\documentclass[a4paper,10pt,twoside]{report}

\usepackage{thesis}
\usepackage[T1]{fontenc}
\usepackage{times}
\usepackage{graphicx}
\usepackage{fancyhdr}
\usepackage{braket}
\usepackage{bbold}
\usepackage{breqn}
\usepackage{dsfont}
\usepackage[numbers]{natbib} 
\usepackage{amssymb}
\newcommand{\shortdoctitle}{Master's Thesis}
\newcommand{\doctitle}{Performing Non-Local Phase Estimation with a Rydberg-Superconducting Qubit Hybrid}

\newcommand{\me}{J.C. Boschero}
\newcommand{\keywords}{Quantum Computing, Quantum Physics, Quantum Algorithms}

\author{\me}

\hypersetup
{
    pdfauthor={\me},
    pdftitle={\shortdoctitle},
    pdfsubject={\doctitle},
    pdfkeywords={\keywords}
}

\begin{document}

%use this include for PDF and distribution versions
\pagenumbering{roman}
% Input from your side:
\newcommand{\thesistitle}{Performing Non-Local Phase Estimation with a Rydberg-Superconducting Qubit Hybrid}			% Title of your MSc thesis.
\newcommand{\yourname}{J.C. Boschero}				% Your name: full initials, and your official surname.
\newcommand{\studentID}{1244356}						% Your student ID.
\newcommand{\defensedate}{October 5, 2023}				% Date of your defence; use the month-day-year format as shown.
\newcommand{\track}{Nano, Quantum and Photonics}								% Track in which you graduate (Fluids, Bio and Soft Matter / Nano, Quantum and Photonics / Plasmas and Beams).
\newcommand{\capacitygroup}{Coherent Quantum Technology}				% Capacity group in which you graduate.
\newcommand{\supervisor}{Dr. Ir., E.J.D, Vredenbregt}		% Your supervisor (titles(s), initial(s) and surname) at TU/e.
\newcommand{\extsupervisor}{MSc. N.M.P. Neumann}	
\newcommand{\extsupervisortwo}{MMath. W.E. van der Schoot}	% External supervisor (titles(s), initial(s) and surname) outside TU/e: leave empty if there is no external supervisor.
\newcommand{\studyload}{45}							% Study load (# of ECTS) of the graduation project (indicate 45 or 60 EC).

% Committee members:
\newcommand{\memberone}{Dr.ir., E.J.D, Vredenbregt}		% Voting member 1, the graduation supervisor, also TU/e examiner, and chair.
\newcommand{\membertwo}{Dr.ir. J. van Dijk}		% Voting member 2, TU/e examiner at least at assistant professor level, not not belonging to the track (FBSM, PB, NQP) of the graduation supervisor.
\newcommand{\memberthree}{Prof.dr. A. Todri-Sanial}	% Voting member 3, TU/e examiner at least at assistant professor level, must be an examiner from the Applied Physics department if member 2 is not from the Applied Physics department.
\newcommand{\memberfour}{title(s), initial(s), surname}		% Optional voting member 4, examiner from TU/e or another university.
\newcommand{\firstadvisor}{MSc. N.M.P Neumann}		% Advisory member 1: for example experts and daily supervisors (e.g. company supervisor, PhD, postdoc), leave empty if there is no first advisor.
\newcommand{\secondadvisor}{MMath. W.E van der Schoot}	% Advisory member 2: leave empty if there is no second advisor.

% External graduation project?
\newcommand{\external}{TNO}					% Give company or institution name if you performed your graduation project outside TU/e, or leave empty.

% Confidential report?
\newcommand{\confidential}{false}	 					% True if your report is confidential, false if public.
\newcommand{\confidentialperiod}{1/2/3/4/5}				% Confidentiality period in years (company may impose a temporary embargo up to 2 years, 
             												% an embargo 2-5 years as requested by  company needs permission from Dean AP, for more info see study guide Graduation Project).
\newcommand{\publicationdate}{January 1, 2024}			% Date of publication after confidentiality period; use the month-day-year format as shown.

%%
%% --- Used packages
%%

%%
%%  --- Pagestyles and page formatting
%%
\parskip            \bigskipamount
\parindent         0mm
\oddsidemargin  0mm
\evensidemargin 0mm
\textwidth        15cm      
\textheight       25cm      
\topmargin      -16mm   
\pagestyle{plain} 

\begin{titlepage}
\includegraphics[width=6cm]{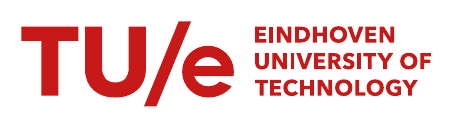} 

\quad\textbf{Department of Applied Physics} \hfill \textbf{CQT2023-15}

\vspace*{4cm}

\begin{center}
    \LARGE{\textbf{\thesistitle}}   
    \vspace*{0.5cm}
    
    \large{by}
    \vspace*{0.5cm}
    
    \Large{\textbf{\yourname}} 
    \vspace*{2cm}
    
    \Large{MSC THESIS}
\end{center}
\vspace{2cm}

\begin{minipage}[t]{6cm}
\textbf{Assessment committee}\medskip

\begin{tabular}{@{}ll}
Member 1 (chair): & \memberone\\
Member 2: & \membertwo\\
Member 3: & \memberthree\\
Advisory member: & \firstadvisor\\
\end{tabular}
\end{minipage}
\qquad\qquad\qquad
\begin{minipage}[t]{5cm}
\textbf{Graduation}\medskip

\begin{tabular}{@{}ll}
Program: & Applied Physics \\
Capacity group: & \capacitygroup\\
Supervisor: & \supervisor\\
Date of defense: &  \defensedate\\
Student ID: & \studentID \\
Study load (ECTS): & \studyload\\
Track: & \track \\
External supervisor(s): & \extsupervisor \\
                        &  \extsupervisortwo
\end{tabular}
\end{minipage}
\vspace*{5mm}

The research of this thesis has been carried out in collaboration with \emph{\external}.\\

This thesis is public and Open Access.

This thesis has been realized in accordance with the regulations as stated in the TU/e Code of Scientific Conduct.

\vfill

Disclaimer: the Department of Applied Physics of the Eindhoven University of Technology accepts no responsibility for the contents of MSc theses or practical training reports.
\end{titlepage}

\normalsize

\clearpage

%Sometimes line numbers are nice, uncomment the next line to enable:
%\linenumbers

%It could be handy to have a list of todos and brainstorms in your thesis
%\chapter*{*General todos*}\todo{remove this chapter}
%\input{chapters/general_todos}

%\chapter*{*Brainstorm results*}\todo{remove this chapter}
%\input{chapters/brainstorm_results}

\chapter*{Abstract}\label{chapter:abstract}
Distributed quantum computation is the key to high volume quantum computation in the NISQ era. This investigation explores the key aspects necessary for the construction of a quantum network by numerically simulating the execution of the distributed phase estimation algorithm in a proposed novel superconducting-resonator-atom hybrid system. The phase estimation algorithm is used to estimate the phase or eigenvalue of a given unitary operator and is a sub-process of many other known quantum algorithms such as Shor's algorithm and the quantum counting algorithm . An entangling gate between two qubits is utilised in the distributed phase estimation algorithm, called an $E2$ gate which provides the possibility to transfer quantum information from one quantum computer to another, which was numerically shown to have a construction time of 17ns at a fidelity of 93\%. This investigation analytically derives the Hamiltonian dynamics as well as the noise sources of each system (flux qubit, resonator and Rydberg atom) and utilizes quantum optimal control (QOC), namely the gradient ascent pulse engineering (GRAPE) algorithm, to minimize fidelity error in the corresponding systems gate construction. The GRAPE algorithm showed very accurate engineering of Rydberg atom single and multi-qubit gates with fidelities higher than 90\% while the flux qubit suffered greatly from noise with multi-qubit gate fidelities lower than 90\%. The manipulation of the C-shunt factor was shown to decrease the noise of the flux qubit which in turn increased the probability of accurately estimating the phase using 4 counting qubits. The study showed a trade off between the number of time steps used in the descent/ascent and the number of GRAPE iterations ran on the optimisation for low C-shunt factors $0<\zeta<1000$. For $\zeta = 1000$, the GRAPE algorithm showed effectiveness for a large number of time steps and large amount of GRAPE iterations by reaching estimation accuracies <90\% . 

\clearpage

%An executive summary if you want:
%\chapter*{Executive summary}\label{chapter:executive_summary}
%\input{chapters/executive_summary}

%\clearemptydoublepage

%\chapter*{Preface}\label{chapter:preface}
%\input{chapters/preface}

%\clearpage

\tableofcontents

\clearpage

\listoffigures

\clearpage

\listoftables

\clearpage

\chapter{Introduction and Purpose of the Investigation}\label{chapter:introduction}
\setcounter{page}{0}
\pagenumbering{arabic}
%from here on, start the 'real' page numbering, from 1, with normal digits
The number of transistors per area on a microchip has been steadily increasing every year as predicted by \citeauthor{Moore_law}, in 1965 \cite{Moore_law} stating that every two years, this value doubles. Currently, the market is looking to produce transistors with a gate pitch of 48nm and a gate length of 24nm \cite{intel_3nm,samsung_3nm}. Although roadmaps exist to reduce transistor size in the coming years \cite{intel_3nm}, continuing this trend requires lithographic wavelengths far beyond extreme ultraviolet that are more expensive and physically harder to reach. Furthermore, transistors require insulating layers to be above a certain size (approximately 1nm) to avoid electrons tunneling through them. Hence, depending solely on classical computing for a sustained exponential boost in computational capacity over the long run is not dependable. Therefore, an alternative approach to achieving significant computational capacity is required.

In 1981, Richard Feynman proposed an alternative to simulating physical systems using classical computation. This alternative, became known as quantum computation \cite{preskill2023quantum}. With his proposal, the amount of quantum computer elements capable of simulating a large physical system are proportional to the space-time volume of the physical system. Classically this requires exponential resources \cite{preskill2023quantum}. Quantum computers use "qubits" describing this quantum computational element \cite{qubit_reference}. 

Feynman's vision of a quantum computer relied purely on its ability to simulate physical systems instead of performing computational tasks. \citeauthor{paul_benioff} possessed a different vision of a quantum computer, whereby he introduced the idea of a quantum computer as a quantum turing machine \cite{paul_benioff}. This was later cemented through the emergence of quantum solutions to oracle-based problems through the likes of Deutsch's algorithm \cite{Deutsch_1985} in 1985, the Bernstein-Vazirani algorithm \cite{Bernstein_vazirani} in 1993 and Simon's algorithm in 1994 \cite{simon_quantum}. Peter Shor famously built on top of these results with his algorithm coined Shor's algorithm \cite{shor_1994} in 1994, where he demonstrated that with a sufficiently large amount of stable qubits, one could break RSA encryption. As of now, however, quantum hardware cannot run these algorithms at a sufficiently high quality. 

Currently, quantum computing is in the noisy intermediate-scale (NISQ) era \cite{Preskill_2018}. The NISQ era is characterized by quantum processors containing up to 1000 noisy qubits \cite{Brooks2019BeyondQS}. These noisy devices can only outperform classical devices on specific tasks. Furthermore, no single architecture exists to build these devices. Different quantum computers have been studied as well as constructed, all with their own set of advantages and disadvantages: trapped ions \cite{Noel_2022} and atoms \cite{Beterov_2018,Saffman_2016}, spin systems \cite{Struck2016} and superconducting circuits \cite{annurev_soa,orlando_flux}. The quantum architectures currently under development include, magnonic qubits~\cite{Lachance_Quirion_2019} and skyrmion qubits~\cite{Psaroudaki_2021}.

IBM have recently announced the construction of a 433 qubit quantum computer~\cite{CHAKRABORTY202353} but with limited circuit depth (the amount of operations a qubit undergoes). Scaling this further poses physical and economical difficulties similar to the difficulties classical computing faced, in which the amount of logic gates capable of fitting on a chip was limited. This was solved by introducing multiprocessor computing in which processes were distributed throughout many processors instead of having a single processor complete one complex task. This same idea is possible in quantum computing called \textit{distributed quantum computing}. Distributed quantum computing involves the collaboration of quantum computers possessing a restricted number of qubits to collectively solve computational tasks that surpass the computational resources achievable within a single quantum device~\cite{caleffi2022distributed}. 

Distributed quantum computing opens the doors to the possibility of a quantum computer network, in which quantum computers can collaborate on executing a task over large distances. Such network has been developing in European research since 2005 and so far progress has been made in the area~\cite{eu2005}. In June 2018, BT announced that it had built a secure quantum network capable of transferring quantum information between Cambridge, UK and BT's laboratory in Ipswich \cite{UKnetwork}. This network was established using quantum computers with the same architecture. However, given the evolution of various quantum architectures, an important question arises: Can two quantum devices with distinct physical architectures collaborate to solve a distributed computational task? In addressing this question, this investigation divides it into three smaller sub-questions.

\noindent\fbox{%
    \parbox{\textwidth}{%
        \textbf{Can two quantum devices with distinct physical architectures collaborate to solve a distributed computational task?}
        \begin{enumerate}
            \item How can one distribute a quantum task or algorithm?
            \item How can a viable connection be established between two disparate quantum architectures?
            \item Is it possible for such distributed system to complete a distributed task accurately?
        \end{enumerate}
    }%
}

The first sub-question will be addressed by distributing the phase estimation algorithm. The quantum phase estimation algorithm was first proposed in 1995 and is used to estimate the phase corresponding to an eigenvalue of a given unitary operator \cite{abelian_stabalizer}. The phase estimation algorithm was chosen because it is a prominent subprocess used in Shor's algorithm \cite{shor_1994}, the quantum algorithm for solving linear systems of equations~\cite{linear_equations}, and the quantum counting algorithm \cite{counting_algorithm}.  

In order to investigate the feasibility of connecting two computers with differing quantum architectures (second sub-question), it is important to understand the coupling (or entanglement) mechanics of one device and then translate it to another \cite{Xiang_2013}. Coupling resonances on different architectures is not a "one-size-fits-all" problem. For this reason, the investigation will focus on two specific quantum architectures, namely, Rydberg atoms and super conducting qubits, which shall be called a Rydberg-superconducting qubit hybrid system.

Rydberg atom computers rely on the interactions of alkali atoms with the valence electron in a highly excited state to execute quantum algorithms. Rydberg atom computers are constructed on an optical table utilizing an array of optical lenses and lasers. Rydberg atom quantum systems as a cloud service have emerged in the last 2 years, with the two notable companies being PASQAL \cite{PASQAL_IEEE} and PennyLane \cite{wang2022predicting}. Furthermore, one of TU Eindhoven's goals for the KAT-1 project, a subsidiary of the Quantum Delta NL ecosystem, is to create a Rydberg atom computer \cite{TUe_KAT1,KAT1}. 

Superconducting (SC) qubit computers were chosen to be the other quantum device because they are arguably the most popular type of qubit. They are constructed using electric circuits and are kept at very low temperatures (sub Kelvin). Because the qubits are built on an electrical chip, the fabrication techniques are analogous to that of classical circuits, consequently making the quantum hardware cheaper and more intuitive when compared to other quantum architectures. The circuit is placed at the bottom of a large structure mainly responsible for its cooling which resembles that of a "golden chandelier". IBM's quantum computers, readily accessed through the cloud by researchers and interested parties, use superconducting qubits \cite{Gambetta_2017}. The SC qubit's popularity is even attributed beyond the realm of the scientific community wherein the image of a quantum computer in pop culture is denoted by this "golden chandelier" \cite{pop_culture}.  

The accuracy of general quantum algorithms depends on the number of operations the qubits undergo. Thus, accurately running a distributed algorithm possessing noisy qubits (third sub-question), requires external optimisation techniques. This is done through Quantum Optimal Control (QOC), in which a laser pulse is tuned to construct a specific quantum state using a noisy state \cite{Ansel_2022}. Algorithms such as CRAB (Chopped RAndom Basis) \cite{Doria_2011}, Krotov \cite{Krotov} and GRAPE (GRadient Ascent Pulse Engineering) \cite{KHANEJA2005296} exist in order to optimize dynamical systems. In this study, GRAPE has been scoped and utilized due to its robustness, efficiency and resource management techniques it provides when coupled to the limited-memory BFGS (L-BFGS) algorithm when optimizing pulses for limited qubit systems.

Chapter \ref{chapter:open_quantum} will answer sub-question (1) by first introducing quantum information theory in Section \ref{sec:qubit_intro} and then describing the steps needed to distribute the phase estimation algorithm, seen in Sections \ref{sec:phase_estimation} and \ref{sec:distributed_algorithm}. Furthermore, its implementation on QuTiP, a Python framework used to model quantum systems \cite{qutip_package}, will be explained in Section \ref{sec:qutip_implementation}.

Chapter \ref{chapter:hybrid_system} will develop the system required to answer sub-question (2). This chapter begins with an introduction of both the Rydberg atom quantum hardware (Section \ref{sec:Rydberg_interaction_main}) and the introduction of the flux qubit, a specific superconducting qubit that is used (Section \ref{sec:flux_qubit_intro}). The dynamics of both systems will be explored and modelled. Once both models have been introduced, Section \ref{sec:hybrid_comp} will denote the specifications of the device required to couple both systems. The chapter finalizes with numerical simulations of the sequence required to establish the communication protocoal between both systems (Section \ref{sec:GHZ_hybrid}).

Chapter \ref{chapter:open_system} details the noise sources that potentially arise from the systems, discussed in chapter \ref{chapter:hybrid_system}, which will be used in Chapter \ref{chapter:GRAPE_sim} to answer sub-question (3). The system will split the noise effects into three parts, the Rydberg atom system, seen in Section \ref{sec:rydberg_noise}, the flux qubit system, seen in Section \ref{sec:flux_qubit_noise}, and the coupled resonator system \ref{sec:hybrid_noise}, seen in Section \ref{sec:hybrid_noise}. 

Chapter \ref{chapter:GRAPE_sim} addresses the optimisation method and its implementation in order to answer the final sub-question. The optimization method, GRAPE, is presented in Section \ref{sec:intro_grape}, and subsequently employed for optimizing quantum gates, as described in Section \ref{sec:process_tomography}. Utilizing the optimised gates, Section \ref{sec:analysis_conclusion} applies the optimised gates in the whole distributed algorithm. The accuracy of the Rydberg-superconducting qubit hybrid system that executes the phase estimation algorithm with GRAPE-optimised gates is subsequently assessed.

Chapter \ref{chapter:conclusions} will include a thorough examination of the model's limitations, along with suggested directions for future research in Section \ref{sec:discussion}. Furthermore, Section \ref{sec:concluding_remarks} will reiterate the results required to answer sub-questions (1)-(3), which will help answer the main question determining if two quantum devices with distinct physical architectures can jointly solve a distributed computational task.

\clearpage

\chapter{Fundamentals of Quantum Information and Quantum Circuit Theory}\label{chapter:open_quantum}
This chapter details the fundamentals of quantum information as well as the local and non-local phase estimation algorithm explored in this research. The goal of this chapter is to introduce the non-local phase estimation algorithm used in the research as well as its numerical implementation. Section \ref{sec:qubit_intro} details the mathematical and physical foundation of what a qubit is and how it propagates through a quantum circuit as well as an explanation on what quantum gates are and how they are represented in a circuit. Section \ref{sec:phase_estimation} details the concepts and intricacies within the phase estimation algorithm such as the quantum Fourier transform as well as the phase estimation algorithm itself. Using the theory discussed on section \ref{sec:phase_estimation}, section \ref{sec:distributed_algorithm} details the construction of non-local circuits while also describing the non-local phase estimation that will be used throughout the entirety of the research. Finally, section \ref{sec:qutip_implementation} highlights the essential components required to implement the phase estimation algorithm using QuTiP.

\section{Qubits and Quantum Circuits}\label{sec:qubit_intro}
 The underlying concept of a qubit is inspired by a classical bit in the sense that they both represent a piece of information, most notably 1 or 0. While a classical bit can only attain the values 0 or 1, a qubit can be in a superposition of both values. Although a qubit will always result in either a 0 or a 1 when measured, the probability of measuring either a 0 or a 1 can be modified. Using Dirac notation (often called bra-ket notation), one can represent a qubit as a wave function $\ket{\psi}$,
\begin{equation}
    \ket{\psi} = \alpha \ket{0} + \beta \ket{1} = 
    \begin{pmatrix}
        \alpha \\
        \beta 
    \end{pmatrix}
\end{equation}
where $\{\alpha,\beta \}  \subset \mathbb{C}$ and $|\alpha|^2+|\beta|^2=1$. Thus, a qubit can take on any value that is a normalized linear combination of the computational basis states $\{\ket{0},\ket{1}\}$. A basis state can represent the different energy levels or states a quantum system can reside in. The normalization of this states allows for a more fruitful representation of the quantum state \cite{nielsen_chuang_2021},
\begin{equation} \label{eq:complex_qubit_state}
    \ket{\psi} = e^{i\gamma}\bigg(\mathrm{cos}\frac{\theta}{2}\ket{0}+e^{i\varphi}\mathrm{sin}\frac{\theta}{2}\ket{1}\bigg),
\end{equation}
where $\gamma$ is the global phase while both $\theta$ and $\varphi$ represent the polar and azimuthal angles, respectively, in spherical coordinates. Visualizing and measuring qubit states in a superposition, for example the one seen in Eq. \eqref{eq:complex_qubit_state}, requires an understanding of the mathematical vector space they are represented by. 

Qubits mathematically inhabit a Hilbert space. A Hilbert space is a vector space $H$ that contains an inner product such that the norm defined by $|f|= \sqrt{\langle f,f \rangle}$ turns $H$ into a complete metric space \cite{stone_2000,weisstein}. Being a complete metric space means that the Hilbert space has a complete inner product. Hence, any point in a Hilbert space can be represented as an infinite sequence of coordinates or as a vector with finite components. In quantum mechanics, the observables such as position, momentum and spin are \textit{self-adjoint} \cite{petrini_pradisi_zaffaroni_2018}. Self adjointness is often exhibited in the representation of quantum information, or the Dirac convention whereby the inner product on a Hilbert space can simply be represented by matrix representation. In order to grasp the self-adjoint principle, it is important to first assume that, $\ket{w} = \sum_i w_i \ket{i}$ and $\ket{v} = \sum_j v_j \ket{j}$ are the representations of vectors $\ket{w}$ and $\ket{v}$ with respect to some orthonormal basis $\ket{i}$, where $\braket{i|j} = \delta_{ij}$, the inner product is said to equal,
\begin{equation}
    \braket{v|w} = \bigg(\sum_i v_i \ket{i},\sum_j w_j \ket{j} \bigg) = \sum_{ij}v_i^{*}w_j\delta_{ij}= \sum_{i}v_i^{*}w_i = \begin{bmatrix}
                            v^*_1 & \cdots & v^*_n 
                        \end{bmatrix}
                        \begin{bmatrix}
                            w_1 \\
                            \vdots \\
                            w_n 
                        \end{bmatrix}.
\end{equation}
Assuming $a$ is a linear operator residing in a Hilbert space, $a$ is said to be self-adjoint if there exists a linear operator $a^{\dag}$ on the Hilbert space such that for the vectors $\ket{v}, \ket{w}$,
\begin{equation}
    (\ket{v},a\ket{w}) = (a^\dag\ket{v},\ket{w}).
\end{equation}
The linear operator $a^\dag$ is known as the adjoint or Hermitian conjugate of the operator $a$ \cite{nielsen_chuang_2021}. To simplify, a linear operator or matrix is called self-adjoint if it is equal to its own adjoint. When analyzing qubits inhibiting a quantum state space (or Hilbert space) the operator $a^\dag$ is commonly referred to as the raising or creation operator while $a$ is the lowering or annihilation operator. 

\begin{figure}
\centering
    \includegraphics[width=0.5\textwidth]{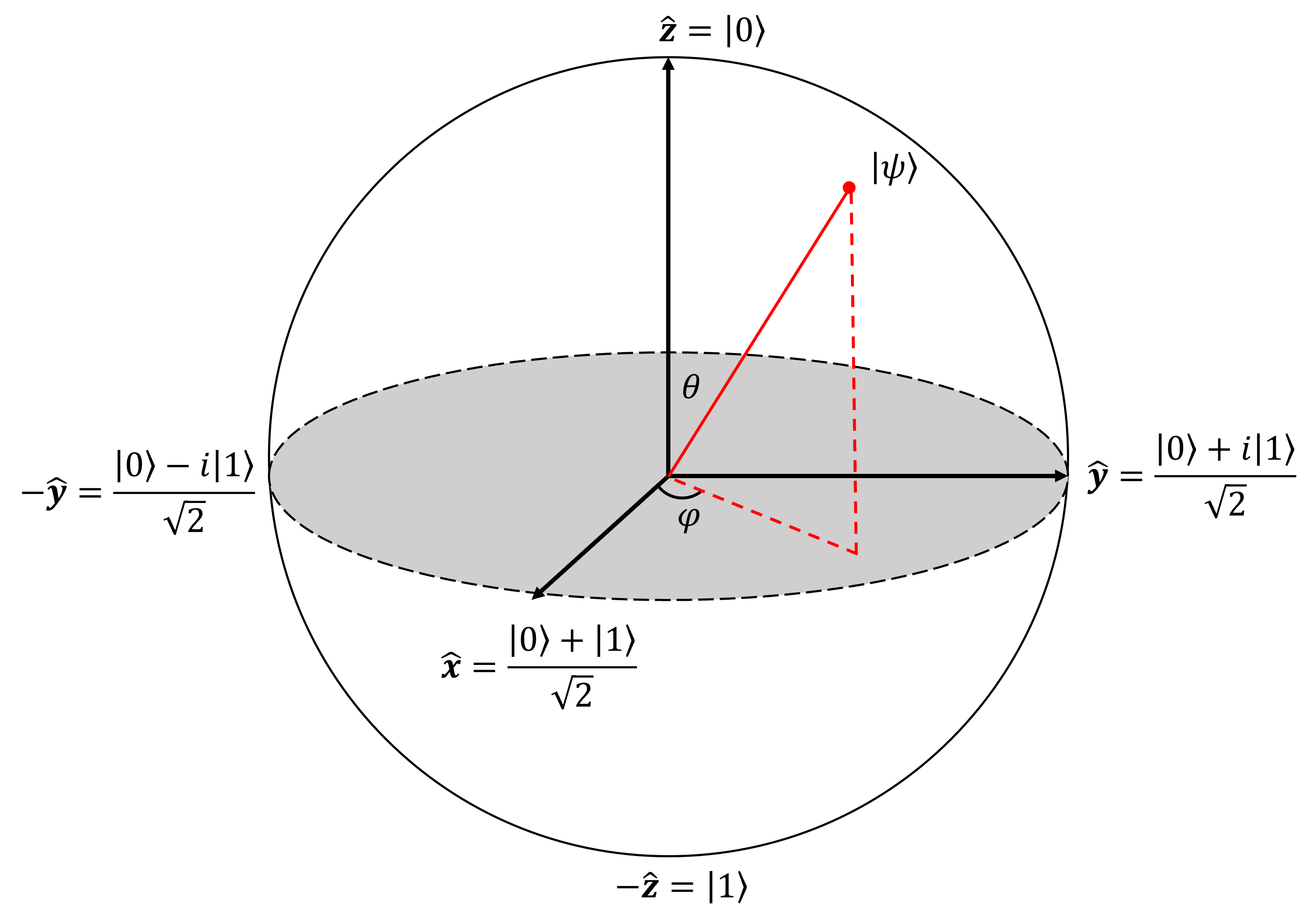}
    \caption{Two-level system represented on a Bloch sphere. The two levels are seen on the $z$ axis where $-z$ is the excited state or computational state $\ket{1}$ and $z$ is the ground state or computational state $\ket{0}$. Given a qubit in a superposition state $\ket{\psi}$, it can be represented as a point on the sphere when plotted. }
    \label{fig:bloch_sphere}
\end{figure}

Measuring of some state $\ket{\psi}$ requires a measurement operator $M_m$. The probabilities of some outcome $m$ occurring are $\bra{\psi}M^\dag_m M_m\ket{\psi}$. Incorporating a global phase to the quantum state, $e^{i\gamma}\ket{\psi}$, and measuring once more yields, $\bra{\psi}e^{-i\gamma}M^{\dag}_m M_m e^{i\gamma}\ket{\psi} = \bra{\psi}M^\dag_m M_m\ket{\psi}$. Therefore, from an observational point of view, the global phase can be ignored in Eq. \eqref{eq:complex_qubit_state}, which leaves the state,
\begin{equation} \label{eq:basic_qubit_state}
    \ket{\psi} = \mathrm{cos}\frac{\theta}{2}\ket{0}+e^{i\varphi}\mathrm{sin}\frac{\theta}{2}\ket{1}.
\end{equation}
The expression acquired in Eq. \eqref{eq:basic_qubit_state} is essential for the creation of qubit gates. Modifying $\theta$ and $\varphi$ such that the phase change alters the superposition of the quantum state, is the basis of quantum computation. A Bloch sphere is a geometric representation of the pure state space of a two-level system. Thus, the modification of phase can be represented using a Bloch sphere. This can be seen depicted in figure \ref{fig:bloch_sphere}. 

The Bloch sphere is also used to represent single qubit gates. The most general single qubit gate is the U-gate, which takes the form,
\begin{equation}\label{eq:general_single_u}
    U(\theta,\varphi,\gamma) = 
    \begin{bmatrix}
        \mathrm{cos}(\frac{\theta}{2}) & -e^{i\gamma}\mathrm{sin}(\frac{\theta}{2})\\
        -e^{i\varphi}\mathrm{sin}(\frac{\theta}{2}) & e^{i(\varphi+\gamma)}\mathrm{cos}(\frac{\theta}{2})
    \end{bmatrix}.
\end{equation}
An example is the Hadamard gate, a gate responsible for placing a single qubit in a superposition state between $\ket{0}$ and $\ket{1}$, which is denoted as, 
\begin{equation}
    H \equiv U(\frac{\pi}{2},0,\pi) = \frac{1}{\sqrt{2}} 
    \begin{bmatrix}
        1 & 1\\
        1 & -1
    \end{bmatrix}.
\end{equation}

\begin{figure}[h]
\centering
    \includegraphics[width=0.5\textwidth]{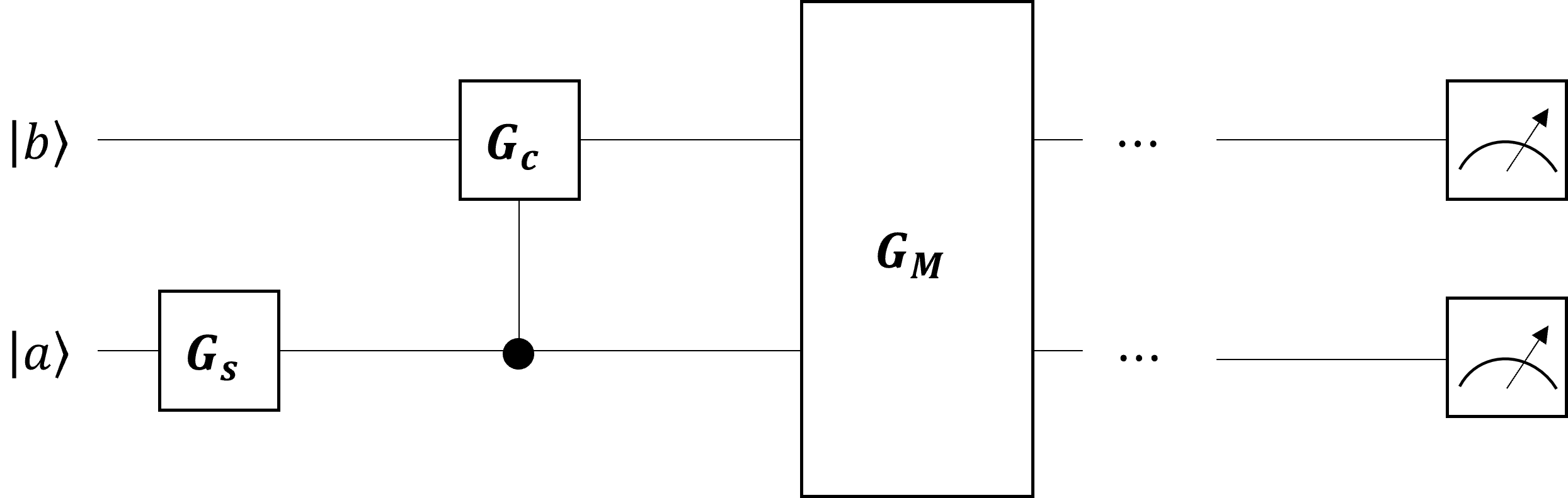}
    \caption{Representation of a general two qubit circuit with initial states $\ket{b}$ and $\ket{a}$. The dot parallel to the square signifies that the corresponding qubit is the control. $G_M$ on the other hand corresponds to an operation involving many gates and spans the qubits that are involved in the operation. The gates at the end of the sequence for each qubit denote the measurement operation for the specific qubit.}
    \label{fig:basic_circuit}
   \end{figure}

A general circuit can be seen represented in figure \ref{fig:basic_circuit} where each qubit travels along an arbitrary synchronous line that ends in a read-out or measurement gate. The gates are presented by squares, either on a single line to denote single qubit gates or spread over many lines denoting a multi-qubit gate. In classical computation, logic gates are physical devices integrated into circuits, where electrons are inputted and outputted.  Within the realm of quantum computing, logic gates are applied to qubits by employing controlled pulses at specific frequencies, as opposed to the physical gates used in classical computing. Although in practice logic gates in quantum computing are very different to those in classical computing, the circuit representations are very similar.

Much like classical computing, quantum computing requires the use of multi-qubit gates. This requires extending the single qubit basis to a multi-qubit basis. Begin by assuming two separated qubits with two states,
\begin{equation}
    \ket{a} = \begin{bmatrix}
        a_0 \\
        a_1
    \end{bmatrix}, \quad
     \ket{b} = \begin{bmatrix}
        b_0 \\
        b_1
    \end{bmatrix},
\end{equation}
the collective state can mathematically be described by applying a tensor multiplication between both states~\cite{pure_state_brenner},
\begin{equation}
    \ket{ba} = \ket{b} \otimes \ket{a} = \begin{bmatrix}
        b_0a_0\\b_0a_1\\b_1a_0\\b_1a_1
    \end{bmatrix}.
\end{equation}

In figure \ref{fig:basic_circuit}, the multi-qubit gate, $G_c$, is shown to be a controlled gate where the qubit in state $\ket{b}$ is the target and the qubit in state $\ket{a}$ is the control. A controlled gate works by only applying a gate to the target qubit if the control qubit is in the excited state or $\ket{1}$ state. This type of gate in particular will be used extensively throughout this research. A list of common single and multi-qubit logic gates can be seen in Appendix \ref{chapter:logic_gate_appendix}. 

\section{The Phase Estimation Algorithm}\label{sec:phase_estimation}
Combining a variety of quantum gates correctly with the combination of a measurement ultimately resulting in an answer (often described in the computational basis). Changing the notion of the measurements, one can extract more complex information from the system. The phase estimation algorithm is an example of such phenomenon. Suppose a unitary operator $U$ has an eigenvector $\ket{u}$ with eigenvalue $e^{2\pi i \varphi}$, where the observable, $\varphi$ is not known \cite{nielsen_chuang_2021}. The goal of the phase estimation algorithm is to estimate the value of $\varphi$. It finds applications in encryption breaking \cite{shor_1994} and many other algorithms. 

\begin{figure}[t]
\centering
    \includegraphics[width=0.8\textwidth]{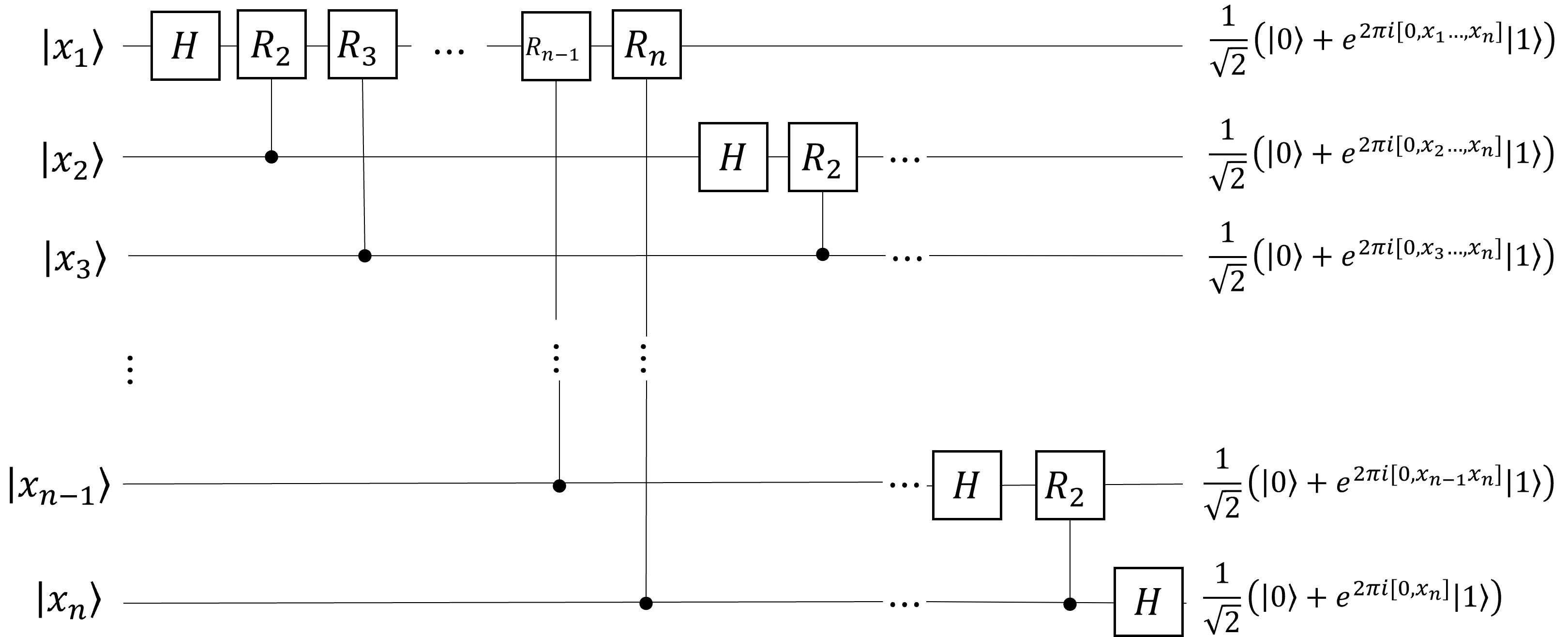}
    \caption{The (inverse) quantum Fourier transform circuit portraying the phase gate positions and final qubit states from eq. \eqref{eq:quantum_fourier} or \eqref{eq:inverse_fourier} with the phase gates described by $R_n$ ($R_n^{-1}$).}
    \label{fig:quantum_fourier}
\end{figure}

\begin{figure}[t]
\centering
    \includegraphics[width=0.8\textwidth]{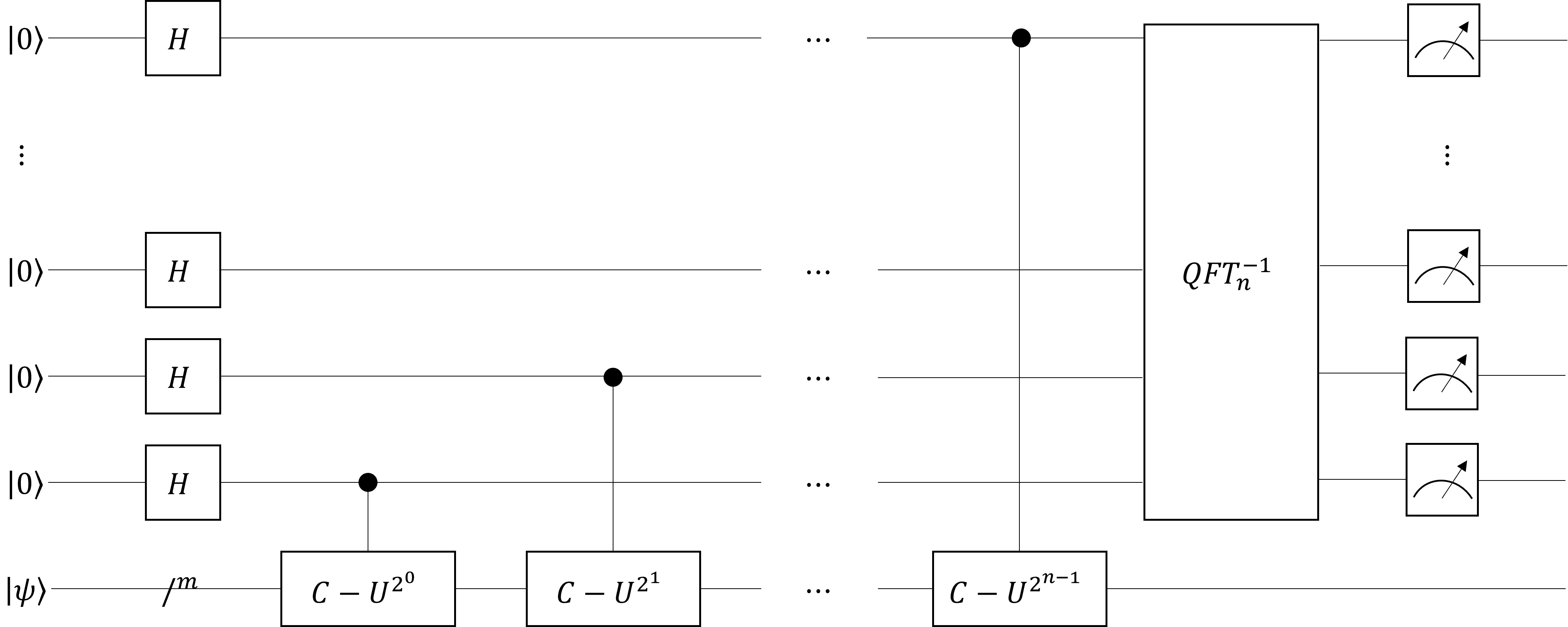}
    \caption{The quantum phase algorithm circuit diagram. The black dots represent the control for the controlled U gates in the $\ket{\psi}$ register \cite{Shende_2006}.}
    \label{fig:phase_estimation_circuit}
\end{figure}

Understanding the phase estimation algorithm requires the understanding of the quantum Fourier transform (QFT), as it constitutes a significant component of the phase estimation algorithm. Parsing multiple qubits through a Fourier transform essentially entangles all qubits at once. The QFT is extremely powerful because it exponentially speeds up factoring algorithms \cite{quantum_computation_information,quantum_fourier}. The classical discrete Fourier transform takes complex components $\{f(0),f(1),...,f(N-1)\}$ and transforms them into $\{\tilde{f}(0),\tilde{f}(1),...,\tilde{f}(N-1)\}$ using \cite{quantum_computation_information}, 
\begin{equation}
    \tilde{f}(j) = \frac{1}{\sqrt{N}}\sum^{N-1}_{k=0}e^{2\pi i \frac{jk}{N}}f(k).
\end{equation}
The quantum Fourier transform works in the same fashion. It is defined on a quantum register of $n$ qubits (with a total of $N=2^n$ states) as the unitary operator $F$ whose action on the states of the computational basis is given by \cite{quantum_computation_information},
\begin{equation}\label{eq:quantum_fourier}
    F(\ket{j}) = \frac{1}{\sqrt{2^n}}\sum^{2^n-1}_{k=0}e^{2\pi i \frac{jk}{2^n}}\ket{k}
\end{equation}
where an arbitrary state $\ket{\psi} = \sum_j f(j)\ket{j}$ transforms into $\ket{\tilde{\psi}} = F\ket{\psi} = \sum^{2^n-1}_{k=0} \tilde{f}(k)\ket{k}$ \cite{quantum_computation_information,quantum_fourier}. The inverse quantum transform is simply given as,
\begin{equation}\label{eq:inverse_fourier}
    F^{-1}(\ket{j}) = \frac{1}{\sqrt{2^n}}\sum^{2^n-1}_{k=0}e^{-2\pi i \frac{jk}{2^n}}\ket{k}.
\end{equation}
Because the implementation of both the quantum Fourier transform and the quantum inverse Fourier transform differs only by a negative phase, the circuit structures for these operations are nearly identical. The goal now is to convert the algebraic representation of the Fourier transform to circuit form as seen in section \ref{sec:qubit_intro}. To do so, it is helpful to write state $\ket{j}$ in binary form, $j = j_1 j_2 \cdots j_n$, or more formally, $j = j_1 2^{n-1} + j_2 2^{n-2} + \cdots + j_n 2^0$. Adopting the notation $0.j_l j_{l+1} \cdots j_m$ to represent the binary fraction $j_l/2 + j_{l+1}/4 +\cdots + j_m/2^{m-l+1}$ allows the quantum Fourier transform to be given by the product representation~\cite{nielsen_chuang_2021},
\begin{equation}\label{eq:fourier_product_representation}
    \ket{j_1,\cdots,j_n} \rightarrow \frac{\big(\ket{0}+e^{2\pi i 0.j_n}\ket{1}\big)\big( \ket{0} + e^{2\pi i0.j_{n-1} j_n)}\ket{1}\big) \cdots \big(\ket{0} + e^{2\pi i 0.j_1j_2 \cdots j_n}\ket{1} \big)}{2^{n/2}}.
\end{equation}
This representation allows the construction of the efficient quantum circuit and a proof as to why the Fourier transform is unitary. The proof as to how the product representation in Eq. \eqref{eq:fourier_product_representation} follows from the formal definition in Eq. \eqref{eq:inverse_fourier} can be seen in Appendix \ref{sec:fourier_transform_product_proof}. 
Figure \ref{fig:quantum_fourier} shows the circuit of the quantum Fourier transform where the $R_n$ gates represent a controlled Z rotation gate seen in eq. \eqref{eq:rot_gate} which follows from the unitary transformations seen in \eqref{eq:fourier_product_representation}. The inverse Fourier transform simply follows by replacing $R_n$ by $R^{-1}_n$,
\begin{equation} \label{eq:rot_gate}
    R_n = 
    \begin{pmatrix}
        1 & 0 & 0 & 0 \\
        0 & e^{-2\pi i/2^n} & 0 & 0\\
        0 & 0 & 1 & 0 \\
        0 & 0 & 0 & e^{2\pi i/2^n}
    \end{pmatrix},
    R^{-1}_n = 
    \begin{pmatrix}
        1 & 0 & 0 & 0 \\
        0 & e^{2\pi i/2^n} & 0 & 0\\
        0 & 0 & 1 & 0 \\
        0 & 0 & 0 & e^{-2\pi i/2^n}
    \end{pmatrix}.
\end{equation}
The inverse Fourier transform can be seen in the context of the phase estimation algorithm in figure \ref{fig:phase_estimation_circuit} as $QFT^{-1}_n$ where it is compartmentalized as a subroutine between $n$ qubits. 

Figure \ref{fig:phase_estimation_circuit} shows the circuit for the phase estimation algorithm. The easiest method of interpreting the diagram is to distinguish two registers, the target register with $n$ qubits denoted as a series of $\ket{0}$ states and the control register with $m$ qubits necessary to store state $\ket{\psi}$. The eigenvector $\ket{\psi}$ is the one targeted for measurement, with the objective of determining its corresponding eigenvalue. 
The $C-U^{2^j}$ gate is a specific type of controlled U gate (seen in Appendix \ref{chapter:logic_gate_appendix}). The action of the gate $C-U^{2^j}$ on a given $\frac{1}{\sqrt{2}}(\ket{0}+\ket{1})\ket{\psi}$ state is given by, 
\begin{equation}
    C-U^{2^j}\frac{1}{\sqrt{2}}(\ket{0}+\ket{1})\ket{\psi} = \frac{1}{\sqrt{2}}(\ket{0}\ket{\psi}+\ket{1}U^{2^j}\ket{\psi})=\frac{1}{\sqrt{2}}(\ket{0}+\ket{1}e^{i2^j \varphi})\ket{\psi},
\end{equation}
 and can be described in matrix form as,
\begin{equation}\label{eq:controlled_cphase}
 C-U^{2^j} =
    \begin{pmatrix}
        1 & 0 & 0 & 0 \\
        0 & 1 & 0 & 0\\
        0 & 0 & 1 & 0 \\
        0 & e^{2\pi i \varphi 2^j}\\
    \end{pmatrix},
\end{equation}
which follows that of a traditional controlled U-gate where global phase is ignored ($\gamma = 0$) and only the $z$ component is taken into account ($\theta,\lambda = 0$). The controlled U-gate described in Eq. \eqref{eq:controlled_cphase} is used to map the phase of the $\ket{\psi}$ state onto the qubits that will then go through the inverse Fourier transform. The output for the quantum phase circuit before the application of the inverse Fourier transform is given by \cite{quantum_computation_information,niels_imperfect,nielsen_chuang_2021},
\begin{equation}\label{eq:phase_estimation_state}
    \frac{1}{2^{t/2}}(\ket{0}+e^{i2\pi 2^{t-1}\varphi}\ket{1})(\ket{0}+e^{2\pi 2^{t-2}\varphi}\ket{1})\cdots (\ket{0}+e^{2 \pi i 2^0\varphi}\ket{1})\ket{\psi} = \frac{1}{2^{t/2}}\sum^{2^t - 1}_{k=0}e^{2\pi i\varphi k}\ket{k}\ket{\psi},
\end{equation}
where $t$ is the amount of counting qubits. To observe how the phase is estimated, it is first important to express $\varphi$ as $\varphi = 0.\varphi_1\cdots\varphi_t$ which leaves Eq. \eqref{eq:phase_estimation_state} as~\cite{nielsen_chuang_2021},
\begin{equation}
    \frac{1}{2^{t/2}}(\ket{0}+e^{2\pi i 0.\varphi_t}\ket{1})(\ket{0}+e^{2\pi i 0.\varphi_{t-1}\varphi_t}\ket{1})\cdots(\ket{0} + e^{2\pi i 0.\varphi_1\varphi_2\cdots \varphi_t} \ket{1}).
\end{equation}
One can see that applying the inverse Fourier transform seen in Eq. \eqref{eq:fourier_product_representation} outputs the product state $\ket{\varphi_1 \cdots \varphi_t}$, which can then be measured in the computational basis to give the exact $\varphi$. 

The description of the special case where $\varphi = 0.a_1\cdots a_m$ can be obtained by applying the inverse QFT, which produces $\ket{a_1,\cdots a_m}$ exactly (and hence $\varphi$). This is because the QFT maps between the state space and the phase space. However $\varphi$ is not in general a fraction of a power of two. For such case, it is important to apply and derive the accuracy of the inverse quantum Fourier transform by giving the best $n$-bit estimate of $\varphi$ with high probability. By conveniently writing $\varphi = 2\pi\Big(\frac{a}{2^n}+\delta\Big)$ where $a = a_{n-1}a_{n-2}...a_0$ and $\delta$ is the phase error defined by $|2^n \delta| \leq 1/2$, it is possible to see that $2\pi a/2^n$ is the best $n$ bit approximation of $\varphi$ \cite{quantum_computation_information,Cleve_1998}.Thus, in the general case, $\delta \neq 0$, the best $n$-bit estimate of $\phi$ is given by $a$. Utilizing the approximations, applying an inverse QFT as seen in Eq. \eqref{eq:phase_estimation_state}, the expression for the best $n$ - bit estimate (with the omitted $\ket{\psi}$ state) is,
\begin{equation}
    \frac{1}{2^n}\sum^{2^n-1}_{k=0}\sum^{2^n-1}_{l=0}e^{2\pi i(a-k)l/2^n}e^{2\pi i\delta l}\ket{k}.
\end{equation}
The next step is to perform a measurement on the $\ket{k}$ state. This is obtained from a standard measurement of the first register containing probability $p_a = |c_a|^2$ where $c_a$ comes from the decomposition $\sum_{k=0}^{2^n-1}c_a \ket{k}$~\cite{quantum_computation_information,Cleve_1998,abelian_stabalizer},
\begin{equation} \label{eq:delta_non_zero}
    c_a = \sum^{2^n-1}_{l=0}e^{2\pi i(a-k)l/2^n}e^{2\pi i\delta l},
\end{equation}
which can finally be evaluated to, 
\begin{equation} \label{eq:geometric_ca}
    |c_a|^2 = \frac{1}{2^n}\bigg[\frac{1-e^{2\pi i\delta 2^n}}{1-e^{2\pi i\delta}}\bigg].
\end{equation}
Employing Euler's form, $|1-e^{2\pi i \delta 2^n}| = 2|\mathrm{sin}(\pi\delta 2^n)|$ and using that $2|\mathrm{sin}(\pi\delta 2^n)| \geq 4|\delta|2^n$ as well as $2|\mathrm{sin}(\pi\delta)| \leq 2\pi |\delta|$ (since for any $z\in[0,1/2]$, $2z \leq \mathrm{sin}(\pi z) \leq \pi z$). These two inequalities can be inserted into eq. \eqref{eq:geometric_ca} to get,
\begin{equation} \label{eq:c_a}
    |c_a|^2 \geq \frac{4}{\pi^2} \approx 0.405.
\end{equation}
This result shows that the best $n$-bit estimate of $\theta$ can be achieved up to this precision \cite{Cleve_1998,abelian_stabalizer,niels_imperfect}. An intuitive way of understanding this value is by first assuming two qubits (thus a precision of 1/4) are used to try to estimate a phase equal to 1/8. Although the qubits lack precision to output the proper result in the computational basis, the results 00 and 01 (0/4 and 1/4 in the computational basis) will have at least 40.5\% probability of being outputted since 1/8 lies between both values. By adding a number of extra qubits in the order of $O(\mathrm{log}(1/\epsilon))$ and truncating the extra qubits, the probability can increase to $1-\epsilon$ \cite{Cleve_1998}. A more general interpretation of this phenomenon is sought by thinking in terms of classical bits. Increasing the amount of bits used to represent a number means increasing the precision the number possesses. Thus, increasing the register $n$, increases the precision of the phase estimation.

\section{Distributed Algorithms and the Non-Local Phase Estimation Algorithm}\label{sec:distributed_algorithm}
\begin{figure}[t]
     \centering
     \begin{subfigure}[t]{0.35\textwidth}
         \centering
         \includegraphics[width=\textwidth]{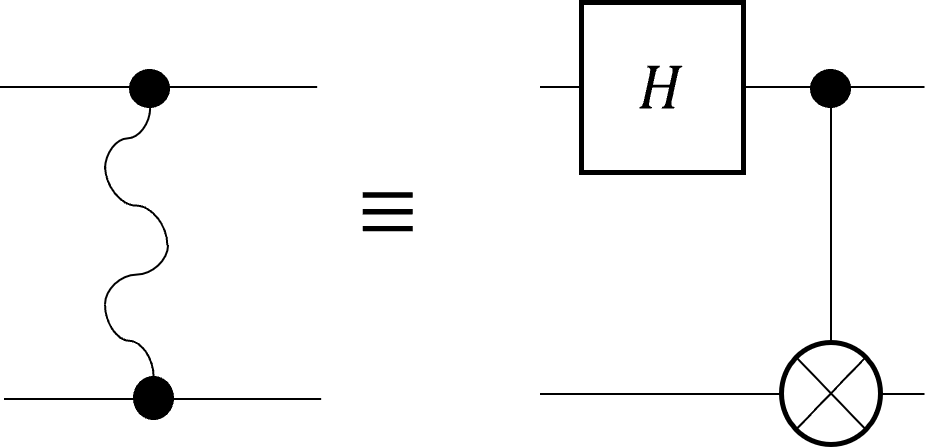}
         \caption{Entangling gate for two qubits ($E_2$ gate).}
         \label{fig: ebit}
     \end{subfigure}
     \hfill
     \begin{subfigure}[t]{0.35\textwidth}
         \centering
         \includegraphics[width=\textwidth]{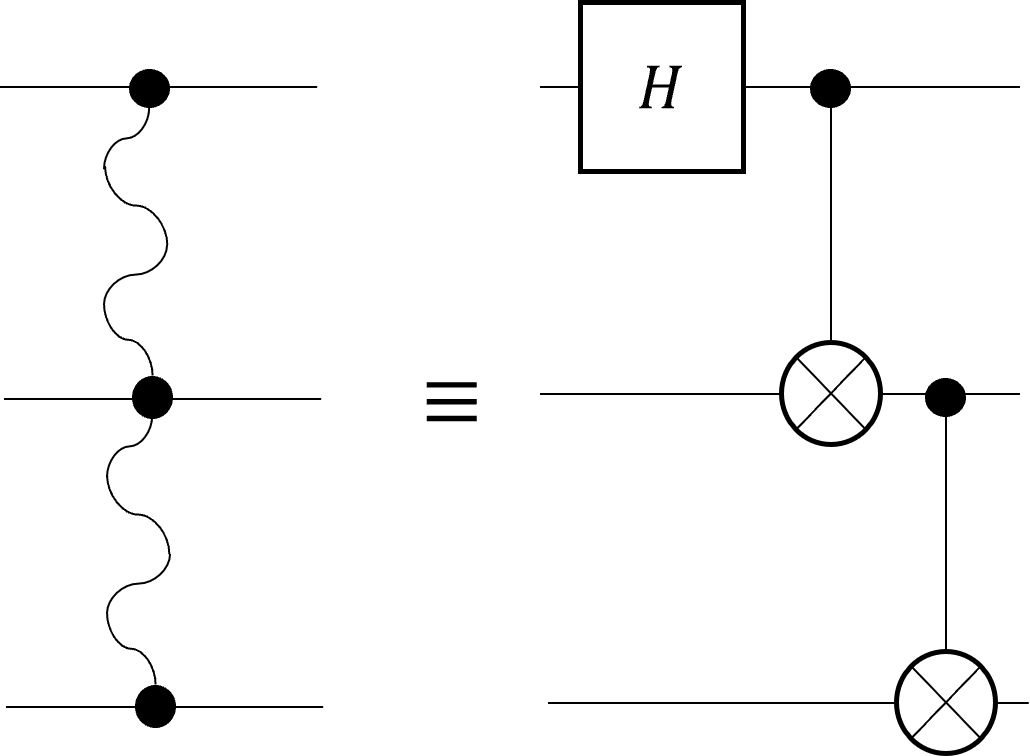}
         \caption{Entangling gate for three qubits ($E_3$ gate).}
         \label{fig: dGHZ}
     \end{subfigure}
     \caption{(a) Non-local circuit diagram of the entangling gate (E2 gate) for two qubits equaling its local counterpart \cite{GHZ_distributed}. (b) Non-local circuit diagram of the entangling gate for three qubits (E3 gate) equaling its local counterpart \cite{GHZ_distributed}. }
\end{figure}

Having revisited both quantum circuit theory and the phase estimation algorithm, it is now possible to expand these concepts into a non-local or distributed quantum computing basis. As opposed to local implementations of quantum algorithms where the interaction between qubits is implied, a non-local representation requires the teleportation or communication of states between channels. These channels are represented by qubits called channel qubits which can be sent back and forth over a network \cite{GHZ_distributed}.  

\citeauthor{Eisert_2000} have shown that setting up these channel qubits relies on the implementation of a distributed CNOT gate \cite{Eisert_2000}. On top of that, \citeauthor{Eisert_2000} shows that the distributed CNOT gate, much like the local CNOT gate seen in Appendix \ref{chapter:logic_gate_appendix}, is a native gate that can create other operations such as controlled gates \cite{Eisert_2000}. Implementing a distributed CNOT gate requires a shared entangled qubit pair (ebit) between two quantum computers and two classical bits (cbits). The generation of the ebit can be seen in figure \ref{fig: ebit}, where an ebit is constructed simply using a Hadamard gate and a CNOT gate. To properly establish an ebit pair, however, one must use an $E_2$-gate which explicitly distributes the ebit pair between two devices. This is done by having the remote computer send a channel qubit to the local machine. The local machine then entangles the channel with one of the local channel qubits. Finally, one qubit of the pair is sent back to the remote machine resulting in a state $\frac{1}{2}(\ket{00}+\ket{11})$ called a \textit{GHZ state} \cite{GHZ_distributed}. However, for simplicity, this investigation will refer to the entangling gate in figure \ref{fig: ebit} as the $E_2$ gate. Figure \ref{fig: dGHZ} shows the same concept but with three qubits which in turn form a GHZ state. 

\begin{figure}[t]
     \centering
     \includegraphics[width=0.7\textwidth]{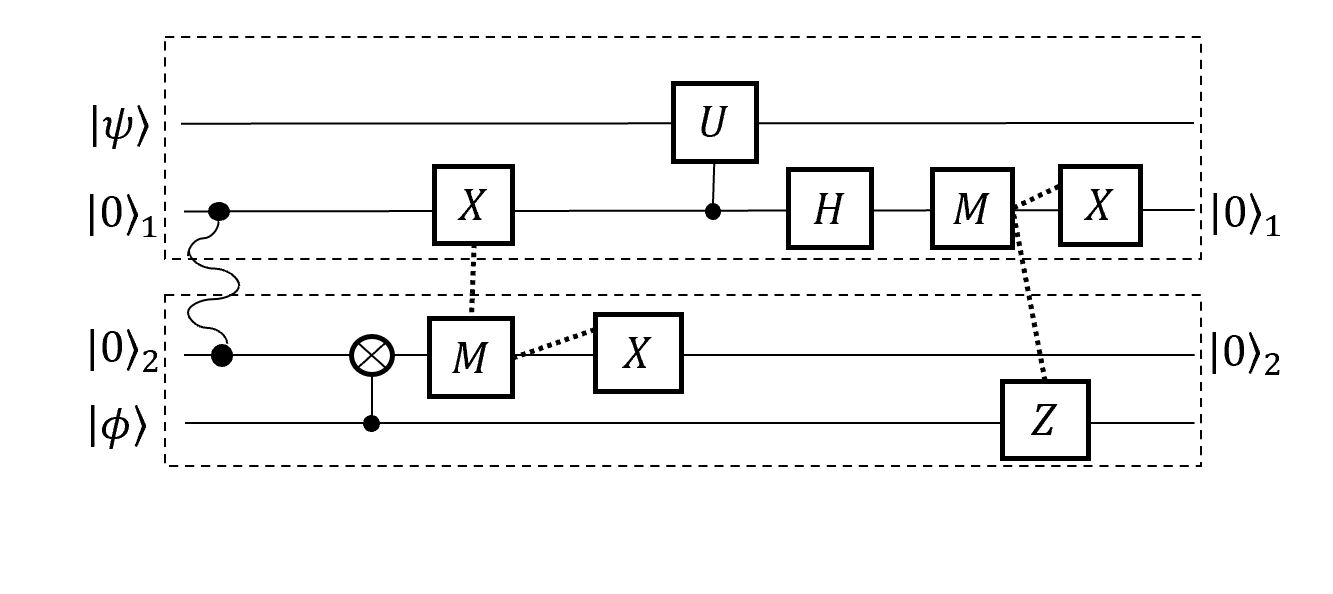}
     \caption{Implementation of a non-local controlled $U$-gate between $\ket{\psi}$ and $\ket{\phi}$ where the dotted box shows the different quantum computers and the dotted lines denote classical communication \cite{niels_imperfect,Yimsiriwattana_2004,GHZ_distributed}.}
     \label{fig: dis_u_gate}
\end{figure}

The subsequent stage in non-local computation involves sharing the information of a qubit from one party with another. \citeauthor{Yimsiriwattana_2004} have identified a set of primitive operations that allow one party to share the information with another \cite{Yimsiriwattana_2004,GHZ_distributed,radzihovsky_espinosa_gim_2021}. Suppose a user, specifically Alice, holds a superposition state and wishes for another party to execute a complex computation on it. In order to share this state, Alice first applies a CNOT gate between her state and the ebit in her system and then measures the corresponding ebit. If the measurement reveals a $\ket{1}$ state, Alice applies a NOT gate (also known as X-gate seen in Appendix \ref{chapter:logic_gate_appendix}) and classically communicates this information to the other party, specifically Bob. Bob then applies a NOT gate to his ebit only if Alice has shown a $\ket{1}$. This procedure is shown as the first five gates in the non-local U-gate implementation seen in figure \ref{fig: dis_u_gate} where Alice's and Bob's quantum computers are shown in dotted black boxes and the dotted lines denote the use of classical communication. In this specific case, Alice is shares the state $\ket{\phi}$ with Bob. 

Bob is now free to use the quantum information on his ebit to perform computations locally using the Alice's shared qubit state. In this case, it is assumed that Bob performs a single controlled U-gate (defined in equation \ref{eq:controlled_cphase}) between the ebit and the qubit with state $\ket{\psi}$ in his system, which can be seen in figure \ref{fig: dis_u_gate} on the $\ket{\psi}$ state. Bob now wishes to share the resulting state with Alice. To do so, Bob first applies a hadamard gate across his ebit and then measures it. If a $\ket{1}$ state is measured, a NOT gate is applied by Bob on his ebit and the result is classically communicated to Alice. Assuming the output is $\ket{1}$, Alice applies a Z gate (seen in Appendix \ref{chapter:logic_gate_appendix})) on the qubit she shared in order to obtain the outcome of Bob's computation. 

This investigation will use the protocol depicted in figure \ref{fig: dis_u_gate} in order to distribute information between two systems. Section \ref{sec:phase_estimation} has emphasized that the precision of the phase estimation algorithm improves as you employ a greater number of counting qubits. Therefore, this study will delve into the scenario in which one quantum computer retains the state requiring estimation of the phase, accompanied by a sole counting qubit. Simultaneously, the other quantum computer will be responsible for managing the counting qubits, thereby enhancing the precision of the estimation process. Thus, the two main distributed processes used in the research will be the distributed QFT and the distributed controlled U-gate.

\begin{figure}[t]
\centering
    \includegraphics[width=0.7\textwidth]{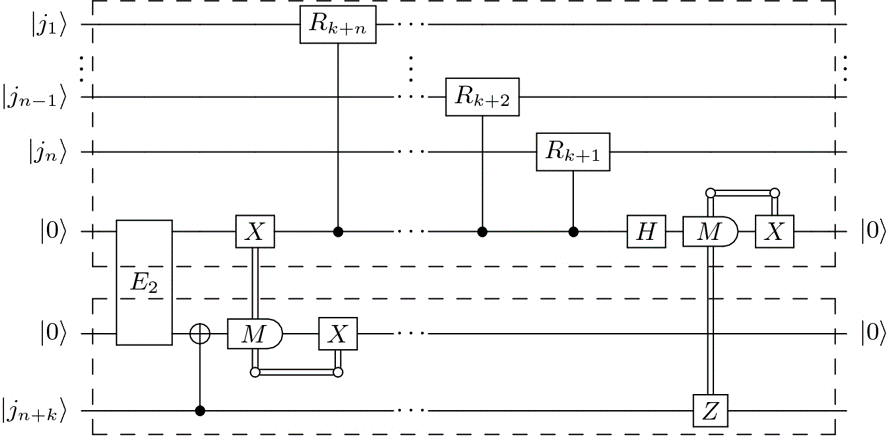}
    \caption{A section of a distributed quantum Fourier transform where $n$-local operations on $n$ qubits are performed using one ebit. The double lines between the gate denote the classical communication channel. The control is $\ket{j_{n+k}}$ acting on the targets $\ket{j_1}...\ket{j_n}$. The remaining $k$ qubits in the second quantum computer are omitted \cite{niels_imperfect}.}
    \label{fig:distributed_fourier}
\end{figure}

The distributed QFT is similar to that of the controlled U-gate. One method of performing the distributed QFT is by replacing each controlled rotation gate with a non-local rotation gate. However, it is also possible to use a single shared entangled state to perform multiple non-local gates, by grouping all operations on one computer that are all controlled by a single qubit from the other computer \cite{niels_imperfect}. This fundamentally leads to the distributed version of the quantum Fourier transform depicted in figure \ref{fig:distributed_fourier} \cite{niels_imperfect}.

\section{QuTiP Implementation of the Non-Local Phase Estimation Algorithm}\label{sec:qutip_implementation}

\begin{figure}[t]
\centering
    \includegraphics[width=0.4\textwidth]{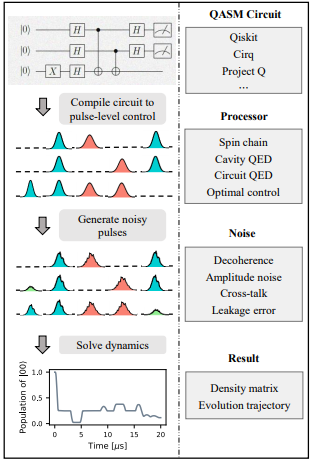}
    \caption{Illustration of the workflow of the pulse-level noisy quantum circuit simulation. First, a quantum circuit is defined in QuTiP (or imported from QASM). Then, the circuit is compiled into control pulses for each control Hamiltonian (blue represents single qubit gates while red represents two-qubit gates). Noise is then generated depending on the chosen processor and finally, the dynamics of the system is solved. This image was taken from \cite{Li_2022}.}
    \label{fig:qutip_workflow}
\end{figure}

The simulation and execution of the quantum algorithms will be constructed on QuTiP \cite{qutip_package}. QuTiP, built on Python, offers many quantum information processing packages. Furthermore, it uses quantum dynamic solvers and control optimization features that will be utilized throughout this research. More specifically, QuTiP can run quantum circuits on simulated quantum processors. The workflow is described in figure \ref{fig:qutip_workflow}.

As seen from figure \ref{fig:qutip_workflow}, the main benefit QuTiP offers when simulating quantum circuits is the ability to simulate the relevant noise depending on the physical architecture. For simplicity, the processor used will be the spin chain model. The spin chain model models the general physics involved in any given system since it describes both a general single qubit interaction and a multi-qubit interaction. The spin chain model acts in accordance with the one dimensional Ising model. The 1D Ising model describes the energy configuration of a lattice containing a finite amount of particles (in this case qubits) allocated in sites that can each be in a spin configuration $\sigma_k \in \{ -1, +1\}$ representing the spin. 

For any two adjacent sites $i,j$ in the lattice, there is an interaction energy with strength $J_{ij}$. Additionally, an extra factor is added to describe the external field interaction $h_j$, which is often manipulated to incorporate quantum gates. The full Ising model can be seen in Eq. \eqref{eq:ising_hamiltonian} described as a Hamiltonian, a function denoting the energy dynamics of a system \cite{Li_2022}.
\begin{equation}\label{eq:ising_hamiltonian}
    H(\sigma) = - \sum_{ij}J_{ij}\sigma_i\sigma_j - \mu \sum_j h_j\sigma_j.
\end{equation}
Knowing the dynamics of the system in use is important because it allows one to determine the native qubit gates. Although this model is very general, it is possible to derive the fact that a controlled U gate is not possible to reproduce using the set of dynamics stated \cite{Huang_2023}. Creating a controlled U gate is possible by extending the model and introducing more parameters, this can be seen in \cite{heinz2023analysis}. In this research, the controlled U-gate will be simplified to controlled phase gate due to the simplicity of the phase in the eigenvalue. The controlled phase gate is similar to the definition used in Eq. \eqref{eq:controlled_cphase}, where $j = 0$ and the $2\pi$ factor in the exponent is omitted, thus leaving, 
\begin{equation}
    \mathrm{CPHASE}(\varphi) = \begin{bmatrix}
        1 & 0 & 0 & 0 \\
        0 & 1 & 0 & 0 \\
        0 & 0 & 1 & 0 \\
        0 & 0 & 0 & e^{i\varphi}
    \end{bmatrix}.
\end{equation}
The CPHASE gate is still not possible to construct in this model, thus, an approximation will be used to create this gate using native gates. The approximation can be constructed using the circuit seen in figure \ref{fig:cphase_deconstruct}. The function for the creation of this gate can be seen in Appendix \ref{sec:distributed_phase_code} as \codeword{cphase_gate}.

\begin{figure}[t]
\centering
    \includegraphics[width=0.9\textwidth]{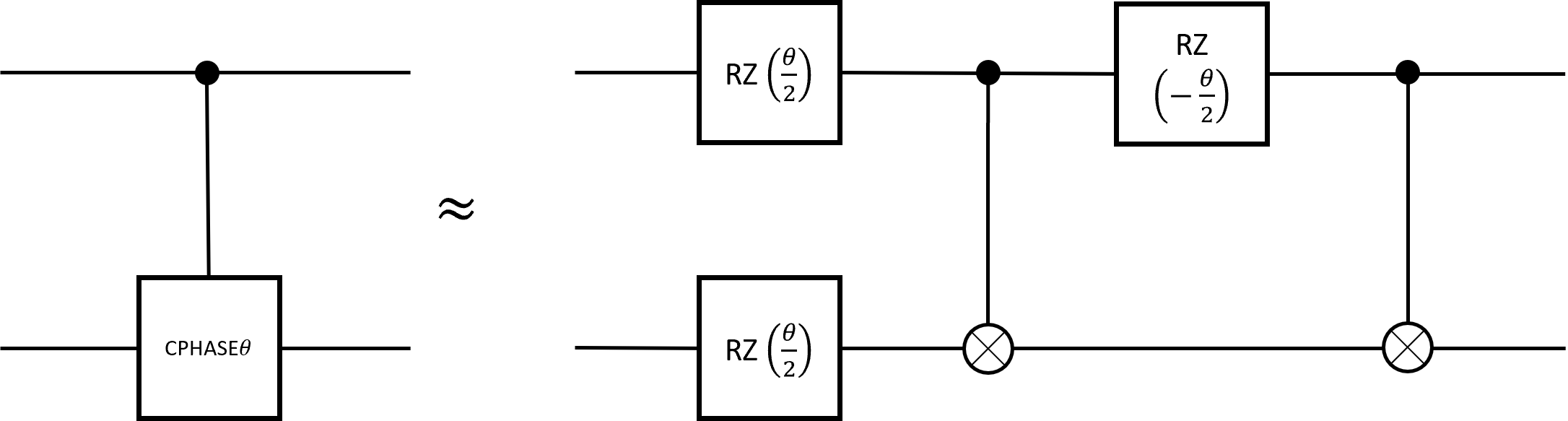}
    \caption{Approximate deconstruction of the CPHASE gate using CNOT gates and Z rotation gates. Inspired by \cite{Fowler_2004}.}
    \label{fig:cphase_deconstruct}
\end{figure}

With the comprehensive description of all gates and processes now provided, it is crucial to gain a clear understanding of the final circuit employed. The research will concentrate on two quantum computers. One of these computers will consist of three qubits, including the qubit containing the phase to be estimated, a single counting qubit, and a communication qubit. The other computer will comprise three counting qubits and a single communication qubit, forming the entangled bit. The total circuit can be seen in figures \ref{fig:final_dpe}, \ref{fig:distributed_CP}, and \ref{fig:distributed_QFT}. The QuTiP functions can be seen in Appendix \ref{sec:distributed_phase_code}. Figure \ref{fig:final_dpe} begins by performing hadamard gates on all the counting qubits, which can be seen in function form as \codeword{perform_hadamards}. 

The following step is to perform the distributed phase gates, or DCP block on the circuit, from the three qubit computer to the four qubit computer. This is seen in figure \ref{fig:distributed_CP} where three non-local controlled phases and a single local controlled phase are performed. This process is described by \codeword{pulse_phase_sequence}, which takes a sub-process responsible for performing the specific non-local gate, \codeword{perform_non_local_phase}. Furthermore, the $E_2$ gate is initialized using \codeword{initialize_E_gate}.

In the non-local functions, it is necessary to measure a single qubit. To do so, one must specify a list of measurement vectors called positive operator-valued measure (POVM). As mentioned in Section \ref{sec:qubit_intro}, given a quantum system $\ket{\psi}$, the probability of an outcome $m$ is given by $p(m) = \bra{\psi}M^\dag M \ket{\psi}$. Defining $E_m = M^\dag M$, it follows that $\sum_m E_m = I$ where $I$ is the identity matrix and $p(m) = \bra{\psi}E_m \ket{\psi}$ which means that the set of operators $E_m$ are sufficient to determine the probabilities of the different measurement outcomes \cite{nielsen_chuang_2021}. The set of $\{E_m\}$ are the POVM which are required to be explicitly defined when making single qubit measurements. This is seen incorporated as \codeword{perform_measurement} which takes a qubit number as input. 

The inverse QFT circuit is then performed, which is expanded upon in figure \ref{fig:distributed_QFT}. As can be observed, only a single non-local operation is required since one of the four total counting qubits are on a separate circuit. This process is described by the recursive function \codeword{iqft_rotations}, which, once again, takes a sub-process responsible for performing the specific non-local gates, \codeword{perform_non_local_iqft}.

\begin{figure}[!htb]
\centering
    \includegraphics[width=0.5\textwidth]{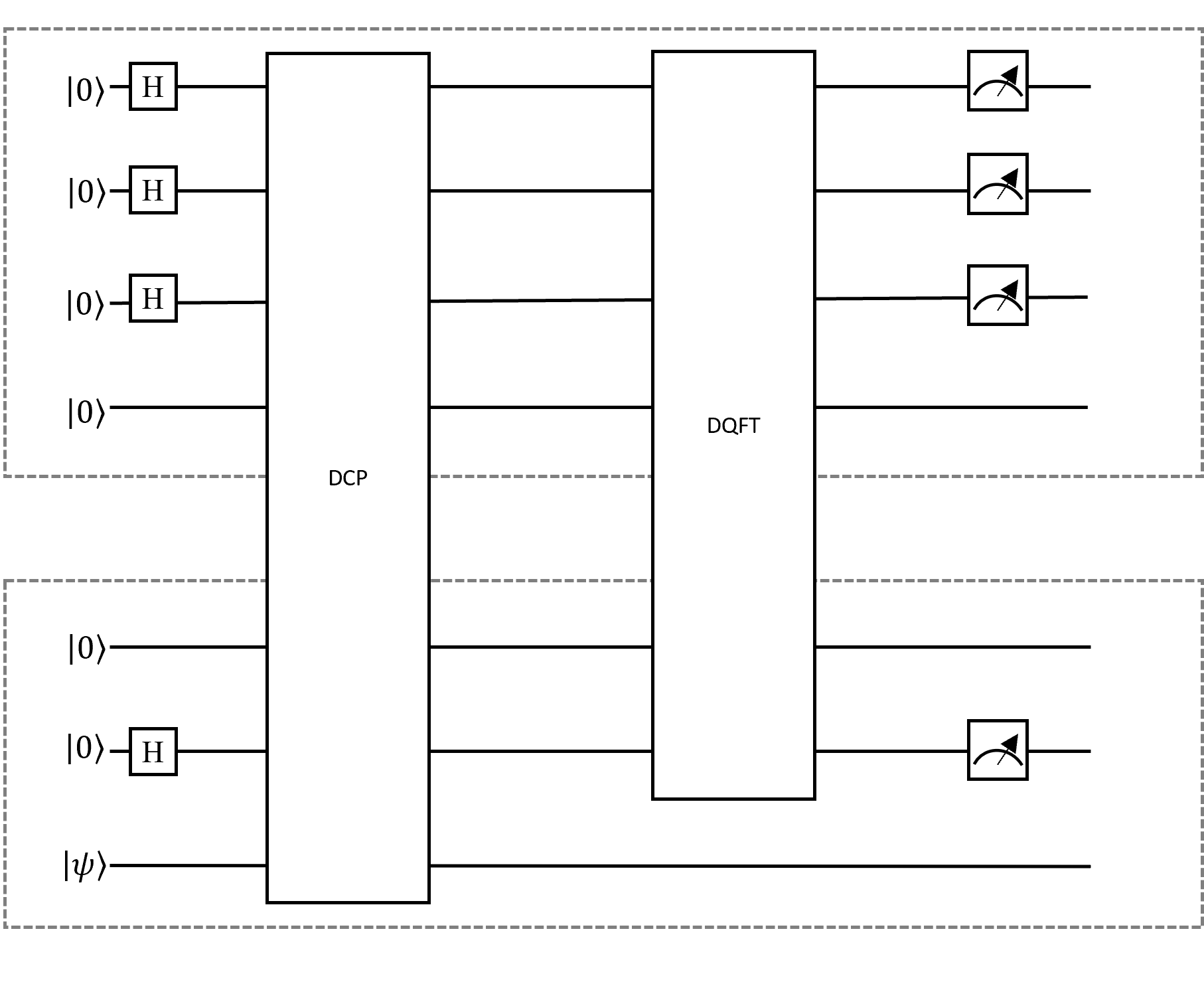}
    \caption{Phase estimation circuit where there are a total of seven qubits across both systems in which four are counting qubits. The computer containing the eigenstate $\ket{\psi}$ contains a single counting qubit while the rest are on the other system. Each system is denoted by gray dotted boxes. }
    \label{fig:final_dpe}
\end{figure}

\begin{figure}[!htb]
\centering
    \includegraphics[width=1\textwidth]{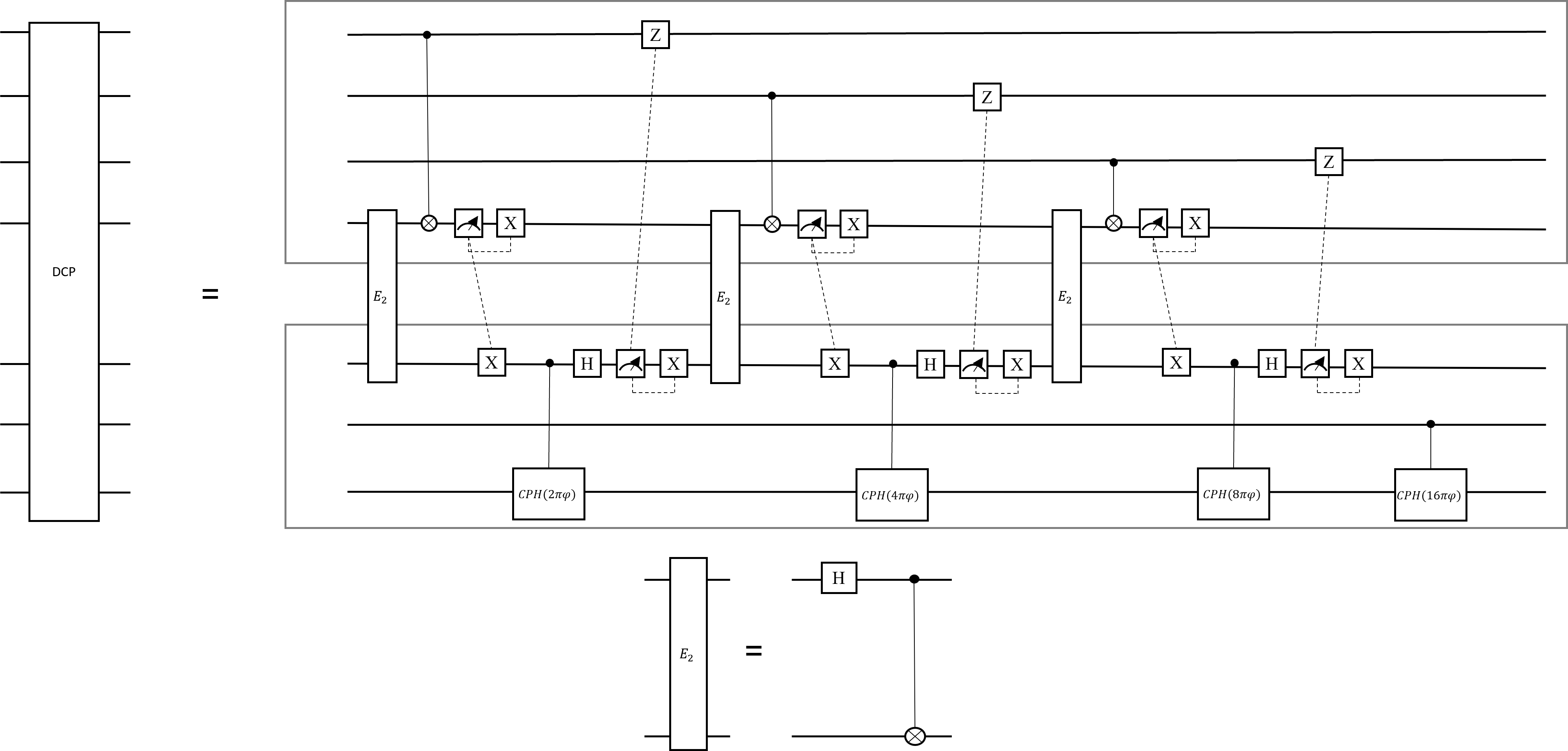}
    \caption{The distributed circuit responsible for applying the controlled phases from the qubit containing the $\ket{\phi}$ state to the counting qubits. The $E_2$ circuit is seen below.}
    \label{fig:distributed_CP}
\end{figure}

\begin{figure}[!htb]
\centering
    \includegraphics[width=1\textwidth]{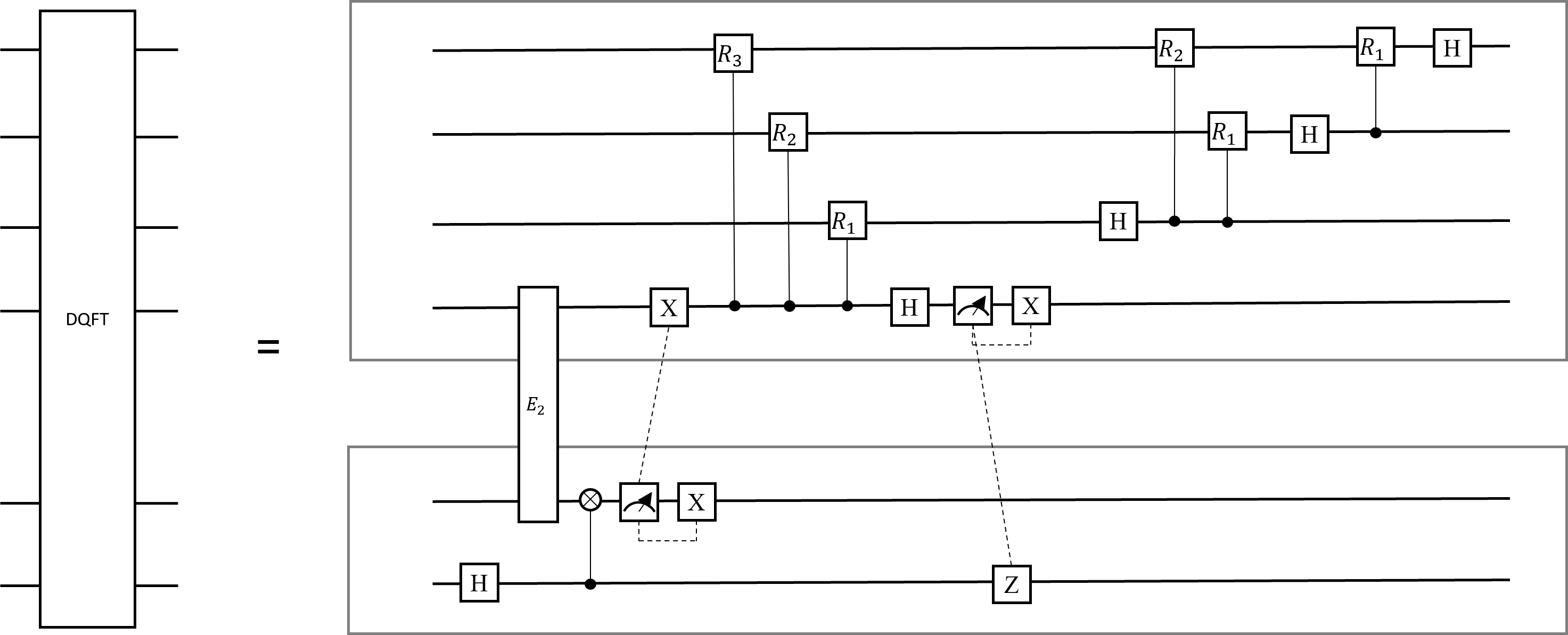}
    \caption{The distributed quantum Fourier transform circuit used on the counting qubits where the $R_i$ gates $i \subset \mathcal{I}$ refer to the $R^{iQFT_n}$ seen in Eq. \eqref{eq:rot_gate}}
    \label{fig:distributed_QFT}
\end{figure}

\begin{figure}[t]
\centering
    \includegraphics[width=0.9\textwidth]{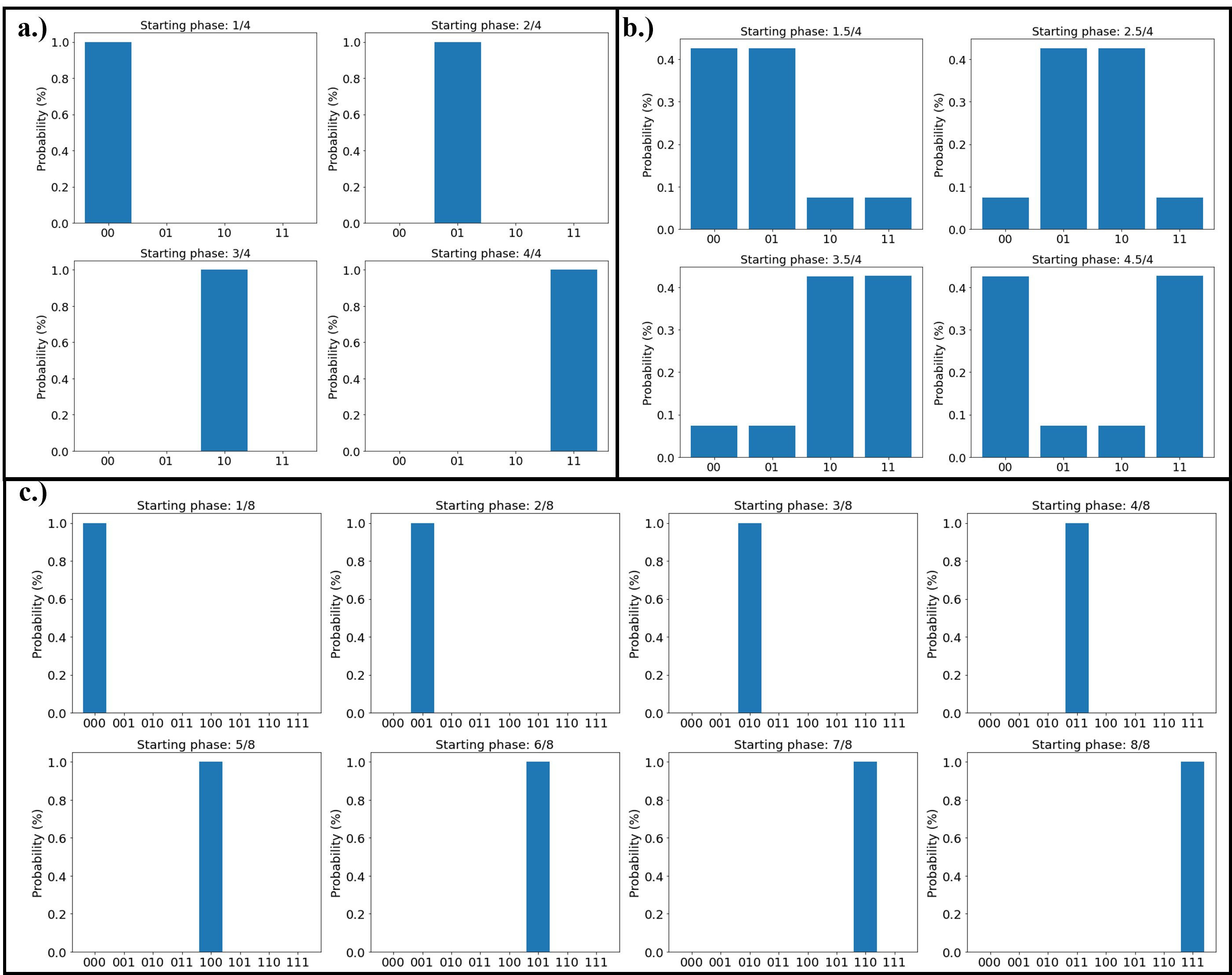}
    \caption{Final computational state probabilities of the distributed quantum phase algorithm with different qubit configurations without noise in the computational basis for different starting phases. The x-axis represents the measured phase in the computational basis, i.e $01 = 1/2^2$. a.) Results of a configuration in which Bob has a qubit that stores the phase, a counting qubit and a communication qubit (3 qubits) whereas Alice has a communication qubit and a counting qubit (2 qubits). b.) Same set up as a. but the starting phase has a $1/2$ factor added. c.) Results of a configuration in which Bob has a qubit that stores the phase, a counting qubit and a communication qubit (3 qubits) whereas Alice has a communication qubit and two counting qubits (3 qubits).}
    \label{fig:complete_no_noise}
\end{figure}

This circuit will be used throughout the entire investigation. However, the results shown in figure \ref{fig:complete_no_noise} shows the QuTiP simulation runs with varying the amount of counting qubits in the computer without the phase register. This is done for simplicity to see the difference in varying initial starting phases. The chosen phase factor $\varphi$ changes each run and the resulting graph denotes the probability of each vector state (measured in the computational basis). Recalling section \ref{sec:phase_estimation}, the final estimate is represented using $\phi/2^N$ where $\phi$ is the measured state in the computational basis and $N$ is the total amount of counting qubits. Thus, figures \ref{fig:complete_no_noise}a and \ref{fig:complete_no_noise}c demonstrate that without noise, the distributed algorithm manages to calculate the phase perfectly for $N = 2$ and $N = 3$ respectively when the phase is divided by the integer $2^N$. However, even when $\delta = 0.5$ (seen in Eq. \eqref{eq:delta_non_zero}), the best estimate of $\varphi$ given by $\phi$ has to have a probability $\mathrm{Pr}(\phi) \geq 4/\pi^2 \approx 40.5\%$. This result can be seen in figure \ref{fig:complete_no_noise}b. where the estimated computational basis states $\phi$ with the highest probabilities are $ \phi = \varphi \pm 0.5$. It is thus shown that the resulting probabilities are in accordance with the inequality observed in Eq. \eqref{eq:c_a}. 

To conclude, the design and implementation of a distributed phase estimation circuit has been realized. Chapters \ref{chapter:hybrid_system} and \ref{chapter:open_system} details the Hamiltonians required execute the noisy version of this algorithm. The execution and implementation of the distributed noisy phase estimation algorithm with four counting qubits using optimization techniques will be seen in Chapter \ref{chapter:GRAPE_sim}.  

\clearpage

\chapter{The Hybrid Quantum System}\label{chapter:hybrid_system}
The future of distributed quantum computation requires the coupling and manipulating of qubits in both the flux qubit and Rydberg atom quantum architectures \cite{quantum_network_hybrid}. The establishment of coupling or entanglement between two qubits possessing distinct physical properties, here referred to as hybrid qubits, becomes essential in enabling the interconnection of quantum computers with varying architectures. Furthermore, a hybrid quantum structure consisting of a variation of different physical realizations has the ability to inherit the advantages of each component \cite{trapped_atomic_ions,Wallquist_2009,Xiang_2013}. Hence, similar to classical computers, which feature multiple chips for executing distributed processes, quantum computers can likewise operate in a comparable manner with the advantage of employing diverse qubits. The goal of this chapter is to introduce the physical characteristics of both quantum computers, which will then be implemented into the circuit seen in Section \ref{sec:qutip_implementation}. Section \ref{sec:rydberg_qubit_intro} introduces the Rydberg qubit, which is then utilized to construct the model of the Rydberg computer. Similarly, Section \ref{sec:flux_qubit_intro} introduces the flux qubit, which is then utilized to construct the model of the flux qubit computer. Section \ref{sec:hybrid_comp} details the coupling of both architectures, which establishes the physical implementation of the $E_2$ gate. This system is then subsequently subjected to numerical evaluation in Section \ref{sec:GHZ_hybrid}, with the objective of gauging the effectiveness of the gate.
\section{The Rydberg Qubit}\label{sec:rydberg_qubit_intro}
Rydberg atoms are atoms that are excited into a high energy level. Specifically, Rydberg atoms possess a single valence electron in the highest energy level. This phenomenon leads to interactions that can be exploited to carry and propagate quantum information between atoms. Hence, Rydberg atom quantum computers use Alkali atoms as mediators of quantum information \cite{Ryabtsev_2016,Saffman_2016}. It is important to note that Sections \ref{sec:qubit_intro}, \ref{sec:intro_flux} and \ref{sec:hybrid_comp} detail three distinct sets of Hamiltonians that will be used in the simulation of the phase estimation.

Rydberg atoms were first discovered in 1885 by J.J Balmer, where he verified that the energy levels $n$ are analogous to that of Hydrogen for $n<16$ \cite{Bohr1913-BOHOTC,briand_kleppner_1987,thompson_salpeter_1977}. The early 1970s saw the advent of tunable lasers and with it further studies on Rydberg atom ensembles, which lead to the exploration of Rydberg atoms in the context of external electric fields. The scaling laws governing Rydberg atoms were revealed to be bizarre in nature as the dipole moment was found to scale as $n^2$. Thus, for high $n$ Rydberg atoms at $n=30$ energy level, the dipole moment was found to equate to $10^3$\r{A} as seen derived in Appendix \ref{sec:ryd_theory} \cite{rydberg_1890,bates_damgaard,quantum_defect_theory,stegun_2013}. The Rydberg atom emerged as a compelling candidate for qubit applications, since the dipole moments at this scale were found to propagate and interact with other atoms at large distances ($\mu$m), as well as the fact that Rydberg atoms could strongly couple to external fields \cite{briand_kleppner_1987}.

The basis of quantum information processes with trapped Rydberg atoms is to encode quantum information in the internal states of single atoms with interactions mediated via the highly excited Rydberg states \cite{Jaksch_2000,Lukin_2001}. Using laser trapping techniques and laser cooling, these atoms can be well isolated and arranged in different geometries that can potentially exaggerate coupling between atoms \cite{Morgado_2021}. In this section, a detailed explanation will be provided on how Rydberg interactions enable atoms to function as qubits, along with a description of the architecture of the Rydberg part in the hybrid quantum computer.

\subsection{Rydberg Interactions} \label{sec:Rydberg_interaction_main}
The main interaction at play when preparing or using Rydberg atoms for quantum computing is the dipole-dipole interaction, which is observed when atoms interact with each other. When an atom interacts with an external electric field, it undergoes a phenomenon known as the Stark effect, resulting in the splitting of its spectral lines. The interaction makes use of the characteristic scaling, which is denoted as $n^* = n -\delta_{nlj}$ where $\delta_{nlj}$ is the quantum defect dependent on the principal quantum number $n$, the orbital angular momentum quantum number $l$ and the total angular momentum quantum number $j$. For the rest of the investigation, the effective principal quantum number $n^*$ and principal quantum number $n$ are interchangeable when describing Rydberg atoms. The mean radius, $r$, of the Rydberg atom scales as $\langle r \rangle \propto n^2 a_0$ ($a_0$ being the Bohr radius).

The single valence electron allows the Rydberg atom to be treated as a dipole which couples strongly to external electric fields. A dipole $\mu$ in an external electric field $E$ has an interaction potential,
\begin{equation}
 V_{dip} = \boldsymbol{\mu} \cdot \boldsymbol{E} = -\mu_zE,   
\end{equation}
where the field is chosen to be in the $z$ direction \cite{ditzhuijzen_2009}. The dipole moment of an atom is given by $\boldsymbol{\mu} = -e\langle \boldsymbol{r} \rangle$ where $\boldsymbol{r}$ is the radial distance of the electron from the atom and $e$ is the electron charge. For pure angular momentum states, the mean radial distance $\langle r \rangle$ is always zero. Since the external electric field perturbates the well defined angular momentum states, $\langle \boldsymbol{r} \rangle \neq 0$. Hence the atomic states are coupled by a transition dipole moment given by $\mu_{1,2}=-e\bra{\psi_1}\boldsymbol{r}\ket{\psi_2}$ which is a second order coupling consequently leading to the second order Stark effect \cite{ditzhuijzen_2009, Szmytkowski_2018},
\begin{equation}
    \Delta E = -\frac{1}{2}\alpha E^2,
\end{equation}
where $\alpha$ is the polarizability. The polarizability as well as the Stark effect can be seen derived in sections \ref{sec:perturb_theory} and \ref{sec:stark_effect} \cite{quantum_theory_mit, JWBHughes_1967,sakurai_napolitano_2021}. Allowing $\boldsymbol{e}_R$ to be the unit vector along the relative coordinate $\boldsymbol{R}$ between two atoms and both $\boldsymbol{r_1}$ and $\boldsymbol{r_2}$ to be the relative displacement from the ion to the Rydberg electron for an atom 1 and atom 2 respectively, the classical dipole-dipole interaction between two Rydberg atoms becomes,
\begin{equation}\label{eq:dipoledipole}
    V_{dd} = \frac{e^2}{4\pi \epsilon_0} \frac{\boldsymbol{r}_1 \cdot \boldsymbol{r}_2 - 3(\boldsymbol{r}_1 \cdot \boldsymbol{e}_R)(\boldsymbol{r}_2 \cdot \boldsymbol{e}_R)}{R^3}.
\end{equation}
Using the fact that Rydberg dipole moments scale $|r| \propto n^2$, the classical interaction strength can be deduced to scale as $V_{dd} \propto n^4$. This however breaks down in the absence of an external electric field. In this scenario, atoms are unpolarized consequently resulting in a vanishing dipole moment due to the spatial symmetry of the wavefunction, $\langle \boldsymbol{r} \rangle = 0$ thus resulting in $\langle V_{dd} \rangle = 0$ \cite{Wu_2021}. The spatial symmetry cannot vanish because $\boldsymbol{r}$ has non-vanishing matrix elements between eigenstates with different parities \cite{Wu_2021}. 

\begin{figure}[t] 
\centering
    \includegraphics[width=0.6\textwidth]{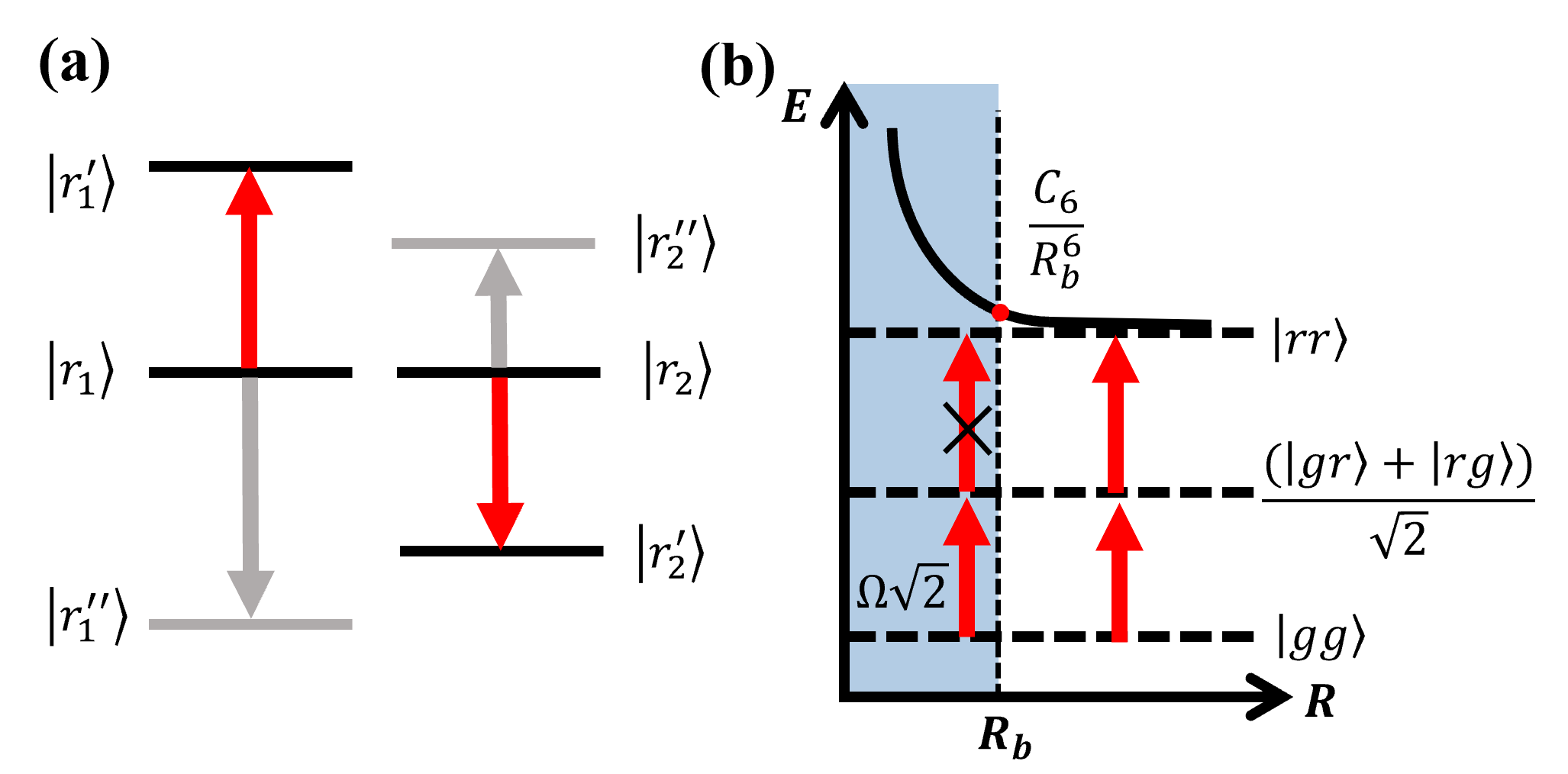}
    \caption{(a) Energy levels with corresponding transition channels contributing to the dipole-dipole interaction. (b) Rydberg blockade for two atoms at distance R. \cite{Wu_2021} }
    \label{fig:rydberg_blockade}
\end{figure}

A more rigorous quantum description is necessary to describe the interaction between different Rydberg states in the absence of an external electric field. In the spherical harmonic basis, the dipole-dipole operator can be expressed as,
\begin{equation}\label{eq:V_dd_extended}
    V_{dd} = \frac{-e^2}{4\pi\epsilon_0 R^3}\sqrt{\frac{24\pi}{5}}\sum_{\mu,\nu}C^{1,1,2}_{\mu,\nu,\mu+\nu}Y^{\mu+\nu}_2(\theta,\phi)^*d_\mu^{(1)}d_\nu^{(2)},
\end{equation}
where $C^{j_1,j_2,J}_{m_1,m_2,M}$ is the Clebsch-Gordan coefficient, $Y^m_l(\theta,\phi)$ is the spherical harmonic function (seen in Appendix \ref{sec:perturb_theory}) and $d_q^{(i)}$ represents the spherical component ($q=\pm 1,0$) of the dipole operator for the $i^{\mathrm{th}}$ atom \cite{Wu_2021}. Utilizing, Eq. \eqref{eq:V_dd_extended} one can derive the energy transfer between two atomic states, known as the F\"{o}rster Resonance Energy Transfer, which is explicitly derived in \cite{Weber_2017}. Knowing this, it is now possible to consider a single transition channel $r_1 + r_2 \rightarrow r_1^' + r_2^'$ in the absence of an external electric field. This transition can be seen on Fig. \ref{fig:rydberg_blockade}(a) depicted as red arrows.

The F\"{o}rster Resonance Energy Transfer is a technique used in Rydberg atom computing to transfer quantum information between Rydberg atoms. The energy transfer is dependent on the potential of both atoms (or the distance between them). This is simply characterized by an energy difference between the initial and final state called the energy defect. The energy defect is defined as $\Delta_F$ and can be seen derived in Appendix \ref{sec:forster_resonance_app} \cite{Ravets_Forster,Beterov_2018}. For simplicity, the dipole-dipole interaction energy will be rewritten to $V_{dd} = C_3(\theta)/R^3$, where $C_3(\theta)$ is the anisotropic interaction coefficient \cite{long_range_dispersion}. Thus, the Hamiltonian of the model system dominated by a single channel in the subspace spanned by $\{\ket{r_1 r_2},\ket{r_1^' r_2^'}\}$ can be written as \cite{zeeman_degeneracy}, 
\begin{equation}
    H = \begin{bmatrix}
    0 & C_3(\theta)/R^3 \\
    C_3(\theta)/R^3 & \Delta_F
    \end{bmatrix}
\end{equation}
The eigenvalues for $H$ are given by,
\begin{equation}
    E_\pm = \frac{\Delta_F}{2} \pm \frac{1}{2}\sqrt{\Delta_F^2+4V^2},
\end{equation}
from which the induced energy shift is $\Delta E_\pm = E_\pm(R) - E_\pm(R\rightarrow\infty)$. For $\Delta_F = 0$, there is strong coupling between the initial and final state since they are degenerate leading to the eigenstates $\ket{\pm} = (\ket{r_1 r_2} \pm \ket{r_1^' r_2^'})/\sqrt{2}$ with eigenenergies proportional to $\pm C_3/R^3$. This is the so called F\"{o}ster resonance. When two atoms are sufficiently close and $\Delta_F \ll V(R)$, the off-diagonal elements of $H$ are still dominant and $\Delta E_{\pm} \approx \pm C_3/R^3$. 

\begin{figure}[t] 
\centering
    \includegraphics[width=0.5\textwidth]{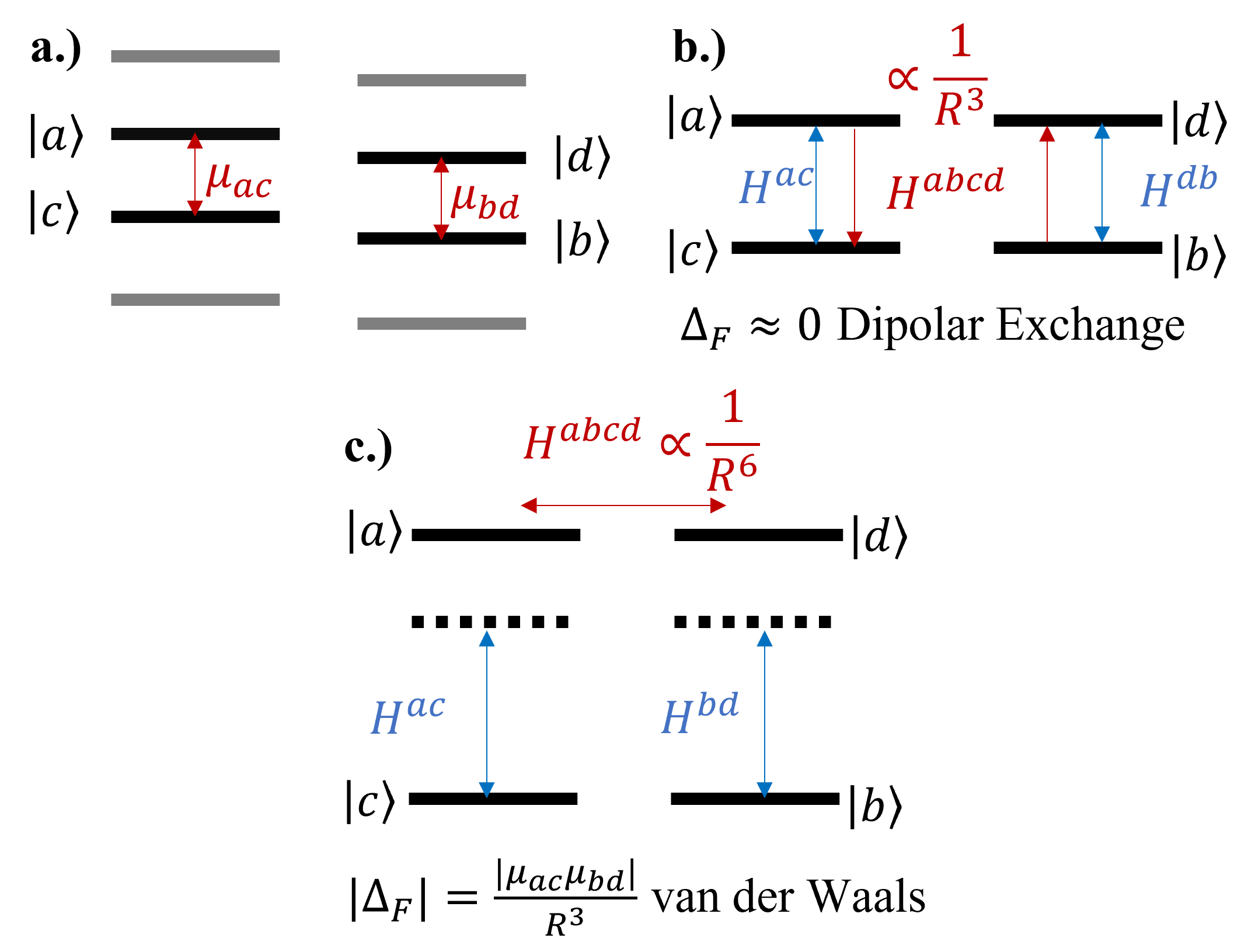}
    \caption{ Rydberg-Rydberg interactions used to construct quantum gates. (a) Each atom possesses a pair dipole coupled states which have transition dipole moments $\mu_{ac}$, $\mu_{bd}$. (b) For $\Delta_F \approx 0$ the Rydberg-Rydberg interactions have a dipolar exchange $\ket{ab}\bra{cd}$ with interaction strength $1/R^3$. (c) When $|\Delta_F|>|\mu_{ac}\mu_{bd}|/R^3$ the interactions are Van de Waals form $\ket{ab}\bra{cd}$ with interaction strength $1/R^6$ \cite{Morgado_2021}. }
    \label{fig:resonance_example}
\end{figure}

This concept can be seen in figure \ref{fig:resonance_example}(a), whereby two dipole coupled Rydberg atoms with states (a,c) and (b,d) have dipole moments $\mu_{ac}$ and $\mu_{bd}$, respectively. The F\"{o}rster defect in this scenario is equal to $\Delta_F = E_c +E_d - (E_a + E_b)$ that offsets the interaction energy between both pair states, where the subscript $i$ in $E_i$ denotes the energy state. When the electrostatic field is tuned to negate this F\"{o}rster defect (make it approximately equal to 0), the Rydberg-Rydberg interactions scales with $1/R^3$ seen in Fig. \ref{fig:resonance_example}(b) \cite{Morgado_2021}. This means that the atomic states assume a two dipole-coupled pair Rydberg state ($a=d, b=c$) and the pair state can be treated as a two level system.

On the other hand, when $\Delta_F \gg V(R)$, the interaction can be treated as a perturbation that just slightly shifts the energies of the bare states, thus giving rise to $\Delta E_- = -C_6/R^6$ for $\ket{r_1 r_2}$ and $\Delta E_+ = C_6/R^6$ for $\ket{r_1^' r_2^'}$ with $C_6 = C_3^2/\Delta_F$ \cite{zeeman_degeneracy, Wu_2021, long_range_dispersion}. As can be seen, the interaction in this regime is proportional to $R^{-6}$, conventionally called the van der Waals interaction. The crossover of the energy shift from $R^{-3}$ to $R^{-6}$ is described by a characteristic radius $R_{vdW} = |C_6/\Delta_F|^{1/6}$ where $V(R_{vdW}) = \Delta_F$.

This concept can be seen in \ref{fig:resonance_example}(c), where the interaction is non-resonant if $|\Delta_F|>|\mu_{ac}\mu_{bd}|/R^3$. Since the dipole interaction is treated using second-order perturbation theory, the $\ket{ab}$ state has a potential proportional to $C_6 = \frac{\mu_{ac}^2\mu_{bd}^2}{\Delta_F}$ \cite{Morgado_2021}. This concept is utilized in the next subsection to make multi-qubit gates between Rydberg atoms. 

\subsection{Rydberg Blockade and Transitions for Information Processing} \label{sec:rydberg_blockade}
As seen in the previous section, the van der Waals interaction scales as $V \propto 1/R^6$ which means that when the interatomic distance between two atoms is large enough, this interaction does not occur. However, at close distances, this interaction governs the system's Hamiltonian. The goal for this specific quantum system is to have minimal interaction between atoms when performing local gates (i.e., little cross-talk) and have large interactions between atoms when performing multi-qubit gates meaning that this phenomenon is highly desired. To fully grasp the phenomena, it is important to assume a two atom system where by each atom possesses two distinct energy levels, a $\ket{g}$ ground state and an excited state $\ket{r}$.

The simplest case of the Rydberg blockade can be observed in figure \ref{fig:rydberg_blockade}(b) where the blockade effect is shown for two atoms. It is assumed that the system starts with the population of both atoms at the ground state, or written as a single state, $\ket{gg}$. A resonant laser excitation with frequency $\Omega\sqrt{2}$ (where $\Omega$ is the \textit{Rabi frequency}) is applied to the system which drives the state to a symmetric single-atom excited state $(\ket{gr}+\ket{rg})/\sqrt{2}$. If the atoms are at a distance $R > R_b$ then another resonant pulse will excite the collective state to the double excited state, $\ket{rr}$. On the other hand, if $R < R_b$ holds, the applied laser pulse will not drive the state to the double excited state since the energy required is much larger. This means that the double excited state is essentially \textit{blocked}.  

This follows the Ising configuration of having Rydberg states adhere to a ground state and a Rydberg excited state \cite{ising_rydberg,Silv_pulser_2022}. This configuration is governed by a transition Hamiltonian of the form,
\begin{equation}\label{eq:pulser_hamiltonian}
    \mathcal{H}(t) = \sum_i \bigg(H_i^D(t) + \sum_{j<i}\frac{C_6}{R_{ij}^6}n_in_j\bigg)
\end{equation}
where $n_i$ denotes the projector $\ket{r}\bra{r}_i$ on the $i^{\mathrm{th}}$ atom. $H^D$ is the driving Hamiltonian between two energy levels. The driving Hamiltonian is defined as,
\begin{equation}
    H^D(t) = \frac{\hbar}{2}\boldsymbol{\Omega}(t)\cdot \boldsymbol{\sigma},
\end{equation}
where $\boldsymbol{\sigma} = (\sigma^x,\sigma^y,\sigma^z)^T$ is the Pauli vector and $\Omega(t) = (\Omega(t)\mathrm{cos}(\varphi),-\Omega(t)\mathrm{sin}(\theta),-\delta(t))^T$ is the rotation vector in which $\delta(t)$ is the detuning and both $\varphi$ and $\theta$ are the azimuthal and polar angle on the Bloch sphere (figure \ref{fig:bloch_sphere}), respectively. It is assumed that the laser targeting the transition is tuned at a resonance frequency $\omega_{ab}= |E_a - E_b|/\hbar$ with detuning $\delta(t) = \omega(t) - \omega_{ab}$ where $a$ and $b$ denote the different energy states. The Pauli vector is defined explicitly as \cite{Silv_pulser_2022},
\begin{equation}
    \sigma^x = \ket{a}\bra{b} + \ket{b}\bra{a},
\end{equation}
\begin{equation}
    \sigma^y = i\ket{a}\bra{b} - i\ket{b}\bra{a},
\end{equation}
\begin{equation}
    \sigma^z = \ket{b}\bra{b} - \ket{a}\bra{a}.
\end{equation}
As can be seen from Eq. \eqref{eq:pulser_hamiltonian}, the Hamiltonian of the system when $R>R_b$, the Hamiltonian can simply be described by the driving Hamiltonian, however, under the Rydberg blockade effect, the Hamiltonian has an added term. At $R>R_b$, \textit{pulsing}, evolving the driving Hamiltonian through means of a laser, a single atom or qubit consequently changes the state of the qubit without influencing the other qubits around it. This is said to be the \textit{digital approach} and is often used to pulse single quantum gates \cite{Silv_pulser_2022}. The digital approach is often encoded on two hyperfine levels of an atom. These are referred to as $\ket{g} = \ket{0}$ and $\ket{h} = \ket{1}$. To evolve the driving Hamiltonian between these states, one needs to define a pulse dictated by the operator,
\begin{equation}
    U(\boldsymbol{\Omega},\tau) = T \mathrm{exp}\Big[-\frac{i}{2} \int_0^\tau \boldsymbol{\Omega}(t) \cdot \boldsymbol{\sigma}\mathrm{d}t\Big],
\end{equation}
where $T$ is the time-ordering operator and $\tau$ is the pulse time. The unitary $U$ describes the rotation around the axis $\boldsymbol{\Omega}$. It is possible to induce rotations around the fixed axis $\boldsymbol{e}(\varphi) = (\mathrm{cos}\varphi,\mathrm{sin}\varphi,0)$ situated on the equator of the Bloch sphere, by pulsing a resonant pulse of phase $\varphi$ ($\delta=0$), there is a rotation angle of \cite{Silv_pulser_2022},
\begin{equation}
    \theta = \int^\tau_0\Omega(t)\mathrm{d}t.
\end{equation}
situated on the equator of the Bloch sphere. The unitary operator is thus,
\begin{equation}
    R_{\boldsymbol{e}(\varphi)}(\theta) = e^{-i\frac{\theta}{2}(\mathrm{cos}(\varphi)\sigma_x-\mathrm{sin}(\varphi)\sigma_y)}=e^{i\frac{\varphi}{2}\sigma_z}e^{i\frac{\theta}{2}\sigma_x}e^{-i\frac{\varphi}{2}\sigma_z} =
    R_z(-\varphi)R_x(\theta)R_z(\varphi).
\end{equation}
Adding an additional $z$-rotation, which can be achieved virtually through a shift in the phase reference frame (more detailed overview in \cite{McKay_2017}), allows the construction of an arbitrary single-qubit gates,
\begin{equation} \label{eq:unitary_rotation}
    U(\theta,\varphi,\gamma) = R_z(\gamma+\varphi)R_{\boldsymbol{e}(\varphi)}(\theta) = R_z(\gamma)R_x(\theta)R_z(\varphi).
\end{equation}
as seen in Eq. \eqref{eq:general_single_u}. To build a set of universal gates, one requires multi-qubit gates. Multi-qubit gates use the manipulation of unitaries provided in equation \eqref{eq:unitary_rotation} and the Rydberg blockade effect. The native multi-qubit gate in Rydberg computing is the CZ gate, which is seen in Appendix \ref{chapter:logic_gate_appendix}. This gate can then be expanded to create other multi-qubit gates. The CZ Rydberg gate turns the initial states into the final states seen in table \ref{table:cz_effect}.

\begin{table}[h]
\centering
\caption{Initial and final states of the CZ pulse sequence in the computational basis.}
\label{table:cz_effect}
\begin{tabular}{ |c|c|c|c|c| } 
 \hline
 Initial state & $\ket{00}$ & $\ket{01}$ & $\ket{10}$ & $\ket{11}$ \\
 \hline
 Final state &   $-\ket{00}$ & $-\ket{01}$ & $-\ket{10}$ & $\ket{11}$ \\ 
 \hline
\end{tabular}
\end{table}

To understand how the two qubit CZ gate is pulsed, it is important to take the computational basis ($\ket{0}$ and $\ket{1}$) that was encoded to the hyperfine states and add an auxiliary Rydberg state $\ket{r}$. Two qubits (target and control) are first physically moved to a distance $R_b$ from each other (seen in figure \ref{fig:rydberg_blockade}) where the van der Waals interaction is observed. Then, a $\pi$ pulse attempts to bring the control qubit in an initial $\ket{0}$ to the auxiliary Rydberg state. If the qubit is in the $\ket{1}$ state, it will not be able to transition to the auxiliary state. A second $2\pi$ pulse is then used to attempt to transition the target qubit to the auxiliary state and back from the $\ket{0}$ state. This transition will be successful if the target is initially in the $\ket{0}$ state and if the control failed to transition to the auxiliary state. This is because only a single qubit is allowed to be in the auxiliary state as seen in figure \ref{fig:rydberg_blockade}. Finally, a third pulse is used to attempt to transition the control qubit back to the $\ket{0}$ state \cite{Silv_pulser_2022}. Applying a global $-1$ factor or a global phase of $\pi$ to the final states in table \ref{table:cz_effect}, leads to the CZ gate. 

\begin{figure}[t] 
\centering
    \includegraphics[width=0.6\textwidth]{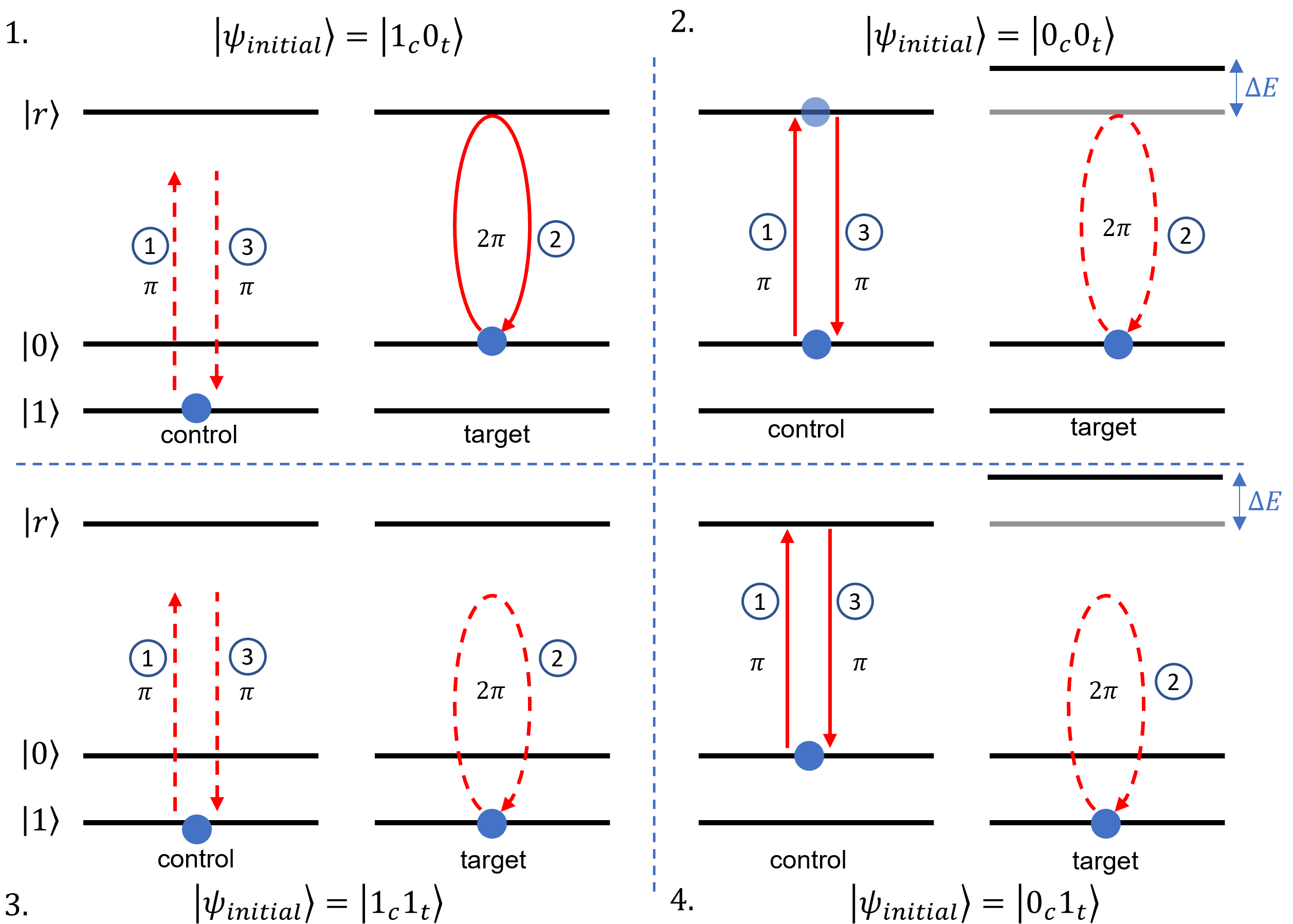}
    \caption{ CZ gate implementation where three pulses are consecutively executed on the qubit. Each panel depicts the different scenarios depending on the initial state. The step number within each panel are shown in blue circles. The bold red lines indicate that the transition was successful while the dashed red lines indicate that the transition was unsuccessful \cite{Silv_pulser_2022}.}
    \label{fig:rydberg_cz}
\end{figure}

\subsection{Rydberg Atom System}\label{sec:rydberg_atom_system}
As described in section \ref{sec:distributed_algorithm}, a distributed quantum algorithm requires the use of an E2 gate. The E2 gate is required to establish quantum communication between two quantum computers, in this case, two different quantum architectures. This subsection will investigate the Rydberg ensemble and how it pairs to the physical E2 gate. A more detailed description of the physical E2 implementation will be investigated in section \ref{sec:hybrid_comp}.

\begin{figure}[t] 
\centering
    \includegraphics[width=0.75\textwidth]{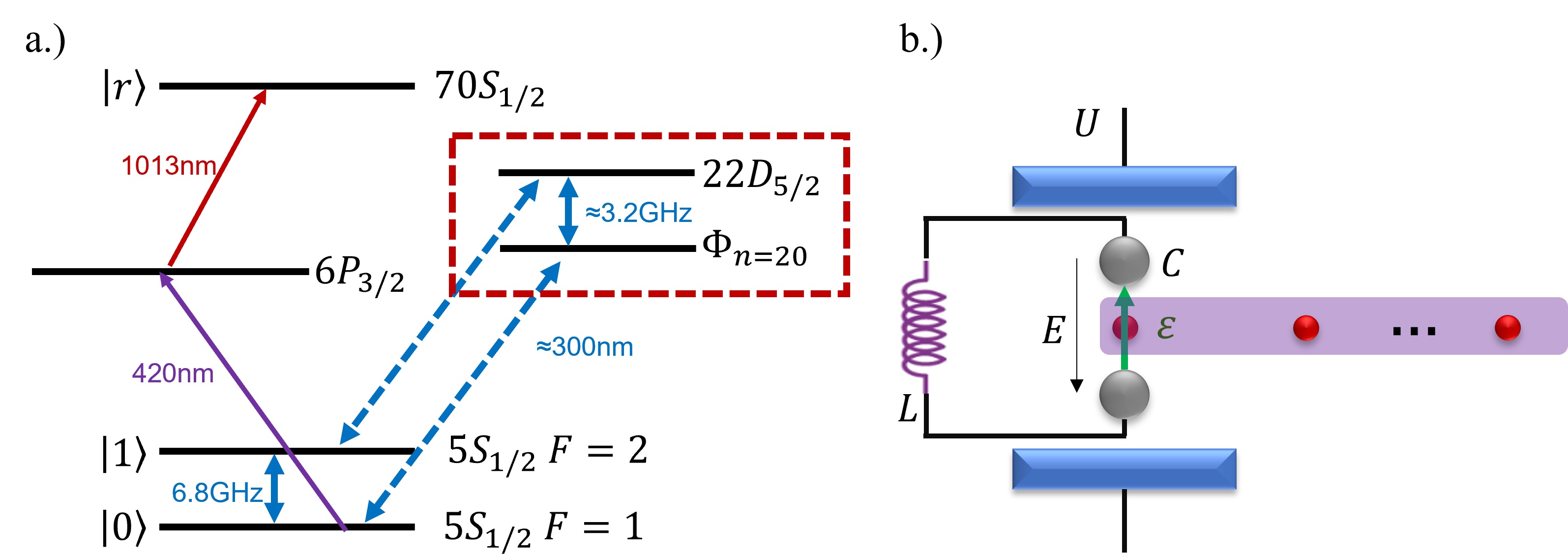}
    \caption{(a) Schematic detailing the Rydberg energy states paired to the respective computational states of the proposed Rydberg computer atom array with the energy states of the communication qubit depicted in the red dotted square. The atoms in the atom array contain two computational levels in the hyperfine $\ket{5S_{1/2},F=1,2}$ states, the Rydberg auxiliary state in the $\ket{70S_{1/2}}$ state and an extra auxiliary $6P_{3/2}$ to facilitate the transfer to the $\ket{70S_{1/2}}$ state. The communication qubit in the red dotted box shows the two energy levels used in the work done by \citeauthor{Yu_atom_flux_2017} whereby two energy states, $\ket{22D_{5/2}}$ and an arbitrary superposition state $\ket{\Phi_{n=20}} = \ket{n=20, l \geq 3, j, m = 5/2}$, are used to denote the computational $\ket{1}$ and $\ket{0}$ states for the $E_2$ gate respectively \cite{Yu_atom_flux_2017}. (b) Schematic of the system whereby an Rubidium-87 atom is held between two spherical capacitors (with capacitance $C$) connected to an LC resonator with inductance $L$. This atomic states are manipulated by an electrostatic field $E$ in the $z$-direction using parallel plates with imposed voltage difference $U$. The atom forms part of Rydberg quantum computer shown as an atom array (red dots in the purple block).}
    \label{fig:forster_atom}
\end{figure}

The Rydberg ensemble requires three energy levels, the ground state $\ket{g}$ (encoded to the computational basis state $\ket{0}$), the hyperfine state $\ket{h}$ (encoded to the computational basis state $\ket{1}$) and the auxiliary excited state $\ket{r}$. For Rubidium-87, the hyperfine levels are chosen to be $\ket{5S_{1/2}, F=1} = \ket{g}$ and $\ket{5S_{1/2}, F=2}= \ket{h}$ \cite{Levine_2019,Wu_2021}. The auxiliary $\ket{r}$ state is encoded to be in the $\ket{70S_{1/2}}$ state, which will be used to initialize the Rydberg blockade. The transition from the $\ket{0}$ to the auxiliary Rydberg state $\ket{r}$ requires frequencies that are optically difficult to achieve (sub 300nm wavelengths). Thus in practice, an auxiliary $\ket{6P_{3/2}}$ state is used to aid the transfer. The hyperfine levels transition with a frequency of 6.8 GHz while the transition from the $\ket{0}$ state to the $\ket{6P_{3/2}}$ state is optically achievable using a 420nm wavelength laser. Finally, the transition from the auxiliary state to the Rydberg auxiliary $\ket{r}$ state is achievable using a 1013nm tuned laser.  A schematic detailing the energy levels in the computational basis can be seen in figure \ref{fig:forster_atom}a.

The apparatus that connects the Rydberg atom system to the flux system follows the work of \citeauthor{Yu_atom_flux_2017}. The study denotes a system in which a Rubidium-87 atom is placed between two spherical capacitors connected to an LC resonator in which an electric field runs in the $z$-direction This system is further elaborated upon in Section \ref{sec:hybrid_comp}. Two states are used in the study, a state $22D_{5/2}$ populated when the Rydberg qubit is in the computational $\ket{1}$ state and an arbitrary $\ket{n=20,l\geq3,j,m=5/2}$ superposition state is populated when the Rydberg qubit is in the $\ket{0}$ computational basis. The transition frequency between the two states is approximately 3.2GHz. This research extends \citeauthor{Yu_atom_flux_2017} by connecting the Rubidium-87 atom in the system to an additional array of Rubidium-87 atoms. 

Transitioning between the atom array to the atom between the capacitor is necessary to bridge both systems. This means that the $\ket{22D_{5/2}}$ state needs to transition to the computational $\ket{1}$ state in the hyperfine $\ket{5S_{1/2}, F=2}$. Similarly, the $\ket{n=20,l\geq3,j,m=5/2}$ (denoted as $\Phi_{n=20}$) state needs to transition to the computational $\ket{0} = \ket{5S_{1/2}, F=1}$ ground state. The optical wavelength to achieve these transitions is approximately 300nm, which equates to a THz frequency. Although, this is theoretically possible, further work should aim to experimentally analyze the transition frequencies between the hyperfine levels of the $\ket{5S_{1/2}}$ state and the $\ket{22D_{5/2},\Phi_{n=20}}$ levels of the communication qubit as well as the effectiveness of the transition. 

For the system to work optimally, it is important that the atoms in the array are far enough apart from each other when processing single qubit operations. This will limit the amount of unwanted interactions from surfacing that may influence the decoherence in the energy states. This means avoiding the dipolar exchange seen in Fig. \ref{fig:resonance_example}b by avoiding a resonant F\"{o}rster defect $\Delta_F \approx 0$. The potential energy of the interaction was seen to be $V_{dd} = C_3/R^3$. Thus, it is important to maximize the interatomic distance $R$ when performing single qubit operations.  

However, for multi-qubit gates, it is important that the atoms are close enough to experience the exchange interaction. In order to implement the universal CZ gate depicted in Fig. \ref{fig:rydberg_cz}, the Rydberg blockade must be in effect. To do so, the F\"{o}rster defect needs to be non-resonant, $\Delta_F > |\mu^2|/R^3| $ where $\mu$ is the transition dipole moment of the transition $\ket{5S_1/2} \rightarrow \ket{70S_1/2}$. The energy shift can thus be calculated using $\Delta E_\pm = \pm C_6/R^6$.  

Using \textit{ARC} \cite{ibali__2017}, the characteristic radius, $R_{vdW}$, was calculated to be $3.5\mu$m which can be seen in Fig. \ref{fig:rybdergblock_system} and Fig. \ref{fig:ryd_val}. The dipole-dipole coefficient was calculated to be $C_3 = 32.45$ GHz $\mu m^{3}$ and the van der Waals coefficient was calculated to be  $C_6 = 801.98$ GHz $\mu m^{6}$ for the pair state. Therefore, when executing single qubit gates, it is important to observe an interatomic distance $R \gtrsim 9\mu$m to avoid substantial dipole-dipole interactions. On the other hand, when processing multi-qubit gates, the interatomic distance is chosen to be $R=3.5\mu$m which observes an energy shift $\Delta E_\pm = 801.98$ GHz capable of achieving the Rydberg blockade effect.  

\section{The Flux Qubit}\label{sec:flux_qubit_intro}
\subsection{Introduction to Flux Qubits} \label{sec:intro_flux}
The Josephson persistent-current qubit or the flux qubit is a type of superconducting qubit which consist of an inductance loop shorted by a Josephson junction \cite{pauw_flux}. Josephson junctions are made up of two superconductors sandwiching a thin non-superconducting layer such that electrons can tunnel through the barrier \cite{Josephson_junction}. This produces what is known as the Josephson effect which is a current that flows continuously across the Josephson junction without supplied current. At temperatures close to zero kelvin, this circuit experiences superconductivity, and in turn, forces the electrons in the circuit to the lowest energy state. In order to not violate Pauli's principle, electrons of opposite spins form a single particle called a Cooper pair, which exhibits the same properties of a boson. The flux qubit's energy states are dependent on the magnetic flux measured in the inductance loop, which is dependent on the amount of Cooper pairs crossing the Josephson junction. The wavefunction of the Cooper pair system as well as the magnetic energy states are described in Appendix \ref{sec:wavefunction_flux}. 

The energetics of this system can be described by the following Hamiltonian, 
\begin{equation}\label{eq:single_JJ_Hamil}
    H = \frac{Q^2_J}{2C_J} - E_J\mathrm{cos}\Big(\frac{2\pi\Phi}{\Phi_0}\Big) + \frac{1}{2}LI_q^2,
\end{equation}
where $Q_J$ is the electric charge on the capacitor, $C_J$ is the self capacitance, $I_q$ is the current circulating in the loop, $L$ is the inductance of the inductor, $\Phi$ is the applied magnetic field, $E_J$ is the energy from the Josephson junction and $\Phi_0$ is the magnetic flux quantum. The Hamiltonian follows that of the quantum LC Hamiltonian seen in Appendix \ref{sec:quantum_oscillator}. Following Eq. \eqref{eq:single_JJ_Hamil}, it is possible to see minimas form at $\Phi = \pm\Phi_0/2$. The two points are called the 'degeneracy points', in which the self-inductance of the Josephson junction is approximately equal to the inductance of the inductive loop $L_J \approx L$ \cite{Birenbaum2014TheCF}. These two values are associated with equal and opposite flux states formed by a persistent current $I_q$ in the loop. Because the magnetic field is seen to rotate clockwise or anti-clockwise around the loop the states are labelled left circle or $\ket{L}$ at $+\Phi_0/2$ and right circle or $\ket{R}$ at $-\Phi_0/2$ \cite{Birenbaum2014TheCF,pauw_flux}.

\begin{figure}[t] 
\centering
    \includegraphics[width=0.6\textwidth]{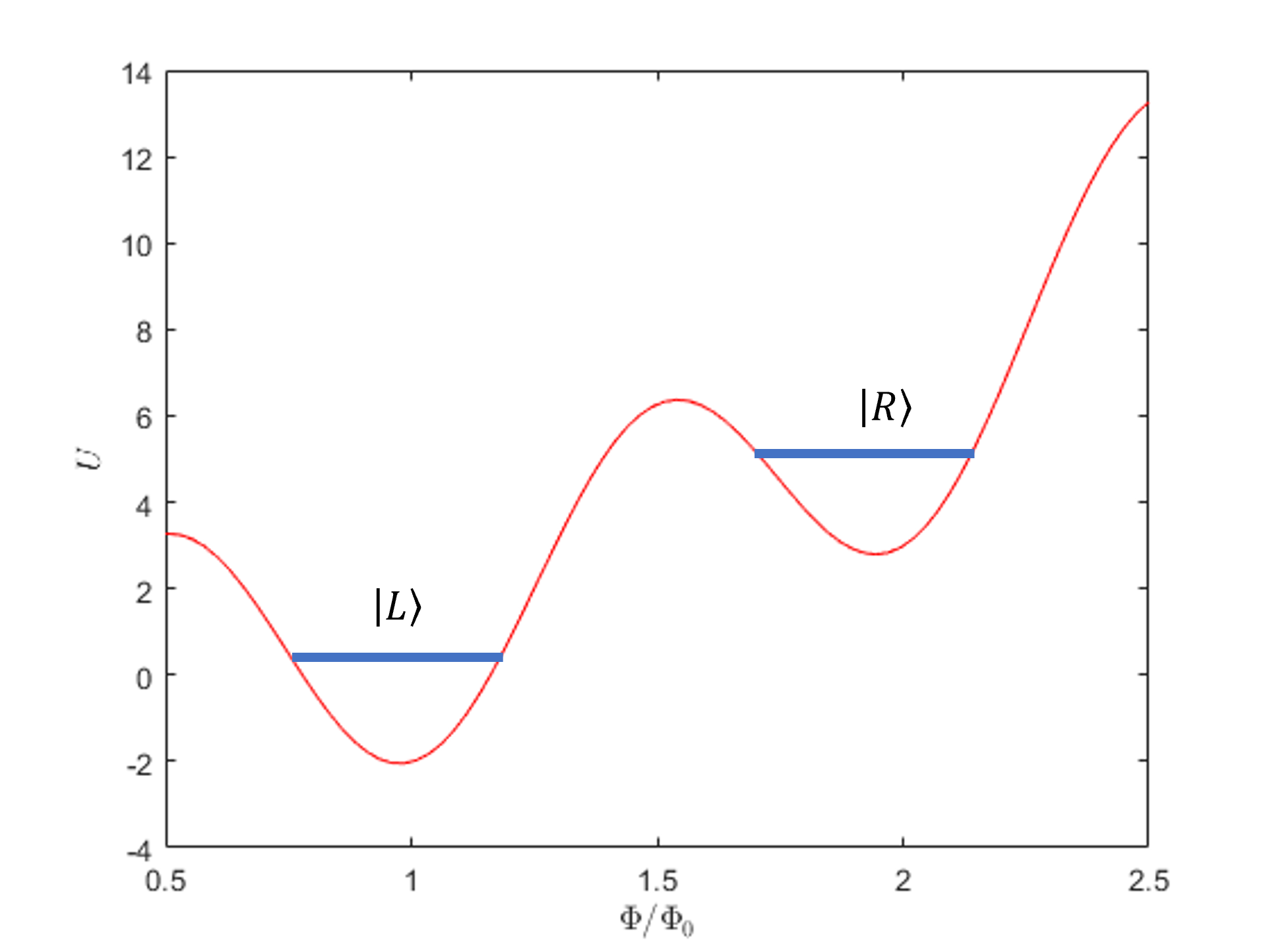}
    \caption{The potential energy $U = \frac{1}{2}L I_q^2 - E_J \mathrm{cos}(\frac{2\pi\Phi}{\Phi_0})$ plotted against $\Phi/\Phi_0$ \cite{single_flux_jj}.} 
    \label{fig:single_JJ}
\end{figure}

The two minimas can be described by analyzing the dependence of $\Phi$ on the potential energy of the system $U$. This relationship is seen in figure \ref{fig:single_JJ} where the strength of the interaction between both states is dependent on the height of the barrier separating them. Both $\ket{R}$ and $\ket{L}$ are classically isolated from each other when the thermal energy is lower than the barrier height, however, due to quantum tunneling, there can be interaction between both states.  Unfortunately, this is very difficult to do with a single junction since the energy splitting (or barrier height) scales exponentially with $L/L_J$. However, to form a double well potential from a single Josephson junction, $L \approx L_J$. The trade-off between the two make it difficult to form a decent qubit. 

\begin{figure}[t] 
\centering
    \includegraphics[width=0.4\textwidth]{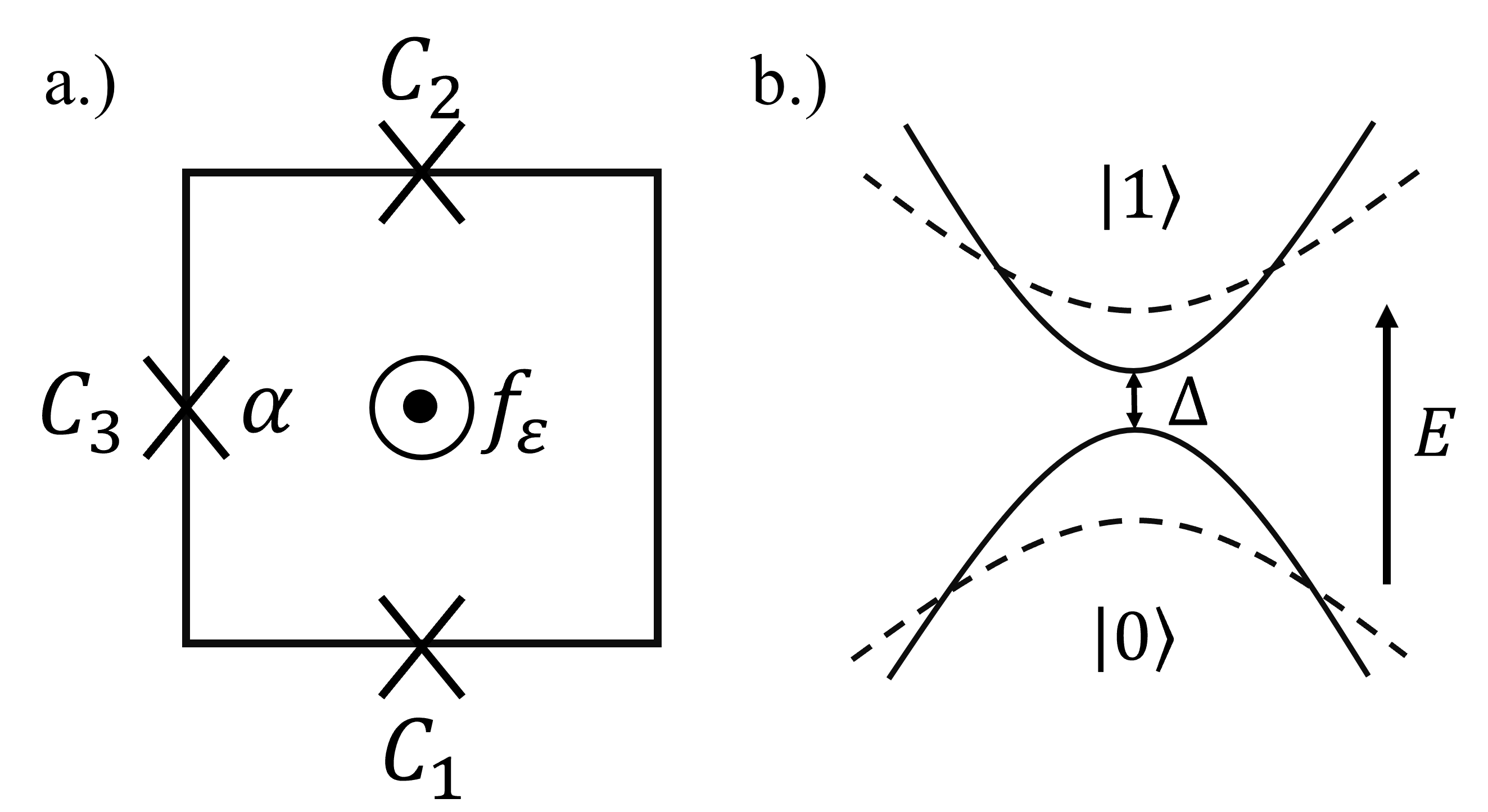}
    \caption{(a) Flux qubit with detailing the three Josephson junctions each denoted by a cross. Each Josephson junction possesses a capacitance $C_i$ that is related to the gauge invariant phase difference of each junction. (b) The ground state $\ket{0}$ and the excited energy state $\ket{1}$ of the qubit versus the frustration near $f_\varepsilon$ = 0.5. As $\alpha$ becomes smaller, $\Delta$ increases while the persistant current $I_p$ increases \cite{pauw_flux}.} 
    \label{fig:flux_qubit}
\end{figure}

The flux qubit was improved by replacing the single Josephson junction with three Josephson junctions. This is called the 3JJ persistent current qubit \cite{pauw_flux}. The purpose of the two junctions is to replace the loop inductance $L$ with the Josephson inductance $L_J$ of the two extra junctions seen in figure \ref{fig:flux_qubit}(a). The ratio $L/L_J$ is now,
\begin{equation}
    \alpha = \frac{2I_3}{I_1+I_2},
\end{equation}
where $I_i$ is the critical currents of each junction, which follow the current-phase relation $I_i = I_0\mathrm{sin}\varphi_i$, where $\varphi_i$ are the gauge-invariant phase of each junction $i$ \cite{orlando_flux}. In this architecture, the ratio of junction sizes is important as opposed to the inductance \cite{Birenbaum2014TheCF}.

With the new 3JJ formulation, it is important to derive the new system dynamics. Figure \ref{fig:flux_qubit}(a) shows an additional variable $f_{\varepsilon}$, called the magnetic frustration, which is simply the magnetic flux measured in units of the flux quantum $\Phi_0$. Two additional parameters of relevance are, $E_J = \Phi_0 I_c/2\pi$ and the charging energy $E_c = (2e)^2/2C$ \cite{pauw_flux}. The energy ratio $E_J/E_c$, determines whether the eigenstates are defined by states of definite charge or flux (or a combination of the two). This ratio also determines whether qubits are more susceptible to charge or flux noise.

Following from \cite{orlando_flux}, the fluxoid quantization around the loop containing the junction is strictly $\varphi_1 - \varphi_2 +\varphi_3 = -2\pi f_{\varepsilon}$. Using the derived parameters, the total Josephson energy $U$ is $U = \sum_i E_{Ji}(1-\mathrm{cos}\varphi_i)$ where each junction has energy $E_{Ji}(1-\mathrm{cos}\varphi_i)$. Combining this with the flux quantization term, the total Josephson energy is,   
\begin{equation}\label{eq:flux_anharmonic}
    \frac{U}{E_J} = 2+\alpha-\mathrm{cos}\varphi_1 -\mathrm{cos}\varphi_2 - \alpha\mathrm{cos}(2\pi f_{\varepsilon}+\varphi_1-\varphi_2)
\end{equation}
Thus the Josephson energy is dependent on the two phases $\varphi_1$ and $\varphi_2$. These can be modified for a range of magnetic frustration $f_{\varepsilon}$. There exists two stable solutions around $f_{\varepsilon}=1/2$, defined by a region $[1/2 - f_c, 1/2+f_c]$, where $f_c$ is derived in Appendix \ref{sec:flux_stability}. The solutions have circulating currents of opposite direction and are degenerate at $f_{\varepsilon}=1/2$. 

\begin{figure}[t] 
\centering
    \includegraphics[width=0.85\textwidth]{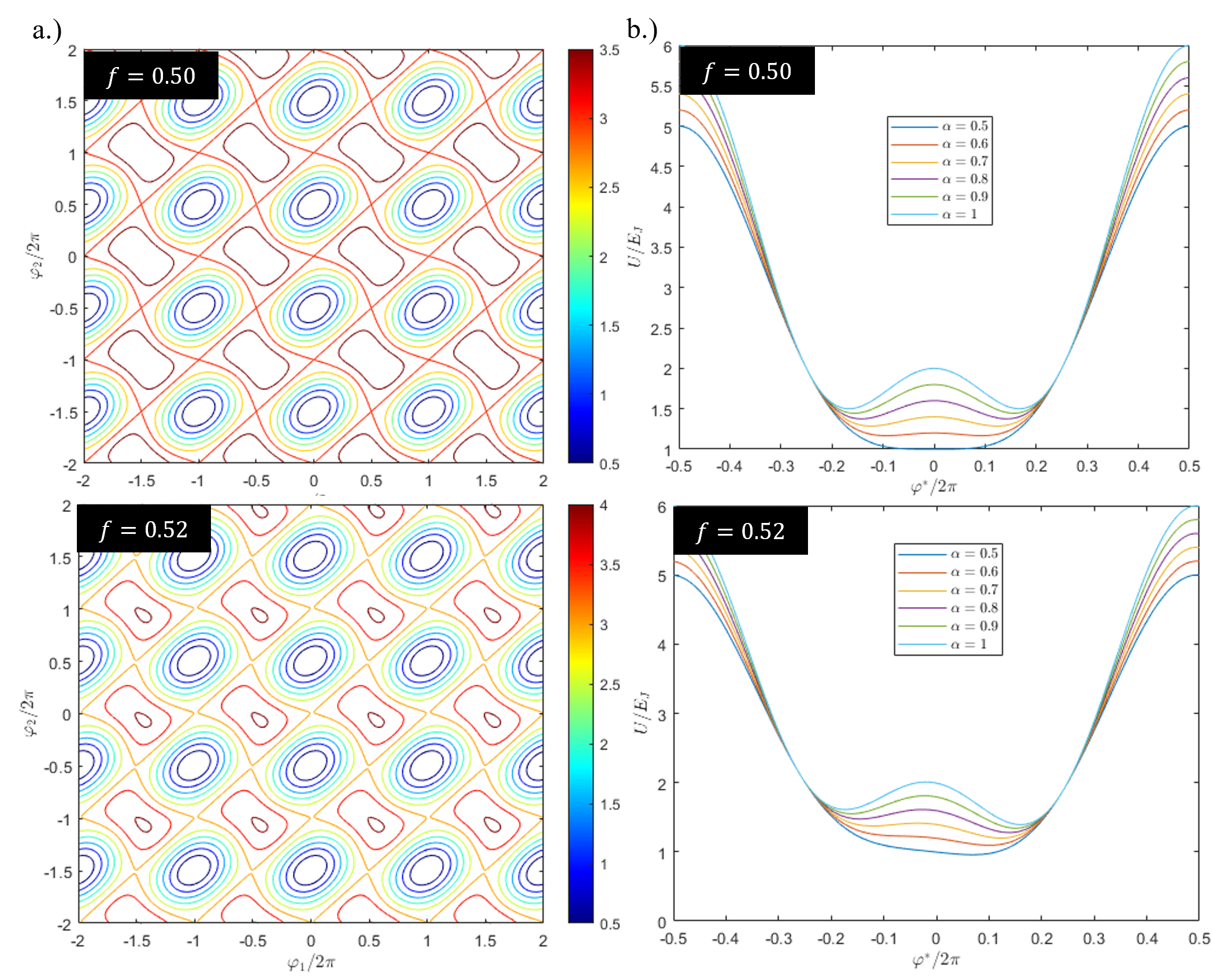}
    \caption{(a) Top: Phase diagram of $\varphi_1$ and $\varphi_2$ with magnetic frustration $f=0.50$ and $\alpha = 0.5$. The circular shapes are the maxima of the potential while the sponge like contours enclose two minima. Bottom: At a magnetic frustration $f = 0.52$, the sponge like contour lines becomes tilted wherein the right side potential is deeper. (b) Top: $U/E_J$ plotted against $\varphi^*$ where $\pm \varphi^* = \varphi_1 = -\varphi_2$ for $f = 0.50$. Bottom: $f = 0.52$.} 
    \label{fig:phase_f}
\end{figure}

It is now crucial to observe the relationship between $f_{\varepsilon}$ and $\alpha$ in order to make suitable qubit eigenstates. At $\alpha \leq 1/2$, $U$ has a single minimum when $\varphi_1 = \varphi_2 = 0$ however, above the critical $\alpha = 1/2$ value, the minimum bifurcates into two degenerate minima, $\pm \varphi^* = \varphi_1 = -\varphi_2$ \cite{orlando_flux,pauw_flux}. Displayed on Fig. \ref{fig:phase_f}(a) the contour of the Josephson energy can be seen as a function of the phase variables for $\alpha = 0.5$. The top graph shows the pattern when $f_{\varepsilon} = 1/2$ (depicted in the figure as $f$) whereby the sponge like pattern shows the stable energy potential minima while the circular patterns show the maxima. When the magnetic frustration diverges from $f_{\varepsilon} = 0.5$ point by a $\delta$, the minimum between both phases becomes biased. The bifurcated minima is seen in the top figure in Fig. \ref{fig:phase_f}(b) for different values of $\alpha$ (for $\alpha \geq 1/2$) while the bottom figure shows the bifurcation effect of a small perturbation ($\delta = 0.2$) in the magnetic frustration. The minimas no longer become degenerate as denoted by the bottom figure in Fig. \ref{fig:phase_f}(a), the potential energy deepens on the right side of the sponge.  

Tunneling through the barrier that separates the two potential minima provides a coupling $\Delta/2$ between the two persistent current states. The tunnel coupling consequently creates an anti-crossing $\Delta$ between the classical energy levels as seen in Fig. \ref{fig:flux_qubit}(b) due to the eigenstates of the Hamiltonian at $f_{\varepsilon}=1/2$ becoming symmetric and anti-symmetric superpositions of the persistent current states \cite{pauw_flux}. To conclude, it has been observed that tuning $\alpha$ is important for varying the barrier height and $f_{\varepsilon}$ is used to control the symmetry between both states when designing a flux qubit. 

Having built upon the fundamental understanding of how $\alpha$ and the magnetic frustration changes the energy dynamics of the qubit, it is now important to derive the two-level Hamiltonian in order to model the system dynamics. In the two-level approximation the qubit Hamiltonian is described in terms of the magnetic energy $E_{LR}$ and the tunnel coupling $\Delta/2$ that creates an anti-crossing \cite{pauw_flux}. In the persistent-current basis ($\ket{L},\ket{R}$) the qubit Hamiltonian is,
\begin{equation}\label{eq:easy_two_level}
    H = -\frac{1}{2}(\varepsilon\sigma_z + \Delta \sigma_x),
\end{equation}
where $\varepsilon = 2E_{R} = 2I_p(f_{\varepsilon}-1/2)\Phi_0$, $I_p$ is the persistent current, and both $\sigma_x$ and $\sigma_z$ are the Pauli matrices \cite{pauw_flux}. The corresponding eigenenergies of the diagonalized Hamiltonian are $E_0,E_1 = \mp\frac{1}{2}\sqrt{\varepsilon^2+\Delta^2}$ where the energy level splitting in the computational basis, $\{0,1\}$, is $E_{01} = E_1 - E_0 =\sqrt{\varepsilon^2+\Delta^2}$.

Dynamically modifying the two level flux qubit system relies on using external magnetic fields. Revising the Bloch sphere (Fig. \ref{fig:bloch_sphere}), the external field is aligned along the $z$-axis. The angle $\theta$ for a flux qubit is determined by $\theta = \mathrm{arctan}(\Delta/\varepsilon)$. The natural precession of the qubit around the $z$-axis depends on the Larmor frequency $\omega_L = E/\hbar$ where $E$ is the qubit energy \cite{pauw_flux}. In order to induce transitions, an oscillating field matching the Larmor frequency has to be applied in the $xy$-plane. To do so, microwaves are usually applied in the lab by means of a microwave frequency magnetic flux in the $z$-direction. This is of the form, $\delta \Phi_{LR} = |\Phi_{LR}|\mathrm{sin}(\omega_{LR}t+\varphi_{LR})$, where $t$ is the time and $\omega_{LR}$ is the applied microwave frequency between the two flux states \cite{pauw_flux}. This modulates the energy bias of the qubit with $\delta \varepsilon_{LR} = 2I_p\delta\Phi_{LR}$. The Hamiltonian describing the two-level flux qubit is thus,
\begin{equation}\label{eq:two_level_flux_driving}
    H^{(01)}_{LR,\varepsilon} = -\frac{\delta \varepsilon_{LR}}{2}\frac{\Delta}{\sqrt{\varepsilon^2+\Delta^2}}
    \begin{pmatrix}
        \varepsilon/\Delta & 1 \\
        1 & -\varepsilon/\Delta 
    \end{pmatrix},
\end{equation}
where the off-diagonal terms determine the strength of driving between the two qubit states \cite{pauw_flux}.

This section has introduced the 3JJ qubit by analyzing the impact of $\alpha$ and $f_{\varepsilon}$, describing the difference between the flux states $\{L,R\}$ and the computational states $\{0,1\}$. The Hamiltonian seen in Eq. \eqref{eq:easy_two_level} with the two level system described in Eq. \eqref{eq:two_level_flux_driving} will be used in the next section to construct the flux qubit architecture. 

\subsection{Flux Qubit System}\label{sec:flux_system}

\begin{figure}[t] 
\centering
    \includegraphics[width=0.8\textwidth]{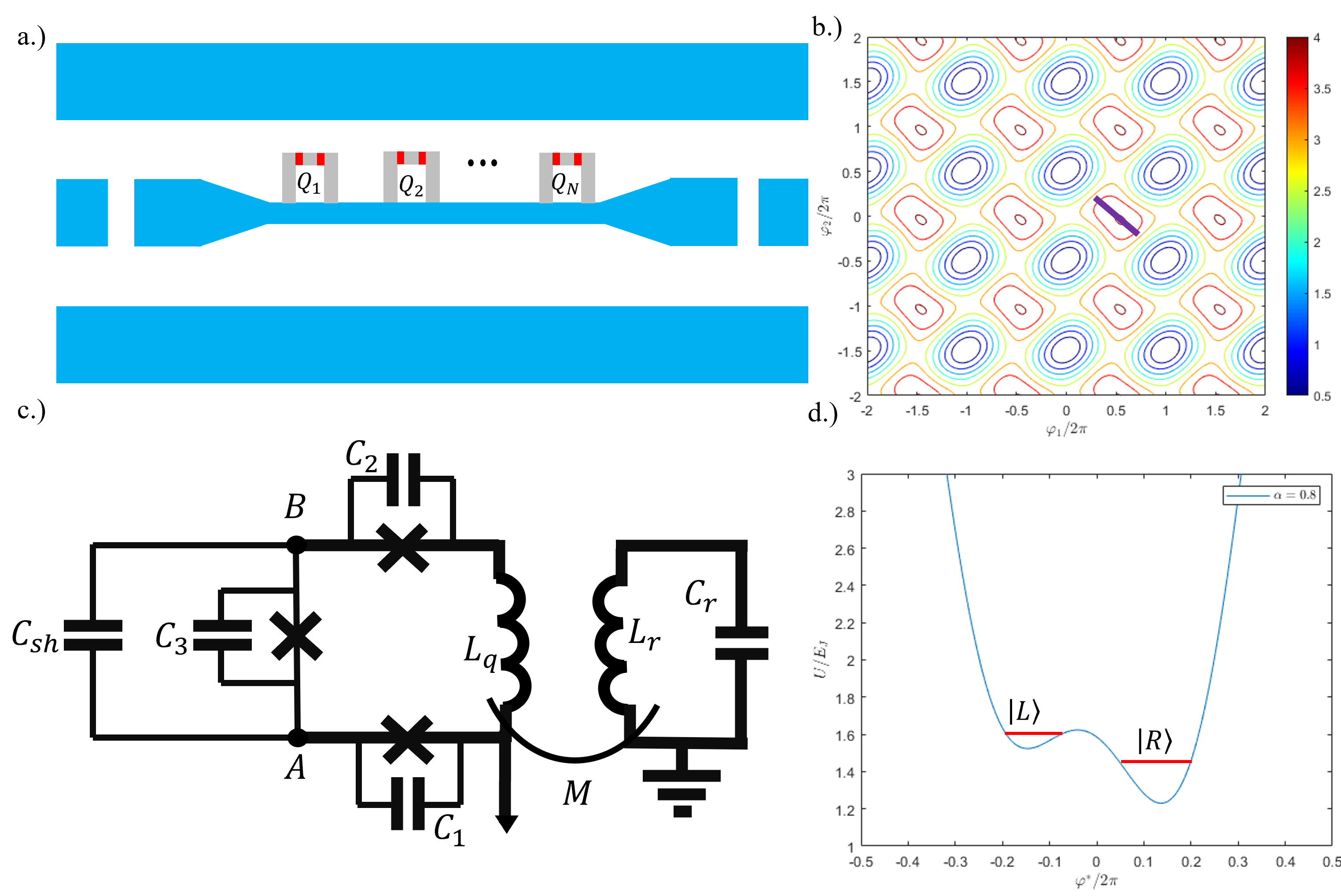}
    \caption{(a) Representation of Flux qubits (each qubit represented by $Q_i$) on a coplanar waveguide (in blue). (b) A contour plot of the potential energy $U/E_J$ at $f_\varepsilon = 0.53$ for $\alpha = 0.8$. The measured potential well is seen highlighted with the purple line. (c) A more specified outline of the C-shunt 3JJ flux qubit (circuit on the left) mutually coupled to a resonator, or in this case, the co-planar wave-guide (circuit on the right). (d) The potential function plotted against $\phi^* = \phi_1 = -\phi/2$ at $\alpha = 0.8$ and $f_\varepsilon = 0.53$ for the potential well seen in (c). The asymmetric bias gives rise to the $\ket{L}$ and $\ket{R}$ states.} 
    \label{fig:flux_system}
\end{figure}

Section \ref{sec:flux_qubit_intro} introduced the flux qubit mainly in the context of a single qubit. Flux qubits coupled together via transmission lines or resonators tend to possess an additional capacitor making them C-shunt flux qubits \cite{Birenbaum2014TheCF,Yan_2016}. This is due to the reduced noise susceptibility that is obtained when introducing an extra capacitor. The goal of this section will be to introduce the C-shunt qubit Hamiltonian, which follows that of Section \ref{sec:flux_qubit_noise} and derive the quasistatics of the system, which will be useful in deriving the noise fluctuations in Chapter \ref{chapter:open_system}. Figure \ref{fig:flux_system}(c) shows a circuit overview of the 3JJ C-shunt flux qubit coupled via a resonator to the CPW. This system of interacting flux qubits will be coupled together through an inductive coupling to the co-planar wave-guide (CPW). Figure \ref{fig:flux_system}(a) shows the flux qubits geometrically on a CPW. 

The two-level-system Hamiltonian for an individual C-shunt flux qubit near flux-degeneracy and coupled to a CPW resonator is,
\begin{equation}\label{eq:approximated_model}
    H = \frac{\hbar}{2}[\varepsilon\sigma_z + \Delta\sigma_x] + \hbar \omega(\hat{a}^\dag \hat{a} + \frac{1}{2}) + \hbar g\sigma_y(\hat{a}^\dag + \hat{a}),
\end{equation}
where the first part of the equation follows Eq. \ref{eq:easy_two_level} and $\hat{a}^\dag (\hat{a})$ is the raising (lowering) operator for photons. This Hamiltonian can be intuitively broken down into three different terms. The three terms are respectively, the qubit, resonator and qubit-resonator Hamiltonians \cite{Yamamoto_2014,Yan_2016}. The coupling strength between the resonator and qubit are dependent on $M$, $L_r$, $C_r$ and $L_q$ as seen in the full derivation, treated in Appendix \ref{sec:flux_resonator_der} \cite{nonlinear_koun,Harris_2007}.

The goal now is to derive the quasistatic Hamiltonian of the C-shunted flux qubit in order to understand the benefits of the C-shunt design and to then derive the noise in Section \ref{sec:flux_qubit_noise}. The C-shunt flux qubit has three features that differentiate it from the normal 3JJ flux qubit. It has a lower critical current, $I_c$, typically $\alpha < 0.5$ and the additional shunting junction has a capacitance $C_{sh} = \zeta C$, where $C$ is the capacitance of the smallest junction and $\zeta \gg 1$. As seen from Appendix \ref{sec:quantum_oscillator}, the 3JJ capacitively-shunted flux qubit can be described by a Hamiltonian with a kinetic and potential part,
\begin{flalign}\label{eq:kinetic_energy_flux}
    H & = T + U, \\
    T & = \frac{1}{2}(\boldsymbol{Q}+\boldsymbol{q})^\intercal \mathbf{C}^{-1}(\boldsymbol{Q}+\boldsymbol{q}), \\
    U & = E_J\{ 2 + \alpha - \mathrm{cos}\varphi_1 - \mathrm{cos}\varphi_2-\alpha\mathrm{cos}(2\pi f_\varepsilon + \varphi_1 - \varphi_2) \}. 
\end{flalign}
The charges $\boldsymbol{Q}$ and $\boldsymbol{q}$ are the charges induced charges on the islands, where the islands refer to the two nodes $A$ and $B$ in figure \ref{fig:flux_system}(c) \cite{Yan_2016}. The matrices can be rewritten to,
\begin{equation}
    \boldsymbol{Q} = -2e\begin{bmatrix}
        \frac{\partial}{\partial \varphi_1} \\
        \frac{\partial}{\partial \varphi_2}
    \end{bmatrix},
    \quad
    \boldsymbol{q} = \begin{bmatrix}
        q_A \\
        q_B
    \end{bmatrix},
    \quad 
    \boldsymbol{C} = C\begin{bmatrix}
        \zeta + 1 + \alpha & -(\zeta + \alpha) \\
        -(\zeta + \alpha) & \zeta + 1 + \alpha
    \end{bmatrix}.
\end{equation}
The system is less sensitive to charge fluctuations due to the shunt capacitor reducing the effective charging energy. Choosing $\varphi_+ = (\varphi_1 + \varphi_2)/2$ and $\varphi_- = (\varphi_1 - \varphi_2)/2$ using the Cooper-pair number operators $\hat{n}_{\sigma} = -i\partial/\partial\varphi_i$ where $i = \{-,+\}$, the reduced Hamiltonian becomes,
\begin{equation}
    H = \frac{1}{2}E_{C,+}\hat{n}^2_+ + \frac{1}{2}E_{C,-}\hat{n}^2_- + E_J\{2+\alpha-2\mathrm{cos}\varphi_-\mathrm{cos}\varphi_+ -\alpha\mathrm{cos}(2\pi f_\varepsilon + 2\alpha_-)\},
\end{equation}
where $E_{C,+} = 2e^2/C$ and $E_{C,-} = (\zeta+\alpha+1)e^2/C$ are the effective charging energy for the $+$-mode and $-$-mode respectively. Ideally, the $+$-mode can be omitted since $\zeta \gg 1$. Thus, the simplified Hamiltonian is \cite{Yan_2016},
\begin{equation}
    H_m = \frac{1}{2}E_{C,-}\hat{n}^2_- + E_J\{-2\mathrm{cos}\varphi_- -\alpha\mathrm{cos}(2\pi f_\varepsilon + 2\alpha_-)\}.
\end{equation}
This will be used in Section \ref{sec:flux_qubit_noise} to derive the flux and charge noise.

To realize the full potential of a C-shunt flux qubit, one would need to choose an $\alpha<0.5$ because it reduces the circuits sensitivity to charge noise and stray fields \cite{Birenbaum2014TheCF,UltrastrongCapacitiveCoupling}. However, in this regime, the qubit resembles a phase qubit where the potential well becomes singular and thus loses anharmonicity between two energy states. In order to realize the hybrid $E_2$ gate, described in Section \ref{sec:hybrid_comp}, each flux qubit requires two potential wells with states $\ket{L}$ and $\ket{R}$ in each \cite{Xiang_2013}. Additionally, the potential wells must have differing depths as seen on figure \ref{fig:flux_system}(b) and \ref{fig:flux_system}(d). Maintaining the anharmonicity between the two levels allows there to be a very strong and a very weak coupling to a resonator depending on the energy level. It is important that the flux qubit system only couples to the Rydberg system when the resonator is in the $\ket{1}$ state, or when the system needs to communicate. Hence, there is a trade-off between noise and the coupling to the hybrid system \cite{Xiang_2013}. Thus, the chosen values for the flux system are $\alpha = 0.8$ and $f_\varepsilon = 0.53$ as seen in figures \ref{fig:flux_system}(b) and \ref{fig:flux_system}(d).

\section{Hybrid Quantum Computer Architecture}\label{sec:hybrid_comp}

Coupling an atomic qubit to a superconducting qubit as of yet remains a theoretical area of study. The current model relies on trapping atoms on a superconducting atom chip \cite{kaiser2021cavity,atom_microwave,Wallquist_2009} or placing the atoms close to a superconducting chip and using a mediator to couple both the superconducting qubits and Rydberg atoms \cite{Yu_resonator_atom_2016,Yu_charge_qubit_2016,Yu_atom_flux_2017}. This mediator is often chosen to be a co-planar wave-guide (CPW) or an LC resonator due to their simplicity as well as their ability to couple many different frequencies \cite{Xiang_2013}. 

This study will mainly focus on a simplistic model whereby a chain of flux qubits are connected to a LC resonator on one side while a chain of Rydberg atoms are connected to the same LC resonator on the other side inspired by \cite{Yu_resonator_atom_2016}. The system dynamics for the atom-resonator-flux qubit, albeit similar to the Hamiltonians described in Sections \ref{sec:rydberg_atom_system} and \ref{sec:flux_qubit_intro}, will use a much simpler model to describe the system dynamics. This is due to the fact that the time scale used to realize the $E_2$ gate is much smaller than many of the modelled effects in previous Hamiltonians \cite{Yu_atom_flux_2017}. This system has been extensively studied in \cite{Yu_atom_flux_2017} and will be used in this research to establish the $E_2$ gate. 

\begin{figure}[t] 
\centering
    \includegraphics[width=0.85\textwidth]{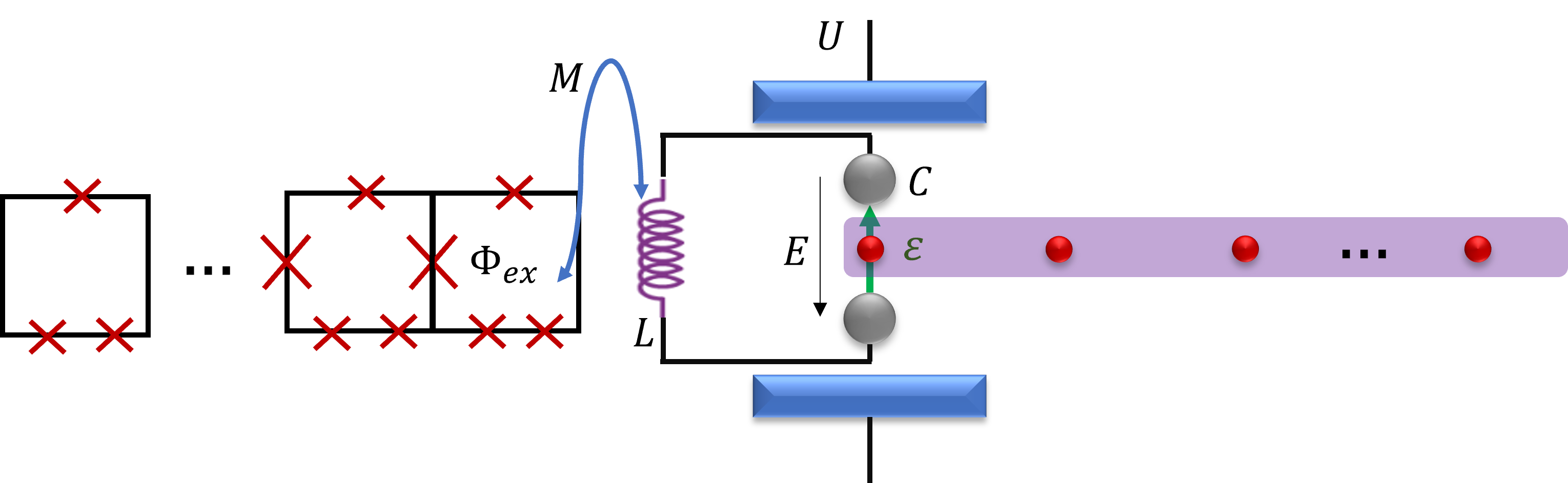}
    \caption{Hybrid system composed of a chain of three-JJ flux qubits, an LC resonator and an atom. The flux qubits have a tunable coupling mitigated by Josephson coupling while the Rydberg atom is Rydberg coupled to a chain of Rydberg atoms. The flux qubit, biased by an external flux $\Phi_{ex}$ is inductively coupled to the resonator with a mutual inductance $M$. The atom is placed in the middle point between two spheres with an oscillating intraresonator electric field with amplitude $\varepsilon$ in the $z$-axis. An additional electrostatic field $E$ runs across the Rydberg qubit in the $z$-axis generated by the parallel-plate capacitors. This model is inspired by \cite{Yu_resonator_atom_2016}.} 
    \label{fig:hybrid_model}
\end{figure}

A diagram of the model can be seen in Fig. \ref{fig:hybrid_model}. The mediator of the model is a superconducting LC resonator operating at mK temperatures. The characteristic frequency of such resonator is $\omega_0 = 1/\sqrt{LC}$ which is chosen to be $2\pi \times 20$ GHz. The capacitor composed of spherical identical spheres has a capacitance $C$ and an inductance $L$. The remaining numerical values as well as their units can be seen in Table \ref{table:hybrid_structure}. The Hamiltonian for the resonator,
\begin{equation}
    H_{LC} = \frac{\phi^2}{2L} + \frac{q^2}{2C} = \hbar \omega_0 (b^\dag b + 1/2),
\end{equation}
where $b^\dag$ and $b$ are the creation and annihilation operators, the magnetic flux $\phi = \sqrt{\hbar/2C\omega_0}(b^\dag + b)$ and $q = i \sqrt{C\hbar\omega_0 /2}(b^\dag - b)$ is the charge operator \cite{Yu_charge_qubit_2016,Yu_atom_flux_2017}. The Rydberg atom is placed in the midpoint between the two spherical capacitors which couples to the local electric field in the $z$-direction resulting in the atom-resonator interaction operator \cite{Yu_resonator_atom_2016,Yu_atom_flux_2017},
\begin{equation}
    V_a = -i D \varepsilon(b^\dag - b).
\end{equation}
$D$ is the atomic dipole moment and $\varepsilon$ is the amplitude of the oscillating electric field inside the resonator. It is assumed that the inhomogeneity of $\varepsilon$ within the atomic wave-pocket is $|\frac{1}{\varepsilon}\frac{\partial \varepsilon}{\partial \alpha}r| < 10^{-3}$ where $\alpha$ are the different dimensions $\alpha = x,y,z$ and $r$ is the radius of the Rydberg state, consequently affecting the atom-resonator coupling \cite{Yu_atom_flux_2017}. Furthermore, the inhomegeneity of the external electric field, $E$, caused by the screening effect of the spheres is also negligible.  

\begin{figure}[h] 
\centering
    \includegraphics[width=0.8\textwidth]{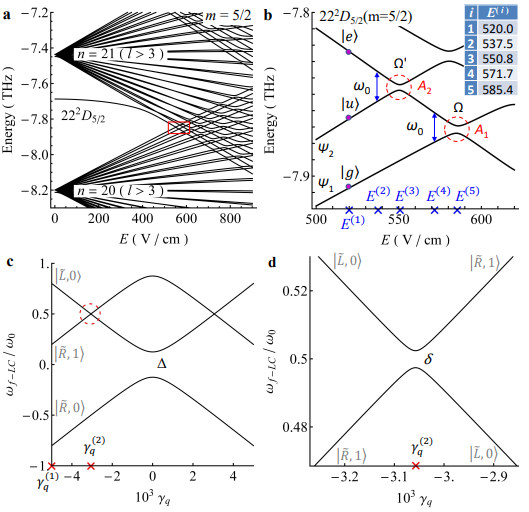}
    \caption{(a) Stark map of Rb around $22^2D_{5/2}$. (b) The atom qubit is formed by the ground state $\ket{g}$ and the excited state $\ket{e}$ at $E^{(1)} = 520.0$ V/cm on two adiabatic energy lines starting with the $22^2D_{5/2}$ and manifold state $\ket{\psi_1}$ composed of a set of $\ket{n=20, l \geq 3, j = l \pm \frac{1}{2}, m = \frac{5}{2}}$ at $E = 0$. The auxiliary state $\ket{u}$ is on another manifold state $\ket{\psi_2}$ yet on the same set as $\ket{n=20, l \geq 3, j = l\pm \frac{1}{2}, m = \frac{5}{2}}$. There are two avoided crossings, $A_1$ and $A_2$ with energy separations $\Omega$ and $\Omega^'$ respectively. (c) Energy spectrum ($\omega_{f-LC}$ of the flux qubit-resonator interface vs. the parameter $\gamma_q$. The interaction between the flux qubit anf the resonator is negligible at $\gamma_q^{(2)} = -5\times10^{-3}$. At $\gamma_q = 0$, an anticrossing exists with spacing $\Delta = 2\pi \times 5$ GHz. Another avoided crossing with spacing $\delta = 2\pi \times 0.1$ GHz induced by the resonant flux qubit-resonator coupling which happens at $\gamma_q^{(1)} = -3.06 \times 10^{-3}$. (d) Anticrossing seen with spacing $\delta$. This image was taken from \cite{Yu_atom_flux_2017}, authored by \citeauthor{Yu_atom_flux_2017}.} 
    \label{fig:hybrid_flux_atom}
\end{figure}

The Rydberg atom coupled to the resonator starts at the excited state $\ket{e}$ on $22^2D_{5/2}(m=5/2)$. The Stark map for $22^2D_{5/2}(m=5/2)$ can be seen in Fig. \ref{fig:hybrid_flux_atom}(a). When the electric field $E$ is around 500 V/cm, the excited state interacts with states in the set $\ket{n=20,l\geq3,j=l \pm \frac{1}{2}, m =\frac{5}{2}}$. These interacting manifold states can be seen interacting with the $\ket{e}$ in \ref{fig:hybrid_flux_atom}(b) for different values of $E$. The anticrossings with manifolds $\psi_2$ and $\psi_1$ can be seen for different values of $E$. The states on the manifolds include an auxiliary state $\ket{u}$ and a ground state $\ket{g}$ with anticrossings labelled A1 and A2. The energy separations are denoted as $\Omega$ and $\Omega^'$. Through adiabatic tuning, one can vary through the states $\ket{\mu = e,g,u}$. The Hamiltonian of the atom is given by,
\begin{equation}
    H_a = \sum_{\mu = e,g,u} \hbar \omega_\mu \ket{\mu}\bra{\mu} +  \frac{\hbar\Omega}{2}(\ket{e}\bra{g} + \ket{g}\bra{e}) + \frac{\hbar\Omega^'}{2}(\ket{e}\bra{u} + \ket{u}\bra{e}),
\end{equation}
where electric-field-dependent energies $\omega_{\mu=e,g,u}$ of atomic states are derived as $\omega_e = -7.81 - (E-500) \times 7.3 \time 10^{-4}$ THz, $\omega_g = -7.92 - (E-500) \times 5.5 \time 10^{-4}$ THz and $\omega_e = -7.88 - (E-500) \times 6.3 \time 10^{-4}$ THz \cite{Yu_atom_flux_2017}. The atom-resonator interaction is expressed as,
\begin{equation}
    V_a = \frac{\hbar g_a}{2}(b^\dag\ket{g}\bra{e} + \ket{e}\bra{g}b) + \frac{\hbar g^'_a}{2}(b^\dag\ket{u}\bra{e} + \ket{e}\bra{u}b),
\end{equation}
with coupling strengths $g_a = |\bra{e}D\ket{g}|\varepsilon/\hbar$ and $|\bra{e}D\ket{u}|\varepsilon/\hbar = 2\pi \times 0.5$ GHz. The atom resonantly interacts with the resonator at $E^{(2)}$ $(\omega_e - \omega_u = \omega_0)$ and at $E^{(4)}$ $(\omega_e - \omega_g = \omega_0)$ \cite{Yu_resonator_atom_2016, Yu_atom_flux_2017}.  

The three-JJ flux qubit is coupled to the LC resonator through means of mutual induction with inductance $M$. $M$ is chosen to be equal to $M = 27 pH$ to obtain a flux strong enough for coupling, furthermore, the flux qubit is biased by an external magnetic flux $\Phi$ to tune the frequency spacing $\varepsilon = 2I_p\Phi_0\gamma_q/\hbar$ with phase bias parameter $\gamma_q = \Phi/\Phi_0 -1/2$ (values introduced in Section \ref{sec:intro_flux}) \cite{Chiorescu_2003,Yu_atom_flux_2017}. The qubit is tuned between the states $\ket{R}$ and $\ket{L}$. Thus, the Hamiltonian for the flux-qubit is similar to Eq. \eqref{eq:easy_two_level} \cite{Yu_charge_qubit_2016,Yu_atom_flux_2017},
\begin{equation}
    H_f = -\frac{\hbar \varepsilon}{2}\sigma_{f,z} - \frac{\hbar \Delta}{2} \sigma_{f,x},
\end{equation}
but with the definitions of the Pauli matrices being, $\sigma_{f,z} = \ket{L}\bra{L} - \ket{R}\bra{R}$, $\sigma_{f,x} = \sigma^\dag_{f,-}+\sigma_{f,-}$ and $\sigma_{f,-} = \ket{R}\bra{L}$. The tunnel splitting is denoted as $\Delta$  with the flux qubit-resonator interaction potential given by,
\begin{equation}
    V_f = -\hbar g_f(b^\dag+b)\sigma_{f,z},
\end{equation}
with the coupling strength given by $g_f = (MI_p/\hbar)\sqrt{\hbar\omega_0/(2L)}$ \cite{Yu_atom_flux_2017}. Fig. \ref{fig:hybrid_flux_atom}(c) shows the varying $\gamma_q$ against the energy spectrum of the flux qubit-resonator interface \cite{Yu_atom_flux_2017},
\begin{equation}
    (H_{LC} + H_f + V_f)\ket{\psi_{f-LC}} = \hbar \omega_{f-LC} \ket{\psi_{f-LC}},
\end{equation}
where $\omega_{f-LC}$ is the eigenvalue and the eigenstate $\ket{\psi_{f-LC}}$ is spanned in the basis $\{\ket{o,n_p},o=L,;n_p=0,1,2,...\}$ where $n_p$ is microwave photon number. In the Hybrid system, the microwave photon number is between $0,1$. The energy separation at the avoided crossing between the states $\ket{R,1}$ and $\ket{L,0}$ at $\gamma_q^{(2)}$ is $\delta = \frac{2V_f\Delta}{\hbar\sqrt{\varepsilon^2+\Delta^2}}$ (in Fig. \ref{fig:hybrid_flux_atom}(d)).

The final system Hamiltonian is,
\begin{equation}\label{eq:final_hamil_hybrid}
    H = H_{LC} + H_a + H_f + V_a + V_f.
\end{equation}
This will be used to model the formation of the GHZ state used in the $E_2$ gate in Section \ref{sec:GHZ_hybrid}. 

\section{Forming the Non-Local GHZ State}\label{sec:GHZ_hybrid}

This section underlines the steps necessary to create the GHZ state. The GHZ state is prepared using the methodology seen in \cite{Yu_atom_flux_2017}. The GHZ state describing the flux-resonator-atom is,
\begin{equation}
    \ket{GHZ} = \frac{1}{\sqrt{2}}(\ket{L,1,e}+\ket{R,0,g}).
\end{equation}

\begin{figure}[t] 
\centering
    \includegraphics[width=0.75\textwidth]{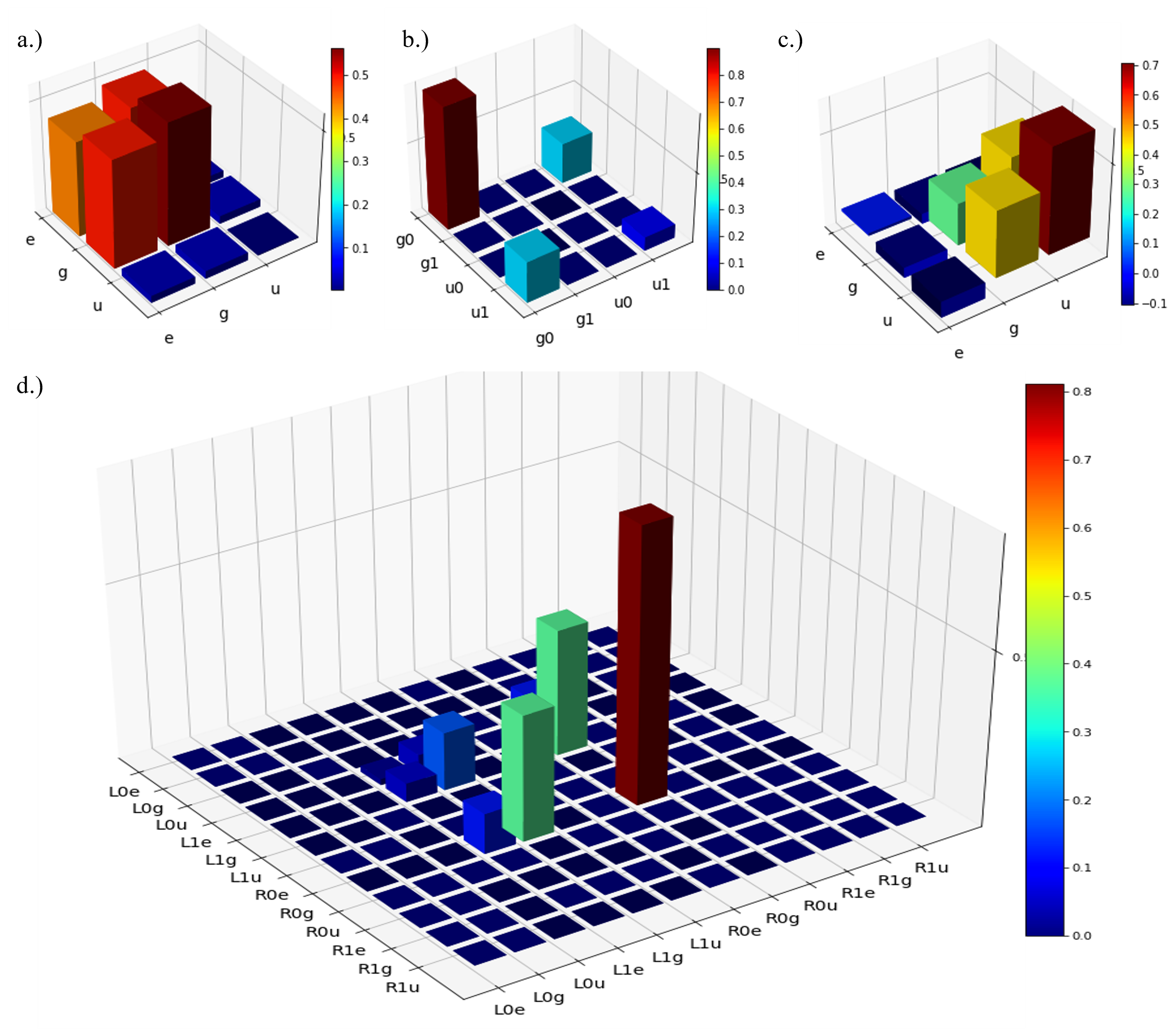}
    \caption{The probability (denoted as a color) of the density matrix of the specific systems $\rho = \ket{\psi}\bra{\psi}$ at different steps of the GHZ implementation. (a) Density state of the atom system after increasing $E$ from $E^{(1)}$ to $E^{(5)}$ where the system transitions to the  $\frac{1}{\sqrt{2}}(\ket{R,0,e}+\ket{R,0,g})$ with 0.99 fidelity. (b) Density state of the atom-resonator system after reducing $E$ from $E^{(5)}$ to $E^{(2)}$ nonadiabatically. After this step, the system is at $\frac{1}{\sqrt{2}}(\ket{R,1,u}+\ket{R,0,g})$ with 0.95 fidelity. (c) Density state of atom system after $E$ is increased to $E^{(3)}$ quickly wherein the final system state is $\frac{1}{\sqrt{2}}(\ket{L,0,e}+\ket{R,0,g})$ with 0.97 fidelity. (d) The final GHZ density state of the system with 0.93 fidelity.}
    \label{fig:GHZ}
\end{figure}

The methodology used to form the $\ket{GHZ}$ in \cite{Yu_atom_flux_2017} follows a five step process. Initially, the electric field and the phase bias parameter are set to $E^{(1)}$ and $\gamma_q^{(1)}$, respectively. Here, both the atom and the flux qubit are far-off-resonantly coupled to the resonator, thus the hybrid system is in the $\ket{R,0,e}$ state. the external electric field, $E$, is increased rapidly to $E^{(5)}$ for a time duration of $\pi/(2\Omega)$ which allows the atom to transit to the $(\ket{e}+\ket{g})/\sqrt{2}$. The density matrix of the result of the simulated transition is visualized in Fig. \ref{fig:GHZ}(a). The graph indicates the amplitude (probability) of the density matrix ($\rho = \ket{\psi}\bra{\psi}$) where the matrix elements are denoted on the axis. The trace of the density matrix $\bra{e}\rho\ket{e} + \bra{g}\rho\ket{g} + \bra{u}\rho\ket{u}$ is equal to 1, where $\bra{u}\rho\ket{u} \approx 0$, $\bra{e}\rho\ket{e} \approx 0.45$ and $\bra{g}\rho\ket{g} \approx 0.55$. The off-diagonal elements denoting the superposition state $(\ket{e}+\ket{g})/\sqrt{2}$, are seen to have a probability of 50\%. This density matrix corresponds to a fidelity of 0.99 relative to the ideal $(\ket{e}+\ket{g})/\sqrt{2}$ state. 

By switching on the resonant interaction between the microwave resonator and the atomic $\ket{e} - \ket{u}$ transition, $E$ is reduced to $E^{(2)}$ for a time of $\pi/g^'_a$ before going back to $E^{(1)}$ which evolves the $\ket{e}$ state to the auxiliary $\ket{u}$ state in order for the hybrid system to transition to state $\frac{1}{\sqrt{2}}(\ket{R,1,u}+\ket{R,0,g})$. The results of this transition is seen depicted in Fig. \ref{fig:GHZ}(b) where the atom and resonator state are combined, thus, $\frac{1}{\sqrt{2}}(\ket{1,u}+\ket{0,g})$. The diagonal terms show $\bra{g,0}\rho\ket{g,0} \approx 0.9$ and $\bra{u,1}\rho\ket{u,1} \approx 0.1$ while the off-diagonal terms show an approximate 30\% probability for the $\bra{u,1}\rho\ket{g,0}$ and $\bra{g,0}\rho\ket{u,1}$ states. This corresponds to a 0.95 fidelity relative to the $\frac{1}{\sqrt{2}}(\ket{R,1,u}+\ket{R,0,g})$ state.  

Moving to the flux qubit side, $\gamma_q$ is then switched to $\gamma_q^{(2)}$ to resonate with the resonator for a time of $\pi/\delta$ which completes the transfer from the resonator to the flux qubit wherein the flux qubit experiences a state flip. Thus, the system state is now at  $\frac{1}{\sqrt{2}}(\ket{L,0,u}+\ket{R,0,g})$. Then, $E$ is set to $E^{(3)}$ for a time length of $\pi/\Omega^'$ consequently evolving the $\ket{u}$ state to the $\ket{e}$. The resulting state is shown in Fig. \ref{fig:GHZ}(c), whereby only the atom system is shown. The diagonal terms show $\bra{u}\rho\ket{u} \approx 0.7$ and $\bra{g}\rho\ket{g} \approx 0.3$ while the off-diagonal terms show an approximate 40\% probability for the $\bra{u}\rho\ket{g}$ and $\bra{g}\rho\ket{u}$ states. This corresponds to a 0.95 fidelity relative to the $\frac{1}{\sqrt{2}}(\ket{L,0,u}+\ket{R,0,g})$ state.

Finally by repeating the nonadiabatic reduction to $E^{(2)}$ for a time length of $\pi/g^'_a$ and then tuning to $E^{(3)}$ for a time length of $\pi/\Omega^'$, the system reaches the GHZ state shown in Fig. \ref{fig:GHZ}(d). The diagonal terms show $\bra{L,1,e}\rho\ket{L,1,e} \approx 0.05$ and $\bra{R,0,g}\rho\ket{R,0,g} \approx 0.8$. It is noticeable that an extra state $\bra{L,1,g}\rho\ket{L,1,g}$ is measured to be around 15\%. The off-diagonal terms show an approximate 15\% probability for the $\bra{L,1,e}\rho\ket{R,0,g}$ and $\bra{R,0,g}\rho\ket{L,1,e}$ states. The states, $\bra{L,1,g}\rho\ket{R,0,g}$ and $\bra{R,0,g}\rho\ket{L,1,g}$, also show an approximate 35\% chance of being in a superposition. This corresponds to a 0.93 fidelity relative to the $\frac{1}{\sqrt{2}}(\ket{L,1,e}+\ket{R,0,g})$ state.

The GHZ state in \cite{Yu_atom_flux_2017} was reported to be completed in 8 ns with a fidelity 0.977. Due to differing methods of solving the master equation and using non-dynamic time steps, the optimized fidelity reached was 0.93 in approximately 17 ns which is still in accordance with the timescale in \cite{Yu_atom_flux_2017}. The code for the numerical simulation can be seen in Appendix \ref{sec:hybrid_code}. The numerical simulation begins by initializing the Hamiltonians of each system as well as the interactions seen in the functions: \codeword{rydberg_hamiltonian}, \codeword{rydberg_resonator_hamiltonian}, \codeword{flux_hamiltonian}, \codeword{flux_resonator_hamiltonian}, and \codeword{resonator_hamiltonian}. The function \codeword{solve_master_equation} takes these Hamiltonians and evolves them over a given time inputted by the user. The output of this function is an array of evolved states at each time step. An additional function, \codeword{find_optimal_time}, takes this array and finds the state with the highest fidelity. Thus, it evaluates the fidelity of the vector state at each time step with the ideal final state and finds the time step with the best fidelity \cite{Li_2022}. The five steps are defined as non-value returning functions, \codeword{step_1}, \codeword{step_2}, \codeword{step_3}, \codeword{step_4}, and \codeword{step_5}. The timescale in the range of nanoseconds is of importance due to the fact that the preparation time is much shorter than the decoherence time, which will be seen in Section \ref{sec:hybrid_noise}. From this, it means that it is indeed possible to establish a feasible $E_2$ gate between the Rydberg atom computer and the flux qubit computer. 

\clearpage

\chapter{Modeling Noise and Decoherence in Quantum Systems}\label{chapter:open_system}
In order to properly simulate the Hybrid system evolving through the quantum phase algorithm, the model requires noise models. The noise models will be dependent on each system, which, like Chapter \ref{chapter:hybrid_system}, will be split into three sections. Section \ref{sec:rydberg_noise} will derive the noise model encountered in the Rydberg atom computer. Section \ref{sec:flux_qubit_noise} will derive the noise fluctuations encountered in the flux qubit computer. Section \ref{sec:hybrid_noise} describes the noise from the flux-resonator-atom apparatus responsible for realizing the connection between both quantum computers, which follows the theory described in \cite{Yu_atom_flux_2017}. The goal of this chapter is to acquire the noise models of the systems for their implementation in the execution of the quantum phase estimation discussed in Chapter \ref{chapter:GRAPE_sim}.  
\section{Rydberg Atom Noise}\label{sec:rydberg_noise}
The Rydberg atom system encounters noise from two main sources, the laser-atom interaction and the energy level dissipation. The goal of this section will be to obtain the \textit{Lindbladians} describing the Rydberg noise. A Lindbladian describes open quantum systems by generalizing the Schr\"{o}dinger equation to a system that is in contact with its environment, which can be seen explained in greater detail in Appendix \ref{sec:open_system} \cite{Viamontes}, that will be used to simulate the Rydberg system dynamics. The Lindbladian describing the noise dynamics with a given initial density matrix, $\rho$, is seen below, 
\begin{equation}\label{eq:lindbladian_form}
    \mathcal{L}[\rho] = i\sum_k \gamma_k \bigg(A_k \rho A_k^\dag - \frac{1}{2} \Big\{ A_k^\dag A_k,\rho\Big \}\bigg),
\end{equation}
where $\{A_k\}$ are the \textit{Lindblad operators} representing the coupling of the system to the environment and thus the dissipative process \cite{joachain_1983,Distributed_Leung,lindblad_1976,nielsen_chuang_2021}. 

In order to derive the Linbaldian for the local phase noise, it is first necessary to derive the full Rydberg interaction Hamiltonian which includes the interaction of the atoms interacting with each other as well as the atoms interacting with the laser. For simplicity, it will be assumed that the atom contains two main energy states, the ground state $\ket{g}$ and the excited state $\ket{e}$. The distinction of the hyperfine states will be ignored because the noise and dissipation of both hyperfine levels are very similar. 

The laser field denoted by $ \mathcal{\boldsymbol{E}}(\boldsymbol{r},t)$, can be described as \cite{rydberg_noise},
\begin{equation}
    \mathcal{\boldsymbol{E}}(\boldsymbol{r},t) = \boldsymbol{E}_0\mathrm{cos}(\omega_Lt+\phi(t) - \boldsymbol{k}\cdot\boldsymbol{r}) = \frac{1}{2}\boldsymbol{\mathcal{E}}_0 \Big(e^{i(\omega_L t+\phi(t) - \boldsymbol{k}\cdot \boldsymbol{r}}) + e^{-i(\omega_L t+\phi(t)-\boldsymbol{k}\cdot\boldsymbol{r})} \Big).
\end{equation}
The amplitude of the field is denoted by $\boldsymbol{\mathcal{E}}_0$ where $t$ denotes the time, $\omega_L$ and $\phi$ are the field frequency and the phase noise, respectively. The Hamiltonian of the Rydberg is considered to have an atomic Hamiltonian whereby the atomic interactions are expressed and a light interaction Hamiltonian. This is of the form,
\begin{equation}\label{eq:general_hamiltonian}
    \mathcal{H} = H_{atom} + H_{int} = \hbar \omega_{eg}\sigma_{ee} + q\boldsymbol{r}\cdot\boldsymbol{E}(\boldsymbol{r},t),
\end{equation}
where $q$ is the electron charge, $q = -e$, and $\sigma_{ij} = \ket{i}\bra{j}$ where $i,j$ denote the energy states. The full interaction Hamiltonian can be obtained by applying the unitary transformation to the Hamiltonian described in Eq. \eqref{eq:general_hamiltonian},
\begin{equation}
    U = e^{-i\hbar\omega_{ge}t\sigma_{ee}}.
\end{equation}
Here, $\omega_{eg}$ is the frequency between both energy levels, $\omega_{eg} = (E_e - E_g)/\hbar$. This leads to, 
\begin{equation}
    \mathcal{H}^I = U^\dag\mathcal{H}U - \hbar \Delta \sigma_{ee}.
\end{equation}
After using the rotating wave approximation and cancelling out negligible terms, the equation becomes,
\begin{equation}\label{eq:manipulated_HI}
    \mathcal{H}^I = -\hbar \Delta\sigma_{ee} + \frac{\hbar \Omega}{2}\bigg(\sigma_{ge}e^{i(\phi(t)-\phi(t^'))}+\sigma_{eg}e^{-i(\phi(t)-\phi(t^'))} \bigg),
\end{equation}
where $\Omega = \boldsymbol{E}_0 \cdot \mu /\hbar$ describes the Rabi frequency where $\mu = e\bra{g}\boldsymbol{r}\ket{e}$ is the dipole matrix and $\Delta = \omega_L - \omega_{eg}$. The instantaneous change in phase noise is described as $e^{i\phi(t-t^')}$. The instantaneous change of phase noise is the so-called frequency noise which corresponds to the energetic shifts of the Rydberg state and is characterized as~\cite{rydberg_noise},
\begin{equation}
    \varepsilon(t) = \dot{\phi} = \frac{\mathrm{d}\phi}{\mathrm{d}t}.
\end{equation}
A final transformation $U = e^{-i\hbar \varepsilon(t)\sigma_{ee}}$ can be applied to Eq. \eqref{eq:manipulated_HI} to remove the time-dependence of the frequency noise in the exponent \cite{one_atom_maser} leaving,
\begin{equation}
     \mathcal{H}^I = -\hbar\Delta\sigma_{ee} - \hbar\varepsilon(t)\sigma_{ee}+ \frac{\hbar\Omega}{2}(\sigma_{ge}+ \sigma_{eg}).
\end{equation}
The full derivation of the interaction Hamiltonian can be seen in Appendix \ref{sec:light_interaction} ~\cite{open_quantum_theory_book,non_markovian_ines,Patsch2022}. In a many atom system, the final Hamiltonian can be formulated to be similar to that seen in Eq. \eqref{eq:pulser_hamiltonian} with an extra noise term,
\begin{equation}\label{eq:total_H_sys}
    \mathcal{H}^I= \frac{\Omega}{2}\sum_i \Big(\sigma_{eg}^{(i)}+\sigma^{(i)}_{ge}\Big) + V_0\sum_{i<j}\frac{\sigma_{ee}^{(i)}\sigma_{ee}^{(j)}}{|i-j|^6} - \sum_i \varepsilon^{(i)}(t)\sigma_{ee}^{(i)},
\end{equation}
where $\hbar = 1$ and the interatomic distance between two atoms is defined as $r_{ij} \equiv |i-j|$. It is important to note that the interaction is possible when $\omega_{eg} = \omega_L$ thus, $\Delta = 0$. 

The frequency noise $\varepsilon(t)$ is dependent on the phase noise $\phi(t)$ which causes fast and slow fluctuations in the laser electric field with a correlation time $\tau_c$ \cite{rydberg_noise}. The phase noise acts as a time-dependent detuning and shares many similarities with Brownian motion which is described by the Langevin equation \cite{one_atom_maser},
\begin{equation}
    \ddot{\phi} = -\gamma\dot{\phi} + F(t),
\end{equation}
where $\gamma$ is the inverse correlation time $\tau_c = 1/\gamma$ and $F(t)$ is a Gaussian distribution denoting a rapidly fluctuating force with zero ensemble average $\overline{F(t)} = 0$ and $\overline{F(t)^2} \neq 0$. It is assumed that $F(t)$ has a very short correlation time compared to the other characteristic time scales of the system, and thus approximately \cite{rydberg_noise,laser_physics},
\begin{equation}
    \overline{F(t)F(t^')} = 2D\delta(t-t^'),
\end{equation}
where $D$ is the magnitude of the fluctuations and $\delta(t-t^')$ is the the Dirac delta function. Together with $\gamma$, $D$ characterizes the spectral width $\Gamma$ of the Lorentzian line shape \cite{one_atom_maser},
\begin{equation}\label{eq:gamma_D}
    \Gamma = \frac{2D}{\gamma^2}.
\end{equation}
The spectral line width, $\Gamma$ is often referred to as the dephasing rate. To dynamically model the dephasing rate using the system Hamiltonian derived in Eq. \eqref{eq:total_H_sys}, one must model the Langevin as a Lindbladian (Eq. \eqref{eq:lindbladian_form}). This follows from the derivation of the Born-Markov master equation \cite{open_quantum_theory_book},
\begin{equation}\label{eq:pre-derived_open}
    \hat{\dot{\rho}} = - \int^t_{-\infty}\mathrm{d}t^'\mathrm{tr}_B \Big[\sum_i \varepsilon^{(i)}(t)\sigma_{ee}^{(i)},\Big[\sum_i \varepsilon^{(i)}(t)\sigma_{ee}^{(i)},\hat{\rho}(t) \Big] \Big],
\end{equation}
where $\mathrm{tr}_B[x]$ is equivalent to the average $\langle x \rangle$. Expanding equation Eq. \eqref{eq:pre-derived_open} gives,
\begin{dmath}
    \hat{\dot{\rho}} = - \int^t_{-\infty}\mathrm{d}t' \sum_{i,j}\mathrm{Tr}_B\Big\{\varepsilon^{(i)}(t)\varepsilon^{(j)}(t')\sigma_{ee}^{(i)}\sigma_{ee}^{(j)}\hat{\rho}+h.c. - \varepsilon^{(i)}(t)\sigma_{ee}^{(i)}\hat{\rho}\varepsilon^{(j)}+h.c. \Big\} = -\sum_{i,j}\int^t_{-\infty}\mathrm{d}t'\mathrm{Tr}_B\Big\{\varepsilon^{(i)}(t)\varepsilon^{(j)}(t^')\Big\}\Big[ \sigma_{ee}^{(i)}\hat{\rho}\sigma_{ee}^{(j)} - \frac{1}{2}\{\sigma_{ee}^{(i)}\sigma_{ee}^{(j)},\hat{\rho} \}\Big] = \mathcal{L}[\hat{\rho}].
\end{dmath}
A local laser is used when addressing qubits in the quantum systems since, as observed in section \ref{sec:rydberg_atom_system}, individual qubits are addressed per operation. Thus, for the local noise, the correlation between the phase noise experienced by atoms $i$ and $j$ at two different times is $\langle \varepsilon^{(i)}(t)\varepsilon^{(j)}(t') \rangle = (\Gamma/2)\delta_{ij}\delta(t-t')$ thus finding the local phase noise superoperator \cite{rydberg_noise, electromagnetic_gamma},
\begin{equation}\label{eq:rydberg_lindblad_1}
    \mathcal{L}[\hat{\rho}]^{l} = \Gamma \sum_{i}\bigg[\sigma_{ee}^{(i)}\hat{\rho}\sigma_{ee}^{(i)} - \frac{1}{2}\{\sigma_{ee}^{(i)}\sigma_{ee}^{(i)},\hat{\rho}\}\bigg].
\end{equation}

The final decoherence effect from the system comes from energy decay or the rate at which the population from an excited state $\ket{e}$ is transferred to a lower lying level $\ket{g}$ via photon emission into the vacuum. Spontaneous emission can be derived through Weisskopf-Wigner theory assuming complete coupling with closely spaced cavity modes with the emission spectrum centered at the atomic transition frequency. Weisskopf-Wigner theory also highlights that phenomenologically, the excited state is only capable of emitting to the ground state but not vice versa \cite{scully_zubairy_1997}. The final superoperator that describes the spontaneous decay from the state $\ket{e}$ reads,
\begin{equation}\label{eq:rydberg_lindblad_2}
    \mathcal{L}[\hat{\rho}]^{e} = \gamma_e\sum_i\Big[\sigma_{ge}^{(i)}\hat{\rho}\sigma_{eg}^{(i)} - \frac{1}{2}\{\sigma_{eg}^{(i)}\sigma_{eg}^{(i)},\hat{\rho}  \} \Big],
\end{equation}
where $\gamma_e$ is the spontaneous emission rate from the excited state. Using ARC \cite{ibali__2017}, the spontaneous decay time for the $\ket{e} = \ket{70S_{1/2}}$ is $375 \mu$s. The Linbladians observed in Eq. \eqref{eq:rydberg_lindblad_1} and \eqref{eq:rydberg_lindblad_2} will be used to model the noise. The implementation will be discussed in Chapter \ref{chapter:GRAPE_sim}. 
\section{Flux Qubit Noise}\label{sec:flux_qubit_noise}
\begin{figure}[t] 
\centering
    \includegraphics[width=0.7\textwidth]{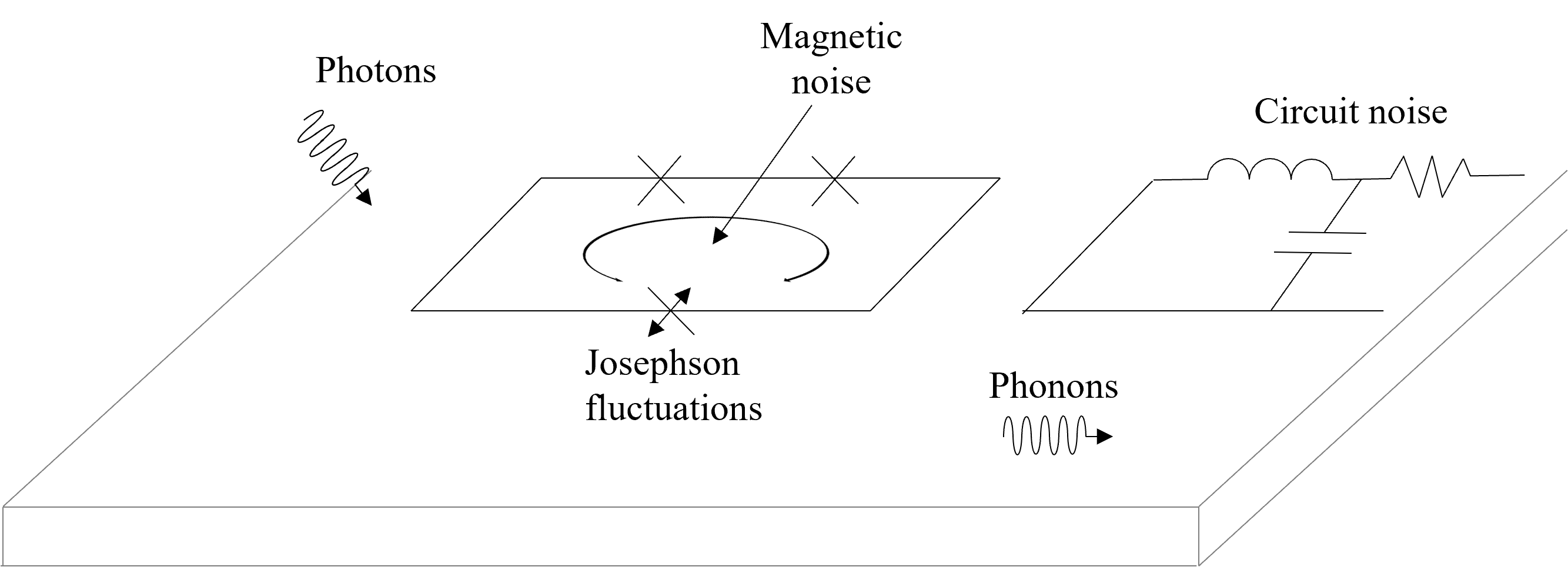}
    \caption{Potential noise sources that can arrise in superconducting qubits. Inspired by \cite{superconducting_plantenberg}.}
    \label{fig:flux_qubit_noise}
\end{figure}

The goal of this section is to describe the flux qubit noise in Linbladian form. Flux qubit noise arises from the general noise in a superconducting circuit which can be seen in figure \ref{fig:flux_qubit_noise}. There exists two categories. The first category of noise is the noise generated from effects on the circuit such as decoherence from qubit coupling effects \cite{pauw_flux}. The second group of noise is the non-circuit related noise such as fluctuations in the magnetic field or the Josephson junction itself.

The noises generated from the circuit (often from defects in the dielectrics) are typically said to cause charge fluctuations, while the noise generated by the magnetic field and the stray fields originating from the Josephson junctions are said to cause flux fluctuations. Flux fluctuations can be described by introducing a perturbation around the magnetic frustration $(\delta f)$ to the anharmonic system described in Eq. \eqref{eq:flux_anharmonic}. The Hamiltonian of the flux fluctuations is thus a first order perturbation of the potential energy in terms of the magnetic frustration. Using the definition for $f_\varepsilon$ and $\varphi_-$ seen in Section \ref{sec:flux_qubit_intro}, as well as, Eq. \eqref{eq:flux_anharmonic}, the flux fluctuation Hamiltonian can be analytically modelled as,
\begin{dmath}
    \delta\mathcal{H} = \delta U = -2\pi\alpha E_J\mathrm{sin}(2\pi f_\varepsilon+2\varphi_-)\delta f = 
    -2\pi\alpha E_J (\mathrm{sin}\phi\mathrm{cos}(2\varphi_{-'})+\mathrm{cos}\phi\mathrm{sin}(2\varphi_{-'}))\delta f \approx -2\pi\alpha E_J(\mathrm{sin}\phi (1-2\varphi_z^2(\hat{a} + \hat{a}^\dag))\delta f,
\end{dmath}
where $\varphi_{-'} = \varphi_- \varphi_-^*$, where $\varphi_-^*$ is the $\varphi$ corresponding to the minimum of the potential well where $\ket{L}$ resides, seen in figure \ref{fig:flux_system}(d), and $\varphi_z = (E_{C,-}/4E_{J,-})^{1/4}$. Furthermore, $E_{C,-}$ is the effective charging energy for the negative minimum and $E_{J,-}$ is the effective Josephson energy in the negative minimum and $\phi = \phi(f_\varepsilon) = 2\pi f_\varepsilon + 2\varphi_-^*(f_\varepsilon)$.

Fluctuations near the frequency $\Omega$ are considered because they have the possibility of inducing transitions between the $\ket{L}$ and $\ket{R}$ states. Thus, in the case $\delta f \propto \mathrm{cos}(\Omega t)$, the system simplifies to two levels,
\begin{equation} \label{eq: basic_hamil_flux}
    \mathcal{H}= \frac{1}{2}(\hbar\omega_q\sigma_z+I_m \Phi_0\delta f\sigma_x)
\end{equation}
where $\omega_q = \omega_q(f_\varepsilon) = \Omega$ and $I_m$, the $f_\varepsilon$-dependent current difference between the parametrised circulating-current states, is defined as $I_m = I_m(f_\varepsilon) = -8\pi\alpha\sigma_z\mathrm{cos}\phi E_J/\Phi_0$, where the parameters are described in Section \ref{sec:flux_system}.

The next step is to derive the charge fluctuation noise on the Hamiltonian. Investigating the charge fluctuations ($\delta q_a$,$\delta q_b$) requires invoking a perturbation via the kinetic energy $T$. Perturbing the kinetic energy given by Eq. \eqref{eq:kinetic_energy_flux} yields,
\begin{equation}
    \delta T = \delta \boldsymbol{q}^\intercal\boldsymbol{C}^{-1}\boldsymbol{Q} = -\frac{e}{C(2\zeta+2\alpha+1)} [-in_z(\hat{a}-\hat{a}^\dag)](\delta q_A - \delta q_B),
\end{equation}
where $n_z = (E_{J,-}/4E_{C,-})^{1/4}$ is the quantum ground-state uncertainty in Cooper-pair number and $\zeta$ is called the shunt factor which relates the capacitance of the smaller junction ($C_3$ seen in figure \ref{fig:flux_system}(c)) to the capacitance of the shunt capacitor through $C_{sh}=\zeta C_3$ \cite{Yan_2016}. This factor is considered to satisfy $\zeta \gg 1$. The charge noise is orthogonal to the flux noise, thus it couples to the Hamiltonian in the $\sigma_y$ basis. The perturbation depends only on the differential mode of the induced charges between the two islands (described by nodes $A$ and $B$ seen in figure \ref{fig:flux_system}(c)) \cite{Yan_2016}. Thus, placing the charge fluctuation back into Eq. \eqref{eq: basic_hamil_flux}, gives,
\begin{equation}\label{eq:flux_fluctuation_final}
    \mathcal{H} = \frac{1}{2}(\hbar \omega_q \sigma_z + I_m \Phi_0 \delta f \sigma_x + n_z E_{C,-}\delta n_- \sigma_y),
\end{equation}
where $\delta n_- = (\delta q_A - \delta q_B)/(-e)$ is the differential electron number fluctuation \cite{Yan_2016}. 

The importance of the C-shunted flux design is observed in the noise sensitivity. Because of this design, the charge noise sensitivity diminishes with larger $\zeta$ due to the relation, $n_z E_{C,-} \propto E^{3/4}_{C,-} \propto \zeta^{-3/4}$. Each one of these fluctuation terms can be applied as a Linbladian to the Hamiltonian seen in Eq. \eqref{eq:approximated_model}. The Lindbladians modelling the noise due to the fluctuations in flux and charge noise in the $\sigma_x,\sigma_y,\sigma_z$ Pauli bases, are thus \cite{Yan_2016},
\begin{equation}\label{eq:flux_linblad_1}
    \mathcal{L}[\hat{\rho}]_z = \frac{1}{2}\omega_q\Big[\sigma_z\hat{\rho}\sigma_z^\dag - \frac{1}{2}\{\sigma_z^\dag\sigma_z, \hat{\rho}\} \Big],
\end{equation}
\begin{equation}\label{eq:flux_linblad_2}
    \mathcal{L}[\hat{\rho}]_x = \frac{1}{2\hbar}I_m\Phi_0\delta f\Big[\sigma_x\hat{\rho}\sigma_x^\dag - \frac{1}{2}\{\sigma_x^\dag\sigma_x, \hat{\rho}\} \Big],
\end{equation}
\begin{equation}\label{eq:flux_linblad_3}
    \mathcal{L}[\hat{\rho}]_y = \frac{1}{2\hbar}n_z E_{C,-}\delta n_-\Big[\sigma_y\hat{\rho}\sigma_y^\dag - \frac{1}{2}\{\sigma_y^\dag\sigma_y, \hat{\rho}\} \Big].
\end{equation}

The final non-circuit noise source comes from the CPW coupled to the flux qubit. The CPW has been simplified to operate similarly to a resonant cavity. Due to this, the system encounters noise in the form of the Purcell effect. The Purcell effect is the enhancement of a quantum system's spontaneous emission rate due to its environment. In this case, the CPW radiates a wave which has been reflected from the environment which consequently excites the resonance out of phase. This is difficult to analytically model due to the vast sources of electromagnetic waves possible, thus, using experimental results~\cite{Yamamoto_2014}, the rate at which the resonator encounters spontaneous emission is modelled by,
\begin{equation}
    \frac{1}{\tau_{photon}} = \frac{\sqrt{A \mathrm{ln} 2}}{\hbar} \Big| \frac{\delta \omega_{LR}}{\delta f_\varepsilon} \Big|,
\end{equation}
where $\omega_{LR}$ is the transition frequency between the states $\ket{L}$ and $\ket{R}$, and $A$ is a fit parameter. According to \cite{Yamamoto_2014}, $1/\tau_{photon}$ was measured to be around 9.19 MHz for $0.5<f_\varepsilon<0.55$. The Lindbladian can, thus, be described as,
\begin{equation}\label{eq:flux_linblad_4}
    \mathcal{L}[\hat{\rho}]_{purcell} = \frac{1}{\tau_{photon}}\Big[\sigma_y\hat{\rho}\sigma_y^\dag - \frac{1}{2}\{\sigma_y^\dag\sigma_y, \hat{\rho}\} \Big].
\end{equation}
The Lindbladians observed in Eq. \eqref{eq:flux_linblad_1}, \eqref{eq:flux_linblad_2}, \eqref{eq:flux_linblad_3} and \eqref{eq:flux_linblad_4} will be used to model the noise. The implementation will be discussed in Chapter \ref{chapter:GRAPE_sim}.

\section{Hybrid System Noise}\label{sec:hybrid_noise}
The goal of this section is to describe the noise (in Linbladian form) of the flux-resonator-atom qubit connection responsible for the construction of the $E_2$ gate seen in Section \ref{sec:hybrid_comp}. Due to the short time scale taken to build the $E_2$ gate, the noise of this system is different to the noise encountered in Sections \ref{sec:rydberg_noise} and \ref{sec:flux_qubit_noise}. It follows the approximations and simplified noise factors seen in ~\cite{Yu_atom_flux_2017}.

Following \citeauthor{Yu_atom_flux_2017}, there are four relevant sources of noise. The flux qubit has been simplified to contain two noise sources, qubit dephasing, given by a dephasing rate, as well as a relaxation loss of a qubit decohering from $\ket{R}$ state to a $\ket{L}$ state (seen in Section \ref{sec:hybrid_comp}), given by a relaxation rate \cite{Yu_atom_flux_2017}. The dephasing rate and relaxation rate are denoted by $\gamma_{relax}$ and $\gamma_\phi$, respectively. The main noise source observed in the hybrid system, following~\cite{Yu_atom_flux_2017}, is the spontaneous decay rate seen in Section \ref{sec:rydberg_noise}. The decay is observed for the states $\mu = e,u$ to the ground state $\ket{g}$. The spontaneous emission rate is assumed to be equal for all $\mu$ because the Rydberg state is close to the capacitor surface which induces extra noise from stray fields and thus drastically reduces the lifetime of the states \cite{Yu_atom_flux_2017}. Finally, the noise of the resonator is characterized by $\kappa = \omega_0/Q$ (assuming a Q-factor value $Q = \mathrm{10^5}$) as the loss rate due to the Purcell effect, as described in Section \ref{sec:flux_qubit_noise}. All the values can be seen on table \ref{table:hybrid_structure}.

The Hybrid system, thus, possesses the following Lindbladians,
\begin{equation}
    \mathcal{L}^f_{relax} = \gamma_{relax}\Big[\sigma_{f,-}\rho\sigma_{f,-}^\dag - \frac{1}{2}\{\sigma_{f,-}^\dag\sigma_{f,-},\rho\}\Big]
\end{equation}
\begin{equation}
    \mathcal{L}^f_{dephase} = \frac{\gamma_\phi}{2}\Big[\sigma_{f,z}\rho\sigma_{f,z}^\dag - \frac{1}{2}\{\sigma_{f,z}^\dag\sigma_{f,z},\rho\}\Big]
\end{equation}
\begin{equation}
    \mathcal{L}^r = \kappa\Big[b\rho b^\dag - \frac{1}{2}\{b^\dag b,\rho\}\Big]
\end{equation}
\begin{equation}
    \mathcal{L}^a = \gamma_{e} \sum_{\mu = e,u}\Big[\sigma_{\mu g}\rho\sigma_{\mu g}^\dag - \frac{1}{2}\{\sigma_{\mu g}^\dag\sigma_{\mu g},\rho\}\Big]
\end{equation}
where $\sigma_{g\mu} = \bra{g}\ket{\mu}$ and the other parameters can be seen in Section \ref{sec:hybrid_comp} \cite{Yu_atom_flux_2017}. The final fidelity with noise is very similar to the fidelity without noise at the same time scale. This is due to the fact that the decoherence timescale is a factor of $\mathrm{10^3}$ larger than the creation of the $E_2$ gate.

\clearpage

\chapter{Quantum Algorithm Simulation and Analysis with Gradient Ascent Pulse Engineering}\label{chapter:GRAPE_sim}
This chapter will detail the simulation of the distributed phase estimation algorithm using the Hamiltonians and noise sources for both the flux and Rydberg qubit computers as seen in Chapters \ref{chapter:hybrid_system} and \ref{chapter:open_system}. As seen in Chapter \ref{chapter:open_quantum}, the distributed phase estimation algorithm, when using four counting qubits, has a circuit depth of around 50 gates. In the case of numerous gates, it becomes imperative that each gate maintains a high level of accuracy since errors in individual gates tend to compound or multiply as they are applied successively. To accomplish accurate gates, the gates will be optimised using the GRAPE (Gradient Ascent Pulse Engineering) algorithm explained in Section \ref{sec:intro_grape}. The accuracy of the gates will then be observed using process tomography, a method of experimentally measuring the fidelity of a gate, in Section \ref{sec:process_tomography}. Finally, the implementation of the gates as well as the GRAPE algorithm in the distributed phase estimation algorithm will be elaborated upon in Section \ref{sec:analysis_conclusion}.
\section{Introduction to GRAPE with BFGS}\label{sec:intro_grape}
Simulating quantum algorithms that require many quantum gates or depth on a noisy open system is unachievable because the system decoheres far too fast. Knowing the noise contributions, one could try analytically solving the master equation and adding the shifts necessary to accomplish the creation of a gate for a certain time. However, this requires solving the Schr\"{o}dinger equation at each step of the evolution which is far too difficult, especially for a system of many qubits.

In the current NISQ era there is a large push towards correcting and optimizing qubit gates through optimal control \cite{Koch_2022}. Quantum optimal control theory (QOCT) refers to a set of methods used to implement shapes of external electromagnetic fields capable of manipulating the quantum dynamical processes at the atomic level in the most efficient way possible. Gradient Ascent Pulse Engineering (GRAPE) algorithm is one of many optimal control methods that was first developed by \citeauthor{KHANEJA2005296} \cite{KHANEJA2005296}. GRAPE has garnered attention among other algorithms because it utilizes direct analytical expressions for the gradient as opposed to difference methods. This means that it can efficiently find suitable solutions in the parameter space with fast convergence speed \cite{chen2022iterative,KHANEJA2005296}. This section will aim to introduce the GRAPE algorithm, which will then be implemented to optimise gates in Sections \ref{sec:process_tomography} and \ref{sec:analysis_conclusion}.

To fully grasp GRAPE, it is crucial to first characterize the system. The system is modelled using the Hamiltonian described in Eq. \eqref{eq:drive_control} where the Hamiltonian is a linear combination of a drift Hamiltonian and a control Hamiltonian. To model the system into an open system, one must first utilize the density operator $\rho (t)$ and characterize the equation of motion as a Liouville-von Neumann equation ( defined in Eq. \eqref{eq:liouville}),
\begin{equation} \label{eq:density_init}
    \dot{\rho}(t) = -\frac{i}{\hbar} \bigg [ \Big(\mathcal{H}_0 + \sum^m_{k=1}u_k(t)\mathcal{H}_k\Big), \rho(t)\bigg],
\end{equation}
where $\mathcal{H}_0$ is the drift Hamiltonian, $\mathcal{H}_k$ are the radiofrequency Hamiltonians that correspond to the available control fields and $u(t) = (u_1(t), u_2(t),\cdots,u_m(t))$ is the control vector or a set of vector amplitudes that can be changed \cite{KHANEJA2005296}. The goal is to find the optimal amplitudes $u_k(t)$ of the RF fields capable of steering a given density operator in initial state $\rho(0) = \rho_0$ in a total time $T$ to a density operator $\rho(T)$ with maximum overlap to some desired target $C$ measured by the standard inner product,
\begin{equation}\label{eq:trace_func}
    \braket{C|\rho(T)} = \mathrm{Tr}\{ C^\dag \rho(T) \}.
\end{equation}
In the case of open quantum systems, the dynamics is described by the Markovian master equation seen in Eq. \eqref{eq:lindbladian_form}. The formal solution to such equation is \cite{Boutin_2017}, 
\begin{equation}
    \rho(t) = \mathrm{exp}\Big\{\int^t_0\mathcal{L}(t^')dt^'\Big\}\rho(0).
\end{equation}
To iterate through time, the transfer time $T$ is discretised in $N$ equal steps of duration $\Delta t = T/N$. It is assumed that at each time step, the control amplitudes $u_k(t)$ are constant. This means that at an arbitrary time step $l$, the amplitude $u_k(t)$ of the $k$th control Hamiltonian is given by $u_k(l)$. Using the Lindbladian formulation described Eq. \eqref{eq:lindbladian_form}, the master equation is,
\begin{equation}
    \dot{\rho} = -i[H,\rho] + \mathcal{L}[\rho],    
\end{equation}
where $\mathcal{L}$ acts as an operator on a density matrix, $\rho$. Using this definition means that one can define a discretised unitary propagator during a time step $l$,
\begin{equation}\label{eq:unitary_grape}
    U_l \boldsymbol{\cdot} = \mathrm{exp} \bigg\{-i\Delta t([H_l, \boldsymbol{\cdot}] + \mathcal{L}[\boldsymbol{\cdot}] \bigg \}.
\end{equation}
A performance index is required for the algorithm to optimise a gate, which can be expressed as the overlap between a final state $\rho(T)$ and the target state, $C$. The performance index is of the form,
\begin{equation}
    \Phi_0 = \mathrm{Tr}\Big\{C U_N\cdots U_1\rho(0)\Big\}.
\end{equation}
The derivative of the performance index takes the form \cite{Boutin_2017},
\begin{equation}
    \frac{\partial \Phi_0}{\partial u_k(l)} = \mathrm{Tr} \bigg\{\lambda_l(C)\frac{\partial U_l}{\partial u_k(l)}\rho_{l-1} \bigg\},
\end{equation}
where,
\begin{equation}
    \rho_l = U_l \cdots U_1 \rho(0),
\end{equation}
is the forward in time evolved density matrix and,
\begin{equation}
    \lambda_j(C) = U_{l+1}^\dag\cdots U_N^\dag C,
\end{equation}
is the backwards in time evolution from the final target state.The first order derivative of the $l^{th}$ time-evolution operator in terms of $\Delta t$ is~\cite{Boutin_2017,Abdelhafez_2019},
\begin{equation}
    \frac{\partial U_l \cdot}{\partial u_l(l)} \approx -i\Delta t [H_k, (U_l \boldsymbol{\cdot})].
\end{equation}
The derivative of the performance index is thus,
\begin{equation}\label{eq:GRAPE_lindblad}
    \frac{\partial\Phi_0}{\partial u_k(l)} = -i \Delta t \mathrm{Tr}\{\lambda_l(C)[H_k,\rho_l]\}.
\end{equation}
The result seen in Eq. \eqref{eq:GRAPE_lindblad} is the core of the GRAPE algorithm ~\cite{Abdelhafez_2019,Boutin_2017}. 

\begin{figure}[t] 
\centering
    \includegraphics[width=0.55\textwidth]{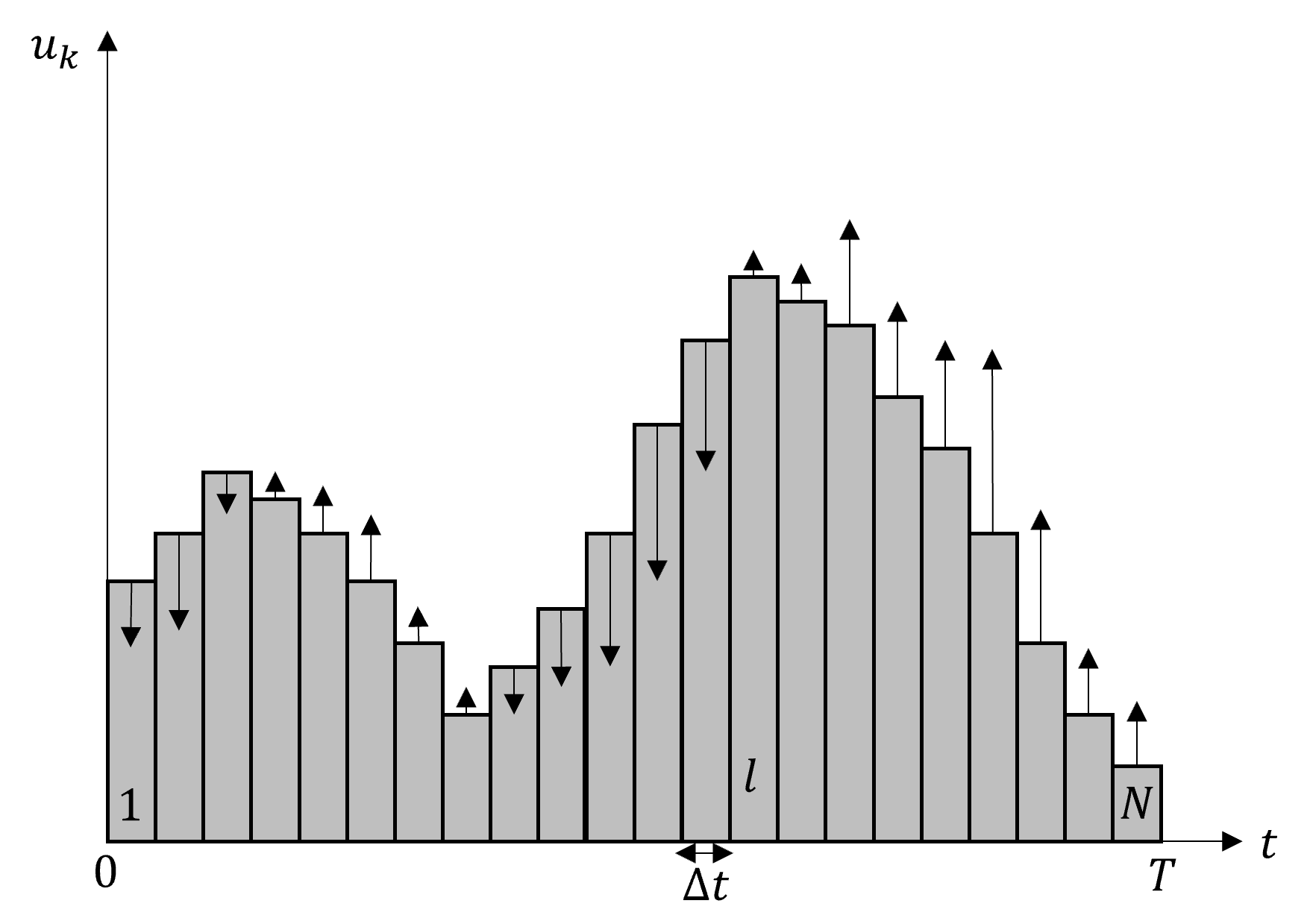}
    \caption{Representation of an arbitrary control amplitude $u_k(t)$ over time $t$ divided in $N$ discrete steps of duration $\Delta t$. During each step $l$, the control amplitude is equal to $u_k(l)$. The vertical arrows represent gradients $\partial\Phi_0/\partial u_k(l)$ which gives an indication of how to modify $u_k(l)$ in the next iteration to improve $\Phi_0$.}
    \label{fig:grape_diagram}
\end{figure}
Fig. \ref{fig:grape_diagram} exhibits how an arbitrary $u_k(t)$ evolves over time. The efficiency of the algorithm is dependent on maximizing the performance gradient $\partial\Phi_0/\partial u_k(l)$ denoted by arrows on Fig. \ref{fig:grape_diagram}. The performance change of the control amplitude is acquired by allowing \cite{chen2022iterative}, 
\begin{equation} \label{eq:u_k_def}
    u_k(l) \rightarrow u_k(l) + \epsilon\ \frac{\delta \Phi_0}{\delta u_k(l)},
\end{equation}
so that the performance function $\Phi_0$ increases when $\epsilon$ is a small enough step size. Tailoring the GRAPE algorithm for qubit optimal control requires modifying the definition of the performance function $\Phi_0$. In the field of qubit optimal control the performance function described by \eqref{eq:trace_func} can be referred to as the fitness function, $f$~\cite{Lu_2017}. This means that the fitness function is the metric that controls the modification such that $\varepsilon = 1 - f$. The closer the algorithm gets to reaching the desired fidelity, the smaller the modification becomes. In a full evolution, there will be $N \times M$ variables where $M$ are the amount of control Hamiltonians. The general algorithm for GRAPE can be seen in Alg. \ref{alg:grape_basic}. 

\begin{algorithm}
\caption{The general GRAPE algorithm.}\label{alg:grape_basic}
\begin{algorithmic}
\Require Get minimum fidelity error $x_{min}$
\Require Guess initial controls $u_k(l)$
\State $\rho \gets \rho_0$
\State $done \gets \mathrm{False}$
\While{not $done$}
\State $l \gets 0$
    \While{$l \leq N$}
        \State $Iterator_{f} \gets 0$
        \While{$Iterator_{f} \leq l$}
            \State $\rho \gets U_{Iterator_{f}} \rho$ 
            \State $Iterator_{f} \gets Iterator_{f} + 1$
        \EndWhile
        \State $\lambda \gets C$
         \State $Iterator_{b} \gets N$
        \While{$Iterator_{b} \leq l+1$}
            \State $\lambda \gets U^\dag_{Iterator_{b}}$
            \State $Iterator_{b} = Iterator_{b} + 1$
        \EndWhile
    \EndWhile
    \State Evaluate $\partial\Phi_0/\partial u_k(l)$ (Eq. \eqref{eq:GRAPE_lindblad}), $\varepsilon=1-f$ and update $m \times N$ control amplitudes $u_k(l)$ according to Eq. \eqref{eq:u_k_def}
    \If{ $\varepsilon \leq x_{min}$ }
        \State $done \gets \mathrm{True}$
    \EndIf
    
\EndWhile

\end{algorithmic}
\end{algorithm}

The main challenge the GRAPE algorithm as well as other optimization algorithms encounter is that the methods have a tendency to get stuck at a local minima. For a system that possesses many control fields $\mathcal{H}_k$, this is usually not a problem since one can define a fidelity with sufficient precision to reach. On the other hand, it is very difficult (or impossible) to know the highest possible fidelity for a system that lacks the required control fields. This consequently forces a trade-off between the number of iterations required to reach the minima and the possibility to overstep a minima. 

An approach one can take to counteract this is to translate the local problem landscape to a parabola which in turn makes it possible to step into the minima without having to guess the step size. For this reason, GRAPE utilizes a quasi-Newtonian gradient descent algorithm called Broyden-Fletcher-Goldfarb-Shanno (BFGS) algorithm which utilizes Hessian based approximation techniques \cite{LBFGS,compareBFGS}. This algorithm efficiently computes the second-order differentials of Hessian matrices that would otherwise pose a challenging calculation. This algorithm is seen explained in Appendix \ref{sec:bfgs_app} ~\cite{LBFGS,Dalgaard_2020,BFGS_practical,wolfe_condition}.

\section{Quantum Process Tomography and Implementation of GRAPE Optimised Gates}\label{sec:process_tomography}

The goal of this section is to analyze the fidelity of the gates using GRAPE. The implementation technique will thus focus on single operations at a time, thus decoherence of atomic states will not be taken into account when the qubit is not being processed. This constraint is reasonably valid since the most noise originates from detuning, magnetic fluctuations and laser effects which are all consequences of qubit state manipulation. Furthermore, when conducting optimal control on hardware, one cannot access the quantum state in the middle of a running execution. For this reason, optimal control is used to create final states from initial states using modelled system noise that has been measured from previous runs \cite{feedback_pulse}.

The initial parameters to consider are those from Eq. \eqref{eq:density_init}. The Pauli matrices, $\sigma_x$, $\sigma_y$ and $\sigma_z$ will be used as the controls. For the Flux system, the drift Hamiltonian will be Eq. \eqref{eq:approximated_model} with the superoperators Eq. \eqref{eq:flux_linblad_1} - \eqref{eq:flux_linblad_4}. For the Rydberg atom system, the drift Hamiltonian will use Eq. \eqref{eq:pulser_hamiltonian} with the superoperators Eq. \eqref{eq:rydberg_lindblad_1} and \eqref{eq:rydberg_lindblad_2}. These master equations will be fed into the GRAPE algorithm that already forms part of the QuTiP optimal control package.

The GRAPE algorithm has many different values that can be modified. This study will focus on varying two, the number of time steps in a specified time $T$ and the number of iterations the algorithm is allowed to run through before reaching the minimum fidelity error ($\mathrm{10^{-10}}$). Allowing the algorithm to run many times for a pulse sequence, allows the GRAPE algorithm to choose different guesses each time in an attempt to lower the fidelity error. The time $T$ chosen was 50 $\mu$s because it is at a small enough timescale to reasonably neglect decoherence effects from state lifetimes. The time steps will vary from 50 to 200 steps which means that the lowest amount of time per step will be 250ns which, according to PASQAL \cite{Silv_pulser_2022}, is enough time to pulse a full rotation gate on a Rydberg atom system. The order of magnitude of the timescale required to form the hybrid GHZ state (in the order of a few nanoseconds) is far too small for GRAPE to be feasible because it would require the ability of pulsing control fields at the picosecond scale. Thus, GRAPE will not be utilized in the creation of the hybrid GHZ gate.   

An additional restriction to be imposed on the algorithm pertains to the allotted CPU time for optimizing a time step before proceeding to the subsequent iteration. This is referred to as wall time or wall clock time because it is counted in seconds and is used as a timeout if the algorithm encounters a local minimum and cannot minimise the error \cite{de_Fouquieres_2011}. This is chosen to be at 200 seconds. 

Appendix \ref{sec:grape_code} shows the code detailing the construction of the GRAPE optimised noisy qubit gates. The function \codeword{single_qubit_hamiltonian} builds the master equation for the single qubit interaction both the flux and Rydberg atom system depending on the argument input. This is also the case for the double qubit interaction using the function \codeword{double_qubit_hamiltonian}. The parameter values can be seen in table \ref{table:values_parameters}. This section will analyze the purple section of the simulation schematic seen in figure \ref{fig:grape_algorithm}. As can be seen in the schematic, the gate optimisation is dependent on the iterating values for both the GRAPE iteration amount and the time step amount. 

An initial examination of how the GRAPE algorithm fine-tunes and constructs both Rydberg and flux quantum gates can be observed through the application of quantum process tomography. Quantum process tomography is the process by which known quantum states are used to probe a quantum process to find how the process or gate is described \cite{nielsen_chuang_2021}. Quantum process tomography stems from quantum state tomography which is the procedure of experimentally determining an unknown quantum state. Since it is not possible to distinguish the superposition states of an unknown state $\rho$ with a single measurement, quantum state tomography allows one to estimate $\rho$ if many copies of $\rho$ are available. This is the case for a quantum gate where the process can be executed many times on a known initial state to generate many copies of $\rho$. 

The goal is to determine a set of operation elements $\{E_i\}$ for $\varepsilon$,
\begin{equation}
    \varepsilon(\rho) = \sum_i E_i\rho E_i^\dag.
\end{equation}
Experimental results however, do not involve operators, they involve numbers. Thus, it is convenient to rewrite $E_i$ to \cite{nielsen_chuang_2021},
\begin{equation}
    E_i = \sum_m e_{im}\tilde{E}_m
\end{equation}
where $\tilde{E}$ is a fixed set of operators and $e_{i}$ is a set of complex numbers. This is often described as the \textit{chi matrix representation} since $\varepsilon$ can be completely described by a complex number matrix $\chi$ defined as, $\chi_{mn} \equiv \sum_i e_{im}e_{in}^*$ given that $\rho$ remains Hermitian with trace one which follows $\sum_i E^\dag_i E_i = I$ \cite{nielsen_chuang_2021}. 

\begin{figure}[!htb] 
\centering
    \includegraphics[width=1.1\textwidth]{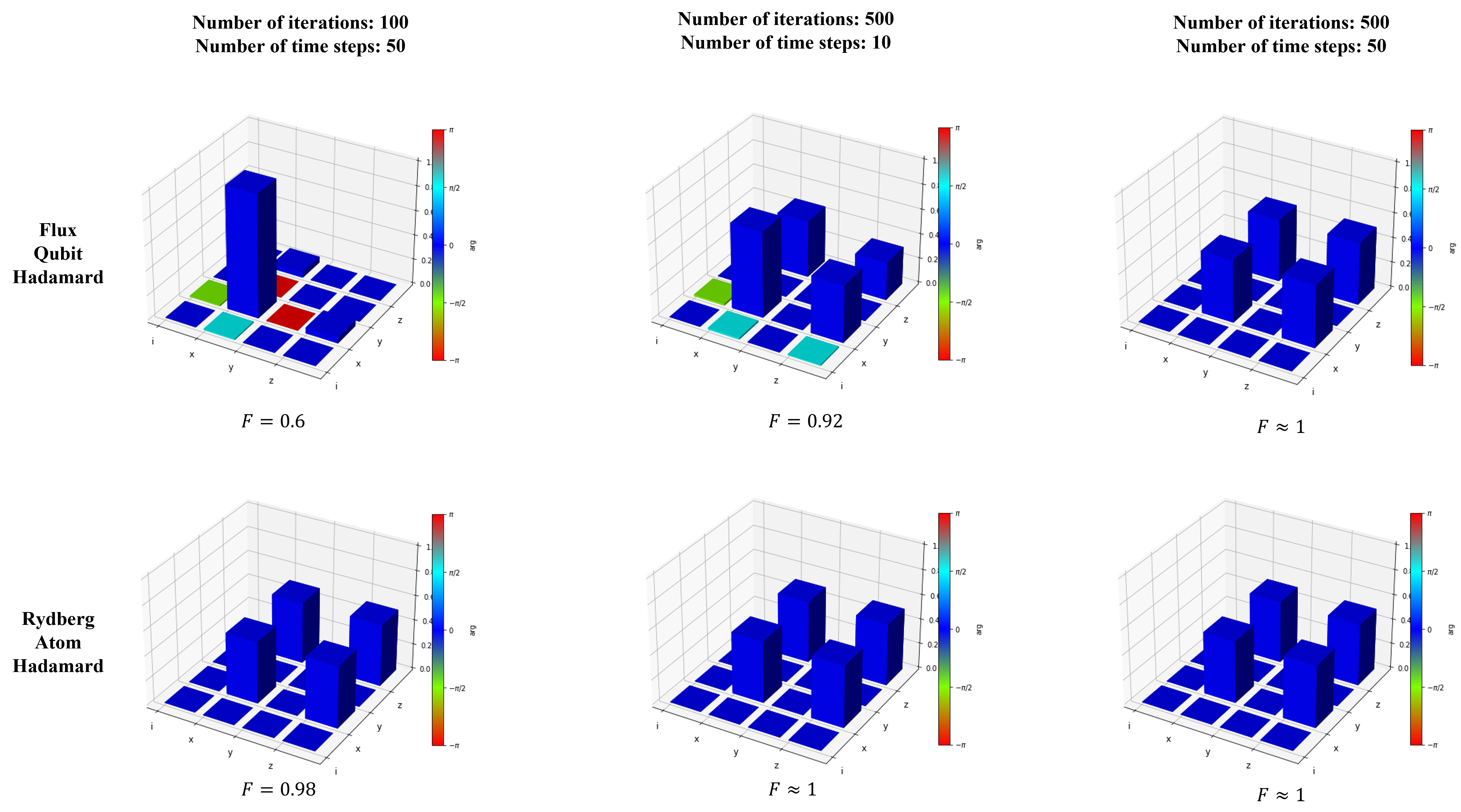}
    \caption{The $\chi_{re}$ matrix evaluating the construction of the Hadamard gate for both the flux and Rydberg systems (each depicted on a row) against GRAPE iterations and time step. Each constructed gate contains the calculated fidelity and the axes represent $i$ as the identity $I$ and $x,y,z$ as the Pauli $\sigma_x,\sigma_y,\sigma_z$, respectively.}
    \label{fig:hadamard_gate_noise}
\end{figure}

\begin{figure}[!htb] 
\centering
    \includegraphics[width=1.1\textwidth]{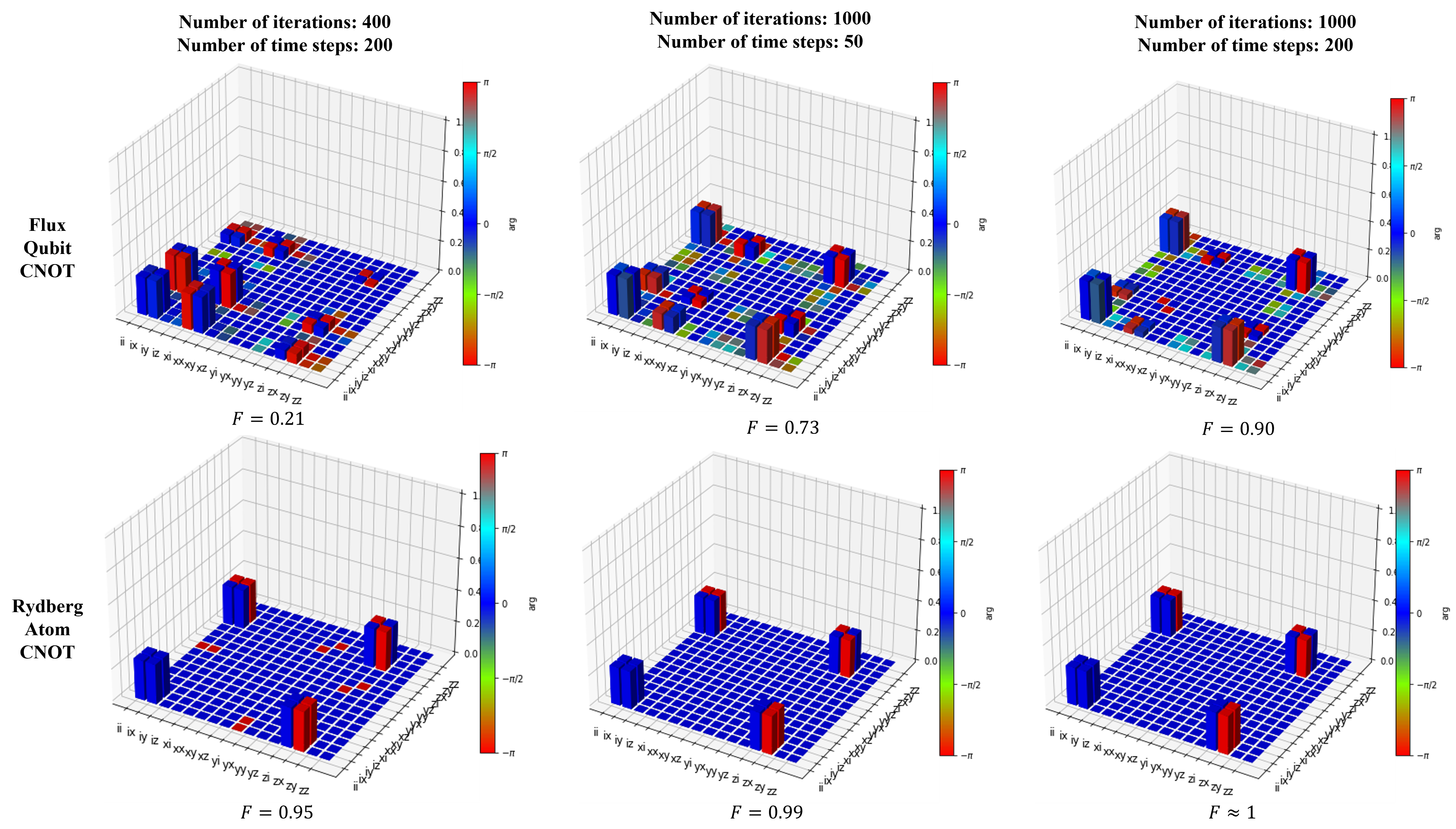}
    \caption{Quantum process tomography evaluating the construction of the CNOT gate for both the flux and Rydberg systems (each depicted on a row) against GRAPE iterations and time step. Each constructed gate contains the calculated fidelity and the axes represent $i$ as the identity $I$ and $x,y,z$ as the Pauli $\sigma_x,\sigma_y,\sigma_z$, respectively. The notation $ii$ refers to a tensor product $I \otimes I$.}
    \label{fig:cnot_gate_noise}
\end{figure}

The input states and measurement projectors must each form a basis for the set of $n$-qubit density matrices, requiring $d^2 = 2^{2n}$ in each set \cite{O_Brien_2004}. This means $\varepsilon$ is,
\begin{equation}
    \varepsilon(\rho) = \sum^{d^2-1}_{m,n=0} \tilde{E}_m \rho \tilde{E}^\dag_n \chi_{mn}.
\end{equation}
Matrix $\chi$ is the transformation matrix that describes how much $\tilde{E}_m \rho \tilde{E}^\dag_n$ contributes to the final state $\epsilon(\rho)$, making it the most important value to analyze. Using $\chi$, the fidelity is then calculated by,
\begin{equation}\label{eq:tomography_equation}
    F(\chi_{th},\chi_{re}) = \frac{\mathrm{Tr}(\chi_{re}\chi_{th}^\dag)}{\sqrt{\mathrm{Tr}(\chi_{th}^\dag \chi_{th})}\sqrt{\mathrm{Tr}(\chi_{re}^\dag\chi_{re})}},
\end{equation}
where $\chi_{th}$ and $\chi_{re}$ are the theoretical and resulting (measured) $\chi$ matrices. The two gates that will be preliminary investigated against the GRAPE iteration number and the time step number are the Hadamard gate and the CNOT gate with the purpose primarily being to analyze both single and multi-qubit interactions. The $\chi_{re}$ matrix is seen plotted in figures \ref{fig:hadamard_gate_noise} and \ref{fig:cnot_gate_noise}.
\begin{figure}[!htb] 
\centering
    \includegraphics[width=1.0\textwidth]{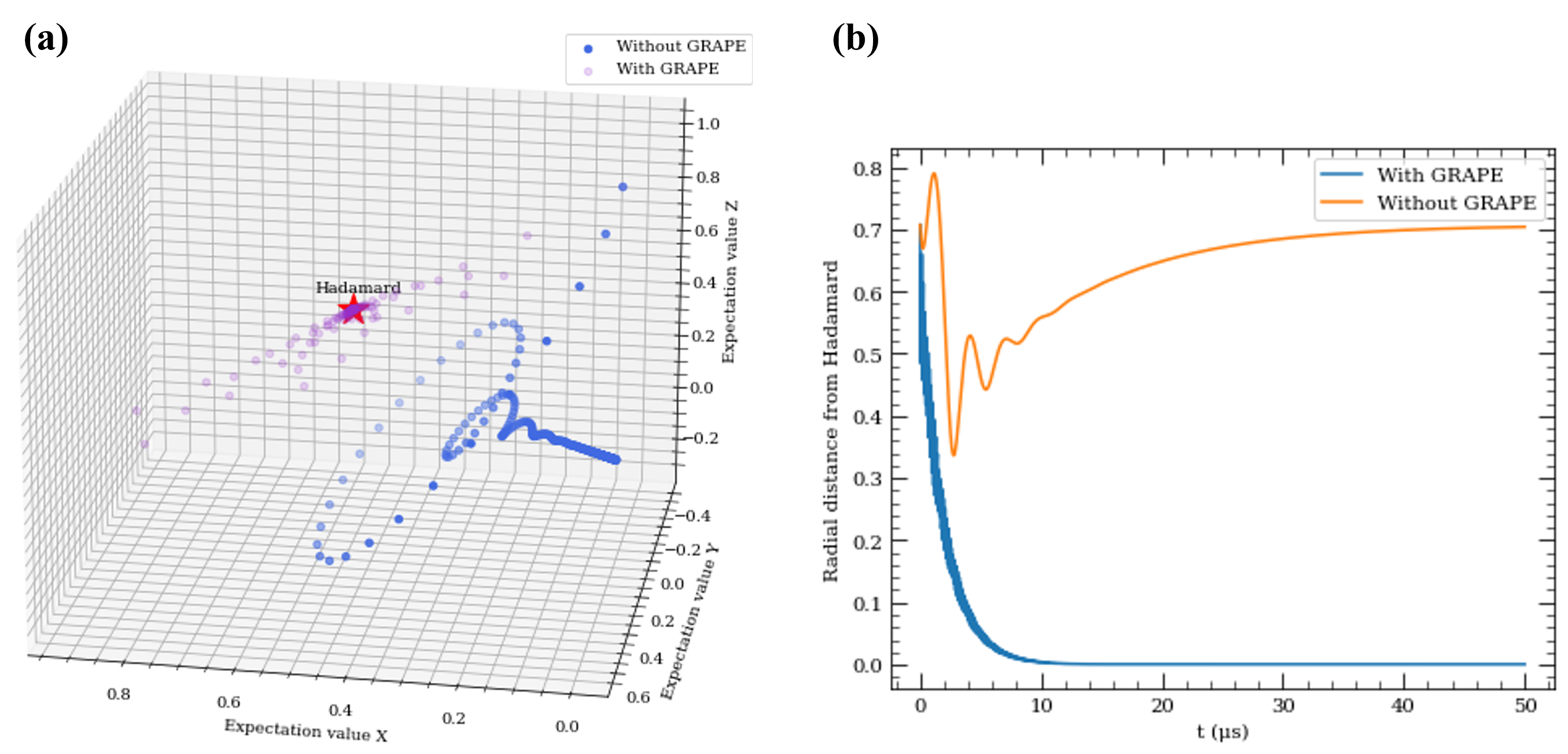}
    \caption{(a) Visualization depicting the expected contributions of $\sigma_x, \sigma_y$ and $\sigma_z$ (labelled as Expectation value $X,Y.Z$) in the evolution of the flux qubit system over time optimised with GRAPE (seen in purple) and without GRAPE (seen in blue). The red star depicts the ideal Hadamard gate.(b) The plot depicting the distance from the ideal Hadamard gate over time in $\mu$s seen in (a), in arbitrary units, for both the GRAPE optimised and none-GRAPE optimised case.}
    \label{fig:flux_hadamard_GRAPE}
\end{figure}
Hadamard gate can be intuitively understood through the contribution of the Pauli bases $\sigma_x$ and $\sigma_z$ as $U^{Hadamard} = \frac{1}{\sqrt{2}}(\sigma_x + \sigma_z)$ \cite{Beterov_2016,IBM_hadamard}. A more formal derivation of the $\chi$ matrix for the ideal Hadamard gate is seen in Appendix \ref{sec:chi_hadamard}. Thus, when calculating the process tomography, the two matrices that have to be measured are the $\sigma_x$ and $\sigma_z$ contributions. This can be seen in figure \ref{fig:hadamard_gate_noise} for the 500 iteration step and 50 time step case for both the flux and Rydberg systems where the fidelity is approximately equal to 1. The color bar on the left hand side of each graph shows the phase, $\theta$, of the phase factor $e^{i\theta}$ multiplied to each $\chi$ element, $\chi_{mn}$ (seen in Appendix \ref{sec:chi_hadamard}). A reduction of the time steps or the iteration steps however, produces fidelity errors in the flux system that are more noticeable than those in the Rydberg system. Furthermore, observing the flux system, it is possible to note that the number of iterations affects the system's fidelity more than the number of steps when reducing each number by a factor of five. 

The GRAPE algorithm's optimisation using the process tomography theory can be visualized in time for a better understanding. Figure \ref{fig:flux_hadamard_GRAPE}(a) depicts a 3D scatter plot with the axes showing the expected contributions of the $\sigma_x, \sigma_y$ and $\sigma_z$ matrices (depicted as $X$, $Y$, and $Z$, respectively). The two sets of points show the evolution over time for the propagation using GRAPE optimisation in purple, using 100 time steps and 500 GRAPE iterations, and the propagation without GRAPE optimisation is in blue. The ideal Hadamard gate is depicted with the red star. Figure \ref{fig:flux_hadamard_GRAPE}(b) shows the distance from the ideal Hadamard gate over time. As seen from figure \ref{fig:flux_hadamard_GRAPE}(a), the GRAPE algorithm utilizes the controls in order to minimize the fidelity error towards the target. With the amount of time steps used, it is possible to see that the GRAPE algorithm has sufficient control at each point in time in order to counter act the system noise, which is seen pulling the system towards a $\sigma_y$ contribution. The fidelity of the Hadamard gate officially calculated by Eq.\eqref{eq:tomography_equation}, can be intuitively thought as the distance away from the ideal gate as seen in figure \ref{fig:flux_hadamard_GRAPE}(b).

The CNOT gate on the other hand is constructed using the Pauli bases $\sigma_x$ and $\sigma_z$ as well as the identity $I$ as $U^{CNOT} = \frac{1}{2}(I \otimes I + I \otimes \sigma_x + \sigma_z \otimes I - Z \otimes X)$. This is seen in \ref{fig:cnot_gate_noise} where at fidelity 1 (seen in the final Rydberg graph), the contributions are indeed $I$, $\sigma_x$, and $\sigma_z$ where the double grouped indices represent a tensor product, e.g $II = I \otimes I$. As expected, the contributions $\sigma_x \otimes \sigma_z$ when paired with $I$ or $\sigma_x$ contain a factor $e^{- i\pi}$ where $\theta = \pi$. The Rydberg system once more produces less fidelity error than the flux system when the parameters are changed. The flux system surprisingly does not converge to an almost perfect fidelity as seen in the single qubit gate example as seen from the incorrect Pauli basis artefacts after 1000 GRAPE iterations and 200 time steps. 

It can be seen that the flux system suffers greater decoherence effects than the Rydberg system. This can be seen analytically from the noise fluctuation Hamiltonian, Eq. \eqref{eq:flux_fluctuation_final} whereby the dynamics are governed by $\sigma_x, \sigma_z$ and, when considering interactions between flux qubits, $\sigma_y$. The magnitude of the noise parameters make it difficult for the GRAPE algorithm to optimise the gates. This is commonly the issue with superconducting qubits, both the charge and the fluctuation noise impact the decoherence heavily. At $E_J/E_C \geq 50$, optimised superconducting C-shunted qubits have a lifetime of 50$\mu$s - 100 $\mu$s \cite{annurev_soa}. The flux qubit used in this study is closely based on the flux qubit used in \cite{Yan_2016} where $E_J/E_C = 65$. On the other hand, Rydberg qubits in the non-excited state have very long coherence times compared to superconducting qubits because atomic states with $L \leq 2$, have 100$\mu$s - 1ms lifetimes, which means that they do not suffer as much from spontaneous emission \cite{PRXQuantum.2.030322}. 

In conclusion, it has been seen that high fidelity single qubit gates are possible for both the flux and Rydberg atom systems but only the Rydberg qubits observe high fidelity in the double qubit gates. The impact of these gates on the phase estimation algorithm will be seen in Section \ref{sec:analysis_conclusion}, where the algorithm will be executed using differing time step and GRAPE iteration amounts. Because the flux qubit system has been observed to perform poorly, differing values for the shunt factor, $\zeta$, will be used in the simulation of the distributed phase estimation algorithm in order to obtain gates with higher fidelities.  

\section{GRAPE Optimised Distributed Phase Estimation}\label{sec:analysis_conclusion}

Section \ref{sec:process_tomography} successfully demonstrated the GRAPE algorithm optimising the single and double qubit gate fidelities for both the flux and Rydberg atom systems. This section will use the optimized gates in order to execute the distributed phase estimation algorithm. Additionally, Section \ref{sec:process_tomography} showed that the flux qubit system performs poorly when compared to the Rydberg atom system. This section will introduce the modification of the shunt factor $\zeta$, previously described in Sections \ref{sec:flux_system} and \ref{sec:flux_qubit_noise}, as a variable in order to analyse potential improvements in the accuracy of the distributed phase estimation algorithm. 

The diagram depicting the simulation can be seen in figure \ref{fig:grape_algorithm}. Section \ref{sec:process_tomography} showed the creation of the gates, which are placed in a gate dictionary (purple area), and Section \ref{sec:qutip_implementation} detailed the distributed phase estimation algorithm, which is seen depicted in orange. As one can see, there is an extra iteration that occurs (depicted as $k$). This is because, during the distributed phase estimation algorithm, one party needs to measure their ebit in order to communicate them to the other party. In the analysis, the expectation value of a specific final state gives insight in the accuracy of the phase estimation algorithm. However, as part of the wavefunction is collapsed by intermediate measurements, ten simulation runs are conducted to calculate the expected outcome. Thus, the orange section of the diagram in figure \ref{fig:grape_algorithm} is executed ten times and the mean outcomes are recorded. 

Firstly, the gates observed in Section \ref{sec:process_tomography} will be used in the simulation whereby, $\zeta = 10$. Although the single qubit gates demonstrate a very high fidelity for the flux qubit, it is important to analyze the impact of the noisy multi-qubit gate seen in figure \ref{fig:cnot_gate_noise}. To do so, the phase to estimate will be chosen to be $\varphi = 3/16$ (or 0011 in the computational basis). Thus, the accuracy will depend on the mean probability (since the algorithm will be run 10 times) of achieving this outcome for a variable GRAPE iteration number and a time step number. 

It is important to note that the GRAPE algorithm will not be run on the communication $E_2$ gate because the GRAPE algorithm is not efficient in these timescales (order of nanoseconds). Because none of the error correction techniques described in this investigation can be employed on the $E_2$ gate, it will be assumed to possess perfect fidelity. This will allow errors arising from GRAPE to be observed more easily. Following figures \ref{fig:hadamard_gate_noise} and \ref{fig:cnot_gate_noise}, the ranges of interest chosen were 50-200 time steps and 100-800 GRAPE iterations due to the computational resources available. This leads to a simulation time of approximately 22 hours on a desktop computer. 

\begin{figure}[!htb]
\centering
    \includegraphics[width=1\textwidth]{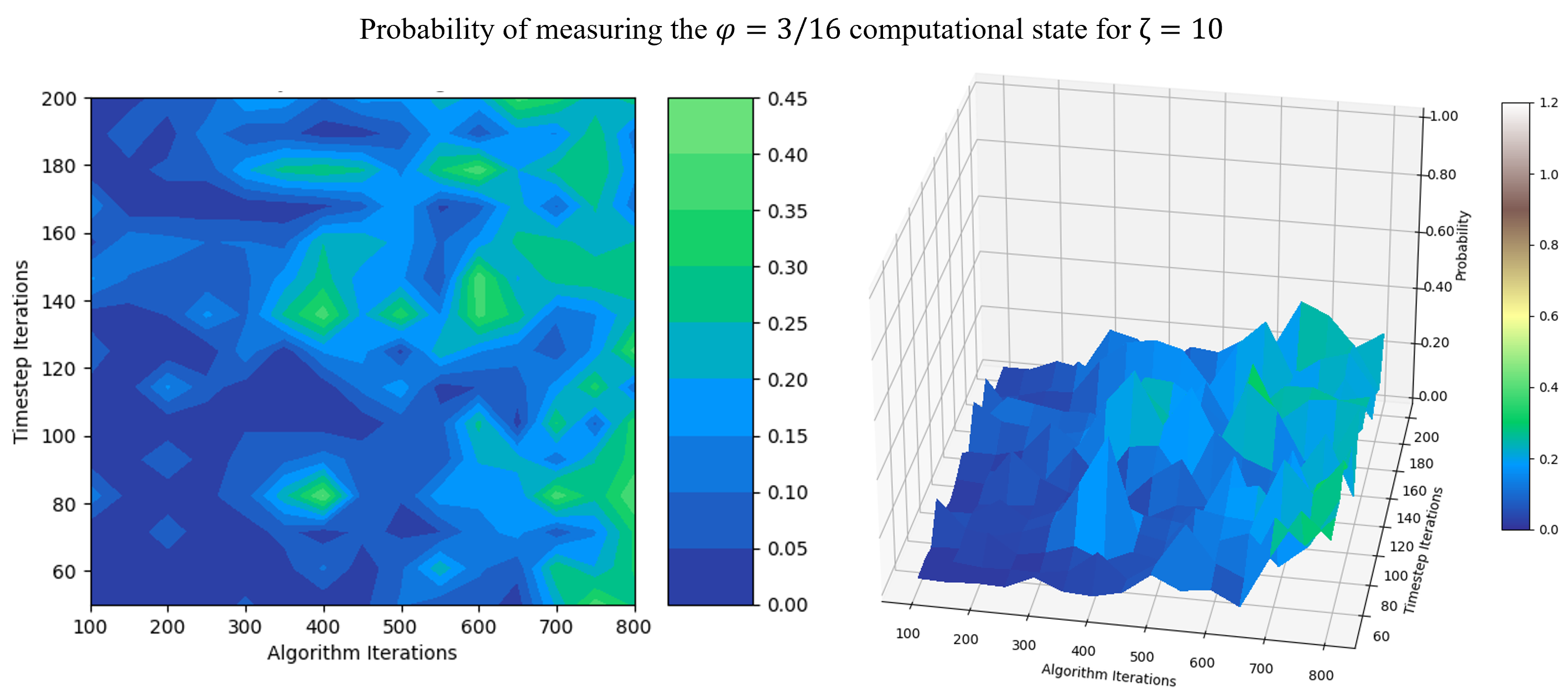}
    \caption{Probability of measuring the $\varphi =3/16$ or $0011$ computational state after 10 executions of the phase estimation algorithm on the noisy hybrid circuit for ranging GRAPE iterations and time steps for a shunt factor of $\zeta = $ 10. Depicted on the left is a graph showing the result in 2D while on the right is the same result on a 3D plane.}
    \label{fig:zeta_10}
\end{figure}

The flux qubit computer was chosen to possess three qubits, one communication qubit, one counting qubit and a qubit to hold the phase to estimate. The Rydberg atom computer was chosen to have four qubits in which, three were counting qubits and one was a communication qubit. The visualization of the entire simulation of a distributed phase estimation algorithm involving this architecture for ranging time steps (50-200) and ranging GRAPE iterations (100-800) can be seen in figure \ref{fig:zeta_10}, where $\zeta = 10$. One can observe that the accuracy of the system at this shunt value, is mainly dependent on the amount of GRAPE iterations. This follows that of the conclusion obtained in Section \ref{sec:process_tomography}, where figure \ref{fig:cnot_gate_noise} showed a better relative performance when changing the GRAPE iteration number for the flux qubit. The highest accuracy obtained for estimating $\varphi = 3/16$ was around 40\%.   

\begin{figure}[!htb]
\centering
    \includegraphics[width=1\textwidth]{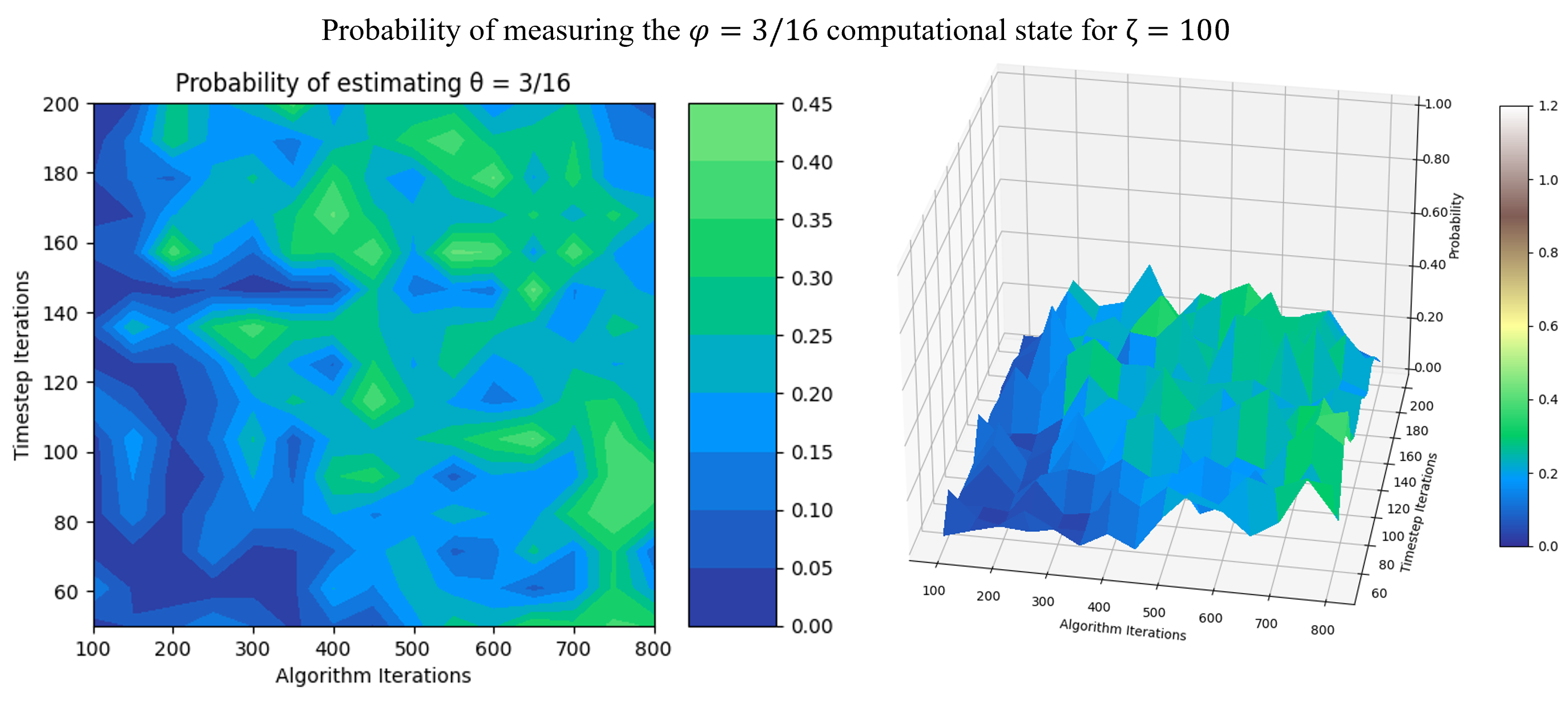}
    \caption{Identical concept to figure \ref{fig:zeta_10} for a shunt factor of $\zeta = $ 100. Depicted on the left is a graph showing the result in 2D while on the right is the same result on a 3D plane.}
    \label{fig:zeta_100}
\end{figure}

Experimentally, the factor $\zeta$ has been modified to be between 1-1000 between different superconducting qubit technologies ~\cite{Birenbaum2014TheCF,Yan_2016}. Thus, the next step is to analyse the performance of the architecture when $\zeta$ is increased by a factor of 10, thus $\zeta = 100$. This is seen in figure \ref{fig:zeta_100}. It can be observed that, although the accuracy does not improve significantly to that of the $\zeta=10$ case for large GRAPE iteration numbers and time step numbers, it does saturate to a probability of 40\% in less GRAPE iterations and time step numbers. Overall, at this effective shunt factor, it is still not able to accurately estimate the phase for the given GRAPE iteration and time step number range.  

\begin{figure}[!htb]
\centering
    \includegraphics[width=1\textwidth]{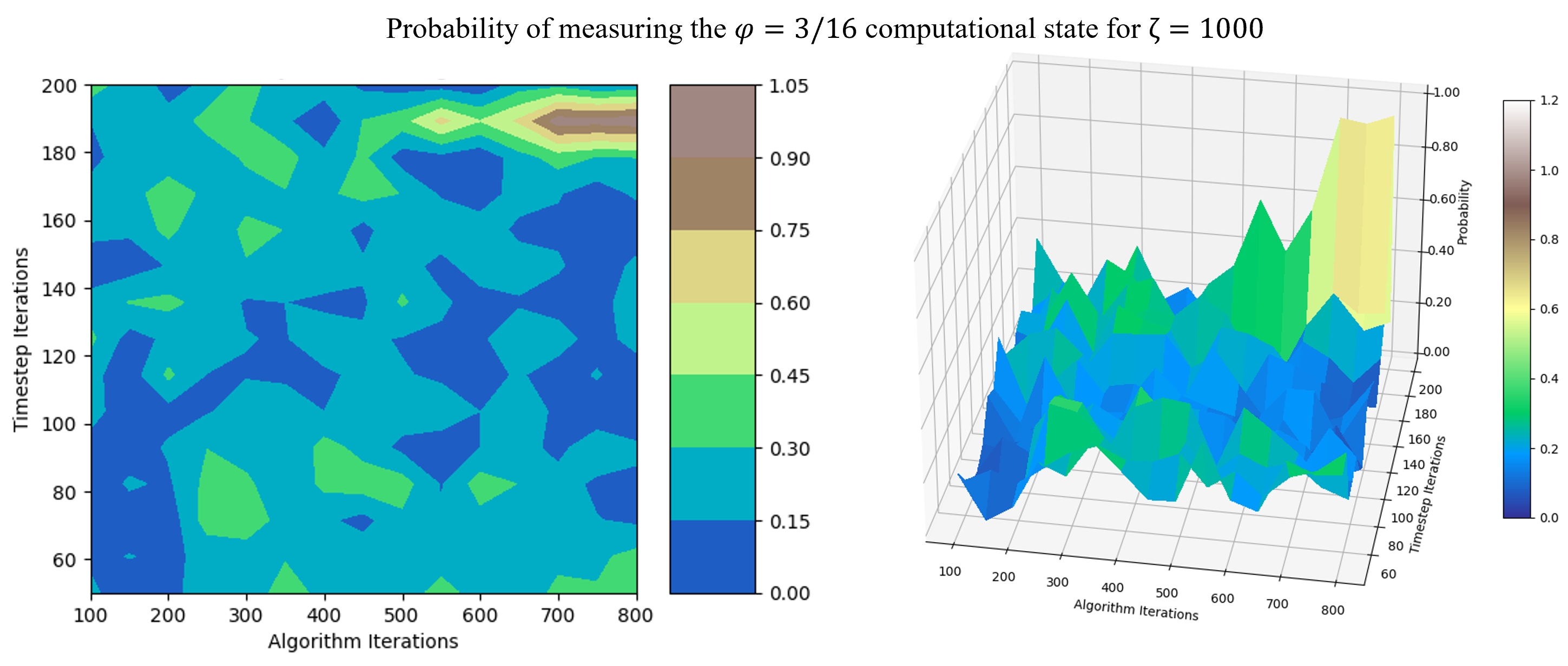}
    \caption{ Identical concept to figure \ref{fig:zeta_10} for a shunt factor of $\zeta = $ 1000. Depicted on the left is a graph showing the result in 2D while on the right is the same result on a 3D plane.}
    \label{fig:zeta_1000}
\end{figure}

A final shunt factor of $\zeta = 1000$ is simulated. The simulation results can be seen in figure \ref{fig:zeta_1000}. It can be observed that for approximately 190 time steps and approximately 700 GRAPE iterations, the algorithm results in an accuracy of around 95\%. The accuracy makes a steep jump from 40\% to 95\% after an increase of only 5 time steps (180 to 185 time steps). Observing the range of very low probability ($<10\%$) between 700-800 GRAPE iterations and 100-160 time steps, it is noticeable that there is a trade off between the time step number and GRAPE iteration number occurring.

\begin{figure}[t]
\centering
    \includegraphics[width=0.5\textwidth]{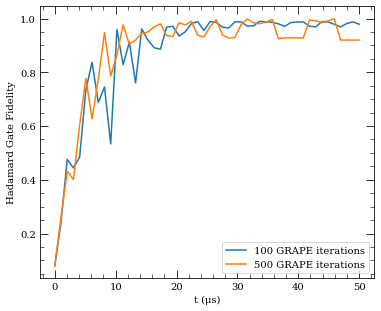}
    \caption{Graph depicting the fidelity of a Hadamard gate for the flux qubit over time for two different GRAPE optimisation configurations. The orange line shows the evolution of the gate with 500 GRAPE iterations while the blue line shows the evolution of the gate with 100 GRAPE iterations. The 100 GRAPE iteration evolution finished with a fidelity of 0.97 while the blue line finished with a fidelity of 0.92.}
    \label{fig:grape_trade_off}
\end{figure}

To describe this trade off, the evolution of flux qubit system aiming to create a Hadamard gate can be observed using two different GRAPE configurations. One configuration is set to use 100 GRAPE iterations while the other configuration is set to use 500 GRAPE iterations. This can be seen in figure \ref{fig:grape_trade_off}. In the 100 GRAPE iteration case, the fidelity reached at the end of the evolution is 0.97 while for the 500 GRAPE iteration case, the fidelity reached is 0.92. The graph shows that, although the final fidelity for the 500 GRAPE iteration case was lower than the 100 GRAPE iteration case at the end of the run, the 500 grape iteration evolution manages to achieve fidelities higher than the 100 GRAPE iteration evolution in the total run. Observing the 500 GRAPE iteration evolution after 30 $\mu$s, one can see that the evolution holds the fidelity around 0.98 but then dips to 0.92. This dip is most likely due to the fact that the system encounters a control landscape whereby the time step required to modify the control amplitudes is not small enough to converge. Thus, at these time steps, a minimum local fidelity is reached until the landscape changes over time as seen at the 40 $\mu$s mark. This concept is further explained in \cite{Larocca_2021}. Observing figure \ref{fig:zeta_1000}, one can see that at 750 GRAPE iterations, the optimisation goes from poor convergence in the control landscape at 170 time steps to proper convergence at 185 time steps.  

In conclusion, performing a distributed phase estimation algorithm using the architecture described in Section \ref{sec:hybrid_comp} is shown to achieve accurate results using GRAPE optimisation configured with a 700-800 GRAPE iteration range and 180-200 time step range (>90\%) when the flux qubit observes a high C-shunt value. 


\clearpage

\chapter{Final Insights and Concluding Remarks}\label{chapter:conclusions}
\section{Conclusion}\label{sec:concluding_remarks}

This investigation aimed to determine whether or not two quantum devices with distinct physical architectures can collaborate to solve a distributed computational task. Three sub-questions were proposed in order to answer this question.

The first sub-question, which explored the feasibility of distributing a quantum algorithm, was addressed by examining the quantum phase estimation algorithm in Chapter \ref{chapter:open_quantum}. Section \ref{sec:distributed_algorithm} demonstrated that, with the use of an $E_2$ gate, a distributed version of the phase estimation algorithm is possible. Section \ref{sec:qutip_implementation} outlined the circuits to distribute the phase estimation algorithm over two quantum computers, one comprising of three qubits and the other comprising of four qubits. The visualization of the QuTiP simulation without noise was presented at the conclusion of the chapter, wherein variations were made in both the number of counting qubits and the phase to be estimated (seen in figure \ref{fig:complete_no_noise}). The simulation proved that the distributed version of the phase estimation algorithm was as accurate as the local version. 

The second sub-question exploring the feasibility of connecting two quantum systems was addressed in Chapter \ref{chapter:hybrid_system}. The quantum hybrid system was broken down into three sub-system, the Rydberg atom system seen in Section \ref{sec:rydberg_qubit_intro}, the flux qubit system seen in Section \ref{sec:flux_qubit_intro} and the flux-resonator-atom system used to couple both architectures inspired by \cite{Yu_atom_flux_2017}, seen in Section \ref{sec:hybrid_comp}. 

Section \ref{sec:rydberg_atom_system} details the Rydberg atom configuration where the Rydberg atom system described in Section \ref{sec:rydberg_blockade} links to the configuration described by \citeauthor{Yu_atom_flux_2017}. The outlined method consists of an atom placed near a resonator (a co-planar waveguide) with an electric field plate and spherical capacitors capable of tuning the atom between a $22D_{5/2}$ state and an arbitrary $\Phi_{n = 20}$ state, which can couple and uncouple from the resonator. Furthermore, an explanation and a diagram was given on how this state could then couple to the other qubits to enable quantum state transfer between the atom and the flux qubit. It was concluded that for the Rydberg atom configuration to operate with the entire hybrid system, the three chosen qubit states were two computational states in the $\ket{0} = 5S_{1/2}, F=1$ and $\ket{1} = 5S_{1/2}, F=2$ as well as the excited state $\ket{r} = 70S_{1/2}$. Using ARC, this gave a dipole-dipole coefficient $C_3 = 32.45$ GHz $\mu \mathrm{m}^3$ and a van der Waals coefficient of $C_6 = 801.98$ GHz $\mu \mathrm{m}^6$ consequently leading to the chosen Rydberg blockade distance and the single gate operating distance to be $R_{b} =$ 3.5 $\mu$m and $R = $ 9 $\mu$m, respectively.  

Section \ref{sec:intro_flux} introduces the key elements describing a flux qubit. The Hamiltonian of the system is shown as well as the anharmoncity of the states depending on the magnetic frustration and the $\alpha$ component. Section \ref{sec:flux_system} uses this to describe the flux qubit in the context of the hybrid system, thus, its coupling to the resonator. The Hamiltonian from section \ref{sec:intro_flux} is modified due to the introduction of the C-shunt as well as by adding the interaction Hamiltonian between the resonator and flux qubit. The primary insight was that a trade-off exists between the $\alpha$ value, where lower $\alpha$ values effectively decrease flux fluctuation noise, while higher values increase the anharmonicity, a factor enabling the system to interact with the resonator. The value was ultimately chosen to be $\alpha = 0.8$ because of its high anharmonicity and its use in literature \cite{Yan_2016}. 

Section \ref{sec:hybrid_comp} explained the flux-resonator-atom connection using the co-planar waveguide. Following the steps described in \cite{Yu_atom_flux_2017}, Section \ref{sec:GHZ_hybrid} depicted the system's ability to evolve into a GHZ state with 0.93 fidelity in 17 nanoseconds as seen in figure \ref{fig:GHZ}. Chapter \ref{chapter:hybrid_system}, thus proved that connecting two quantum computers is feasible. 

In order to address the the third sub-question regarding the possibility of a distributed system distributing an algorithm with accuracy, a noise model had to be created on top of the driving Hamiltonians previously described in Chapter \ref{chapter:hybrid_system}. Chapter \ref{chapter:open_system} addresses the noise, in terms of Lindbladian equations, for the three different systems, the Rydberg system seen in Section \ref{sec:rydberg_noise}, the flux system noise seen in Section \ref{sec:flux_qubit_noise} and the flux-resonator-atom coupling device seen in Section \ref{sec:hybrid_noise}. The Linbladians described in Sections \ref{sec:rydberg_noise} and \ref{sec:flux_qubit_noise} are then used in Chapter \ref{chapter:GRAPE_sim} to model the noise of the flux and Rydberg atom computers, however, the Linbladian to establish the connection or GHZ gate for the flux-resonator-atom system was shown to have no effect in Section \ref{sec:hybrid_noise}. Thus, it was concluded that the noise would be neglected for the flux-resonator-atom system.  

Answering the third sub-question requires the use of GRAPE optimised gates, which were introduced in Section \ref{sec:intro_grape}. Section \ref{sec:process_tomography} described the implementation of the master equations involving the driving Hamiltonians (derived in Chapter \ref{chapter:hybrid_system}) and the Lindbladians (derived in Chapter \ref{chapter:open_system}) for use in the GRAPE algorithm. This section showed that both the Rydberg and flux systems were able to create single-qubit high fidelity ($F \approx 1$) gates when 500 GRAPE iterations and 50 time steps were used. For the double-qubit gates, the Rydberg atom managed to reach a high fidelity ($F \approx 1$) CNOT gate with 1000 GRAPE iterations and 200 time steps, but the flux qubit gate only managed to reach a fidelity of $F = 0.90$. These optimised gates were then used in Section \ref{sec:analysis_conclusion} to run the distributed phase estimation algorithm. The initial result (figure \ref{fig:zeta_10}) showed that the flux system was too noisy to complete a distributed phase estimation algorithm. However, changing the shunt effectiveness to $\zeta = 1000$, resulted in an estimated phase with an accuracy of over $90\%$ already at 190 time steps and 700 GRAPE iterations (configured to be $\varphi = 3/16$ or $\ket{0011}$ in the computational basis). The chapter concluded that a distributed quantum system is capable of executing a distributed quantum algorithm with high accuracy.

Overall, with the results having shown the possibility of answering the three sub-questions posed, the investigation has successfully concluded that two quantum devices with distinct physical architectures can collaborate to solve a distributed computational task. 

\section{Discussion on Limitations and Further Research}\label{sec:discussion}

The simulations of this investigation mainly used QuTiP. QuTiP proved to be a powerful tool due to the package's ability of easily simulating open quantum systems, more specifically, open quantum gates with configurable Lindbladians. However, due to the fact that QuTiP is not yet equipped to handle non-local processes, the construction of the non-local gates were built using local assumptions. A consequence of non-locality in distributed quantum computing is asynchronicity which occurs when two operations are not accomplished sequentially. The $E_2$ gate was simply assumed to be two sequential operations with an immediate response from both parties over a classical network. Experimentally, this would not work using current classical guidelines. Both \citeauthor{vanbrandwijk2016asynchronous} have shown that the TCP/IP structure of classical networks would require extra quantum operations from both parties \cite{vanbrandwijk2016asynchronous}. This would far exceed the nanosecond timescale gate creation time set in section \ref{sec:GHZ_hybrid}, and would ultimately lead to the noise described in Section \ref{sec:hybrid_noise} to be non-negligible \cite{Yu_atom_flux_2017}.

Investigating the TCP/IP effect on the $E_2$ gate requires the use of a different framework for the simulations. QooSim \cite{vanbrandwijk2016asynchronous} addresses the TCP/IP issue, but cannot simulate noisy circuits. On the other hand, different methods of establishing quantum networks are being studied. For example, \citeauthor{virtual_path} proposed a protocol in which an additional \textit{virtual path} is established between entangled qubits or nodes connected to repeaters that are responsible for optimising the fidelity by iterating through links between the qubits \cite{virtual_path}. Further research must evaluate the decoherence loss on hardware due to classical networks and whether or not a different strategy for classical communication between quantum computers is worth implementing. 

The proposed quantum devices to couple in this investigation were Rydberg atom computers and superconducting computers. Overall, it is unlikely that a specific qubit will be standardised throughout quantum computing \cite{DiVincenzo_2000}. Thus, the selection of both quantum devices will likely remain valid for the foreseeable future. However, implementing a coupling between both of these devices can be done in a variety of different forms. In hybrid qubit coupling, there exists two main methods of entangling two hybrid qubits, \textit{direct} or \textit{indirect} coupling \cite{Xiang_2013}. Direct coupling refers to a coupling in which the magnetic and electric fields resonate at similar resonances meaning that two qubits physically next to each other can couple or entangle without the need of an apparatus in the middle. Indirect coupling is when an apparatus mitigates the resonances of both qubits, this is seen in this study by the co-planar waveguide. Direct coupling leads to less decoherence and is, thus, more desired but is sometimes impossible.

Focusing on the system proposed in this investigation, some limitations of the system need to be further investigated. The indirect coupling utilized comes from \cite{Yu_atom_flux_2017} whereby the $22D_{5/2}$ state was researched numerically as the coupling state between a flux qubit-resonator-atom system. Other system architectures may couple better than the system introduced in this investigation, for example, \citeauthor{Thiele_2014} experimentally showed that a cloud of meta-stable helium atoms are potential candidates for coupling to superconducting surfaces \cite{Thiele_2014} because they are not affected by stray fields generated by the superconducting surface. Although \cite{Yu_atom_flux_2017} theorized that the hybrid entanglement would not be affected by a single flux qubit's stray field, an experiment would need to be conducted to verify if that remains valid. 

On top of this, for distributed quantum computation to work in its fullest extent, the quantum computers should also be able to communicate over longer distances. The current scenario requires both computers to be approximately in the same vicinity. It has been shown that qubits can be transferred via fibre optic cables by indirectly opto-mechanically coupling phonons to photons \cite{Fiaschi_2021,Lago_Rivera_2023}. This is already partially the case within the co-planar wave guide used in the investigation, however, further optimisation for photon trapping within the resonator can be accomplished \cite{Kellner_2023}. To do so, each system would need a single qubit to be attached to an opto-mechanical resonator and these resonators have to be attached to a transmission or optical fiber cable.  

ARC was utilized to calculate the parameters in the Rydberg atom qubit configuration. However, ARC does not take into consideration the calculation of hyperfine states. As seen in figure \ref{fig:forster_atom}(a), it was assumed that the RF transition between a $\Phi_{n=20}$ state and a $5S_{1/2} F=1$ state was possible. This would need to be further researched experimentally. It may be the case that an auxiliary state needs to be provided, which would lead to an increase in decoherence. 

The flux qubit was shown to suffer more decoherence than the Rydberg atom qubit. However, because the hybrid system from \cite{Yu_atom_flux_2017} was chosen, it put further strain on the flux qubit. For the system to work, the anharmonic energy levels in figure \ref{fig:flux_system}(d) are required in order for the system to properly couple to the resonator without affecting the rest of the flux qubit system. C-shunt flux qubits operate best (with an approximate decoherence of 100MHz) when $\alpha < 0.5$. Unfortunately, this cannot be done on the system used because there is a large drop in anharmonicity when $\alpha$ drops below 0.5 as seen in figure \ref{fig:phase_f}(b). This would mean that the $\ket{R}$ and $\ket{L}$ states would disappear and the flux qubit system would constantly be coupled to the resonator, thus it would generalize the states to a $\ket{0}$ and $\ket{1}$ energy state, which forbids the construction of the GHZ state seen Section \ref{sec:GHZ_hybrid}.

\begin{figure}[!htb]
\centering
    \includegraphics[width=1\textwidth]{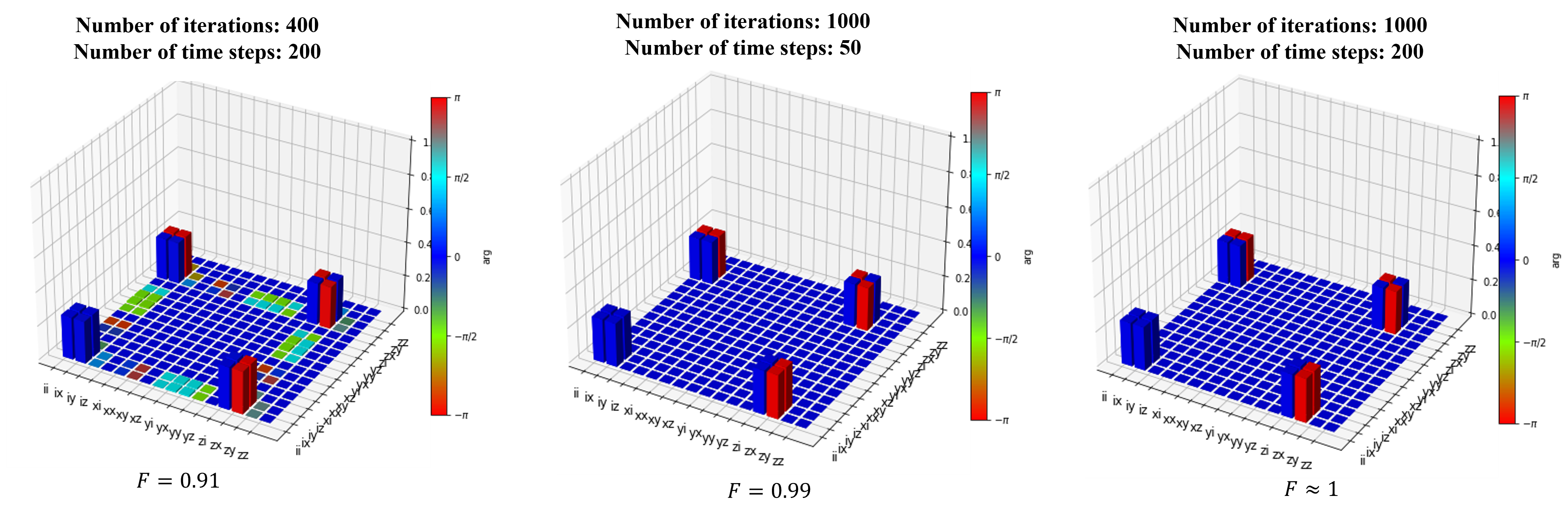}
    \caption{Quantum process tomography evaluating the construction of the CNOT gate for the flux qubit with $\alpha = 0.49$ GRAPE iterations and changes in time step number. Each constructed gate contains the calculated fidelity and the axes represent $i$ as the identity $I$ and $x,y,z$ as the Pauli $\sigma_x,\sigma_y,\sigma_z$, respectively. The notation $ii$ refers to a tensor product $I \otimes I$.}
    \label{fig:alpha_0.49}
\end{figure}

\begin{figure}[!htb]
\centering
    \includegraphics[width=1\textwidth]{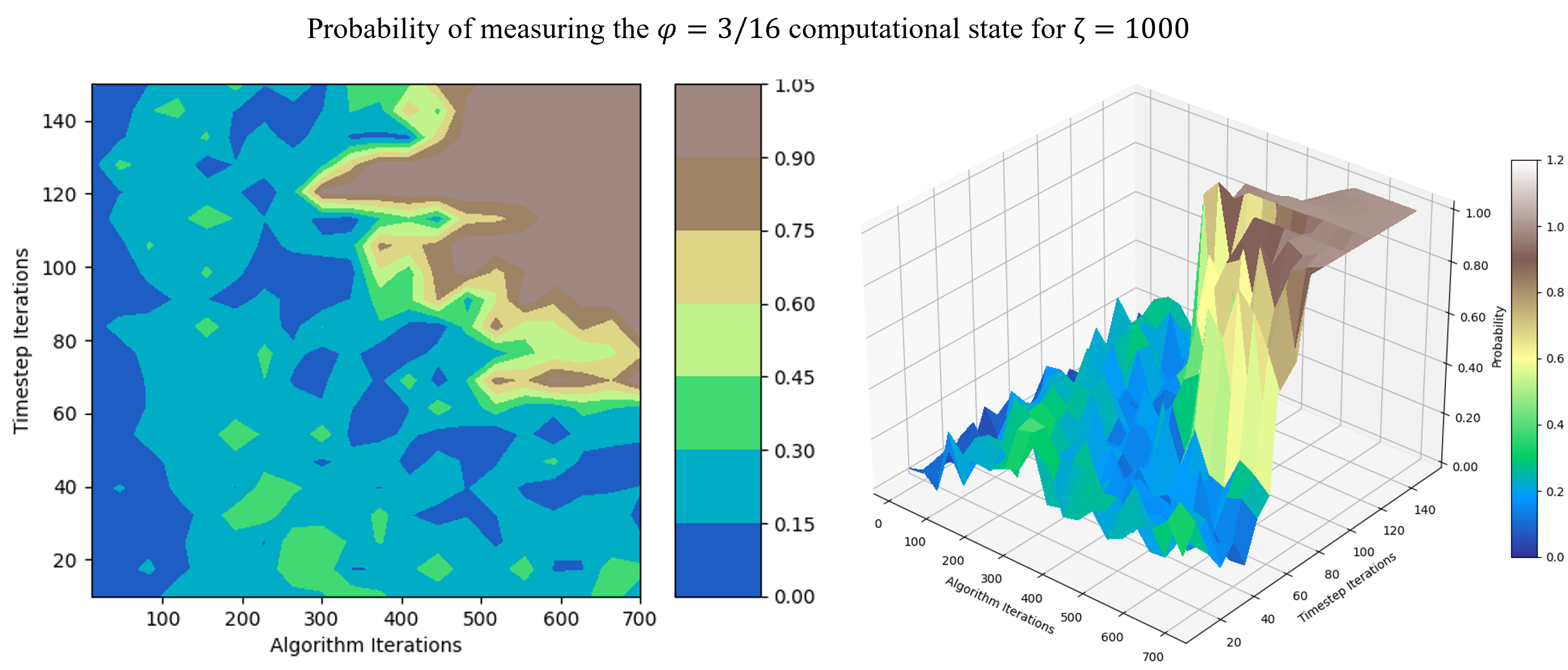}
    \caption{Probability of measuring the $\varphi =3/16$ or $0011$ computational state after 10 executions of the phase estimation algorithm on the noisy hybrid circuit with the flux qubit modifed per total GRAPE iteration number per time step number for a shunt factor of $\zeta = $ 1000 and $\alpha = 0.49$. Depicted on the left is a graph showing the result in 2D while on the right is the same result on a 3D plane.}
    \label{fig:alpha_0.49_total}
\end{figure}

In order to show the effect a C-shunt flux qubit with $\alpha = 0.49$ has on noise, the process tomography for the construction of a CNOT gate is visualized in figure \ref{fig:alpha_0.49}. The gate constructed using GRAPE with 1000 GRAPE iterations and 200 time steps has a 0.09 higher fidelity than the gate obtained when $\alpha = 0.8$, depicted in figure \ref{fig:cnot_gate_noise}. Utilizing the optimised gates with a flux qubit $\alpha = 0.49$ in the distributed phase algorithm is seen in figure \ref{fig:alpha_0.49_total}. As observed, the distributed phase estimation algorithm, estimates the correct state with 90\% accuracy, at the 400 iteration mark with 120 time steps and at the 500 iteration mark with 60 time steps. This occurs in less time steps/GRAPE iterations when compared to the $\alpha = 0.8$ case (seen in figure \ref{fig:zeta_1000}). Using $\alpha = 0.49$, however, is not theoretically feasible since it would introduce further noise effects caused by lack of anharmonicity. Thus, a method in which the connection does not rely on the qubit's anharmonicity must be researched in order to utilize C-shunt flux qubits with $\alpha < 0.5$ values. 

Although the GRAPE algorithm was mainly used in this investigation, it is worth noting that other algorithms for optimal control exist. The two notable ones are Krotov \cite{Krotov} and CRAB \cite{Doria_2011}. Krotov was initially tested but it was far too computationally expensive when optimising pulses for a multi-qubit gate. The main strength Krotov possesses is its ability to converge to small fidelity errors when optimising systems comprised of many qubits \cite{Krotov}. CRAB on the other hand, did not converge at the given fidelity error rate at all because it is a gradient-free optimisation and not robust enough to minimise the fidelity using the controls. \citeauthor{Krotov} states that GRAPE must be used when the control parameters are discrete and the derivative of error rate with respect to the control parameters are known and Hermitian while Krotov is used when the control parameters are near-continuous without the restriction of Hermitian Hamiltonians \cite{Krotov}. In the case used in this research, it is apparent that the best choice was indeed GRAPE.

Instead of using optimal quantum control, one can use variational quantum algorithms (VQAs) to optimise quantum circuits. Instead of pulsing a quantum gate at pulse level, one could replace this with a gate or series of gates capable of manipulating the quantum state to match the process of a perfect quantum gate. Although GRAPE cannot be utilized in VQAs, other algorithms or machine learning techniques offer a less intrusive method of engineering gates \cite{Magann_2021}. This would defeat the need to offer custom control parameters assuming both systems can generate Pauli basis states. Variational quantum algorithms and quantum optimal control can be used in parallel, offering an interesting expansion of this study \cite{Magann_2021}. Integrating VQAs through machine learning could simulate circuit noise and subsequently rectify it, thus enhancing the examination of the original assumption, which solely introduced noise during qubit manipulation.

%Choose a good bibliography style, plain would do often, but these might be nice too
\bibliographystyle{plainnat}
\bibliography{thesis}

\clearpage

\appendix
\addcontentsline{toc}{chapter}{Appendix}

\chapter{Quantum Logic Gates}\label{chapter:logic_gate_appendix}

\section{Single Qubit Gates}

\begin{figure}[h]
\centering
    \includegraphics[width=0.4\textwidth]{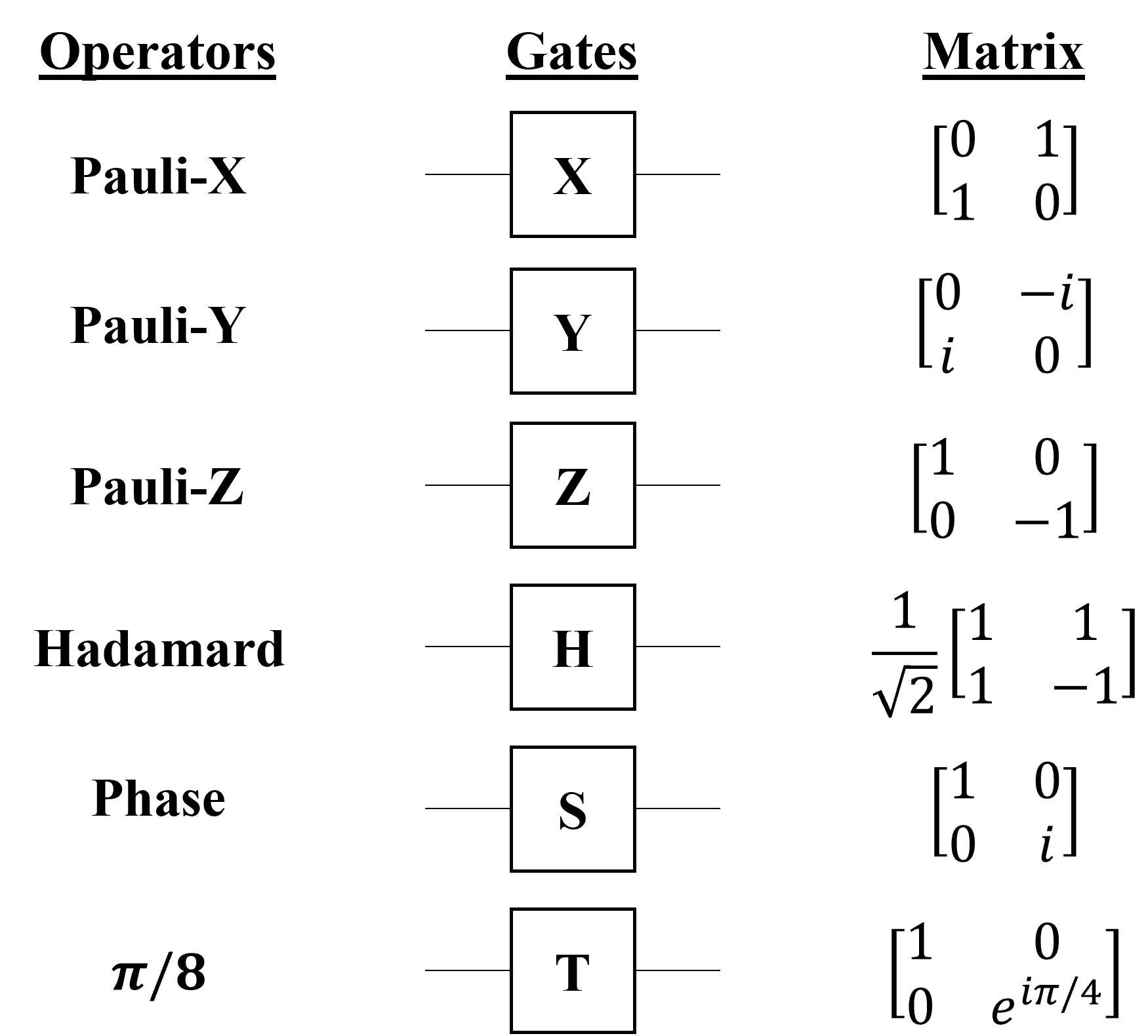}
\end{figure}

\section{Multiple Qubit Gates}

\begin{figure}[h]
\centering
    \includegraphics[width=0.9\textwidth]{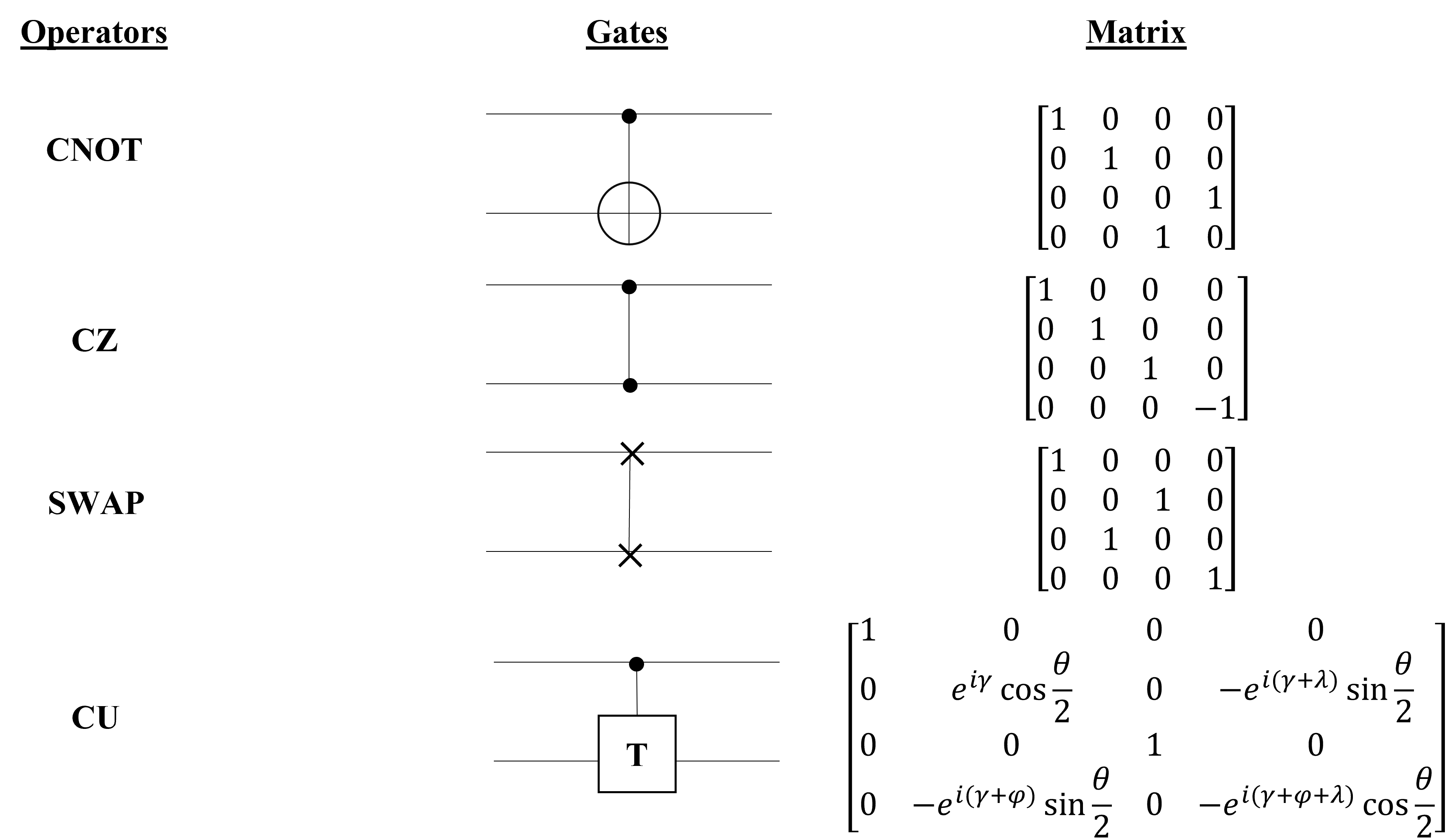}
\end{figure}

\chapter{Proofs and Concepts in Quantum Computation}

\section{Quantum Fourier Transform Product Representation} \label{sec:fourier_transform_product_proof}

The equivalence of the product representation seen in Eq. \ref{eq:fourier_product_representation} and the formal definition seen in Eq. \ref{eq:inverse_fourier} follows from,
\begin{flalign}
\ket{j}  & \rightarrow \frac{1}{2^{n/2}}\sum^{2^n-1}_{k=0}e^{2\pi i jk/2^n}\ket{k} \\
    & = \frac{1}{2^{n/2}} \sum^1_{k_1=0} \cdots \sum^1_{k_n=0}e^{2\pi i j \big(\sum^n_{l=1}k_l 2^{-l}) }\ket{k_1 \cdots k_n} \\
    & = \frac{1}{2^{n/2}} \sum^1_{k_1=0} \cdots \sum^{1}_{k_n=0} \otimes^n_{l=\mathbb{1}}e^{2\pi i jk_l2^{-l}}\ket{k_l}\\
    & = \frac{1}{2^{n/2}} \otimes^n_{l=1} \sum^1_{k_l=0} e^{2\pi i j k_l 2^-l}\ket{k_l} \\
    & = \frac{1}{2^{n/2}} \otimes^n_{l=1} \Big[ \ket{0} + e^{2\pi i j 2^{-l}} \ket{1} \Big] \\
    & = \frac{\big(\ket{0}+e^{2\pi i 0.j_n}\ket{1}\big)\big( \ket{0} + e^{2\pi i0.j_{n-1} j_n}\ket{1}\big) \cdots \big(\ket{0} + e^{2\pi i 0.j_1j_2 \cdots j_n}\ket{1} \big)}{2^{n/2}}.
\end{flalign}

\section{Introduction to Open Quantum Systems and the Lindbladian Derivation}\label{sec:open_system}
The first postulated quantum equation was introduced by Erwan Schr\"{o}dinger which describes the behaviour of an isolated or closed system. By definition, a closed system is one that does not exchange information with another system. Assuming the system is in a pure state $\ket{\psi(t)} \in \mathcal{H}$ at time $t$ where $\mathcal{H}$ describes the Hilbert space of the system, the time evolution is,
\begin{equation}
    \frac{d}{dt} \ket{\psi(t)} = -\frac{i}{\hbar}H(t)\ket{\psi(t)}
\end{equation}
where $H(t)$ is the Hamiltonian operator of the system. An important property of of the Schr\"{o}dinger equation is that it does not change the norm of the states,
\begin{dmath}\label{eq:schrodinger_proof}
   \frac{d}{dt}\bra{\psi(t)}\ket{\psi(t)}=\bigg(\frac{d\bra{\psi(t)}}{dt}\bigg)\ket{\psi(t)} + \bra{\psi(t)}\bigg(\frac{d\ket{\psi(t)}}{dt}\bigg) = i\bra{\psi(t)}H^{\dag}(t)\ket{\psi(t)}-i\bra{\psi(t)}H(t)\ket{\psi(t)} = 0.
\end{dmath}
This is due to the Hamiltonian being self-adjoint, $H(t) = H^\dag(t)$ \cite{Rivas_2012}. Exploiting the linearity of the Schr\"{o}dinger equation, one can describe its solution using an evolution family, $\ket{\psi(t)}=U(t-t_0)\ket{\psi(t_0)}$, in which $U(t_0,t_0) = \mathbb{1}$.
Equation \ref{eq:schrodinger_proof} implies that the evolution operator is isometric. Thus, if the system possesses finite dimensions, the system can be considered a unitary operator \cite{Rivas_2012}. The specific form of the evolution operator relies on the properties of the system Hamiltonian. In the simplest case, whereby the Hamiltonian is time independent, the solution to Schr\"{o}dinger's equation is obtained using,
\begin{equation}
    U(t,t_o) = e^{-\frac{i}{\hbar}(t-t_0)H},
\end{equation}
however, for a time-dependent system, the unitary evolves as a Dyson expansion \cite{joachain_1983},
\begin{equation}
    U(t,t_0) = \mathcal{T} e^{\int^t_{t_0}H(t')dt'},
\end{equation}
where $\mathcal{T}$ is the time ordering operator. From this one can deduce that due to the self-adjoint nature $H(t)$ possesses, $-H(t)$ is also self adjoint suggesting that the every evolution has a physical inverse. Furthermore, the unitary also possesses an inverse, for example in the time-independent case, $U(t_1,t_0) = U(t-t_0)$. This can be generalized to $U^{-1}(\tau)=U(-\tau)=U^\dag(\tau)$ such that $U^{-1}(\tau) U(\tau) = U(\tau) U^{-1}(\tau) = \mathbb{1}$. For time dependent Hamiltonians, the subtlety is confined within the evolution operator's definition, $U(t,s)$. The inverse is shown to be unitary by  $U^{-1}(s,t) = U(s,t)=U^\dag(s,t)$ \cite{Rivas_2012}.

Inherited from classical computing, a quantum circuit is an abstracted model of a set of operations performed on a single or many two level states. The execution of a circuit on a quantum state is given by,
\begin{equation} \label{eq:unitary_gates_example}
    \ket{\psi_f} = U_k U_{K-1}\cdots U_2U_1\ket{\psi_i},
\end{equation}
where $\ket{\psi_f}$ and $\ket{\psi_i}$ are the final and initial state respectively and $U_k$ with $k \in {1,2,\cdots,K}$ the quantum gates. Simulating quantum circuits is implemented by representing each the gates and states of the system as complex matrices and vectors. This is referred to as a gate level quantum circuit description because it is a quantum algorithm at an abstract level \cite{Viamontes}. The set of unitaries described in eq. (\ref{eq:unitary_gates_example}) can be thought of a process multiplied to a state, which in turn applies the process to the state. 

In order to understand how unitaries evolve a system it is first important to understand time dependent Hamiltonians. One can define a Hamiltonian as a function of time $t$ as a set of two Hamiltonians. A non-controllable drift Hamiltonian $H_d$, which describes the system evolution over time and a control Hamiltonian $H_c$, which can control the evolution of the system in time \cite{Distributed_Leung}. This may be depicted as,
\begin{equation}\label{eq:drive_control}
    H(t) = H_d(t) + H_c(t) = H_d(t) + \sum_j c_j(t)H_j
\end{equation}
where $H_j$ describes the effects of physical controls on the system modulated by time-dependent control coefficients $c_j(t)$ used to realize desired unitary gates \cite{qutip_package,Distributed_Leung}. The unitary is a solution of the Schr\"{o}dinger equation,
\begin{equation} \label{unitary_schrodinger}
    i \hbar \frac{\partial U(t)}{\partial t} = H(t)U(t).
\end{equation}
Implementing the desired unitaries can be obtained by choosing $H(t)$, equally, the choice of the solver depends on the parametrization of the control coefficients $c_j(t)$. Parameters within $c_j(t)$ can be determined through theoretical models or through cost optimisation \cite{qutip_package}.   

\begin{figure}[t] 
\centering
    \includegraphics[width=0.6\textwidth]{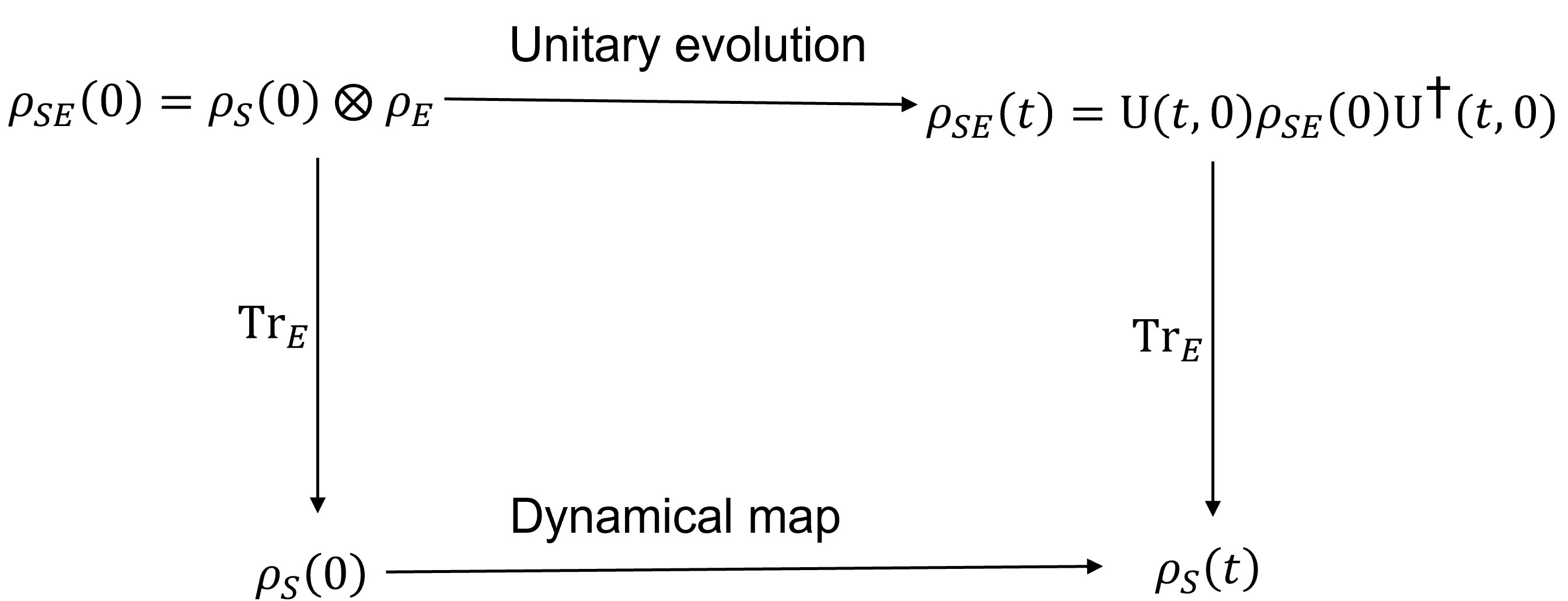}
    \caption{Diagram demonstrating how tracing out the environment from the unitary evolution of the total state leads to $\mathcal{E}(t,0)$ \cite{Rivas_2012,open_quantum_Heinz,phd_goerz}.}
    \label{fig:dynamic_map}
\end{figure}

In order to evolve the Hamiltonian through the unitaries one needs to define the respective density matrix. A density matrix is a generalization of a wave function which describes the physical state of a quantum system. Density formalism allows a mathematical representation of decoherence on a state vector $\ket{\psi}$. A density matrix is defined as,
\begin{equation}
    \rho = \sum_j p_j \ket{\rho_j}\bra{\rho_j}.
\end{equation}
A density matrix is said to describe a pure state if the sum has a single term. If not, it is called a mixed state which describes a classical ensemble of quantum states. The evolution of $\rho$ in time $t$ can be depicted using unitaries, $\rho(t_1) = U(t_1,t_0)\rho(t_0)U^\dag(t_1,t_0)$ \cite{Rivas_2012}. For a pure state, the density matrix is equivalent to the state space vector in Hilbert Space, i.e $\rho = \ket{\psi}\bra{\psi}$ and can be substituted into the Schr\"{o}dinger equation, forming the so called Liouville-von Nuemann equation \cite{phd_goerz},
\begin{equation} \label{eq:liouville}
    i\hbar\frac{\partial}{\partial t}\rho = [H,\rho].
\end{equation}
The density matrix is also a key parameter to the analysis of quantum systems. For example, considering a larger system composed by two quantum subsystems in Hilbert space $\mathcal{H} = \mathcal{H_A} + \mathcal{H_B}$, the density matrix of subsystem A is given by \cite{nielsen_chuang_2021,Rivas_2012},
\begin{equation}
    \rho_A = \mathrm{Tr}_B(\rho).
\end{equation} 
Provided that the total evolution is not factorized $U(t_1,t_0) = U_A(t_1,t_0)\otimes U_B(t_1,t_0)$, both quantum systems $A$ and $B$ are interchanging information, ergo converting the system to an open system \cite{Rivas_2012}. Even in such system, the density matrix of an independent subsystem at time $t$ can still be acquired, for example, for system $A$,
\begin{equation}\label{eq:dynamic_map_example}
    \rho_A(t_1) = \mathrm{Tr}_B[U(t_1,t_0)\rho(t_0)U^\dag(t_1,t_0)].
\end{equation}
This is a so called dynamical map \cite{phd_goerz,open_quantum_Heinz}. In order to fully establish a true open quantum system, the environment and the quantum system must entangle over time. Allowing $\rho_S$ and $\rho_E$ to depict the qubit and the environment density matrices, respectively, the full quantum system can be initialized as a separable state $\rho_S \otimes \rho_E$ that entangles over later time $t$ \cite{phd_goerz}. Knowing that both the time evolution of in this composite Hilbert space is unitary with eq. \ref{eq:liouville} and that the state of the qubit can be obtained by taking a partial trace over the environment, the equation of motion can be derived,
\begin{equation}\label{eq:equation_motion_density}
    i\hbar\frac{\partial}{\partial t}\rho_S = \mathrm{Tr}_E[H,\rho].
\end{equation}
The solution of $\rho_S(t)$ is thus the dynamical map \cite{phd_goerz},
\begin{equation}
    \rho_S(t) = \mathcal{E}(t,0)\rho_S(0) = \mathrm{Tr}_E\Big[U(t,0)\rho_{SE}(0)U^\dag(t,0)\Big].
\end{equation}
Fig. \ref{fig:dynamic_map} details the relationship between the unitary evolution and the dynamical map. 
The operation of a dynamical map is considered independent of any prior evolution in the total Hilbert space when the separation or distinguishability between two initial states is assumed to not grow during its evolution \cite{open_quantum_Heinz}. If this is satisfied, a dynamical map can be split into two time intervals $t_1$ and $t_2$ as,
\begin{equation}\label{eq:dynamic_map_independence}
    \mathcal{E}(t_1+t_2,0) = \mathcal{E}(t_2,0)\mathcal{E}(t_1,0).
\end{equation}
Eq. \ref{eq:dynamic_map_independence} can be rewritten into the following form \cite{open_quantum_Heinz},
\begin{equation}
    \mathcal{E}(t,0) = e^{-\frac{i}{\hbar}\mathcal{L}t},
\end{equation}
where $\frac{i}{\hbar}$ has been factored out of $\mathcal{L}$, a generator that follows $\mathcal{L}(\cdot) = -i[H(t),\cdot]$ \cite{Rivas_2012}. Substituting this relation into Eq. \ref{eq:equation_motion_density} leads to \cite{phd_goerz},
\begin{equation}
    i\hbar\frac{\partial}{\partial t}\rho_S(t) = \mathcal{L}[\rho_S(t)] \equiv \lim_{t\to 0}\frac{1}{t}(\mathcal{E}(t,0)\rho_S-\rho_S).
\end{equation}
Choosing a complete basis $\{R_i\}$ of $N_S^2$ operators of dimensions $N_S$ in Hilbert space such that $\Big[ R_{N^2_S} \Big] = 1$ and $\mathrm{Tr}[R_i]=0$ \cite{phd_goerz}. Using the expansion of Kraus operators \cite{open_quantum_Heinz}, the equation of motion can then be reduced to,
\begin{equation}
    \mathcal{L}[\rho_S] = [\mathrm{H},\rho_S] + i\sum^{N_S^2-1}_{i,j=1} a_{ij}\bigg( R_i\rho_S R_j^\dag - \frac{1}{2}\Big\{ R_j^\dag R_i,\rho_S \Big\}\bigg),
\end{equation}
where $\{\cdot,\cdot\}$ denotes an anti-commutator, $\mathrm{H}$ is a Hermitian operator constructed from $\{ R_i \}$ and $a_{ij}$ is a given coefficient matrix. Rotating the basis to a new basis $\{ A_k \}$ such that the coefficient matrix is diagonalized with eigenvalues $\gamma_k$, one arrives at the master equation in Lindblad form \cite{lindblad_1976},

\begin{equation}
    \mathcal{L}[\rho_S] = [\mathrm{H},\rho_S] + i\sum_k \gamma_k \bigg(A_k \rho_S A_k^\dag - \frac{1}{2} \Big\{ A_k^\dag A_k,\rho_S\Big \}\bigg),
\end{equation}

where $\{A_k\}$ are the \textit{Lindblad operators} representing the coupling of the system to the environment and thus the dissipative process \cite{nielsen_chuang_2021,lindblad_1976}. The Lindbladian is not a unitary, however, it does still satisfy the property of being trace-preserving and positive for any initial condition. In this investigation, the $[H,\rho_S]$ term will be omitted to only describe the noise dynamics of a system, thus,
\begin{equation}
    \mathcal{L}[\rho_S] = i\sum_k \gamma_k \bigg(A_k \rho_S A_k^\dag - \frac{1}{2} \Big\{ A_k^\dag A_k,\rho_S\Big \}\bigg),
\end{equation}

\section{Measuring Expected Values with the $\chi$ Matrix for a Hadamard Gate} \label{sec:chi_hadamard}

The final state, $\varepsilon(t)$, can be expressed as, 
\begin{equation}
    \varepsilon(t) = \rho_{out} \sum_{m,n = 0} \tilde{E}_m \rho \tilde{E}_n^\dag \chi_{mn} =U\rho_{in},
\end{equation}
where $U$ is a propagator for density matrix in superoperator form. One can then identify,
\begin{equation}
    U = \sum_{mn}\chi_{mn}\tilde{E}_m \tilde{E}_n^\dag
\end{equation}
The interest lies in forming the $\chi_{mn}$ matrix originating from a process, which is a $N^4 \times N^4$ matrix, where $N$ is the amount of qubits involved. The product $\tilde{E}_m \tilde{E}_n^\dag$ can be written as a $N^4 \times N^4$ matrix called $M$. This means,
\begin{equation}
    U_I = \sum^{N^4}_J M_{IJ}\chi_J,
\end{equation}
with the solution for $\chi$,
\begin{equation}
    \chi = M^{-1}U.
\end{equation}
For an ideal Hadamard gate, the $U$ superoperator is $U = (G_H \otimes I) \times (I \otimes G_H^\dag)$ where $G_H$ is the Hadamard gate seen below,
\begin{equation}
    \frac{1}{\sqrt(2)} \begin{bmatrix}
        1 & 1 \\
        1 & -1
    \end{bmatrix}.
\end{equation}
Next, the list of operators that define the basis $\{\tilde{E}_i\}$ are defined in the form of list operators for each composite system. For the Hadamard gate, the Pauli gates are used to define the basis, thus one can define a matrix $\Pi$, such that,
\begin{equation}
    \Pi = \begin{bmatrix}
        I & X & Y & Z
    \end{bmatrix} \otimes \begin{bmatrix}
        I \\
        X \\ 
        Y \\
        Z
    \end{bmatrix} = \begin{bmatrix}
    II & IX & IY & IZ \\
    XI & XX & XY & XZ \\
    YI & YX & YY & YZ \\
    ZI & ZX & ZY & ZZ \\
    \end{bmatrix}.
\end{equation}
The $\chi$ matrix is computed by,
\begin{equation}
    \chi = \frac{1}{N^2}\begin{bmatrix}
        \mathrm{Tr}(II \times U) && \mathrm{Tr}(IX \times U) &&
        \mathrm{Tr}(IY \times U) && \mathrm{Tr}(IZ \times U) \\
        \mathrm{Tr}(XI \times U) && \mathrm{Tr}(XX \times U) &&
        \mathrm{Tr}(XY \times U) && \mathrm{Tr}(XZ \times U) \\
        \mathrm{Tr}(YI \times U) && \mathrm{Tr}(YX \times U) &&
        \mathrm{Tr}(YY \times U) && \mathrm{Tr}(YZ \times U) \\
        \mathrm{Tr}(ZI \times U) && \mathrm{Tr}(ZX \times U) &&
        \mathrm{Tr}(ZY \times U) && \mathrm{Tr}(ZZ \times U)
        \end{bmatrix}.
\end{equation}
For the ideal Hadamard gate this equals,
\begin{equation}
    \chi_{Hadamard} = \begin{bmatrix}   
        0 & 0 & 0 & 0 \\
        0 & 0.5\cdot e^{i\cdot0} & 0 & 0.5\cdot e^{i\cdot0} \\
        0 & 0 & 0 & 0 \\
        0 & 0,5\cdot e^{i\cdot0} & 0 & 0.5\cdot e^{i\cdot0} \\
    \end{bmatrix},
\end{equation}
which is seen plotted in figure \ref{fig:hadamard_gate_noise} for the $F\approx 1$ cases. The colour in of the elements depicts the phase, $\theta$, of the phase factor ($e^{i\theta}$) multiplied to the $\chi_{mn}$ element. 
\chapter{Code \& Functions}

\section{Distributed Phase Estimation}\label{sec:distributed_phase_code}

\begin{tcolorbox}[title=\textbf{Apply initial hadamard gates function}]
\begin{mintedbox}{python}
def perform_hadamards(circuit):
    print("Performing Hadamards:")
    #Add Hadamard to Flux qubit,
    for i in range(0,N_rydberg-1):
        circuit.add_gate("SNOT", targets = i)
        print("q"+str(i))
    #Add Hadamard to Rydberg qubits
    for i in range(N_rydberg+1,N_rydberg+N_flux-1):
        print("q"+str(i))
        circuit.add_gate("SNOT", targets = i)   
    return circuit
\end{mintedbox}
\end{tcolorbox}

\begin{tcolorbox}[title=\textbf{Establish the E2 gate (GHZ state) between the two systems}]
\begin{mintedbox}{python}
def initialize_E_gate(circuit,state):
    if (len(circuit.gates) > 0):
        [circuit,state] = get_state(circuit,N_rydberg+N_flux,state)
    # Flux circuit initialisation
    circuit = QubitCircuit(N=N_flux+N_rydberg)
    circuit.add_gate("SNOT",targets=N_rydberg-1)
    circuit.add_gate("CNOT",targets=N_rydberg, controls=N_rydberg-1)
    [circuit,state] = get_state(circuit,N_rydberg+N_flux,state)
    return [circuit,state] 
\end{mintedbox}
\end{tcolorbox}

\begin{tcolorbox}[title=\textbf{Implementation of the CPHASE gate}]
\begin{mintedbox}{python}
 def cphase_gate(circuit,control,target,angle):
    if (abs(angle) > 0):
        circuit.add_gate("RZ",targets = target, arg_value = angle/2)
        circuit.add_gate("RZ",targets = control, arg_value = angle/2)
        circuit.add_gate("CNOT",targets = target, controls = control)
        circuit.add_gate("RZ",targets = target, arg_value = -angle/2)
        circuit.add_gate("CNOT",targets = target, controls = control)

    return circuit
\end{mintedbox}
\end{tcolorbox}

\begin{tcolorbox}[title=\textbf{Performing a measurement on a specific qubit}]
\begin{mintedbox}{python}
 def perform_measurement(circuit,qubit,state):
    [circuit,state] = get_state(circuit,N_rydberg+N_flux,state)
    circuit = QubitCircuit(N=N_flux+N_rydberg)
    measurement_basis = []
    measurement_matrix = qeye(2)
    for i in range(N_flux + N_rydberg-1):
        if(i == (qubit-1)):
            measurement_matrix = tensor(measurement_matrix, basis(2,0) * basis(2,0).dag())
        else:
            measurement_matrix = tensor(measurement_matrix,qeye(2))
    measurement_basis.append(measurement_matrix)
    measurement_matrix = qeye(2)
    for i in range(N_flux + N_rydberg-1):
        if(i == (qubit-1)):
            measurement_matrix = tensor(measurement_matrix,basis(2,1) * basis(2,1).dag())
        else:
            measurement_matrix = tensor(measurement_matrix,qeye(2))
    measurement_basis.append(measurement_matrix)
    measured_flux_basis = qutip.measurement.measure_povm(state,measurement_basis)[-1]
    if ((measured_flux_basis.dag() * measurement_basis[1] * measured_flux_basis)[0][0][0] > 0.9):
        measured_flux = 1
        return [circuit,measured_flux,measured_flux_basis]
    measured_flux = 0
    return [circuit,measured_flux,measured_flux_basis]
\end{mintedbox}
\end{tcolorbox}

\begin{tcolorbox}[title=\textbf{Applying the non-local controlled phase gates seen in figure \ref{fig:distributed_CP}}]
\begin{mintedbox}{python}
def perform_non_local_phase(circuit,angle,control_flux,target_flux,
                            control_rydberg,target_rydberg,state):
    print("CNOT:")
    circuit.add_gate("CNOT", targets = target_rydberg, controls = control_rydberg)
    measurement_rydberg_results = perform_measurement(circuit,target_rydberg,state)
    measurement_rydberg = measurement_rydberg_results[1]
    circuit = measurement_rydberg_results[0]
    state = measurement_rydberg_results[2]
    if measurement_rydberg:
        print("Applying X gates to "+ "q"+ str(target_rydberg) + " and " + "q"+ str(control_flux))
        circuit.add_gate("X", targets = control_flux)
        circuit.add_gate("X", targets = target_rydberg)
    print("CPHASE from q" + str(control_flux) + " to q" + str(target_flux) + " with angle 2pi*" + str(angle))
    circuit = cphase_gate(circuit,control_flux,target_flux,
                            (2*np.pi)*angle)
    circuit.add_gate("SNOT", targets = control_flux)
    measurement_flux_results = perform_measurement(circuit,control_flux,state)
    measurement_flux = measurement_flux_results[1]
    circuit = measurement_flux_results[0]
    state = measurement_flux_results[2]
    if measurement_flux:
        print("Applying Z gate and X gate to: " + "q"+ str(control_rydberg) + " and " + "q"+ str(control_flux))
        circuit.add_gate("Z", targets = control_rydberg)
        circuit.add_gate("X", targets = control_flux)
        [circuit,state] = get_state(circuit,N_rydberg+N_flux,state)
    
    return [circuit,state]

\end{mintedbox}
\end{tcolorbox}

\begin{tcolorbox}[title=\textbf{Applying the DCP circuit seen in figure \ref{fig:distributed_CP}}]
\begin{mintedbox}{python}
def pulse_phase_sequence(circuit,angle,state):
    print("Pulse phase sequences:")
    for i in range(N_rydberg-1):
        phase_angle = angle*(2**i)
        [circuit,state] = initialize_E_gate(circuit,state)
        control = i
        target = N_rydberg-1
        print("Perform phasegate from " + "q"+str(control)+"_r to q" + str(N_flux-1)+"_f with angle:" )
        print(str(2)+"π"+str(angle)+"*"+str((2**i)))
        [circuit,state] = perform_non_local_phase(circuit,phase_angle,N_rydberg,
                            N_rydberg+N_flux-1,control,target,state)
    for i in range(N_rydberg+1,N_rydberg+N_flux-1):
        phase_angle = angle*(2**(i-2))
        print("Perform phasegate from " + "q"+str(i-N_rydberg)+"_f " + "to " + "q"+str(N_flux-1)+"_f" + " with angle:" )
        print(str(2)+"π"+str(angle)+"*"+str((2**(i-2))))
        circuit = cphase_gate(circuit,i,N_flux+N_rydberg-1,
                                (2*np.pi)*phase_angle)
    [circuit,state] = get_state(circuit,N_rydberg+N_flux,state)

    return [circuit,state]
\end{mintedbox}
\end{tcolorbox}

\begin{tcolorbox}[title=\textbf{Performing the non-local section of the inverse Fourier transform circuit seen in figure \ref{fig:distributed_QFT}}]
\begin{mintedbox}{python}
def perform_non_local_iqft(circuit, n, counting_qubit_arr, channel_ryd, channel_flux,state):
    circuit.add_gate("CNOT", targets = channel_flux, controls = counting_qubit_arr[n])
    measurement_rydberg_results = perform_measurement(circuit,channel_flux,state)
    measurement_rydberg = measurement_rydberg_results[1]
    circuit = measurement_rydberg_results[0]
    state = measurement_rydberg_results[2] 
    if measurement_rydberg:
        print("Applying X gates to "+ "q"+ str(channel_ryd) + " and " + "q"+ str(channel_flux))
        circuit.add_gate("X", targets = channel_ryd)
        circuit.add_gate("X", targets = channel_flux)
    for qubit in range(n):
        phase = -np.pi/2**(n-qubit)
        control = channel_ryd
        target = counting_qubit_arr[qubit]
        circuit = cphase_gate(circuit,control, target, phase)
        print([target,control])
    circuit.add_gate("SNOT",targets = channel_ryd)
    measurement_flux_results = perform_measurement(circuit, channel_ryd,state)
    measurement_flux = measurement_flux_results[1]
    circuit = measurement_flux_results[0]
    state = measurement_flux_results[2]
    if measurement_flux:
        print("Applying Z gate and X gate to: " + "q"+ str(counting_qubit_arr[n]) + " and " + "q"+ str(channel_ryd))
        circuit.add_gate("X", targets = channel_ryd)
        circuit.add_gate("Z", targets = counting_qubit_arr[n])
        [circuit,state] = get_state(circuit,N_rydberg+N_flux,state)
    return [circuit,state]
\end{mintedbox}
\end{tcolorbox}

\begin{tcolorbox}[title=\textbf{Execute the inverse Fourier transform circuit in figure \ref{fig:distributed_QFT}}]
\begin{mintedbox}{python}
def iqft_rotations(circuit, name_arr, counting_qubit_arr, n, state):
    if n == 0: # Exit function if circuit is empty
        [circuit,state] = get_state(circuit,N_rydberg+N_flux,state)
        return [circuit,state]
    n -= 1
    circuit.add_gate("SNOT", targets = counting_qubit_arr[n])
    print(circuit.gates)
    [circuit,state] = get_state(circuit,N_rydberg+N_flux,state)
    non_local = False
    for qubit in range(n):
        target = counting_qubit_arr[n]
        control = counting_qubit_arr[qubit]

        print(control,target)
        if (name_arr[n] == name_arr[qubit]):
            circuit = cphase_gate(circuit,n,qubit,-np.pi/2**(n-qubit))
            [circuit,state] = get_state(circuit,N_rydberg+N_flux,state)
            
        else:
            non_local = True
            break

    if non_local:
        [circuit,state] = initialize_E_gate(circuit,state)
        [circuit,state] = perform_non_local_iqft(circuit, n, counting_qubit_arr, N_rydberg-1, N_rydberg,state)
        
    return iqft_rotations(circuit, name_arr,counting_qubit_arr, n,state)
\end{mintedbox}
\end{tcolorbox}

\begin{tcolorbox}[title=\textbf{Distributed phase estimation algorithm for estimating 16 different phases}]
\begin{mintedbox}{python}
N_flux = 3
N_rydberg = 4 
nsteps = 10000000
N_phase = 2**(N_flux-2+N_rydberg-1)

list_of_qubit_types = []
list_of_qubit_num = []

for i in range(N_rydberg-1):
    list_of_qubit_types.append('r')
    list_of_qubit_num.append(i)
for i in range(1,N_flux-1):
    list_of_qubit_types.append('f')
    list_of_qubit_num.append(i+N_rydberg)
    
for current_phase in range(0,N_phase):
    start_phase = current_phase/N_phase
    total_results =[]
    final_string = []
    qc = perform_hadamards(qc)
    [qc,current_state] = pulse_phase_sequence(qc,start_phase,current_state)
    [qc,current_state] = iqft_rotations(qc,list_of_qubit_types,list_of_qubit_num, len(list_of_qubit_types),current_state)
\end{mintedbox}
\end{tcolorbox}

\section{Flux-Resonator-Atom Hybrid Numerical Simulation}\label{sec:hybrid_code}

\begin{tcolorbox}[title=\textbf{Constructor of the Rydberg hamiltonian}]
\begin{mintedbox}{python}
def rydberg_hamiltonian(E,Omega,Omega_d,omega_e,omega_g,omega_u):
    basis_e = basis(3,0)
    basis_g = basis(3,1)
    basis_u = basis(3,2)
    
    if omega_e != 0:
        omega_e = (-7.81-(E-500)*7.3*10**-1)
    if omega_g != 0:
        omega_g = (-7.92+(E-500)*5.5*10**-1)
    if omega_u != 0:
        omega_u = (-7.88+(E-500)*6.6*10**-1)
    
    H_a_basis = omega_e*(tensor(basis_e,basis_e.dag()))+
                omega_g*(tensor(basis_g,basis_g.dag()))+
                omega_u*(tensor(basis_u,basis_u.dag()))+
                Omega/2*(tensor(basis_e,basis_g.dag())+
                tensor(basis_g,basis_e.dag()))+
                Omega_d/2*(tensor(basis_e,basis_u.dag())+
                tensor(basis_u,basis_e.dag()))
    return to_dim(H_a_basis,[[3],[3]])
\end{mintedbox}
\end{tcolorbox}

\begin{tcolorbox}[title=\textbf{Constructor of the Rydberg-Resonator Hamiltonian}]
\begin{mintedbox}{python}
def rydberg_resonator_hamiltonian(g_a,g_a_d):
    basis_e = basis(3,0)
    basis_g = basis(3,1)
    basis_u = basis(3,2)
    
    V_a = g_a/2 * (create(3) * to_dim(tensor(basis(3,1),basis(3,0).dag()), [[3],[3]]) + to_dim(tensor(basis(3,0),basis(3,1).dag()), [[3],[3]]) * destroy(3)) + g_a_d/2 * (create(3) * to_dim(tensor(basis(3,2),basis(3,0).dag()),[[3],[3]]) + to_dim(tensor(basis(3,0),basis(3,2).dag()),[[3],[3]]) * destroy(3))
    return to_dim(V_a,[[3],[3]])
\end{mintedbox}
\end{tcolorbox}

\begin{tcolorbox}[title=\textbf{Constructor of the Flux Hamiltonian}]
\begin{mintedbox}{python}
def flux_hamiltonian(gamma_q,Delta):
    basis_L = basis(2,0)
    basis_R = basis(2,1)
    
    I_p = 0.8*10**-6
    h = 6.63*10**-34
    e = 1.9**10**-16
    flux_quantum = h/(2*e)
    eps = (2*I_p*flux_quantum*gamma_q/(h/2*np.pi))*10**9
    
    sigma_fz = to_dim(tensor(basis_L,basis_L.dag()) + tensor(basis_R,basis_R.dag()),[[2],[2]])
    sigma_fx = to_dim(tensor(basis_R,basis_L.dag()).dag(),[[2],[2]]) + to_dim(tensor(basis_R,basis_L.dag()),[[2],[2]])
    
    H_f = -eps/2*sigma_fz - Delta/2*sigma_fx
    return H_f
\end{mintedbox}
\end{tcolorbox}

\begin{tcolorbox}[title=\textbf{Constructor of the Flux Resonator Hamiltonian}]
\begin{mintedbox}{python}
def flux_resonator_hamiltonian(g_f):
    basis_L = basis(2,0)
    basis_R = basis(2,1)
    sigma_fz = to_dim(tensor(basis_L,basis_L.dag()) + tensor(basis_R,basis_R.dag()),[[2],[2]])
    
    V_f = -g_f*(create(2)+destroy(2))*sigma_fz
    
    return V_f
\end{mintedbox}
\end{tcolorbox}

\begin{tcolorbox}[title=\textbf{Constructor of the Resonator Hamiltonian}]
\begin{mintedbox}{python}
def resonator_hamiltonian(omega_0):
    H_LC = to_dim(omega_0*(create(2)*destroy(2)+1/2),[[2],[2]])
    return H_LC
\end{mintedbox}
\end{tcolorbox}

\begin{tcolorbox}[title=\textbf{Function to simulate a hamiltonian for a given time.}]
\begin{mintedbox}{python}
def solve_master_equation(H,times,ref,c_ops_arr,psi0):
    options = Options(nsteps=10000)
    result = mesolve(H,psi0,times,c_ops_arr,ref)
    return result
\end{mintedbox}
\end{tcolorbox}

\begin{tcolorbox}[title=\textbf{Function to maximize the fidelity depending on evolution time.}]
\begin{mintedbox}{python}
def find_optimal_time(result,gate):
    highest_score = 0
    total_time = 0
    for i in range(len(result.states)):
        if (highest_score < fidelity(result.states[i],gate)):
            highest_score = fidelity(result.states[i],gate)
            total_time = result.times[i]
            
    print("Highest fidelity: " + str(highest_score))
    print("Time: " + str(total_time))
    return total_time
\end{mintedbox}
\end{tcolorbox}

\begin{tcolorbox}[title=\textbf{First step of the Flux-Resonator-Atom system}]
\begin{mintedbox}{python}
def step_1():
    E = 585.4
    Omega = 2*np.pi*4.6
    Omega_d = 0
    Ht = rydberg_hamiltonian(E,Omega,Omega_d,1,1,0) 
    times = np.linspace(0,np.pi/(2*Omega),10000)
    result = solve_master_equation(Ht, times, [], [Gamma_decay * qeye(3)], basis(3,0))
\end{mintedbox}
\end{tcolorbox}

\begin{tcolorbox}[title=\textbf{Second step of the Flux-Resonator-Atom system}]
\begin{mintedbox}{python}
def step_2():
    E = 537.5
    Omega = 2*np.pi*4.6
    Omega_d = 0

    g_a = 2*np.pi
    g_a_d = np.pi
    Ht_atom = rydberg_hamiltonian(E,Omega,Omega_d,0,0,1) + rydberg_resonator_hamiltonian(g_a,g_a_d)
    times = np.linspace(0,np.pi/(g_a_d),1000)
    result1 = solve_master_equation(Ht_atom, times, [], [[Gamma_decay * qeye(3)]], result.states[-1])

    E = 520.0
    Omega = 2*np.pi*4.6
    Omega_d = 2*np.pi*3.2
    Ht = rydberg_hamiltonian(E,Omega,Omega_d,0,1,1)
    times = np.linspace(0,10,10000)
    result_expect = solve_master_equation(Ht, times, [], [Gamma_decay * qeye(3), kappa_decay * qeye(3)], result1.states[-1])
    final_time = find_optimal_time(result_expect, np.sqrt(1/2) * (basis(3,1) + basis(3,2)))
    times = np.linspace(0,final_time,10000)
    result = solve_master_equation(Ht,times,[],
            [Gamma_decay8 * qeye(3),kappa_decay * qeye(3)],result1.states[-1])
    x_label = ['e','g','u']
    matrix_histogram(result.states[-1] * result.states[-1].dag(),x_label,x_label)

    flux_noise = [to_dim(gamma_relax * tensor(sigmaz(),sigmaz()),[[4],[4]]),
    to_dim(gamma_phi * tensor(sigmaz(),sigmaz()),[[4],[4]])]
\end{mintedbox}
\end{tcolorbox}

\begin{tcolorbox}[title=\textbf{Third step of the Flux-Resonator-Atom system}]
\begin{mintedbox}{python}
def step_3():
    gamma_q = -3.06*10**-3
    g_f = 2*np.pi*0.2
    Delta= 2*np.pi*50.0
    delta = 2*np.pi*0.1
    omega_0 = 2*np.pi*20
    Ht_flux = flux_resonator_hamiltonian(g_f) + flux_hamiltonian(gamma_q,Delta)
    Ht = to_dim(tensor(Ht_flux,qeye(2)), [[4],[4]]) + to_dim(ket2dm(basis(4,0)+basis(4,3)), [[4],[4]]) * to_dim(tensor(qeye(2), resonator_hamiltonian(omega_0)), [[4],[4]])
    times = np.linspace(0, np.pi/delta,10000)
    result_coupling = solve_master_equation(Ht, times,[], flux_noise, basis(4,3))
    final_time = find_optimal_time(result_coupling, basis(4,0))
    times = np.linspace(0, final_time, 10000)
    result_coupling = solve_master_equation(Ht,times,[], flux_noise ,basis(4,3))
    x_label = ['g0','g1','u0','u1']
    matrix_histogram(result_coupling.states[-1] * result_coupling.states[-1].dag(), x_label, x_label)
\end{mintedbox}
\end{tcolorbox}

\begin{tcolorbox}[title=\textbf{Fourth step of the Flux-Resonator-Atom system}]
\begin{mintedbox}{python}
def step_4():
    E = 550.8
    Omega = 2*np.pi*4.6
    Omega_d = 2*np.pi*3.2
    Ht = rydberg_hamiltonian(E,Omega,Omega_d,1,0,0)
    times = np.linspace(0,10,10000)
    result_expect = solve_master_equation(Ht,times,[],[Gamma_decay * qeye(3)],result.states[-1])
    final_time = find_optimal_time(result_expect,np.sqrt(1/2) * (basis(3,0)+basis(3,1)))
    times = np.linspace(0,final_time,10000)
    result = solve_master_equation(Ht,times,[],[Gamma_decay * qeye(3)],result.states[-1])
    x_label = ['e','g','u']
    matrix_histogram(result.states[-1] * result.states[-1].dag(), x_label, x_label)

    psi_0 = (qeye(12) - ket2dm(basis(12,1)-basis(12,7))) * to_dim(tensor(result_coupling.states[-1], 
            result.states[-1]),[[12],[1]])
\end{mintedbox}
\end{tcolorbox}

\begin{tcolorbox}[title=\textbf{Fifth step of the Flux-Resonator-Atom system}]
\begin{mintedbox}{python}
def step_5():
    E = 537.5
    Omega = 2*np.pi*4.6
    Omega_d = 0

    g_a = 2*np.pi
    g_a_d = np.pi
    Ht_atom = rydberg_hamiltonian(E,Omega,Omega_d,1,0,1) + rydberg_resonator_hamiltonian(g_a,g_a_d)
    times = np.linspace(0,np.pi/(g_a_d),1000)
    result1 = solve_master_equation(Ht_atom,times,[],[Gamma_decay * qeye(3), kappa_decay * qeye(3)], result.states[-1])

    E = 520.0
    Omega = 2*np.pi*4.6
    Omega_d = 2*np.pi*3.2
    Ht = rydberg_hamiltonian(E,Omega,Omega_d,0,1,1)
    times = np.linspace(0,10,10000)
    result_expect = solve_master_equation(Ht,times,[],[Gamma_decay * qeye(3)],result1.states[-1])
    final_time = find_optimal_time(result_expect,np.sqrt(1/2) * (basis(3,1)+basis(3,2)))
    times = np.linspace(0,final_time,10000)
    result = solve_master_equation(Ht,times,[],[Gamma_decay * qeye(3)],result1.states[-1])

    psi_0 = (qeye(12)-ket2dm(basis(12,5)-basis(12,7))) * to_dim(tensor(basis(4,1), result.states[-1]),[[12],[1]])

    E = 520.0
    Omega = 2*np.pi*4.6
    Omega_d = 2*np.pi*3.2
    Ht = to_dim(tensor(qeye(2), rydberg_hamiltonian(E, Omega, Omega_d, 0, 1, 1), qeye(2)), [[12],[12]])
    times = np.linspace(0,10,10000)
    result_expect = solve_master_equation(Ht,times,[],[Gamma_decay * qeye(12)],psi_0)
    final_time = find_optimal_time(result_expect,np.sqrt(1/2) * (basis(12,4)+basis(12,7)))
    times = np.linspace(0,final_time,10000)
    result = solve_master_equation(Ht,times,[],[Gamma_decay * qeye(12)],psi_0)
    x_label = ['L0e', 'L0g', 'L0u', 'L1e', 'L1g', 'L1u', 'R0e', 'R0g', 'R0u', 'R1e', 'R1g', 'R1u']

    fig,ax = matrix_histogram(result.states[-1] * result.states[-1].dag(), x_label, x_label)
    fig.set_size_inches(18, 6, forward=True)
\end{mintedbox}
\end{tcolorbox}

\begin{tcolorbox}[title=\textbf{Executing the E2 (GHZ) connection.}]
\begin{mintedbox}{python}
def perform_GHZ():    
    Gamma_decay = 2*np.pi*0.15*10**-3
    kappa_decay = 2*np.pi*0.2*10**-3
    gamma_relax = 2*np.pi*0.03*10**-3
    gamma_phi = 2*np.pi*0.1*10**-3
    
    step_1()
    step_2()
    step_3()
    step_4()
    step_5()
    
    return result
perform_GHZ().states[-1]
\end{mintedbox}
\end{tcolorbox}

\section{Creation of GRAPE optimised gates}\label{sec:grape_code}

\begin{tcolorbox}[title=\textbf{Fabrication of the RZ gate.}]
\begin{mintedbox}{python}
def rz_gate(arg_value):
    # controlled rotation X
    mat = np.zeros((2, 2), dtype=complex)
    mat[0, 0] = np.exp(-1j*arg_value/2)
    mat[1, 1] = np.exp(1j*arg_value/2)
    return Qobj(mat, dims=[[2], [2]])
\end{mintedbox}
\end{tcolorbox}

\begin{tcolorbox}[title=\textbf{Fabrication of the CNOT gate.}]
\begin{mintedbox}{python}
def cnot_gate():
    # controlled rotation X
    mat = np.zeros((4, 4), dtype=complex)
    mat[0, 0] = mat[1, 1] = 1.
    mat[2:4, 2:4] = x_gate()
    return Qobj(mat, dims=[[2, 2], [2, 2]])
\end{mintedbox}
\end{tcolorbox}

\begin{tcolorbox}[title=\textbf{Function to establish the drift and controls for the single qubit case.}]
\begin{mintedbox}{python}
def single_qubit_hamiltonian(is_ryd):
    if is_ryd:
        lt_70 = 1/(3.75 * 10**(-4)) * 10**(-7)
        gamma_d1 = lt_70*2
        H_d = Omega*(sigmax()+sigmay())+sigmaz()
        c_ops = [np.sqrt(gamma_d/2)*sigmaz(), Gamma*sigmaz()]
    else:
        E_J = 65
        E_C = 1
        E_cm = E_C*1/(zeta+alpha+1/2)
        E_jm = 2*(1-2*alpha)*E_J
        E_Cp = 2*E_C
        theta_z = (E_cm/(4*E_jm))**(-1/4)
        n_z = (E_jm/(4*E_cm))**(-1/4)
        l = (8*alpha-1)/(8*(1-2*alpha))
        omega = np.sqrt(8*E_jm*E_cm) - l*E_cm
        theta = 2*np.pi*fe+2*1.5
        I_m = -8* np.pi* alpha* theta_z* np.cos(theta)* E_J
        epsilon = 6.7
        delta = 0.82
        g = 2
        d_n = 4*np.pi/g
        H_d = -1/2*(epsilon*sigmaz() + delta*sigmax())
        c_ops = [1/2*( (omega)* sigmaz() + I_m*omega* sigmax() + n_z* E_cm* d_n* 0.092* sigmay() )]

    # The (single) control Hamiltonian
    H_c = [sigmax(), sigmay(), sigmaz()]

    # start point for the map evolution
    E0 = identity(2)

    return [E0, H_d, c_ops, H_c]
\end{mintedbox}
\end{tcolorbox}

\begin{tcolorbox}[title=\textbf{Function to establish the drift and controls for the double qubit case.}]
\begin{mintedbox}{python}
def double_qubit_hamiltonian(is_ryd):
    if is_ryd:
        lt_70 = 1/(3.75 * 10**(-4)) * 10**(-7)
        gamma_d1 = lt_70*2
        H_d = Omega* (tensor( sigmax(), sigmax()) + tensor(sigmay(), sigmay())) + tensor(sigmaz(),sigmaz()) + 801.98 / (3.5**6)
        c_ops = [ tensor( qeye(2), np.sqrt(gamma_d1/2)* sigmaz() ), Gamma* tensor (sigmaz(), qeye(2)) ]
    else:
        E_J = 65
        E_C = 1
        E_cm = E_C*1/(zeta + alpha + 1/2)
        E_jm = (1 - 2* alpha)* E_J
        E_Cp = 2* E_C
        theta_z = (E_cm/ (4* E_jm) )**(-1/4)
        n_z = (E_jm/ (4*E_cm) )**(-1/4)
        l = (8*alpha - 1) / (8*(1 - 2*alpha))
        omega = np.sqrt(8*E_jm*E_cm) - l*E_cm
        theta = 2*np.pi*fe+2*1.5
        I_m = -8*np.pi* alpha* theta_z* np.cos(theta)* E_J
        g = 2
        epsilon = 6.7
        delta = 0.82
        d_n = 4*np.pi/g
        H_d = -1 / 2 * (epsilon*tensor(sigmaz(), sigmaz()) + delta*tensor(sigmax(), sigmax())) + omega* (tensor(qeye(2), qeye(2)).dag() + tensor(qeye(2), qeye(2))) + 10*tensor(sigmay(), sigmay())* ( tensor(qeye(2), qeye(2) ).dag() + tensor(qeye(2), qeye(2)) ) 
        c_ops = [ (omega* tensor(sigmaz(), qeye(2)) + I_m*omega*tensor(sigmax(), qeye(2)) + n_z*E_cm*d_n*tensor(sigmay(), qeye(2))) ]

    H_c = [tensor(sigmax(), identity(2)),
         tensor(sigmay(), identity(2)),
         tensor(sigmaz(), identity(2)),
         tensor(identity(2), sigmax()),
         tensor(identity(2), sigmay()),
         tensor(identity(2), sigmaz()),
         tensor(sigmax(), sigmax()) +
         tensor(sigmay(), sigmay()) +
         tensor(sigmaz(), sigmaz())]
    # start point for the gate evolution
    U_0 = identity(4)
    return [U_0, H_d, c_ops, H_c]
\end{mintedbox}
\end{tcolorbox}

\begin{tcolorbox}[title=\textbf{Function to create single qubit optimised gates.}]
\begin{mintedbox}
    def create_single_gates(gate_dict):
        U_targ = x_gate()
        single_qubit = single_qubit_hamiltonian(1)
        H_0 = mesolve(single_qubit[1], ket2dm(basis(2,0)), times, [ single_qubit[2][0].unit() ],[]).states
        result = get_result(single_qubit[0], U_targ,H_0, single_qubit[3], algorithm)
        result = result.evo_full_final
        gate_dict.update({"x_ryd": result})
    
        U_targ = x_gate()
        single_qubit = single_qubit_hamiltonian(0)
        H_0 = mesolve(single_qubit[1], ket2dm(basis(2,0)), times, [single_qubit[2][0].unit()],[]).states
        result = get_result(single_qubit[0], U_targ, H_0, single_qubit[3], algorithm)
        result = result.evo_full_final
        gate_dict.update({"x_flux" : result})
    
        U_targ = z_gate()
        single_qubit = single_qubit_hamiltonian(1)
        H_0 = mesolve(single_qubit[1], ket2dm(basis(2,0)), times, [single_qubit[2][0].unit()],[]).states
        result = get_result(single_qubit[0], U_targ, H_0, single_qubit[3], algorithm)
        result = result.evo_full_final
        gate_dict.update({"z_ryd": result})
    
        U_targ = z_gate()
        single_qubit = single_qubit_hamiltonian(0)
        H_0 = mesolve(single_qubit[1], ket2dm(basis(2,0)), times, [single_qubit[2][0].unit()],[]).states
        result = get_result(single_qubit[0], U_targ, H_0, single_qubit[3],algorithm)
        result = result.evo_full_final
        gate_dict.update({"z_flux": result})
    
        U_targ = snot()
        single_qubit = single_qubit_hamiltonian(1)
        H_0 = mesolve(single_qubit[1], ket2dm(basis(2,0)), times, [single_qubit[2][0].unit()],[]).states
        result = get_result(single_qubit[0], U_targ, H_0, single_qubit[3], algorithm)
        result = result.evo_full_final
        gate_dict.update({"snot_ryd": result})
    
        U_targ = snot()
        single_qubit = single_qubit_hamiltonian(0)
        H_0 = mesolve(single_qubit[1], ket2dm(basis(2,0)), times, [single_qubit[2][0].unit()],[]).states
        result = get_result(single_qubit[0], U_targ, H_0, single_qubit[3], algorithm)
        result = result.evo_full_final
        gate_dict.update({"snot_flux": result})
        return gate_dict
\end{mintedbox}
\end{tcolorbox}

\begin{tcolorbox}[title=\textbf{Function to create double qubit optimised gates.}]
\begin{mintedbox}
    def create_double_gates(gate_dict):
        U_targ = cnot_gate()
        double_qubit = double_qubit_hamiltonian(1)
        H_0 = mesolve(double_qubit[1], ket2dm(tensor(basis(2,0), basis(2,0))), times, [double_qubit[2][0].unit()], []).states
        result = get_result(double_qubit[0], U_targ, H_0, double_qubit[3], algorithm)
        result = (result.evo_full_final/ result.evo_full_final[0,0])
        result = to_dim( result, [[2, 2], [2, 2]])
        gate_dict.update({"cnot_ryd": result})
    
        U_targ = cnot_gate()
        double_qubit = double_qubit_hamiltonian(0)
        H_0 = mesolve(double_qubit[1], ket2dm(tensor(basis(2,0),basis(2,0))), times,[ double_qubit[2][0].unit() ], []).states
        result = get_result(double_qubit[0], U_targ, H_0, double_qubit[3], algorithm)
        result = (result.evo_full_final/ result.evo_full_final[0,0])
        result = to_dim(result, [[2, 2], [2, 2]])
        gate_dict.update({"cnot_flux": result})
\end{mintedbox}
\end{tcolorbox}

\begin{tcolorbox}[title=\textbf{Creating the optimised gates using GRAPE.}]
\begin{mintedbox}{python}
def create_gate_dict():

    gate_dict = {}
    gate_dict = create_single_gates(gate_dict)
    gate_dict = create_double_gates(gate_dict)

    return gate_dict
\end{mintedbox}
\end{tcolorbox}

\begin{tcolorbox}[title=\textbf{Function executing the GRAPE optimiser.}]
\begin{mintedbox}{python}
def get_result(U_0,U_targ,H_d,H_c,algorithm): 

    # pulse type alternatives: RND|ZERO|LIN|SINE|SQUARE|SAW|TRIANGLE|
    p_type = 'RND'

    U_targ = to_dim(U_targ, U_0.dims)

    result = cpo.optimize_pulse_unitary(H_d, H_c, U_0, U_targ, n_ts, evo_time, 
                    fid_err_targ=fid_err_targ, min_grad=min_grad, 
                    max_iter=max_iter, max_wall_time=max_wall_time,
                    init_pulse_type=p_type, alg = algorithm,
                    log_level=log_level, gen_stats=True)

    global N_depth 
    N_depth = N_depth + 1
    return result
\end{mintedbox}
\end{tcolorbox}

\clearpage

\section{Values for Parameters}
\begin{table}[h] \label{table:values_hybrid}
\centering
\begin{tabular}{ |c|c|c| } 
 \hline
 Parameters & Values \\
 \hline
 $E_J$ (GHz) & 65 \\
 $E_C$ (GHz) & 1 \\
 $\Delta$ (GHz) & 0.82 \\
 $\alpha$ & 0.8 \\
 $\varepsilon$ (GHz) & 6.7 \\
 $f_\varepsilon$ & 0.53 \\
 $\zeta$ & 10,100,1000 \\
 $g$ (GHz) & 2 \\
 $\delta n_m$ & 1\\
 $\Omega$ (GHz) & $2\pi \times 6.8$\\
 $\Gamma$ (MHz) & $2\pi \times 470$\\
 $\gamma_e$ (MHz) & $2 \times \pi 0.03$\\
 \hline
\end{tabular}
\caption{List of physical parameters used for the master equations.}
\label{table:values_parameters}
\end{table}

\section{Diagram of Simulation Processes}
\begin{figure}[t] 
\centering
    \includegraphics[width=1.1\textwidth]{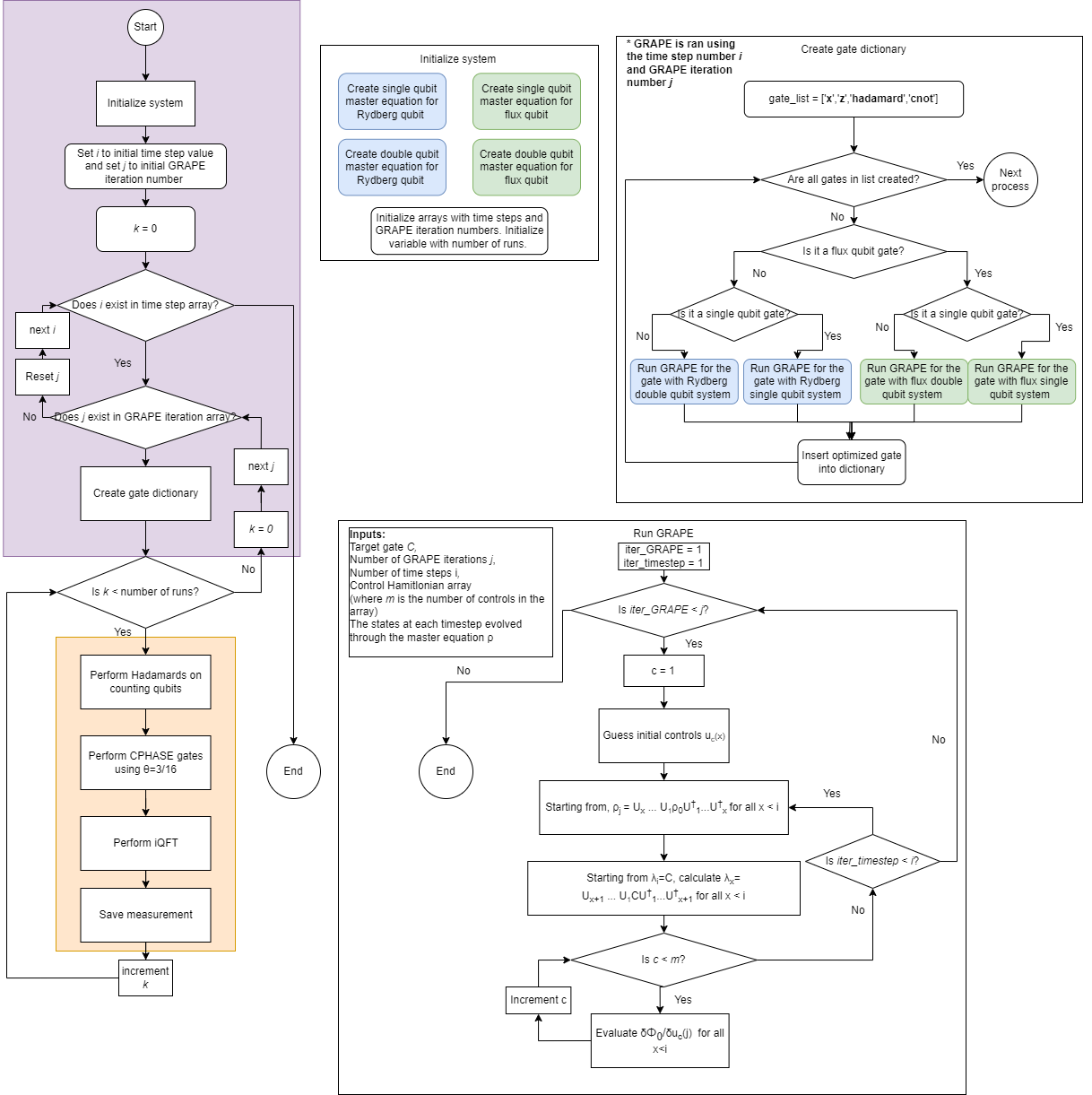}
    \caption{Detailed schematic of the steps involved when simulating the distributed phase estimation with GRAPE optimised gates.}
    \label{fig:grape_algorithm}
\end{figure}
\chapter{Quantum Principles in Rydberg Systems}

\section{Wavefunctions for Alkali Atoms}\label{sec:ryd_theory}

Alkali atoms are the names given to the chemical elements Li, Na, K, Rb and Cs. These atoms are found in the first group in the periodic table along with Hydrogen. Although alkali atoms obviously differ from hydrogen, both hydrogen and alkali metals have an outermost electron in an s-orbital. Thus it possible to derive the wavefunction for a hydrogen atom to derive the analytical wavefunction for alkali atoms. In order to do so, one needs to solve the time-independent Schr\"{o}dinger equation,

\begin{equation}
    -\frac{1}{2}\Delta^2\psi(\boldsymbol{r})-\frac{1}{r}\psi(\boldsymbol{r}) = E\psi(\boldsymbol{r}),
\end{equation}

where the Coulomb potential is exhibited by $V=-\frac{1}{r}$ where $r$ is the distance of the electron to the nucleus. Due to the spherical symmetric nature of an atom, one can separate the Schr\"{o}dinger equation into a radial and angular part using spherical coordinates,

\begin{equation}
    \psi(\boldsymbol{r}) = R(r)Y_{lm}(\theta,\phi).
\end{equation}

\begin{figure}[t] 
\centering
    \includegraphics[width=0.6\textwidth]{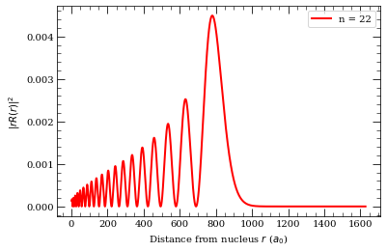}
    \caption{The probability of the radial part of the wavefunction for the state $\ket{22D_{5/2},j=5/2, s = 1/2}$ as a function of the atomic distance.}
    \label{fig:d_wavefunction}
\end{figure}

The angular part $Y_{lm}(\theta,\phi)$ has a solution \cite{thompson_salpeter_1977},

\begin{equation}
    Y_{lm}(\theta,\phi) = \sqrt{\frac{(2l+1)}{4\pi}\frac{(l-m)!}{(l+m)!}}P_l^m(\mathrm{cos}\theta)e^{im\phi}
\end{equation}

where $P^m_l$ the associated Legendre function \cite{stegun_2013}. $l$ is the orbital angular momentum quantum number and $m$ is the magnetic quantum number.
Solving the radial part of the Schr\"{o}dinger equation requires one to rewrite $u(r) = rR(r)$ to obtain, 
\begin{equation}
    \bigg(-\frac{1}{2}\frac{\mathrm{d}^2}{\mathrm{d}r^2}+\frac{l(l+1)}{2r^2}-\frac{1}{r}\bigg)u(r) = Eu(r).
\end{equation}
The solution to $R(r)$ can be generalized to \cite{thompson_salpeter_1977},
\begin{equation}\label{eq:radial_hydrogen}
    R_{nl}(r) = \sqrt{\bigg( \frac{2}{n} \bigg)^3 \frac{(n-l-1)!}{2n(n+l)!}}e^{-r/n} \bigg(\frac{2r}{n}\bigg)^l L^{2l+1}_{n-l-1} \bigg(\frac{2r}{n}\bigg),
\end{equation}
with energy,
\begin{equation}\label{eq:energy_hydrogen}
    E_n = -\frac{1}{2n^2},
\end{equation}
where $n$ is the principal quantum number, $l$ is the angular momentum quantum number which cannot be greater than $n-1$. Eq. \eqref{eq:energy_hydrogen} can be adjusted for alkali atoms by adding a term to the orbital quantum number known as a quantum defect $\delta_{nlj}$ \cite{ditzhuijzen_2009} where,
\begin{equation}
    \delta_{nlj} = \delta_0 + \frac{\delta_2}{(n-\delta_0)^2}.
\end{equation}
In this formulation, both $\delta_0$ and $\delta_2$ are dependent on $l$ but the spin orbit coupling extends the dependence to an additional spin orbit coupling term $j$. The discrepancy in energy between the energy of a hydrogen atom and an alkali metal arises from the potential near the core which in turn changes the energy. Hence, the energy can be rewritten as, 
\begin{equation}
    E_{nlj} = -\frac{1}{2(n-\delta_{nlj})^2}.
\end{equation}
This defect was originally measured and postulated by Johannes Rydberg in 1980 \cite{rydberg_1890}. In order to solve the radial wavefunction for alkali metals, equation \eqref{eq:radial_hydrogen} must be rewritten using the new definitions for the principal quantum number $n$ and the orbital momentum quantum number $l$. The new principal quantum number can simply be rewritten as $n^* = n-\delta_{nlj}$. For the orbital quantum number however, $l* = l - \delta_{nlj}+I(l)$ where $I(l)$ is an integer responsible for fixing the number of radial nodes as seen in \cite{bates_damgaard,quantum_defect_theory}. The restriction applied to $I(l)$ is that $\delta_{nlj}-l-\frac{1}{2} < I(l) \leq n_{min} - l - 1$ where $n_{min}$ is the principal quantum number of the ground state. Setting $I(l) = \delta_{nlj}$ establishes a satisfactory agreement for the lifetime values 5s-5p \cite{ditzhuijzen_2009}. 
\begin{equation}\label{eq:numerical_radial}
    R_{n^*l^*}(r) = \sqrt{\bigg(\frac{2}{n^*}\bigg)^2\frac{(n^*-l^*-1)!}{2n^*(n^*+l^*)!}}e^{-r/n^*}\bigg(\frac{2r}{n^*}\bigg)^{l^*}L^{2l^*+1}_{n^*-l^*-1}\bigg(\frac{2r}{n^*}\bigg).
\end{equation}
Fig. \ref{fig:d_wavefunction} shows the absolute value of Eq. \eqref{eq:numerical_radial} multiplied by $r$ given in $a_0$ squared as a function of $r$ for the state  $\ket{22D,j=5/2, s = 1/2}$ for $\mathrm{Rb^{87}}$. This shows the probability of finding an electron from the nucleus at distance $r$. 

\section{Non-Degenerate Perturbation Theory}\label{sec:perturb_theory}

Solving physical problems in large Hilbert spaces, much like analyzing the interaction of radiation with matter, is analytically too complex to solve exactly. A method of solving such complex problems can be solved using an approximate method.

Assume a Hamiltonian in the form, 
\begin{equation}\label{eq:starting_pertub}
    \mathcal{H} = \mathcal{H}_0 - \varepsilon V
\end{equation}
where the eigenvalue for the unperturbed Hamiltonian $\mathcal{H}_0$ is known and $\varepsilon$ is a value varying between 0 and 1 responsible for the expansion. The solution for $\varepsilon \rightarrow 1$ is the desired solution to \eqref{eq:starting_pertub}. Knowing the eigenvalues and eigenkets of the uperturbed Hamiltonian, $\mathcal{H}_0 \ket{k} = E_k^{(0)}\ket{k}$ allows one to rewrite the complete Hamiltonian as, 
\begin{equation} \label{eq:perturb_ket}
    (\mathcal{H}_0+\varepsilon V)\ket{\varphi_k}_\varepsilon = E_k(\varepsilon)\ket{\phi_k}_\varepsilon,
\end{equation}
this is assuming that at first that the energy spectrum is not degenerate (that is, all $E_k^{(0)}$ are different). Expanding the eigenket where $\ket{\phi^{(0)}_k} = \ket{k}$,
\begin{equation}
    \ket{\phi_k} = \ket{\phi_k^{(0)}}+\varepsilon \ket{\phi_k^{(1)}}+\varepsilon^2 \ket{\phi_k^{(2)}}+\dots,
\end{equation}
and
\begin{equation}
    E_k = E_k^{(0)}+\varepsilon E^{(1)}+\varepsilon^2 E_k^{(2)}+\dots.
\end{equation}
When $\varepsilon$ is small, the energy shift due to the perturbation is given by $\Delta_k = E_k = E_k^{(0)}$. This means that equation \eqref{eq:perturb_ket} can be rewritten to $(E^{(0)}_k - \mathcal{H}_0)\ket{\varphi_k} = (\varepsilon V-\Delta_k)\ket{\varphi_k}$. Then, projecting this onto $\bra{k}$, $\bra{k}(E^{(0)}_k - \mathcal{H}_0)\ket{\varphi_k} = \bra{k}(\varepsilon V-\Delta_k)\ket{\varphi_k}$ leaves \cite{quantum_theory_mit, sakurai_napolitano_2021},
\begin{equation}
    \Delta_k = \varepsilon \frac{\bra{k}V\ket{\varphi_k}}{\braket{k|\varphi_k}},
\end{equation}
since $\bra{k}\mathcal{H}_0\ket{\varphi_k} = \bra{k}E_k^{(0)}\ket{\varphi_k}$ and $\bra{k}\varepsilon V - \Delta_k \ket{\varphi_k} = 0$. 
It is possible to introduce the relation $\braket{k|\varphi_k} = 1$, however this implies that the perturbed state $\ket{\varphi_k}$ is not normalized. It can be shown that by defining, 
\begin{equation}
    \ket{\psi_k} = \frac{\ket{\varphi_k}}{\sqrt{\braket{\varphi_k | \varphi_k}}}
\end{equation}
such that $\braket{k|\psi_k} = 1/\sqrt{\braket{\varphi_k|\varphi_k}}$, $\braket{k|\phi_k}$ can be approximated to 1 \cite{quantum_theory_mit}. Calculating the perturbed normalization factor,
\begin{equation}
    \braket{\varphi_k | \varphi_k} = \braket{k + \varepsilon \varphi_k^{(1)} + \dots | k +\varepsilon \varphi_k^{(1)} + \dots} = 1 + \varepsilon \braket{k|\varphi^{(1)}_k} + \dots,
\end{equation}
where $\varepsilon \braket{k|\varphi^{(1)}_k} = 0$. Thus, $\braket{\varphi_k | \varphi_k} \approx 1$ meaning that $\braket{k|\phi_k} \approx 1$.

Using the approximation $\braket{k|\phi_k} \approx 1$, it follows that $\Delta_k = \varepsilon \braket{k|V|\varphi_k}$. Replacing, $\Delta_k$ by $\varepsilon E^1_k+\varepsilon^2 E^2_k+\dots$ and $\ket{\varphi_k}$ by it's expansion and equating terms of the same order in $\epsilon$, it is possible to obtain,
\begin{equation}
    E^n_k = \braket{k|V\varphi_k^{(n-1)}}.
\end{equation}
This result shows that by knowing only the eigenstates of lower orders, one can find the energy of all orders. The question is now finding the value of $\varphi_k^{(n-1)}$. 
Due to the fact that $(E^{(0)}_k - \mathcal{H}_0)^{-1}\ket{k}$ possesses a singularity, it is not possible to invert the operator in the expansion $\ket{\varphi_k} = (E^{(0)}_k - \mathcal{H}_0)^{-1}(\varepsilon V - \Delta_k)\ket{\varphi_k}$. To ensure $(E^{(0)}_k - \mathcal{H}_0)^{-1}$ is never applied to eigenstates of the unperturbed Hamiltonian, $\ket{\psi_k}$ needs to equal  $(\varepsilon V - \Delta_k)\ket{\varphi_k} \neq \ket{k}$ for any $\ket{\varphi_k}$ \cite{quantum_theory_mit}. By defining a projector $P_k = \mathds{1}-\ket{k}\bra{k}=\sum_{h\neq k}\ket{k}\bra{k}$, one ensures that $\forall \ket{\psi}$ the projected state is $\ket{\psi}' = P_k\ket{\psi}$ is such that $\braket{k|\psi'} = 0$ since $\bra{k} P_k \ket{\psi} = \braket{k|\psi} - \braket{k|k}\braket{k|\psi}=0$ \cite{sakurai_napolitano_2021,quantum_theory_mit}. Using the projector, $(E_k^{(0)}-\mathcal{H_0})$ can be multiplied by $P_k\ket{\psi}$, thus giving, 
\begin{equation}
    P_k(E_k^{(0)}-\mathcal{H}_0)\ket{\varphi_k},
\end{equation}
which is now solvable since $(E_k^{(0)}-\mathcal{H_0})^{-1}\ket{P_k}$ is no longer ill defined. Thus, taking $(E_k^{(0)}-\mathcal{H}_0)\ket{\varphi_k} = (\varepsilon V-\Delta_k) \ket{\varphi_k}$, multiplying it by $P_k$ and rearranging for $P_k\ket{\varphi_k}$ gives, 
\begin{equation}
    P_k\ket{\varphi_k} = (E_k^{(0)}-\mathcal{H}_0)^{-1}P_k (\varepsilon V-\Delta_k)\ket{\varphi_k}
\end{equation}
Noting that $P_k\ket{\varphi_k} = \ket{\varphi_k} - \ket{k}\braket{k|\varphi_k} = \ket{\varphi_k} - \ket{k}$, one can obtain \cite{sakurai_napolitano_2021,quantum_theory_mit},
\begin{equation}
    \ket{\varphi_k} = \ket{k}+ (E^{(0)}_k - \mathcal{H}_0)^{-1}P_k(\varepsilon V -\Delta_k)\ket{\varphi_k}.
\end{equation}
In order to simplify the expression, it is possible to define an operator $F_k$ as,
\begin{equation}
    F_k = (E_k^{(0)}-\mathcal{H}_0)^{-1}P_k = \sum_{h\neq k} \frac{\ket{h}\bra{h}}{E^0_k - E^0_h}.
\end{equation}
By using the expansion, 
\begin{equation}
    \ket{k} +\varepsilon\ket{\varphi^{(1)}_k}+\dots = \ket{k} + F_k \varepsilon(V-E_k^1-\varepsilon E^2_k - \dots)(\ket{k}+\varepsilon \ket{\varphi_k^{(1)}}+\dots)
\end{equation}
one can solve to obtain the first order term,
\begin{equation}
    \ket{\varphi_k^{(1)}} = F_k (V-E_k^1)\ket{k} = F_k(V-\bra{k}V\ket{k})\ket{k} = F_k V\ket{k}
\end{equation}
and the second order term,
\begin{equation}
    E^2_k = \bra{k}V\ket{\varphi_k^{(1)}} = \bra{k}V F_k V \ket{k} = \bra{k} V \bigg(\sum_{h\neq k} \frac{\ket{h}\bra{h}}{E^0_k - E^0_h}\bigg) V\ket{k}
\end{equation}
which is explicitly equal to \cite{sakurai_napolitano_2021,quantum_theory_mit},
\begin{equation}
    E^2_k = \sum_{h\neq k} \frac{|V_{kh}|^2}{E^0_k - E^0_h}.
\end{equation}

\section{Stark Effect}\label{sec:stark_effect}

Suppose the atom under study has a Hamiltonian in the form $\mathcal{H} = \mathcal{H}_0 + \mathcal{H}_1$ where $\mathcal{H}_0$ is the unperturbed part of the Hamiltonian and $\mathcal{H}_1 = e|\boldsymbol{E}|z$ is the perturbed part of the Hamiltonian. This problem can be studied using non-degenerate perturbation theory assuming that the unperturbed states are non-degenerative. A change in energy of the eigenstates characterized by quantum numbers $n$,$l$, and $m$ in the presence of a small electric field is given by,
\begin{equation}
    \Delta E_{nlm} = e|\boldsymbol{E}|\bra{n,l,m}z\ket{n,l,m} + e^2 |\boldsymbol{E}|^2 \sum_{n',l',m'=n,l,m} \frac{|\bra{n,l,m}z\ket{n',l',m'}|^2}{E_{nlm}-E_{n'l'm'}}
\end{equation}
Trivially, it is possible to see that in almost all cases, $\bra{n,l,m}z\ket{n',l',m'}$ is zero. Thus, it is important to determine a set of rules for these matrix elements to be non-zero. These rules are referred to as selection rules. Knowing that the z orbital angular momentum operator commutes with z ($[L_z,z] = 1$), it follows that \cite{sakurai_napolitano_2021,quantum_theory_mit},
\begin{equation}
    \bra{n,l,m}[L_z,z]\ket{n',l',m'}= \bra{n,l,m}L_z z - zL_z \ket{n',l',m'} = \hbar (m-m')\bra{n,l,m}z\ket{n',l',m'} = 0.
\end{equation}
This shows two clear selection rules. The first being that $m'=m$ has to hold in order for $\bra{n,l,m}z\ket{n',l',m'}$ to be non-zero. The second being that $l'=l \pm 1$ which can be calculated by using the commutation of the total angular momentum $L^2$ with $z$ \cite{quantum_theory_mit}. 
\begin{equation}
    \Delta E_{nlm} = e^2|\boldsymbol{E}|^2\sum_{n',l'=l\pm 1}\frac{|\bra{n,l,m}z\ket{n',l',m'}|^2}{E_{nlm}-E_{n'l'm}}.
\end{equation}
This is thus the definition of the quadratic Stark effect. Since, $\Delta E = -\frac{1}{2}\alpha |\boldsymbol{E}|^2$, it follows that,
\begin{equation}
    \alpha_{nlm} = 2e^2 \sum_{n',l'=l\pm 1}\frac{|\bra{n,l,m}z\ket{n',l',m'}|^2}{E_{nlm}-E_{n'l'm}}.
\end{equation}
For states that are degenerate at zero field, the Stark effect will cause a splitting of the states. This means that at these states, the energy shift is linear meaning that they have a permanent dipole moment \cite{ditzhuijzen_2009}. This occurs in alkalis for all higher angular momentum states with equal $n$. Following \cite{JWBHughes_1967}, it is convenient to convert the quantum number $l$ to $k$, the parabolic quantum number, using the parabolic coordinate system. The state $\ket{nkm}$ is thus expressed as \cite{ditzhuijzen_2009,JWBHughes_1967},
\begin{equation}
    \ket{nkm} = \sum_l C_{kl}\ket{nlm},
\end{equation}
where the Clebsch-Gordan coefficient is,
\begin{equation}
    C_{kl} = (-1)^m\sqrt{2l+1}
    \begin{pmatrix}
    \frac{1}{2}(n-1) & \frac{1}{2}(n-1) & l \\
    \frac{1}{2}(m+k) & \frac{1}{2}(m-k) & m
    \end{pmatrix}.
\end{equation}
The parabolic quantum number $k$ is restricted to $n-|m|-1 < k < -n+|m|+1$ in steps of 2. The Stark shift resulting in \cite{thompson_salpeter_1977},
\begin{equation}
    \Delta E = \frac{3}{2}nkE
\end{equation}
\section{Anti-Crossing}

Assume a two level state with levels $h$ and $k$ with energies $E^0_k$ and $E^0_h$ which is perturbed by an applied perturbation $V$ connecting the two states ($\bra{l}V\ket{j}=0$) while still different to zero for the transition from $h$ to $k$ ($\bra{h}V\ket{k} \neq 0$). The perturbed state energies, assuming the perturbation is small, are \cite{quantum_theory_mit},
\begin{equation}
    E_k^{(2)} = \sum_{j \neq k}\frac{|V_{kj}|^2}{E_k^0-E^0_j} = \frac{|V_{kh}|^2}{E_k^0-E^0_h},
\end{equation}
and,
\begin{equation}
    E_k^{(2)} = \sum_{j \neq h}\frac{|V_{hj}|^2}{E_h^0-E^0_j} = \frac{|V_{kh}|^2}{E_h^0-E^0_k}=-E_k^{(2)}.
\end{equation}
In the absence of a perturbation, these two energy levels would cross when $E_k^0 = E^0_h$. However, with an added perturbation, both these energy levels repel each other with opposite energy shifts. This phenomenon is called an anti-crossing and is more noticeable when the energies $E_k^0$ and $E_h^0$ are closer to each other \cite{sakurai_napolitano_2021}.  

\section{F\"{o}rster Energy Defect} \label{sec:forster_resonance_app}

Consider the electric dipole-dipole interaction between an atom A and an atom B, described by the following,
\begin{equation}
\begin{split}\label{eq: Vdd_small}
    V_{dd} = \frac{1}{4\pi\varepsilon_0 R^3}\bigg[\mathcal{A}_1(\theta) \Big(d_{1+}d_{2-}+d_{1-}d_{2+}+2d_{1z}d_{2z} \Big) + \\ 
    \mathcal{A}_2(\theta) \Big(d_{1+}d_{2z}-d_{1-}d_{2z}+d_{1z}d_{2+}-d_{1z}d_{2-} \Big) - \mathcal{A}_3(\theta) \Big(d_{1+}d_{2+}+d_{1-}d_{2-}\Big) \bigg],
    \end{split}
\end{equation}
where $d_{d,\pm} = \mp (d_{k,x} \pm id_{k,y}/\sqrt{2}$ are the components of the dipole operator in the spherical basis with prefactors $\mathcal{A}_1(\theta) = (1-3\mathrm{cos}^2\theta)/2$, $\mathcal{A}_2(\theta) = 3\mathrm{sin}\theta \mathrm{cos}\theta/\sqrt{2}$ and $\mathcal{A}_3(\theta) = 3\mathrm{sin}^2\theta/2$ \cite{Ravets_Forster}. The operator $V_{dd}$ couples two-atom states where the total magnetic quantum number $M = m_1 + m_2$ changes by $\Delta M = 0, \pm 1 $ and $\pm 2$. The matrix element of the $V_{dd}$ operator for the transition between atom states $\ket{r_a r_b} \rightarrow \ket{r_s r_t}$ in which each atomic state $\ket{r} = \ket{nljm_j}$, can be expressed as,
\begin{equation} \label{eq: Vdd_large}
    \begin{split}
        \bra{n_s m_s l_s j_s; n_t m_t l_t j_t}V_{dd}\ket{n_a m_a l_a j_a; n_b m_b l_b j_b} = \frac{e^2}{4\pi\varepsilon_0 R^3} \bigg\{ \mathcal{A}_1(\theta) \Big[2C^{j_s m_s}_{j_a m_a 10}C^{j_t m_t}_{j_b m_b 10} + C^{j_s m_s}_{j_a m_a 11}C^{j_t m_t}_{j_b m_b 1-1} \\
        + C^{j_s m_s}_{j_a m_a 1-1}C^{j_t m_t}_{j_b m_b 11}\Big] + \mathcal{A}_2(\theta)\Big[ \Big(C^{j_s m_s}_{j_a m_a 11}-C^{j_s m_s}_{j_a m_a 1-1} \Big) C^{j_t m_t}_{j_b m_b 10} + C^{j_s m_s}_{j_a m_a 10}\Big(C^{j_s m_s}_{j_a m_a 11}-C^{j_s m_s}_{j_a m_a 1-1} \Big)\Big] - \\
        \mathcal{A}_3(\theta) \Big[C^{j_s m_s}_{j_a m_a 11}C^{j_t m_t}_{j_b m_b 11} + C^{j_s m_s}_{j_a m_a 1-1}C^{j_t m_t}_{j_b m_b 11} \Big] \bigg \} \times \sqrt{max(l_a,l_s)}\sqrt{max(l_b,l_t)}\sqrt{(2j_a+1)(2j_b+1)} \times \\
        \bigg\{\frac{l_a j_a}{2j_s l_s}\bigg\}\bigg\{\frac{l_b j_b}{2j_t l_t}\bigg\}(-1)^{l_s + \frac{l_a+l_s+1}{2}} \times (-1)^{l_s + \frac{l_b+l_t+1}{2}} (-1)^{j_a+j_b} R^{n_s l_s}_{n_a l_a} R^{n_t l_t}_{n_b l_b}.
    \end{split}
\end{equation}

$R$ is the interatomic distance and both $R^{n_s l_s}_{n_a l_a}$ and $R^{n_t l_t}_{n_b l_b}$ are the radial matrix elements for $\ket{n_a l_a} \rightarrow \ket{n_s l_s}$ and $\ket{n_b l_b} \rightarrow \ket{n_t l_t}$ transitions, respectively \cite{Ravets_Forster,Beterov_2018}. The F\"{o}rster energy defect for channel $k$ is the difference between the initial collective state $\ket{r_a r_b}$ and of the final state $\ket{r_s r_t}$,
\begin{equation}
    \hbar \delta_k = [U(r_a)-U(r_s)] + [U(r_b)-U(r_t)].
\end{equation}

\section{Rydberg Light Interaction Hamiltonian}\label{sec:light_interaction}
The Hamiltonian of the Rydberg system is thus given by,
\begin{dmath}
    \mathcal{H} = H_{atom} + H_{int} = \hbar\omega_{eg}\sigma_{ee}+q\boldsymbol{\hat{r}}\cdot\boldsymbol{\mathcal{E}}(\boldsymbol{\hat{r}},t) = \hbar \omega_{eg}\sigma_{ee}+\frac{1}{2}\boldsymbol{\hat{r}}\cdot \boldsymbol{E}_0 \Big(e^{i(\omega_L t+\phi(t) - \boldsymbol{k}\cdot \boldsymbol{r}} + e^{-i(\omega_L t+\phi(t)-\boldsymbol{k}\cdot\boldsymbol{r})} \Big),
\end{dmath}
where $q = -e$ (the electron charge). It is possible to express $q\boldsymbol{\hat{r}}\mathrm{exp}(\pm i\boldsymbol{k}\cdot \boldsymbol{\hat{r}})$ in terms of the energy state $\ket{g}$ and $\ket{e}$,
\begin{dmath}\label{eq:wave_laser}
    q\boldsymbol{\hat{r}}\mathrm{exp}(\pm i\boldsymbol{k}\cdot \boldsymbol{\hat{r}}) = - \bra{g}e \boldsymbol{\hat{r}}\mathrm{exp}(\pm i\boldsymbol{k}\cdot \boldsymbol{\hat{r}})\ket{e}\ket{g}\bra{e}-\bra{e}e \boldsymbol{\hat{r}}\mathrm{exp}(\pm i\boldsymbol{k}\cdot \boldsymbol{\hat{r}}) \ket{g}\ket{e}\bra{g} - \bra{g}e \boldsymbol{\hat{r}}\mathrm{exp}(\pm i\boldsymbol{k}\cdot \boldsymbol{\hat{r}})\ket{g}\ket{g}\bra{g} - \bra{e} e \boldsymbol{\hat{r}}\mathrm{exp}(\pm i\boldsymbol{k}\cdot \boldsymbol{\hat{r}}) \ket{e}\ket{e}\bra{e}.
\end{dmath}
Eq. \eqref{eq:wave_laser} can be simplified by applying $\bra{g}\boldsymbol{\hat{r}}\ket{g} = \bra{e}\boldsymbol{\hat{r}}\ket{e} = 0$ since the laser does not couple to the same energy states. One can neglect the spatial dependence of the electrical field (known as the dipole approximation) to further simplify the equation. This is possible to do because on a Rydberg atom, state $\ket{e}$ is many Bohr radii away from state $\ket{g}$ which is the same order of magnitude of the laser wavelength. However if the scalar products in Eq. \eqref{eq:wave_laser} are written as integrals of the form \cite{rydberg_noise}, 
\begin{equation}
    \int \mathrm{d}^3 r \psi_g(\boldsymbol{r}) e \boldsymbol{r} \mathrm{exp}(\pm i \boldsymbol{k}\cdot\boldsymbol{r}) \psi_e(\boldsymbol{r}),
\end{equation}
the significant contributions come from $\psi_g(\boldsymbol{r})\psi_e(\boldsymbol{r})$. Since the ground state is close to the nucleus of the atom, $\psi_g(\boldsymbol{r})\psi_e(\boldsymbol{r})$ has a size smaller than the wavelength $(\boldsymbol{k}\cdot\boldsymbol{r}\ll1)$. Hence, it is possible to apply the dipole approximation \cite{rydberg_noise},
\begin{equation}
    e \boldsymbol{\hat{r}}\mathrm{exp}(\pm i\boldsymbol{k}\cdot \boldsymbol{\hat{r}}) \approx e\boldsymbol{\hat{r}} = \bra{g}e\boldsymbol{\hat{r}}\ket{e}\ket{g}\bra{e}+\bra{e}e\boldsymbol{\hat{r}}\ket{g}\ket{e}\bra{g} = \hat{\mu} (\sigma_{ge}+\sigma_{eg}),
\end{equation}
where $\hat{\mu} = e\bra{g}\boldsymbol{\hat{r}}\ket{e}$ is the dipole matrix element and the operators $\sigma_{\alpha \beta} = \ket{\alpha}\bra{\beta}$. The Hamiltonian in Eq. \eqref{eq:general_hamiltonian} becomes,
\begin{equation}
    \mathcal{H} = \hbar \omega_{eg}\sigma_{ee} + \frac{1}{2}\boldsymbol{E}_0\cdot\hat{\mu}(\sigma_{eg} + \sigma_{ge})\Big( e^{i(\omega_L t +\phi(t))} + e^{-i(\omega_L t +\phi(t))} \Big).
\end{equation}
which can be transformed to the interaction Hamiltonian by applying the unitary \cite{rydberg_noise},
\begin{equation}
    U = e^{-i\hbar\omega_{ge}t\sigma_{ee}}.
\end{equation}
One can thus obtain the Hamiltonian,
\begin{dmath}
    \mathcal{H}^I = U^\dag\mathcal{H}U - \hbar \Delta \sigma_{ee} = -\hbar\Delta\sigma_{ee}+\frac{\hbar \Omega}{2}\Big(\sigma_{eg} e^{i[(\omega_L+\omega_{eg}) t +\phi(t)]} + \sigma_{ge} e^{i(\Delta t + \phi(t))}+\sigma_{eg} e^{-i(\Delta t + \phi(t))}+\sigma_{ge} e^{-i[(\omega_L+\omega_{eg}) t +\phi(t)]} \Big),
\end{dmath}
where $\Delta = \omega_L - \omega_{eg}$ is the detuning and $\Omega = \boldsymbol{E}_0 \cdot \mu/\hbar$ is the Rabi frequency. To eliminate the time dependence, another unitary transformation $U =  e^{-\hbar\Delta\sigma_{ee}t}$ can be applied,
\begin{dmath}\label{eq:H_I}
    \mathcal{H}^I = U^\dag\mathcal{H}U - \hbar \Delta \sigma_{ee} = -\hbar\Delta\sigma_{ee}+\frac{\hbar \Omega}{2}\Big(\sigma_{eg} e^{i[(\omega_L+\omega_{eg}) t +\phi(t)]} +  \sigma_{eg} e^{-i\phi(t)} + \sigma_{ge} e^{i\phi(t)}+\sigma_{ge} e^{-i[(\omega_L+\omega_{eg}) t +\phi(t)]} \Big),
\end{dmath}
Using the Hilbert space basis $(\ket{e},\ket{g})$, so that $\ket{\psi} = c_g\ket{g}+c_e\ket{e}$, the solution of the Schr\"{o}dinger equation becomes \cite{rydberg_noise},
\begin{equation}
    i\hbar\frac{\partial}{\partial t}  \ket{\psi(t)} = \mathcal{H}^I\ket{\psi(t)},
\end{equation}
with initial condition $\ket{\psi(0)} = \ket{g}$ \cite{rydberg_noise}. A further approximation can be utilized by assuming that $\omega_L$ oscillates at a very high frequency or short timescale. This means that $(\omega_L + \omega_{eg})^{-1} \ll \Delta^{-1},\Omega^{-1}$ which means that the exponential terms with $(\omega_L + \omega_{eg})$ can be neglected. This is known as the rotating wave approximation (RWA). Furthermore, defining the phase noise at an initial condition $t=0$ as $e^{i\phi(0)}$ means that the instantaneous change in phase noise can be defined as $e^{-i\phi(t-t^')}$. Thus applying the new definition of the phase noise with the RWA to Eq. \eqref{eq:H_I} gives \cite{rydberg_noise},
\begin{equation} 
    \mathcal{H}^I = -\hbar \Delta\sigma_{ee} + \frac{\hbar \Omega}{2}\bigg(\sigma_{ge}e^{i(\phi(t)-\phi(t^'))}+\sigma_{eg}e^{-i(\phi(t)-\phi(t^'))} \bigg).
\end{equation}

\chapter{Flux Qubit Calculations}\label{chapter:flux_appendix}

\section{Collective Wavefunction in Superconducting Qubits}\label{sec:wavefunction_flux}
Electrons in superconducting materials are cooled below the transition temperature $T_c$. At this temperature, electrons in a lattice with opposing spins form a single spin-0 particle, or boson, called Cooper pairs. This means that the Cooper pairs can occupy the same state without without violating the Pauli principle. Thus, the Cooper pairs form a highly-degenerate condensate which can be effectivley modelled by a single wavefunction~\cite{Birenbaum2014TheCF},
\begin{equation}
    \psi(r) = [n(r)]^{1/2}e^{i\theta(r)}.
\end{equation}
In the equation, $n(r)$ represents the fractional density of Cooper pairs while $\theta(r)$ represents the phase collective wavefunction. Knowing that $\theta(r)$ is a single valued function expressing only capable of changing by a multiple of $2\pi$ when traversing any closed contour within the superconductor, it can be mathematically written as,
\begin{equation}\label{eq: theta_r_init}
    \oint \nabla \theta(r) \cdot dl = 2\pi m,
\end{equation}
where $m$ is an integer. Ginzburg-Landau theory states that in the interior of a superconductor ~\cite{Birenbaum2014TheCF},
\begin{equation}
    \hbar \nabla \theta = q \boldsymbol{A}
\end{equation},
where $\boldsymbol{A}$ is the magnetic vector potential and $q$ is the charge of a Cooper pair. Substituting this into Eq. \eqref{eq: theta_r_init}, one obtains,
\begin{equation}
    \oint \boldsymbol{A} \cdot d\boldsymbol{l} = \frac{hm}{2e}.
\end{equation}
With the use of Stoke's theorem, one can rewrite this as, 
\begin{equation}
    \int \int \boldsymbol{B}\cdot \boldsymbol{A} \equiv \Phi = m \frac{h}{2e},
\end{equation}
where the substitution $\boldsymbol{B} = \nabla \times \boldsymbol{A}$ was made \cite{Birenbaum2014TheCF}. From this expression, it is possible to see that the flux threading a superconducting ring is quantized in integer multiples of the magnetic flux quantum $\Phi_0 = h/2e$.

\section{Quantum LC Oscillator}\label{sec:quantum_oscillator}
The Hamiltonian for an LC oscillator can be expressed using the canonical conjugates $Q$ and $\Phi$ where $Q = CV$ is the electric charge on the capacitor and $\Phi = LI$ is the generalized flux variable \cite{Birenbaum2014TheCF}. The Hamiltonian is thus,
\begin{equation}
    H = \frac{1}{2}\frac{Q^2}{C} + \frac{1}{2}\frac{\Phi^2}{L}.
\end{equation}
The voltage across the capacitor and the inductor must be equal, thus one can see that $Q$ and $\Phi$ satisfy the relations for canonical variables \cite{Birenbaum2014TheCF},
\begin{equation}\label{eq: canonical_hamiltonian}
    \frac{\partial H}{\partial Q} = \frac{Q}{C} = -L \frac{\partial I}{\partial t} = -\dot{\Phi},
\end{equation}
\begin{equation}
    \frac{\partial H}{\partial \Phi} = \frac{\Phi}{L} = I = \dot{Q}.
\end{equation}
One can see that the capacitive term $Q^2/2C$ acts as the kinetic energy term $p^2/2m$ where the mass of the 'phase particle' is the capacitance \cite{Birenbaum2014TheCF}. Similarly, the inductive term $\Phi^2/2L$ acts as the potential energy term $m\omega^2x^2/2$ by rewriting the inductive enrgy term as $C\omega^2\Phi^2/2$ with $\omega = (LC)^{-1/2}$. The Hamiltonian described in Eq. \eqref{eq: canonical_hamiltonian}, can be transformed to a quantum version by replacing the canonical variables with their respective operators. Thus, the commutation relation, $[\hat{\Phi},\hat{Q}] = i\hbar$, holds. The solutions of the derived Hamiltonian for the quantum LC oscillator must be the same as the quantum harmonic oscillator. From this, it is possible to see that the eigenenergies must form an equally spaced ladder expressed as $(n+\frac{1}{2})\hbar \omega$~\cite{Birenbaum2014TheCF}.

An equally spaced energy ladder however, is not favorable when designing a qubit. Driving transitions between two levels on an equally spaced ladder will result in unintended resonant transitions between all energy levels, since the energy spacing is always $\hbar \omega$ \cite{Birenbaum2014TheCF}. An anharmonic potential yields an uneven spread of energy levels, consequently allowing one to isolate two energy levels. Josephson junctions possess an inductance,
\begin{equation}
    L_J = \frac{L_{J_0}}{\sqrt{1-(I/I_c)^2}},
\end{equation}
where, $I_c$ is the critical current, or the maximum electric current density a superconducting material can carry before switching into the normal state, and $L_{J_0} \equiv \frac{\Phi_0}{2\pi I_c}$. It can be observed that this inductance is non-linear, meaning that Josephson junctions function as non-linear resonators. 

\section{Potential Energy Stability}\label{sec:flux_stability}

The purpose of this section will be to derive the two stable solutions of the three-junction potential around $f = 1/2$ starting at the equation, 
\begin{equation}
    \frac{U}{E_J} = 2 + \alpha - \mathrm{cos}\alpha_1 -\mathrm{cos}\alpha_2 - \alpha \mathrm{cos}(2\pi f+\varphi_1-\varphi_2).
\end{equation}

To find the minimum energy phase estimation, one must first take the derivative of the potential in terms of $\phi_1$ and $\phi_2$ and equate them to 0 assuming $\xi = U/E_J$,
\begin{equation}
    \frac{\partial \xi}{\partial \varphi_1} = \mathrm{sin}\varphi_1 + \alpha \mathrm{sin}(2\pi f + \varphi_1 - \varphi_2) = 0,
\end{equation}
and,
\begin{equation}
    \frac{\partial \xi}{\partial \varphi_2} = \mathrm{sin}\varphi_2 - \alpha \mathrm{sin}(2\pi f + \phi_1 - \varphi_2) = 0.
\end{equation}

The characteristic solutions $(\varphi_1^*,\varphi_2^*)$ are chosen such that they comply with $\mathrm{sin}\phi_1^* = -\mathrm{sin}\varphi_2^* = \mathrm{sin}\varphi^*$ \cite{orlando_flux}. Assuming, $\mathrm{sin}\varphi^* = -\alpha \mathrm{sin}(2\pi f + 2\varphi^*)$, it is possible to check the characteristic solution by computing the eigenvalues of the stability matrix, $\partial^2\xi/\partial\varphi_i\partial\varphi_j$, where, 
\begin{equation}
    \frac{\partial^2 \xi}{\partial \varphi_1^2} = \mathrm{cos}\varphi_1 + \alpha\mathrm{cos}(2\pi f + \varphi_1 - \varphi_2),
\end{equation}
\begin{equation}
    \frac{\partial^2 \xi}{\partial \varphi_2^2} = \mathrm{cos}\varphi_2 + \alpha\mathrm{cos}(2\pi f + \varphi_1 - \varphi_2),
\end{equation}
\begin{equation}
    \frac{\partial^2 \xi}{\partial \varphi_1 \varphi_2} = - \alpha\mathrm{cos}(2\pi f + \varphi_1 - \varphi_2).
\end{equation}
For states with $\mathrm{cos}\varphi_1^*=\mathrm{cos}\varphi_2^*=\mathrm{cos}\varphi^*$, the eigenvalues are, $\lambda_1 = \mathrm{cos}\varphi^*$ and $\lambda_2 = \mathrm{cos}\varphi^* + 2\alpha \mathrm{cos}(2\pi f+2\varphi^*)$. Both eigenvalues are larger than zero which assures the minimum energy condition \cite{orlando_flux}.

The next step is to calculate the critical values of the external field for the coexistence of two minimum energy phase configurations. This is easily done by restricting the analysis around the $f=1/2$ point, thus, $[0.5-f_c,0.5+f_c]$. These extrema values of the field correspond to solutions for which one of the values is a finite positive number and the other is zero. At $f = 1/2 \pm f_c$, $\lambda_1 = 0$ implying that $\varphi^* = \mp \pi/2 \mathrm{mod} 2\pi$. Substituting $\varphi^* = \mp \pi/2$, into the characteristic equation, $\mathrm{sin}\varphi^* = -\alpha \mathrm{sin}(2\pi f + 2\varphi^*)$, one arrives at $\pm1 = \pm\alpha\mathrm{sin}(2\pi f_c)$, where once rearranged,
\begin{equation}
    f_c = \frac{1}{2\pi} \mathrm{arcsin}\frac{1}{\alpha}.
\end{equation}

It is now possible to calculate $f_c$ when $0.5\leq\alpha\leq 1.0$ \cite{orlando_flux}. Now the eigenvalue to equal zero is $\lambda_2$ where the system of equations to solve are,

\begin{equation}
    \mathrm{sin}\varphi^* = -\alpha\mathrm{sin}(2\pi f+2\varphi^*) = \alpha \mathrm{sin}(\pm 2 \pi f_c + 2\varphi^*),
\end{equation}
\begin{equation}
    \mathrm{cos}\varphi^* = -2\alpha\mathrm{cos}(2\pi f+2\varphi^*) = 2\alpha \mathrm{sin}(\pm 2 \pi f_c + 2\varphi^*).
\end{equation}

Allowing $\Delta = \pm 2\pi f+ 2\varphi^*$, it is possible to write $1 = \alpha^2 + 3\alpha^2 \mathrm{cos}^2 \Delta$ \cite{orlando_flux}. Then,
\begin{equation}
    \mathrm{cos}\Delta = \sqrt{\frac{1-\alpha^2}{3\alpha^2}},
\end{equation}
\begin{equation}
    \mathrm{cos}\varphi^* = 2\sqrt{\frac{1-\alpha^2}{3}}\varphi^*.
\end{equation}
Using this, one arrives to the solution for $f_c$ for when $\mathrm{cos}(\varphi^*) \geq 0$,
\begin{equation}
    f_c = \frac{1}{2\pi}\bigg[ 2\mathrm{arccos}\Big(2\sqrt{\frac{1-\alpha^2}{3}} \Big)  - \mathrm{arccos}\Big(2\sqrt{\frac{1-\alpha^2}{3\alpha^2}} \Big) \bigg].
\end{equation}

\section{Flux-Resonator Hamiltonian}\label{sec:flux_resonator_der}

The addition of the extra capacitor changes the 3JJ qubit Hamiltonian to \cite{Yamamoto_2014,Yan_2016},
\begin{equation} \label{eq:hamiltonian_extended_flux}
    \mathcal{H}_{q} = \boldsymbol{n} \cdot 4\boldsymbol{n}\cdot \boldsymbol{E}_C + \alpha E_{J}[1-\mathrm{cos}\Big(\frac{2\pi \Phi_\varepsilon}{\Phi_0} +\varphi_1 - \varphi_2 \Big)] + E_J[2-\mathrm{cos}\varphi_2 - \mathrm{cos}\varphi_1],
\end{equation}
where $\boldsymbol{n} \equiv \{n_i\}$ is the vector node charge operator and $\boldsymbol{E}_C \equiv e^2/2 \cdot \boldsymbol{C}^{-1}$ is the inverse charging energy matrix and $\varphi_L$ is the phase across the loop inductance. It can be seen that now the capacitors are treated in matrix form. Assuming $C_1 = C_3$, the capacitance matrix $\boldsymbol{C}$ becomes,
\begin{equation}
    \boldsymbol{C} = \begin{bmatrix}
        C_2+C_1+C_{sh} & -C_1-C_{sh} & 0 \\
        -C_1 - C_{sh} & C_{2}+C_{1}+C_{sh} & -C_2 \\
        0 & -C_{2} & C_{2}
    \end{bmatrix}.
\end{equation}
It is now important to model the flux qubit interacting with the resonator. The Hamiltonian responsible for the inductive coupling of the 3JJ qubit circuit to the LC resonator is as follows \cite{Yamamoto_2014},
\begin{equation}\label{eq:resonator_hamiltonian_flux_full}
    \mathcal{H}_{r} = E_r(\hat{a}^\dag\hat{a} + \frac{1}{2}) + \frac{\pi\alpha\delta}{(1+2\alpha)\beta_L}\frac{E_J}{E_M}\sqrt{E_r E_{Lr}}\sqrt{E_r E_{Lr}}(\hat{a}^\dag+\hat{a}),
\end{equation}
where $E_M = \Phi^2_0/(2M)$, $E_{Lr} = \Phi^2_0/(2L_r)$, $E_r = \hbar/\sqrt{L_r C_r}$, $\delta$ is the phase across the loop inductance of the qubit, and $\beta_L = 2\pi I_0\alpha L_q/ [(2\alpha+1)\Phi_0$]. The coupling of both these systems is dependent on the persistent current across both circuits since both $\beta_L$ and the loop inductance phase depends on the current as seen in Section \ref{sec:flux_qubit_intro} \cite{Harris_2007}. To simplify this interaction, one can envision this coupling as an interaction Hamiltonian between a qubit and a resonator, 
\begin{equation}
    \mathcal{H}_{r} = \hbar\omega (\hat{a}^\dag \hat{a} + \frac{1}{2}) + \hbar g \sigma_y (\hat{a}^\dag + \hat{a}) ,
\end{equation}
where the coupling strength (the factor in the second term) in \eqref{eq:resonator_hamiltonian_flux_full} is generalized to $g$ \cite{UltrastrongCapacitiveCoupling,nonlinear_koun,Yan_2016}.
\section{BFGS Algorithm}\label{sec:bfgs_app}
BFGS aims to minimize a function $f(x)$ using an initial estimate $\boldsymbol{x}_0$ where $\boldsymbol{x}$ is a vector $\mathcal{R}^n$ and $f$ is a differentiable scalar function \cite{LBFGS}. The descent or ascent direction $\boldsymbol{p}_l$ at a stage $l$ is given by the solution of the Newton equation,
\begin{equation}
    B_l \boldsymbol{p}_l = -\nabla f(\boldsymbol{x}_l),
\end{equation}
where $B_l$ is an approximation of the Hessian matrix at $\boldsymbol{x}_l$ which is updated iteratively at each stage and $\nabla f(\boldsymbol{x}_l)$ is the gradient of the function at $\boldsymbol{x}_l$. Incorporating a line search to find the next point $\boldsymbol{x}_{l+1}$ leads to a quasi-Newton condition which can be imposed on the updated $B_l$ \cite{LBFGS},
\begin{equation}
    B_{l+1}(\boldsymbol{x}_{l+1}-\boldsymbol{x}_l) = \nabla f(\boldsymbol{x}_{l+1})-\nabla f(\boldsymbol{x}_l).
\end{equation}
To avoid the need for step size estimation, the algorithm employs the secant equation \cite{BFGS_practical}. The secant equation is defined as, 
\begin{equation}
    B_{l+1}\boldsymbol{s}_l = \nabla f(\boldsymbol{x}_{l+1}) - \nabla f(\boldsymbol{x}_l),     
\end{equation}
where $\boldsymbol{s}_k = \boldsymbol{x}_{l+1} - \boldsymbol{x}_{l}$. The curvature condition $\boldsymbol{s}_l^\intercal (\nabla f(\boldsymbol{x}_{l+1}) - \nabla f(\boldsymbol{x}_{l})) > 0$, should be satisfied for when $B_{l+1}$ 
is a positive definite. If this is not satisfied, it means the function is not strongly convex. When this is the case, the algorithm explicitly enforces the step length to find a point $\boldsymbol{x}_l$ where the inequalities below hold,
\begin{enumerate}[i]
\centering
\item $f(\boldsymbol{x}_l+\alpha_l\boldsymbol{p}_l)\leq f(\boldsymbol{x}_l) + c_1\alpha_l\boldsymbol{p}_l^\intercal\nabla f(\boldsymbol{x}_l)$,
\item $-\boldsymbol{p}_l^{\intercal}\nabla f(\boldsymbol{x}_l+\alpha_l\boldsymbol{p}_l) \leq -c_2\boldsymbol{p}_l^\intercal\nabla f(\boldsymbol{x}_l)$
\end{enumerate}
which are called the Wolfe conditions where $\alpha_l$ is the step length in a restricted direction $\boldsymbol{p}_l$ 
with coefficients $0<c_1<c_2<1$ \cite{wolfe_condition}. In order to maintain the symmetry and positive definiteness of $B_{l}$, the updated $B_{l+1}$ is chosen to follow,
\begin{equation}\label{eq:B_update}
 B_{l+1} = B_l+\alpha \nabla (f(\boldsymbol{x}_{l+1}) - \nabla f(\boldsymbol{x}_{l}))(\nabla f(\boldsymbol{x}_{l+1}) - \nabla f(\boldsymbol{x}_{l}))^\intercal + \beta B_l\boldsymbol{s}_l(B_l\boldsymbol{s}_l)^\intercal,   
\end{equation}
where,
\begin{equation}
    \alpha = \frac{1}{\nabla f(\boldsymbol{x}_{l+1}) - \nabla f(\boldsymbol{x}_{l})^\intercal \boldsymbol{s}_l},
\end{equation}
and
\begin{equation}
    \alpha = \frac{1}{\boldsymbol{s}_l^\intercal B_l \boldsymbol{s}_l}.
\end{equation}
Substituting the values of $\alpha$ and $\beta$ into Eq. \eqref{eq:B_update} leads to the final updated $B_{l+1}$ \cite{BFGS_practical},
\begin{equation}\label{eq:final_BFGS}
    B_{l+1} = B_l + \frac{(f(\boldsymbol{x}_{l+1}) - \nabla f(\boldsymbol{x}_{l}))(f(\boldsymbol{x}_{l+1}) - \nabla f(\boldsymbol{x}_{l}))^\intercal}{(f(\boldsymbol{x}_{l+1}) - \nabla f(\boldsymbol{x}_{l}))^\intercal\boldsymbol{s}_k} - \frac{B_l \boldsymbol{s}_l \boldsymbol{s}_l^\intercal B_l^\intercal}{\boldsymbol{s}_l B_l\boldsymbol{s}_l}.
\end{equation}

The step direction can be determined by defining the infidelity to be a function of the control vector $J = \varepsilon(u_k)$, following Eq. \eqref{eq:u_k_def}, consequently leading to $\boldsymbol{p}_l = -\nabla J(u_k)$. The steepest descent only uses information about the first derivative which means that it suffers from having no information about the curvature of the landscape. By defining the Hessian,

\begin{equation}
    \boldsymbol{H} = \begin{bmatrix}
        \frac{\partial^2 J}{\partial u^2_{1,1}} & \frac{\partial^2 J}{\partial u^2_{1,1}\partial u_{2,1}} & \cdots & \frac{\partial^2 J}{\partial u_{1,1} \partial u_{N,M}} \\
        \frac{\partial^2 J}{\partial u_{2,1}\partial u_{1,1}} & \frac{\partial^2 J}{\partial u^2_{2,1}} & \cdots & \frac{\partial^2 J}{\partial u_{2,1} \partial u_{N,M}} \\
        \vdots & \vdots & \ddots  & \vdots \\
        \frac{\partial^2 J}{\partial u_{N,M}\partial u_{1,1}} & \frac{\partial^2 J}{\partial u_{N,M} \partial u_{2,1}} & \cdots & \frac{\partial^2 J}{\partial u^2_{N,M}},
    \end{bmatrix}
\end{equation}
where $M$ are the total amount of control Hamiltonians and $N$ are the total amount of amplitude segments, one can derive the standard search direction as $\boldsymbol{p}_l = -[\boldsymbol{H}(u_k(l))]^{-1}\nabla J(u_k(l))$ \cite{Dalgaard_2020}. The Hessian-approximation scheme allows one to approximate $B \approx \boldsymbol{H}$ thus allowing the iterative use of the BFGS form written in Eq. \eqref{eq:final_BFGS} \cite{BFGS_practical}. 
\chapter{Hybrid System}
\begin{figure}[h]
    \centering
    \includegraphics[width=1\textwidth]{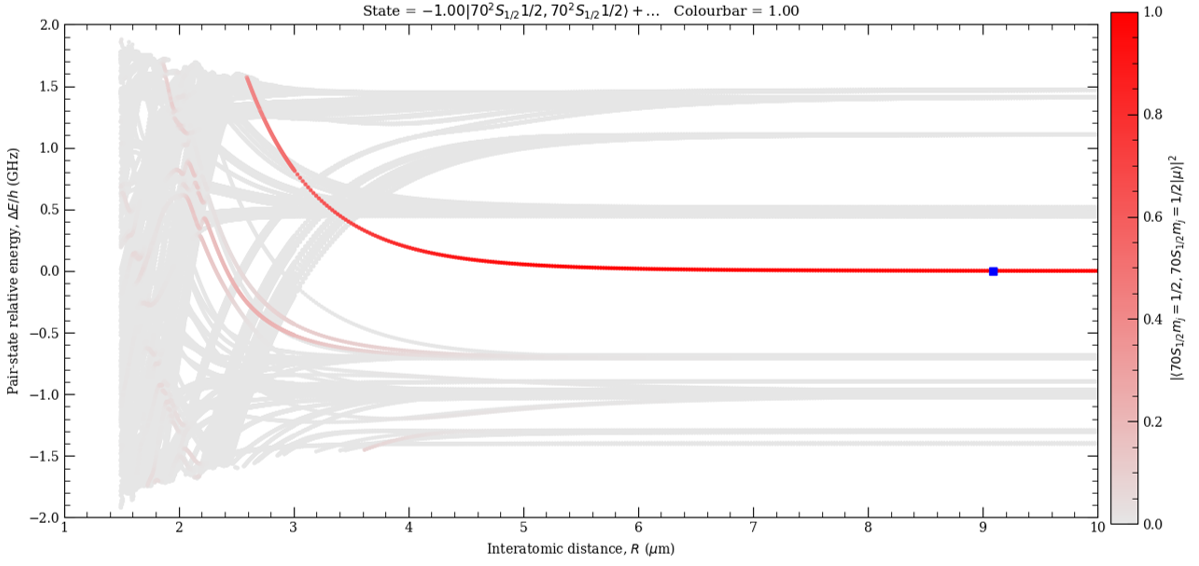}
    \caption{The interaction strength as a function of the pair-state relative energy between two atoms in state $\ket{70S_{1/2}}$. The Rydberg blockade effect can be seen before the 4.1$\mu m$ distance where the phenomena causes the mixing of states resulting in an exponential relative energy. This image was produced using \cite{ibali__2017}.}
    \label{fig:rybdergblock_system}
\end{figure}

\begin{figure}[h]
    \centering
    \includegraphics[width=1\textwidth]{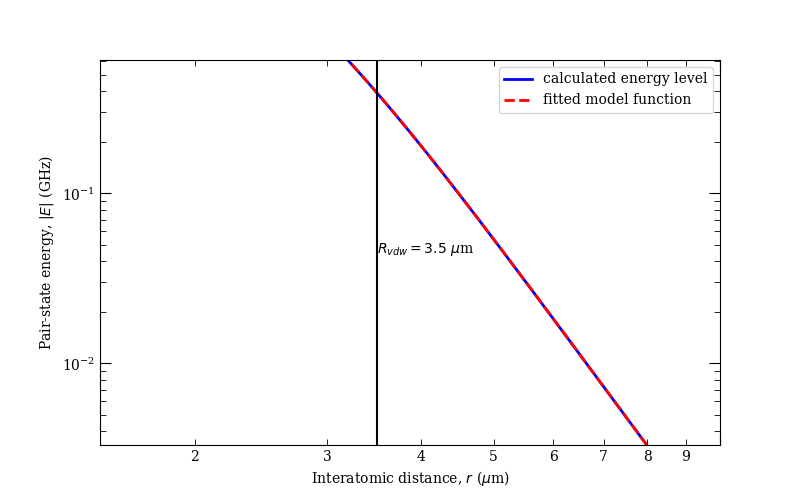}
    \caption{Fitted model function of the interaction strength as a function of the pair-state relative energy between two atoms in state $\ket{70S_{1/2}}$ in the vicinity of the characteristic radius. The black line shows the calculated characteristic radius $R_{vdW}$ equal to $3.5\mu$m. This image was produced using \cite{ibali__2017}.}
    \label{fig:ryd_val}
\end{figure}

\begin{table}[h] \label{table:values_hybrid}
\centering
\begin{tabular}{ |c|c|c| } 
 \hline
 Physical parameters (units) & Symbols & Values \\
 \hline
 Capacitance (aF) & $C$ & 256 \\
 Inductance (nH) & $L$ & 247 \\
 Resonator frequency (GHz) & $\omega_0/(2\pi)$ & 20\\
 Value of electric field at $i = 1$ (V/cm) & $E^{(1)}$ & 520.0\\
 Value of electric field at $i = 2$ (V/cm) & $E^{(2)}$ & 537.5\\
 Value of electric field at $i = 3$ (V/cm) & $E^{(3)}$ & 550.8\\
 Value of electric field at $i = 4$ (V/cm) & $E^{(4)}$ & 571.7\\
 Value of electric field at $i = 5$ (V/cm) & $E^{(5)}$ & 585.4\\
 Value of flux-bias parameter at $i=1$ & $\gamma_q^{(1)}$ & $-5 \times 10^{-3}$\\  
 Value of flux-bias parameter at $i=1$ & $\gamma_q^{(1)}$ & $-3.06 \times 10^{-3}$\\
 Anticrossing A1 (GHz) & $\Omega/(2\pi)$ & 4.6 \\
 Anticrossing A2 (GHz) & $\Omega^'/(2\pi)$ & 3.2 \\
 Energies of atomic states $\ket{\mu = e,g,u}$ & $\omega_{e,g,u}$ & \\
 Atom $(\ket{g} - \ket{e})$ resonator coupling (GHz) & $g_a/(2\pi)$ & 1.0 \\
 Atom $(\ket{u} - \ket{e})$ resonator coupling (GHz) & $g_a^'/(2\pi)$ & 0.5 \\
 Mutual inductance (pH) & $M$ & 27 \\
 Tunnel splitting of flux qubit (GHz) & $\Delta/(2\pi)$ & 5.0 \\ 
 Relaxation rate of flux qubit (MHz) & $\gamma_{relax}/(2\pi)$ & 0.03\\
 Anticrossing between $\ket{R,1}$ and $\ket{L,0}$ (GHz) & $\delta/(2\pi)$ & 0.1 \\
 Flux qubit-resonator coupling (GHz) & $\gamma_{\phi}/(2\pi)$ & 0.1 \\
 Loss rate of resonator (MHz) & $\kappa/(2\pi)$ & 0.2 \\
 Decay rate of Rydberg states (MHz) & $\Gamma/(2\pi)$ & 0.15 \\ 
 \hline
\end{tabular}
\caption{List of physical parameters mentioned in the three-qubit hybrid structure \cite{Yu_atom_flux_2017}.}
\label{table:hybrid_structure}
\end{table}

\end{document}